\documentclass[a4,11pt]{article}
\usepackage{jheppub1}
\usepackage{amsmath,bm}
\usepackage{amssymb}
\usepackage{slashed}
\usepackage{graphicx}
\usepackage{color}
\usepackage[autostyle]{csquotes}
\usepackage{lscape}
\usepackage{subfig}
\usepackage{setspace}
\usepackage{bbm}
\usepackage{breqn}
\usepackage{wrapfig}
\usepackage{hyperref}
\usepackage{mathtools}
\usepackage{physics}
\usepackage{simpler-wick}

\newcommand{\be}{\begin{equation}}
\newcommand{\ee}{\end{equation}}
\newcommand{\bea}{\begin{eqnarray}}
\newcommand{\eea}{\end{eqnarray}}
\newcommand{\beas}{\begin{eqnarray*}}
\newcommand{\eeas}{\end{eqnarray*}}

\begin{document}

\title{A single-cut discontinuity for cosmological correlators from unitarity and analyticity}

\author[\spadesuit]{Shibam Das,}
\author[\clubsuit]{Debanjan Karan,}
\author[\clubsuit]{Babli Khatun,}
\author[\spadesuit]{ and Nilay Kundu}
\affiliation[\spadesuit]{Department of Physics \\ Indian Institute of Technology Kanpur, Kalyanpur, Kanpur 208016, India}
\affiliation[\clubsuit]{International Centre for Theoretical Sciences (ICTS-TIFR) \\ Tata Institute of Fundamental Research,
Shivakote, Hesaraghatta, Bangalore 560089, India.}

\emailAdd{shibamdas23@iitk.ac.in}
\emailAdd{debanjan.karan@icts.res.in}
\emailAdd{babli.khatun@icts.res.in}
\emailAdd{nilayhep@iitk.ac.in}

\abstract{We derive discontinuity relations, also known as cutting rules, and explore the analytic properties of cosmological correlators, fundamental observables of the primordial universe. Our emphasis is on how these relations arise from unitarity and hermitian analyticity in interacting quantum field theories on de Sitter space-time. Instead of analyzing wave-function coefficients, we apply these relations directly to cosmological correlators. By studying conformally coupled and massless scalar fields with $\phi^n$ self-interactions, we demonstrate that the discontinuity of a cosmological correlator can be expressed as a sum of products of lower-point discontinuities, stemming from a single-cut of one internal line in the corresponding tree-level exchange Witten diagram. Notably, beyond lower-point correlators, the decomposition of the discontinuities of cosmological correlators includes contributions from auxiliary elements that consist of both the real and imaginary parts of the lower-point wave-function coefficients, which have not been reported in the existing literature. Interestingly, depending on whether $n$ is even or odd in a $\phi^n$ interaction, these different lower-point discontinuities contribute as the leading or sub-leading piece in the late-time limit to the discontinuity relations. Additionally, our single-cut discontinuity relation leads to a decomposition rule for the residue of the cosmological correlators at partial energy singularities, incorporating contributions from these auxiliary objects. Through explicit calculations in several models, we confirm that our discontinuity relations are consistent with results from the in-in formalism. While primarily developed using tree-level exchanges with polynomial interactions, we also demonstrate that our framework can be extended to include loop corrections and cases with derivative interactions.}


\maketitle
\section{Introduction} \label{sec:intro}

Cosmological correlators are natural observables to study the early universe. They contain information about the quantum correlation between cosmological fluctuations produced during the exponential expansion of our universe in the inflationary phase, which are encoded in the CMB radiation and are among the very few directly measurable quantities available to learn about the physics of the early universe. 

The geometry of the space-time during that phase was de Sitter (dS) (approximately dS, to be precise). The analytic calculation of cosmological correlators is formulated in an interacting quantum field theory (QFT) on de Sitter (dS), where one needs to calculate the correlation functions of the fields/observables inserted on the late-time slice of dS. The standard approach to computing cosmological correlators is known as the in-in method (also referred to as the Schwinger-Keldysh method) \cite{Schwinger:1960qe, Keldysh:1964ud, Weinberg:2005vy}. For cosmological correlators, we need to compute an equal-time correlation function at the end of inflation (or the late-time slice of an approximate de Sitter space-time), which is inherently different from calculating an S-matrix. Knowing the state in which one needs to compute this correlator is non-trivial, and by using the in-in method, one prepares the state at the late-time slice of de Sitter, starting with an appropriate ground state (generally taken as the Bunch-Davies state) at a far past time. 

In an alternative approach \cite{Maldacena:2002vr, Anninos:2014lwa}, one can also first compute the wave-function of the quantum fields at the late-time slice of de Sitter as a function of the boundary values of the quantum fields (which are observables), following a path-integral setup in the bulk. The cosmological correlator can then be computed by knowing the probability amplitude, which is given by the modulus square of the wave-function. The wave-function method also led to a model-independent study of cosmological correlators based on symmetries of de Sitter and using techniques of momentum space CFT \cite{Mata:2012bx, Ghosh:2014kba, Kundu:2014gxa, Kundu:2015xta,Creminelli:2011mw,Liu:1998ty,Dey:2025kci}. There also exist closely related methods to compute cosmological correlators, for example, the bootstrap method \cite{Arkani-Hamed:2018kmz, Baumann:2019oyu, Baumann:2020dch,Ansari:2025nsf}.

Although, as algorithms, the methods mentioned above are, in principle, relatively straightforward, the computation becomes increasingly complicated once we start considering higher-point correlators or even lower-point correlators with an increasing number of vertices (which we will also refer to as sites) in the bulk, or move beyond tree-level and examine loop diagrams. This can be attributed, for example, in the in-in method, to the increasing number of bulk conformal time integration associated with multiple vertices, also involving various bulk-to-boundary and bulk-to-bulk propagators, rendering analytic calculation next to impossible. Similarly, in the wave-function approach, to obtain a higher-point cosmological correlator, one must know all the lower-point data (known as the wave-function coefficients, which we will define later). Therefore, it is reasonable to seek a fundamentally different reconceptualization that enables us to go beyond merely computing cosmological correlators to limited orders in perturbation theory. 

One conceptually deep idea that has been tremendously fruitful in this regard is to use fundamental principles, such as the unitarity and analyticity properties of a QFT in de Sitter space-time, as axioms to start with and reconstruct arbitrary higher-point cosmological correlators from a few lower-point data recursively. This method is compelling, as it replaces the off-shell Lagrangian or Hamiltonian-based approach to computing correlators, which employs tedious methods such as the in-in or wave-function approach, where unitarity and other properties can only be verified on a case-by-case basis for each given Hamiltonian. On the contrary, in this new approach, it is fascinating to realize that the constraints obtainable from the starting physical principles - such as unitarity, analyticity - on the analytic structure of cosmological correlators are powerful enough that by exploiting them, the correlators can self-consistently reconstruct themselves  \cite{Arkani-Hamed:2018kmz, Baumann:2019oyu, Baumann:2020dch,Ansari:2025nsf, Goodhew:2020hob, Cespedes:2020xqq, Baumann:2021fxj, Benincasa:2022omn, Jazayeri:2021fvk, Melville:2021lst, Goodhew:2021oqg, Bonifacio:2021azc, Meltzer:2021zin, Hogervorst:2021uvp, DiPietro:2021sjt, Hillman:2021bnk, Baumann:2022jpr, Salcedo:2022aal, AguiSalcedo:2023nds, Stefanyszyn:2023qov, Lee:2023kno, Chowdhury:2023arc, Stefanyszyn:2024msm, Liu:2024xyi, Goodhew:2024eup, Lee:2024sks, Chowdhury:2025ohm, Cespedes:2025dnq, Melville:2024ove,Arkani-Hamed:2023kig,Werth:2024mjg,Anninos:2024fty}. Recent progress has been done from the holographic point of view \cite{Dey:2024zjx, Nanda:2023wne, Moitra:2022glw, Afshordi:2017ihr, Antoniadis:1996dj, Larsen:2002et, McFadden:2010vh} and it would  therefore be interesting to understand the connections of the cosmological correlators with the physical principles.

The elegance and effectiveness of this new approach towards studying cosmological correlators may not be surprising at all, as one finds similar developments in the field of research in scattering amplitudes. A paradigm shift happened with the advent of the so-called on-shell program for amplitudes based on the BCFW recursion relations \cite{Britto:2004ap, Britto:2005fq} and the unitary methods \cite{Eden:1966dnq, Benincasa:2013faa, Taylor:2017sph, Correia:2020xtr}. Undoubtedly, the cosmological correlators are inherently different from the scattering amplitudes. However, in recent times, several studies of cosmological correlators have followed a similar approach. They have placed the constraints coming from unitarity and analyticity at the forefront as discussed above, and developed cutting or Cutkosky rules, optical theorems following from unitarity and analyticity for a QFT in de-Sitter space-time, borrowing from the ideas well-developed in Minkowski space-time \cite{Goodhew:2020hob, Cespedes:2020xqq, Baumann:2021fxj, Melville:2021lst, Goodhew:2021oqg}. These relations are expressed in terms of certain discontinuity relations defined in momentum space, such that the magnitude of the momentum is analytically continued in the complex plane. The discontinuity relations are well-known mathematical tools in complex variables and have been applied to various branches of physics, namely the Kramers-Kronig relations. 

However, since we are interested in learning about the cosmological correlators, it is crucial to understand on which objects these discontinuity relations are being applied. In majority of recent studies on the cutting rules or optical theorems in cosmological setup, the final results are written in terms of the wave-function coefficients \cite{Goodhew:2020hob, Cespedes:2020xqq, Baumann:2021fxj, Melville:2021lst, Goodhew:2021oqg, Salcedo:2022aal, Donath:2024utn, Ema:2024hkj, Stefanyszyn:2024msm,Anninos_2015}, which are, in some sense, the primitive building blocks of the wave-function of the quantum fields at the late-time de Sitter slice. That the discontinuity relations are only written in terms of the wave-function coefficients and not directly in terms of the cosmological correlators is not very satisfactory. However, in a few of them - such as \cite{Goodhew:2020hob, Donath:2024utn, Ema:2024hkj, Stefanyszyn:2024msm}, we can find the cutting rules for the correlators. Our primary goal in this paper is to address this issue and derive the discontinuity relations explicitly written for cosmological correlators, which will hold in a broader class of cases more generally, and provide a comprehensive understanding of how these relations actually work.

Next, let us briefly discuss the techniques we employed to achieve our goals. In this paper, we focus on interacting scalar fields denoted by $\phi$ in rigid de Sitter space-time, where the scalar field can have arbitrary polynomial self-interaction of the form $\phi^n$. The standard mode-expansion of the scalar field obtainable from the classical solution of the Klein-Gordon equation in de-Sitter will enable us to write down the various bulk-to-boundary and bulk-to-bulk propagators for the scalar field. These propagators, both in the in-in method and the wave-function method, are the basic ingredients to compute the scalar $n$-point correlators with the scalar field operator inserted at the late-time de Sitter slice. Next, we investigate the consequences of physical principles such as unitarity and hermitian analyticity and obtain specific identities that the bulk-to-boundary and bulk-to-bulk propagators must satisfy. These identities include operations involving the analytically continued complex magnitude of the $3$-momentum, which appear as arguments of the propagators. As we will see, these propagator identities are naturally tailor-made for obtaining discontinuity relations for the correlators by directly manipulating them. 

For concreteness and convenience, we will initially focus on cases with conformally coupled massive scalar fields and also on tree-level diagrams with an arbitrary number of interaction vertices in the bulk. Our discontinuity relation will naturally decompose into a sum of products of discontinuities of lower-point data upon cutting the internal line. For the relation to hold, we will only need to cut one single internal line for tree-level diagrams. Therefore, we will refer to our identities as the single-cut discontinuity relations. As it is clear from the discussion above, these relations will also serve as the statements about the cutting rules, or the optical theorems \cite{Goodhew:2020hob, Melville:2021lst, Goodhew:2021oqg}. That the discontinuity relations are equivalent to cutting rules is not surprising, since previous studies have already established this equivalence for the wave-function coefficients, and we are now explicitly writing these relations for cosmological correlators. Although our analysis initially will be restricted to conformally coupled scalars, we will explicitly show that our formalism can also be extended and generalized to other situations, such as the massless scalar with polynomial self-interaction, correlators with derivative interactions in the bulk, and also to loop diagrams. For loop diagrams, our single-cut rule involves cutting all the internal lines successively between two adjacent vertices in the bulk Witten diagram. 

A related issue highlighted in the existing literature about the discontinuity relations involving wave-function coefficients concerns the factorization aspect. When a single internal line is cut in a tree-level Witten diagram corresponding to an $n$-point wave-function coefficient with multiple vertices, the diagram separates into two parts: the left block and the right block. The factorization principle implies that the discontinuity of the original Witten diagram can be factored into the discontinuities of these left and right blocks that arise from cutting the internal line. These blocks correspond to the lower-point objects obtained after attaching the cut internal line to the boundary and treating it as an external leg.

It is interesting to ask if our discontinuity relation for cosmological correlators manifests the factorization property mentioned above. Instead of producing the product of two discontinuities of lower point correlators, our discontinuity relation is expressed as a sum of the products of discontinuities from both the left and right blocks after cutting the internal lines. Consequently, there is an additional contribution that is not expressable precisely as a lower-point correlator. This additional piece in the discontinuity relation involves the imaginary components of the wave-function coefficients for lower-point diagrams, which have not been previously reported in earlier studies. That the discontinuity relation for a real correlator involves the imaginary part of the wave-function coefficients might be confusing. However, we have observed that these imaginary parts of the lower-point wave-function coefficients are always paired with another imaginary part from a similar lower-point coefficient. As a result, the contribution to the discontinuity of the cosmological correlator from these additional terms is always real.

It is worth mentioning at this point about one recent work \cite{Donath:2024utn} where a similar discontinuity relation for cosmological correlators was written down using an approach different from what we follow. While comparing our discontinuity relation for cosmological correlators with \cite{Donath:2024utn}, we have found instances, such as for conformally coupled $\phi^n$ self-interactions with $n$ being an odd integer, where the cutting rule mentioned in that paper, to the best of our understanding, fails to hold. We have also understood the reason behind this - there is a subtle interplay between leading and sub-leading contributions to the discontinuity relations as the late-time limit is taken, which was not taken into account in the above-mentioned work. More interestingly, the problematic cases mentioned above are consistently dealt with in our discontinuity relations - precisely due to those extra contributions involving imaginary pieces of the wave-function coefficients discussed previously. Therefore, this justifies the novelty of our result. We provide several instances of explicit checks working with varied types of interactions, which verify the consistency of the formal derivation of our discontinuity relations.

As a related application of our discontinuity relation for cosmological correlators, we extract the residue at partial-energy poles, which occur when the total energy flowing through a sub-diagram at an interaction vertex vanishes \cite{Maldacena:2011nz, Raju:2012zr}, \cite{Goodhew:2020hob, Baumann:2022jpr}. We demonstrate that the residue of the scalar cosmological correlators at a partial energy singularity can be decomposed into two distinct contributions: one that factorizes into a product of lower-point correlators, and another that factorizes into a product involving the real and imaginary parts of the corresponding lower-point wave-function coefficients. This structure provides new insights into the analytic properties of cosmological correlators. Notably, the decomposition at partial-energy poles emphasizes the role of the imaginary part of the lower-point wave-function coefficient. 

The rest of the paper is organised as follows. In the next section \S\ref{basicsetup}, we establish the basic framework, providing the background that will be useful for the calculations that follow in the later sections. Next, in section \S\ref{ininpropagatorfromunitarity}, we will work out the relations involving the bulk-to-boundary and the bulk-to-bulk propagators starting from the statements of unitarity and hermitian analyticity. Following that, in \S\ref{confscalartreelevel} we will derive the discontinuity relations for the $n$-point correlator, working with conformally coupled scalar fields with polynomial self-interaction and focusing on tree-level diagrams with two, and finally with generic $r$-site interaction vertices. In \S\ref{sec:generalization_cases} we generalise our results for cases with massless scalar fields, derivative interactions, and for loop diagrams. Next in \S\ref{sec:Explicit checks}, we work out several examples and justify the consistency of our results derived so far. In \S\ref{partial energy}, using our single-cut discontinuity rule, we have discussed the decomposition of the residue of the cosmological correlators at the partial energy singularity in terms of lower point data. Following that, we conclude with a summary and outlook in \S\ref{conclusions}. Finally, we present various supplementary materials and technical details in Appendices. 

\textbf{Note added:} While this manuscript was in preparation we came across \cite{MarinMacedo:2025jco} and an online seminar \footnote{The link is available at - 
\href{https://www.youtube.com/watch?v=KXUq9-8KomQ}{(link: https://www.youtube.com/watch?v=KXUq9-8KomQ)}} where related results have been obtained. We comment on the similarities and differences with \cite{MarinMacedo:2025jco} in relevant places in the subsequent sections.

\section{Basic set-up and summary of results} \label{basicsetup}


In this section we will discuss the basic set-up briefly reviewing the background material required for one to get an idea about how cosmological correlators or late time de Sitter correlators are calculated analytically. Starting with some basic description of the properties of a QFT with scalar fields in rigid de Sitter space-time, we will mention the primary ingredients of the technical toolbox for both in-in and wave-function method for computing the late-time de Sitter correlators. This will be very useful for the technical computations that we will perform in the following sections. 

The de Sitter space-time will be written in the Poincare patch (expanding) coordinates as follows
\begin{equation}
ds^2 = \frac{- d\eta^2 + d \vec{x}^2 }{H^2 \eta^2} ,
\end{equation}
where the conformal time $\eta$ will span the range $\eta \in \left( - \infty, 0 \right)$, such that $\eta = - \infty$ will be at far past and $\eta = 0$ will correspond to the future de Sitter boundary or the late-time slice of de Sitter. For our calculations we will use $\eta = \eta_0$ as the late-time slice, with an understanding that, at the end $\eta_0 \to 0^-$ limit will be taken. Also, $H$ is known as the Hubble parameter and it will be constant for us since we are considering rigid de Sitter space-time. 

Let's recall the dynamics of a free scalar field in the expanding Poincare patch of a $(d+1)$-dimensional de Sitter space-time \footnote{For detailed discussion on this topic, readers may look into the references \cite{Akhmedov:2013vka, Spradlin:2001pw, Galante:2023uyf}}. From the following action 
\begin{equation}
S = - \frac{1}{2} \int d^{d+1} x \, \sqrt{-g} \left( (\partial \phi)^2 + m^2 \phi^2\right) \, ,
\end{equation}
we can check that the classical solutions for the Fourier space mode function $\phi_k (\eta)$ \footnote{Our convention for the Fourier transform is the following 
$  \phi(\eta, \vec{x}) = \int \frac{d^d \vec{k}}{(2\pi)^d} e^{i \vec{k}.\vec{x}} \phi_{\vec{k}}(\eta)$ .
}
will be given by $f_k^-(\eta)$ and $f_k^{+}(\eta)$ corresponding to positive and negative frequency modes respectively \cite{Goodhew:2020hob,Goodhew:2021oqg}
\begin{equation} \label{mode f+-}
\begin{split}
    & f_k^-(\eta) = \frac{-i\sqrt{\pi}}{2} H^{\frac{d-1}{2}} (-\eta)^{\frac{d}{2}} e^{\frac{i\pi}{2}\left( \nu + \frac{1}{2} \right)} \mathbb{H}^{(1)}_\nu (- k \eta) \, ,
    \\
    & f_k^+(\eta) = \frac{i\sqrt{\pi}}{2} H^{\frac{d-1}{2}} (-\eta)^{\frac{d}{2}} e^{-\frac{i\pi}{2}\left( \nu + \frac{1}{2} \right)} \mathbb{H}^{(2)}_\nu (- k \eta) \, .
\end{split}
\end{equation}
Here $\mathbb{H}_\nu^{(1)}$ and $\mathbb{H}_\nu^{(2)}$ are the Hankel functions of the first and second kind with the parameter $\nu$ given by $\nu = \sqrt{(d/2)^2- (m/H)^2}$. 

Specific choices for $m^2$ are given by fixing the index $\nu$. Of particular interest to us will be the conformally coupled scalar field with $m^2= (d-1)(d+1)H^2/4$ and the massless case. For the rest of the discussion in this paper we will restrict ourselves to the $(3+1)$-dimensional de Sitter space-time. The mode functions for the choices of $\nu=1/2$ and $\nu=3/2$, which describe conformally coupled and massless scalar fields respectively, have the expressions as follows
\begin{equation} \label{mode functions}
\begin{split}
    & f_k^-(\eta) = \frac{i H\eta}{\sqrt{2k}} e^{-ik\eta},~~ f_k^+(\eta) = \frac{-i H\eta}{\sqrt{2k}} e^{ik\eta}~~~~~~~~~~~~~~~~~~~~~~~~~~:~\text{conformally coupled scalar} \, ,
    \\
    & f_k^-(\eta) = \frac{H}{\sqrt{2k^3}} (1+i k\eta) e^{-ik\eta}, ~~f_k^+(\eta) = \frac{H}{\sqrt{2k^3}} (1-i k\eta) e^{ik\eta}~~:~\text{massless scalar} \, .
\end{split}
\end{equation}
\textbf{In-In method:} Next we turn to describing the techniques used for calculating the late-time de Sitter correlators. We start with the so-called in-in (also known as Schwinger-Keldysh \cite{Schwinger:1960qe,Keldysh:1964ud}) formalism. Using this method, one addresses the following question: given some initial state at an early time, what is the expectation value of an operator (or a product of operators all inserted at equal $\eta_0$) in the evolved state, say $\vert \Omega \rangle$, at a later time $\eta_0$. This formally defines the in-in late-time de Sitter correlator (or the cosmological correlator) in the limit $\eta_0 \rightarrow 0^-$, 
\begin{equation} \label{inincorrdef}
\begin{split}
    \langle \Omega \vert \mathcal{O}(\eta_0) \vert \Omega \rangle &= \langle 0 \vert\,  \bar{T}  \left[ \text{exp} \left( i \int_{- \infty_+}^{\eta_0} d\eta' H_{\text{int}}(\eta') \right)\right] \, \mathcal{O}(\eta_0)  \, T  \left[ \text{exp} \left( - i \int_{-\infty_-}^{\eta_0} d\eta' H_{\text{int}}(\eta') \right)\right] \vert 0 \rangle \, ,
\end{split}
\end{equation}
where $\pm \infty_{\pm} \equiv \pm \infty \left(1 \pm i\epsilon \right)$. In eq.\eqref{inincorrdef} all the operators are in the interaction picture and $H_{\text{int}}$ denotes the interacting Hamiltonian in the interaction picture. $T$ and $\bar{T}$ represent the time-ordering and anti time-ordering respectively. Also, the particular $ \pm i \epsilon$ principle ensures that the state at very far past is projected onto the so-called Bunch-Davies vacuum denoted by $\vert 0 \rangle$. 

With this formula in hand, one can expand it perturbatively in powers of $H_{\text{int}}$ and obtain an expression for the cosmological correlator. Expanding up to first and second order in perturbation (i.e. at $\mathcal{O}(H_{\text{int}})$ and $\mathcal{O}(H_{\text{int}}^2)$) one gets
\begin{equation}
\begin{split}
\langle \Omega \vert \mathcal{O}(\eta_0) \vert \Omega \rangle \equiv& \langle \mathcal{O}(\eta_0) \rangle = i \int_{-\infty}^{\eta_0} d\eta \langle 0 \vert \left[ H_{\text{int}}(\eta) , \mathcal{O}(\eta_0) \right] \vert 0 \rangle
    \\
    &  - \int_{-\infty}^{\eta_0}  d\eta \int_{-\infty}^{\eta} d\eta' \langle 0 \vert \left[ H_{\text{int}}(\eta'),\left[ H_{\text{int}}(\eta) , \mathcal{O}(\eta_0) \right]\right] \vert 0 \rangle + \mathcal{O}(H_{\text{int}}^3) \, ,
\end{split}
\end{equation}
which will serve as the main working formula for computing cosmological correlators using in-in method. 

In practice one can compute the in-in correlators by calculating Feynman diagrams using certain rules. First one needs to draw the late time boundary $\eta_0 \to 0$ where the fields are inserted and then consider all diagrams contributing to the relevant process. In order to evaluate those diagrams, one needs to use the following vertices, and different types of propagators: 
\begin{itemize}
\item Vertices: Each vertex can be of two types: $``+"$ and $``-"$. 
A $``+"$-vertex will contribute to a factor $+i V$ whereas a $``-"$-vertex will contribute to a factor $-i V^\dagger$. $V$ is the functional differentiation of the interaction Hamiltonian corresponding to each vertex.
\item Propagators: Lines connecting the bulk fields with the boundary fields are represented by ``bulk-to-boundary" propagators which are essentially the Wightman functions 
\begin{equation}\label{in in bulk-bndry prop}
\begin{split}
    K_{k}^{+}(\eta_0, \eta) = f_k^-(\eta_0) f_k^+(\eta) \, , \quad 
    K_{k}^{-}(\eta_0, \eta) = f_k^+(\eta_0) f_k^-(\eta) \, ,
\end{split}
\end{equation}
where $K_k^+$ appears when the boundary fields are contracted to the bulk fields at a vertex of $``+"$ type whereas $K_k^-$ appears when they are connected to the bulk fields at a vertex of $``-"$ type. 
Lines contracting two bulk fields are represented by ``bulk-to-bulk" propagators
\begin{equation}\label{in in bulk to bulk prop}
\begin{split}
    & G_p^{++}(\eta_1, \eta_2) = f_p^-(\eta_1) f_p^+(\eta_2) \theta(\eta_1 - \eta_2) +(\eta_1 \leftrightarrow \eta_2)
    \\
    & \quad = \frac{1}{P_p(\eta_0)}\bigg[K_p^-(\eta_0,\eta_1) K_p^+(\eta_0,\eta_2) \theta(\eta_1-\eta_2)+(\eta_1 \leftrightarrow \eta_2) \bigg] \, ,
    \\
    & G_p^{--}(\eta_1, \eta_2) = f_p^+(\eta_1) f_p^-(\eta_2) \theta(\eta_1 - \eta_2) +(\eta_1 \leftrightarrow \eta_2)
    \\
    & \quad = \frac{1}{P_p(\eta_0)}\bigg[K_p^+(\eta_0,\eta_1) K_p^-(\eta_0,\eta_2) \theta(\eta_1-\eta_2)+(\eta_1 \leftrightarrow \eta_2) \bigg] \, ,
    \\
    & G_p^{+-}(\eta_1, \eta_2) = f_p^+(\eta_1) f_p^-(\eta_2) = \frac{K_p^+(\eta_0,\eta_1) K_p^-(\eta_0,\eta_2)}{P_p(\eta_0)} \, ,\\
    &     G_p^{-+}(\eta_1, \eta_2) = G_p^{+-}(\eta_2, \eta_1)= f_p^-(\eta_1) f_p^+(\eta_2)\, ,
\end{split}
\end{equation}
where $P_p(\eta) = f^-_p(\eta)f^+_p(\eta)$, is the power spectrum.
\item Each vertex will be accompanied by an integral $\int d\eta/(H \eta)^4$. In case of the loop diagrams, one needs to integrate over the loop momenta.
\end{itemize}

Here we give the explicit form of the bulk-to-bulk and bulk-to-boundary propagators for both the conformally coupled and massless scalar theory using the mode functions written in eq.\eqref{mode functions}. 
In case of conformally coupled scalar theory the bulk-to-boundary propagator is given by
\begin{equation}\label{bulk-boundary prop}
  \begin{split}
       K_{k}^{+}(\eta_0, \eta) = \frac{H^2 \eta_0 \eta}{2k} e^{-ik(\eta_0-\eta)} \, , \quad  K_{k}^{-}(\eta_0, \eta) = \frac{H^2 \eta_0 \eta}{2k} e^{ik(\eta_0-\eta)} \, .
  \end{split}  
\end{equation}
And the bulk-to-bulk propagators are given by
\begin{equation}\label{bulk-bulk prop}
 \begin{split}
   & G_p^{++}(\eta_1, \eta_2) = \frac{H^2 \eta_1 \eta_2}{2p} \bigg(e^{-ip(\eta_1-\eta_2)}    \theta(\eta_1 - \eta_2) +(\eta_1 \leftrightarrow \eta_2) \bigg)\, ,
    \\
    & G_p^{--}(\eta_1, \eta_2) =\frac{H^2 \eta_1 \eta_2}{2p} \bigg(e^{-ip(\eta_2-\eta_1)}    \theta(\eta_1 - \eta_2) +(\eta_1 \leftrightarrow \eta_2) \bigg)\, ,
    \\
    & G_p^{+-}(\eta_1, \eta_2) =  \frac{H^2 \eta_1 \eta_2}{2p} e^{ip(\eta_1-\eta_2)}    \, ,
    \qquad G_p^{-+}(\eta_1, \eta_2) = \frac{H^2 \eta_1 \eta_2}{2p} e^{-ip(\eta_1-\eta_2)}    \, .
 \end{split}   
\end{equation}
Similarly for massless scalar case, the bulk-to-boundary and bulk-to-bulk propagator is given by the following
\begin{equation}\label{bulk-boundary prop1}
  \begin{split}
       K_{k}^{+}(\eta_0, \eta) = \frac{H^2  }{2k^3} (1+ik\eta_0) (1-ik\eta) e^{-ik(\eta_0-\eta)}\, , \quad 
     K_{k}^{-}(\eta_0, \eta) =  \frac{H^2  }{2k^3} (1-ik\eta_0) (1+ik\eta) e^{ik(\eta_0-\eta)} \, .
  \end{split}  
\end{equation}
And the bulk-to-bulk propagators are given by
\begin{equation}\label{bulk-bulk prop1}
 \begin{split}
   & G_p^{++}(\eta_1, \eta_2) = \frac{H^2 }{2p^3} \bigg((1+ip\eta_1) (1-ip\eta_2)e^{-ip(\eta_1-\eta_2)}    \theta(\eta_1 - \eta_2) + \eta_1 \leftrightarrow \eta_2\bigg) \, ,
    \\
    & G_p^{--}(\eta_1, \eta_2) =\frac{H^2 }{2p^3} \bigg((1-ip\eta_1) (1+ip\eta_2)e^{ip(\eta_1-\eta_2)}    \theta(\eta_1 - \eta_2) + \eta_1 \leftrightarrow \eta_2 \bigg)\, ,
    \\
    & G_p^{+-}(\eta_1, \eta_2) =   \frac{H^2}{2p^3}(1-ip\eta_1) (1+ip\eta_2) e^{ip(\eta_1-\eta_2)}    \, ,
    \\
    & G_p^{-+}(\eta_1, \eta_2) = \frac{H^2}{2p^3}(1+ip\eta_1) (1-ip\eta_2) e^{-ip(\eta_1-\eta_2)}    \, .
 \end{split}   
\end{equation}
\textbf{Wave-function method:} In an alternative method, the late-time de Sitter correlators can be computed with the knowledge of the wave-function of the quantum fields (denoted by $\Phi$) in the bulk. A path-integral over the field configurations $\Phi$, given the initial condition at far past (i.e., at $\eta \rightarrow - \infty$) which we choose to be the Bunch-Davies one, formally defines the wave-function as a functional of the late time values of the fields $\phi = \Phi(\eta=\eta_0)$. Furthermore, in a semi-classical limit the path-integration receives major contribution when the action gets evaluated at its saddle-point (i.e., an on-shell action) and the wave-function, at tree-level, can be approximated as
\begin{equation}
\Psi[\phi]=\int_{\Phi(-\infty_{+})=0}^{\Phi(0)=\phi} \mathcal{D}\Phi~ e^{i S[\Phi]} \quad \xRightarrow[\text{approximation}]{\text{Saddle-point}} \quad \Psi[\phi] \sim e^{i \, S_{\text{on-shell}}[\phi]}  \,. 
\end{equation}
In an perturbative expansion in the interaction, the semi-classical wave-function can be formally expressed as 
\begin{equation} \label{Psi-wvfncoeff}
\begin{split}
    \Psi[\phi] = \exp\bigg (&-\frac{1}{2} \int\frac{ d^3 \vec{k}_1 d^3 \vec{k}_2 }{(2\pi)^6} \, \Psi_2(\vec{k}_1,\vec{k}_2) \phi(\vec{k}_1) \phi(\vec{k}_2) 
    \\
    &  +\sum_{n=3}^{\infty} \frac{1}{n!} \int\frac{ d^3 \vec{k}_1 \, d^3 \vec{k}_2 \, \cdots d^3 \vec{k}_n}{(2\pi)^{3n}} \, \Psi_n(\vec{k}_1,\vec{k}_2,\cdots,\vec{k}_n) \, \phi(\vec{k}_1) \, \phi(\vec{k}_2)\cdots \phi(\vec{k}_n) \bigg) \, ,
\end{split}    
\end{equation}
where $\Psi_n(\vec{k}_1,\vec{k}_2,\cdots,\vec{k}_n)$ are known as the wave-function coefficients, which are evidently the ingredients that build the wave-function.  

Once, we know the wave-function the late-time de Sitter correlators can in principle be defined through another path-integration performed at 
\begin{equation}\label{correlator from wavefunction def}
\begin{split}
    \big{\langle} \phi_1 \phi_2 \cdots \phi_n \big{\rangle}=\int \mathcal{D} \phi~ \phi_1 \phi_2 \cdots \phi_n \, \Big{\vert}  \Psi[\phi]  \Big{\vert}^2 \, ,
\end{split}    
\end{equation}
where we have used the following short-hand notation $\phi_i = \phi(\vec{k}_i)$ for $i=1,\, 2, \cdots , n$. 

It is obvious from the expansion in the exponential on the RHS of eq.\eqref{Psi-wvfncoeff} that $\Psi_2$, as the coefficient of the quadratic piece in $\phi$, captures information about the inverse propagator in free theory. Similarly, the wave-function coefficients $\Psi_n$ for $n \ge 3$ carries information about the interactions. Therefore, in an perturbative expansion in the interactions, one can substitute eq.\eqref{Psi-wvfncoeff} in eq.\eqref{correlator from wavefunction def} and further manipulations imitating Wick contractions would yield the following expressions for the cosmological correlators in terms of the wave-function coefficients
\begin{equation} \label{cosmo-cor-wvfn-def}
\begin{split}
     \langle \phi_1 \phi_2 \rangle = & \,  \frac{1}{2 \, \mathbb{R}e \psi_2 (\vec{k})} \, ,
    \\
     \langle \phi_1 \phi_2 \phi_3 \rangle = & \, \frac{ 2 \mathbb{R}e \psi_3 (\vec{k}_1, \vec{k}_2, \vec{k}_3)}{\prod_{i=1}^3 2 \mathbb{R}e\psi_2 (\vec{k}_i)} \, ,
    \\
     \langle \phi_1 \phi_2 \phi_3 \phi_4 \rangle = & \, \frac{2}{ \prod_{i=1}^4 2\mathbb{R}e  \psi_2 (\vec{k}_i) } \left( \mathbb{R}e \psi_4(\vec{k}_1,\vec{k}_2, \vec{k}_3, \vec{k}_4 ) + \frac{\mathbb{R}e \psi_3 (\vec{k}_1, \vec{k}_2, \vec{p})  \, \mathbb{R}e \psi_3 (\vec{k}_3, \vec{k}_4, -\vec{p})}{\mathbb{R}e\psi_2(\vec{p})}\right) \, , \\
     \cdots & \cdots \, .
\end{split}
\end{equation}
where by $\vec{p}$ we denote the momentum flowing within the bulk to bulk vertices \footnote{In eq.\eqref{cosmo-cor-wvfn-def} we are writing the $s$-channel expression, ignoring the other channels. Also, in subsequent analysis, we will be focussing on $s$-channel diagrams only unless explicitly mentioned.}.

From eq.\eqref{cosmo-cor-wvfn-def}, it is clear that to know an $m$-point cosmological correlator all the lower point wave-function coefficients \footnote{In writing $\psi_n$ from $\Psi_n$, we have stripped off a three momentum conserving delta function.} $\psi_n$ such that $n \le m$ have to be calculated, which can be done by evaluating Witten diagrams using Feynman rules similar to those for in-in correlators. In this case we can have only one type of bulk-to-boundary and bulk-to-bulk propagators. The bulk-to-boundary propagator is given by following
\begin{equation}\label{wvfnbulkbndry}
K^{\psi}_k( \eta_0,\eta) = \frac{f^+_k(\eta)}{f^+_k(\eta_0)} \, ,
\end{equation}
and the bulk-to-bulk propagator is expressed as
\begin{equation}\label{wvfnbulkbulk}
  G_{p}^\psi(\eta_1, \eta_2) =f_p^-(\eta_1) f_p^+(\eta_2) \theta(\eta_1 - \eta_2) + f_p^+(\eta_1) f_p^-(\eta_2) \theta(\eta_2 - \eta_1)-\frac{f_p^-(\eta_0)}{f_p^+(\eta_0)} f_p^+(\eta_1) f_p^+(\eta_2) \, .
\end{equation}
\textbf{Relating in-in and wave-function propagators:} The propagators required to calculate wave-function coefficients can be related to that of the in-in correlators. Using eq.\eqref{in in bulk-bndry prop} we could therefore write the following relation between them
\begin{equation}\label{bulkboundary WF and inin}
 \begin{split}
  K^{\psi}_k( \eta_0,\eta) =  \frac{f^+_k(\eta)}{f^+_k(\eta_0)} =\frac{K^{+}_k( \eta_0,\eta)}{P_k(\eta_0)} \, ,
 \end{split}   
\end{equation}
where $P_k(\eta) = f^-_k(\eta)f^+_k(\eta)$, is the power spectrum. At this point it is also useful to write the power spectrum in terms of the wave-function-coefficients as well as the in-in propagator as follows 
\begin{equation} \label{defpowerspec}
P_k(\eta_0) =  f^-_k(\eta_0)f^+_k(\eta_0) = \frac{1}{2 \mathbb{R}e \psi_2(k)} \, .
\end{equation}
To establish a similar relationship between the bulk-to-bulk propagators, we can rewrite eq.\eqref{wvfnbulkbulk} using eq.\eqref{in in bulk-bndry prop} and eq.\eqref{in in bulk to bulk prop} as
\begin{equation}
     G_{p}^\psi(\eta_1, \eta_2) =G_p^{++}(\eta_1, \eta_2)-\frac{K^{+}_p( \eta_0,\eta_1) K^{+}_p( \eta_0,\eta_2)}{P_p(\eta_0)} \, .
\end{equation}
Equivalently, using eq.\eqref{bulkboundary WF and inin} we can also write the following relation which we will use in many places in the derivation of the cutting rules
\begin{equation} \label{bulkbulkreln}
 \begin{split}
 &G_p^{++}(\eta_1, \eta_2)=    G_{p}^\psi(\eta_1, \eta_2) +P_p(\eta_0) K^{\psi}_p(\eta_0,\eta_1) K^{\psi}_p(\eta_0,\eta_2) \, , \\
 & G_{p}^{--} (\eta_1, \eta_2) = G_{p}^\psi(\eta_1, \eta_2)^* + P_{p}(\eta_0) K^\psi_{p}(\eta_0,\eta_1)^* K^\psi_{p}(\eta_0,\eta_2)^* \, .
 \end{split}  
\end{equation}
\subsection*{Overview of notation and summary of results} 
We will now summarise the main results derived in this paper. Before that, we will give a brief overview of various notations used throughout the paper. See appendix \ref{notation_convention} for a detailed summarised list of the notations and conventions. 

We will denote a correlator by stripping off the three-momentum-conserving delta function. For example, any $n$-point correlator will be defined as follows 
\begin{equation}
    \big{\langle} \phi(\vec{k}_1) \, \phi(\vec{k}_2)  \cdots \phi(\vec{k}_n) \big{\rangle} = (2\pi)^3 \, \delta^3 (\vec{k}_1 + \vec{k}_2+ \cdots + \vec{k}_n) \, \mathcal{B}(\lbrace k_i \rbrace; \lbrace p_i \rbrace) \, .
\end{equation}
In subsequent sections, our calculations will primarily use the object $\mathcal{B}(\lbrace k_i \rbrace; \lbrace p_i \rbrace)$ defined above. Let us elaborate on the significance of its notation. In $\mathcal{B}(\lbrace k_i \rbrace; \lbrace p_i \rbrace)$, the argument $\lbrace k_i \rbrace$ corresponds to the magnitude of all the external momenta, and $\lbrace p_i \rbrace$ signifies the magnitude of the internal momenta in the associated Witten diagram considered to compute this correlator. It should be noted that the internal momenta are not independent data for evaluating the $n$-point correlator, as they are all determined in terms of the external momenta. For loops, also, the internal momenta will be integrated and have no significance for the correlator. However, we are using this notation to highlight the internal momenta, as we need to indicate which internal momentum will be cut when writing our cutting rules. 

For us it will not be of much importance to keep track of the number of field operators inserted at the late-time slice $\eta_0^-$, i.e., we will not be explicitly writing the `$n$' in an $n$-point cosmological correlator. Instead, we will use a notation $\mathcal{B}^{(2)}\left( \lbrace \mathbf{k}_L, \mathbf{k}_R \rbrace ;p \right)$ where the superscript `$^{(2)}$' will be used to denote that we are looking at a $2$-site correlator with $2$ interaction vertices, with $\lbrace \mathbf{k}_L, \mathbf{k}_R \rbrace$ signifying the set of external momenta at the left and right vertices and $p$ denotes the modulus of the bulk-bulk momentum. 
\begin{figure}[h]
	\centering
	\includegraphics[width=0.65\textwidth]{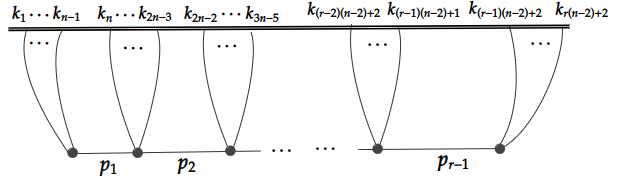}
	\caption{}
	\label{rsite tree}
\end{figure}
 Similarly, $\mathcal{B}^{(r)}\left( \lbrace \mathbf{k}_L,\mathbf{k}_{M_1},...,\mathbf{k}_{M_{r-2}}, \mathbf{k}_R \rbrace ; \lbrace p_1,..,p_{r-1} \rbrace \right)$ will denote a correlator with $r$-sites or vertices. 
  In the similar fashion as before, $\lbrace \mathbf{k}_L, \mathbf{k}_R \rbrace$ signifies the set of external momenta at the left-most and right-most vertices at the tree-level diagram. Also note that, in a $r$-site diagram, except the left and right-most vertices we have $(r-2)$ number of other ``middle" vertices. $\lbrace \mathbf{k}_{M_1},...,\mathbf{k}_{M_{r-2}} \rbrace$ signify the collection of external momenta at these vertices starting from the left. To be precise, e.g. $\lbrace \mathbf{k}_{M_1} \rbrace$ denotes the collection of external lines at the next to left most vertex. This can also be understood from the tree level diagram in Fig-\ref{rsite tree} with a polynomial interaction of type $\phi^n$. From this diagram we can understand that 
 \begin{equation*}
 \begin{split}
 	& \lbrace \mathbf{k}_L \rbrace \equiv \lbrace \vec{k}_1,\cdots, \vec{k}_{n-1} \rbrace \, , ~ \lbrace \mathbf{k}_{M_1} \rbrace \equiv \lbrace \vec{k}_n,\cdots, \vec{k}_{2n-3} \rbrace \,,  \cdots \, , \\
 	& \lbrace \mathbf{k}_{M_{r-2}} \rbrace \equiv \lbrace \vec{k}_{(r-2)(n-2)+2},\cdots, \vec{k}_{(r-1)(n-2)+1} \rbrace\, , \lbrace \mathbf{k}_R \rbrace \equiv \lbrace \vec{k}_{(r-1)(n-2)+2},\cdots, \vec{k}_{r(n-2)+2} \rbrace \, ,
 \end{split}
 \end{equation*}
 and the internal momenta are denoted by $\lbrace \vec{p}_1, \cdots, \vec{p}_{r-1} \rbrace$ and their magnitude are written without the vector symbol. Also, $\mathcal{B}^{(1)}( \lbrace \mathbf{k}_{L} \rbrace ,p)$ will denote a contact correlator at the left vertex of the original multi-site correlator. Keeping eq.\eqref{cosmo-cor-wvfn-def} in mind, we can also express the $r$-site correlators $\mathcal{B}^{(r)}$ in terms of wave-function coefficients, 
\begin{equation}\label{corr 1 2 def}
\begin{split}
&\mathcal{B}^{(1)}(k_1,\ldots,k_{n})
= \frac{2\,\mathbb{R}e\,\psi_n(\vec{k}_1,\ldots,\vec{k}_{n})}
{\left(\prod_{i=1}^{n} 2\,\mathbb{R}e\,\psi_2(\vec{k}_i)\right)
} \, ,
\\[6pt]
&\mathcal{B}^{(2)}(k_1,\ldots,k_{2n-2};p)
= \frac{2}{\prod_{i=1}^{2n-2} 2\,\mathbb{R}e\,\psi_2(\vec{k}_i)}
\Bigg[
\mathbb{R}e\,\psi_{2n-2}(\vec{k}_1,\ldots,\vec{k}_{2n-2};\vec{p})
\\
&\qquad\qquad\qquad\qquad\qquad\qquad
+ \frac{
\mathbb{R}e\,\psi_n(\vec{k}_1,\ldots,\vec{k}_{n-1},\vec{p})\,
\mathbb{R}e\,\psi_n(-\vec{p},\vec{k}_n,\ldots,\vec{k}_{2n-2})
}
{\mathbb{R}e\,\psi_2(\vec{p})}
\Bigg] \,,\\ 
& \ldots\ldots \, ,
\end{split}
\end{equation}
upto possible additional contributions on the RHS due to different channels. 

We will be deriving a discontinuity relation for the cosmological correlators. Therefore, it is very important to define what we mean by the discontinuity operation. Our definition of the discontinuity operation denoted by `$\text{Disc}_p$' is \footnote{It is worth noting that this definition is different than what appeared in \cite{Melville:2021lst}, see eq.(3.3). The main difference lies in the fact that in our definition the internal energy $p$ is analytically continued as $p \to -p$, but the external energies are untouched. However, in the definition used in \cite{Melville:2021lst} it was opposite, where the external energies and the spatial momenta got reversed in sign, but the internal energies remained same. However, when they appear in the discontinuity relations, it is possible to argue that content-wise both definitions of `$\text{Disc}_p$' are ``equivalent".}
\begin{equation} \label{defdisc}
\begin{split}
    & \text{Disc}_{p} f(p,k_1,...,k_n) = f(p,k_1,...,k_n) - f(-p,k_1,...,k_n) \, .
\end{split}
\end{equation}
We will also use another related discontinuity operation denoted by `$\widetilde{\text{Disc}}_p$' defined as follows 
\begin{equation} \label{defdisct}
\begin{split}
    &\widetilde{\text{Disc}}_{p} f(p,k_1,...,k_n) = f(p,k_1,...,k_n) + f(-p,k_1,...,k_n) \, .
\end{split}
\end{equation}
Note that the two definitions given above in eq.\eqref{defdisc} and eq.\eqref{defdisct} differ by a relative sign between the two terms on the RHS's. As we will see these definitions of the discontinuity operation will lead us to the desired discontinuity relations involving the cosmological correlators for conformally coupled and massless scalars with polynomial interactions. However, one might wonder why such definitions are needed, and are they related to standard notions of discontinuity familiar in complex analysis, such as $f(p + i \, \epsilon) - f(p - i \, \epsilon)$. Moreover, are these notions of discontinuity extend to cases involving scalars with generic masses? We address these issues using analytic continuation and properties of Hankel functions in Appendix~\ref{genericmassscalar}.

With these notations explained above, let us write our final result for the single-cut discontinuity of a $2$-site tree-level cosmological correlator with scalar operators inserted at the late-time slice of de Sitter:
\begin{equation}\label{2siteDiscFinal Result I}
\begin{split}
     {\text {Disc}}_{p} \mathcal{B}^{(2)}( \lbrace \mathbf{k}_L, \mathbf{k}_R \rbrace ;p) = \frac{1}{2 P_p(\eta_0)} \Bigg(& \text{Disc}_p \mathcal{B}^{(1)}( \lbrace \mathbf{k}_{L} \rbrace ,p)  \times \text{Disc}_p \mathcal{B}^{(1)}(\lbrace \mathbf{k}_{R} \rbrace ,p) 
     \\
     & - \widetilde{\text{Disc}}_{p} \widetilde{\mathcal{B}}^{(1)}( \lbrace \mathbf{k}_{L} \rbrace ,p) \times \widetilde{\text{Disc}}_{p} \widetilde{\mathcal{B}}^{(1)}( \lbrace \mathbf{k}_{R} \rbrace ,p) \Bigg) \, .
\end{split}
\end{equation}

For $r$-site tree level correlator we will use a convention that our single-cut will be applied to the right-most internal line, and the result is the following: 
\begin{equation}\label{r site correlator I}
\begin{split}
    & {\text {Disc}}_{p_{r-1}} \mathcal{B}^{(r)}( \lbrace \mathbf{k}_L,..., \mathbf{k}_R \rbrace ; \lbrace p_1, ..., p_{r-1} \rbrace ) = 
    \\
    &\frac{1}{2 P_{p_{r-1}} (\eta_0)} \bigg[ \text{Disc}_{p_{r-1}} \mathcal{B}^{(r-1)}\left(  \lbrace \mathbf{k}_L,...,\mathbf{k}_{M_{r-2}} \rbrace, p_{r-1} ; \lbrace p_1,..,p_{r-2} \rbrace \right)  \times  \text{Disc}_{p_{r-1}} \mathcal{B}^{(1)}( \lbrace \mathbf{k}_{R} \rbrace, p_{r-1})\\
    & ~~ - \widetilde{\text{Disc}}_{p_{r-1}} \widetilde{\mathcal{B}}^{(r-1)}\left(  \lbrace \mathbf{k}_L,...,\mathbf{k}_{M_{r-2}} \rbrace, p_{r-1} ; \lbrace p_1,..,p_{r-2} \rbrace \right)  \times  \widetilde{\text{Disc}}_{p_{r-1}} \widetilde{\mathcal{B}}^{(1)}( \lbrace \mathbf{k}_{R} \rbrace, p_{r-1}) \bigg] \, .
\end{split}
\end{equation}
where the object $\widetilde{\mathcal{B}}$ takes the following forms at various sites
\begin{equation}\label{corrt 1 2 def}
	\begin{split}
		& \widetilde{\mathcal{B}}^{(1)}(k_1,\ldots,k_n)
		= \frac{2i\,\mathbb{I}m\,\psi_n(\vec{k}_1,\ldots,\vec{k}_{n})}
		{\left(\prod_{i=1}^{n} 2\,\mathbb{R}e\,\psi_2(\vec{k}_i)\right)} \, ,
		\\[6pt]
		& \widetilde{\mathcal{B}}^{(2)}(k_1,\ldots,k_{2n-2};p)
		= \frac{2i}{\prod_{i=1}^{2n-2} 2\,\mathbb{R}e\,\psi_2(\vec{k}_i)}
		\Bigg[
		\mathbb{I}m\,\psi_{2n-2}(\vec{k}_1,\ldots,\vec{k}_{2n-2};\vec{p})
		\\
		& \qquad \qquad \qquad \qquad + \frac{
			\mathbb{R}e\,\psi_n(\vec{k}_1,\ldots,\vec{k}_{n-1},\vec{p})\,
			\mathbb{I}m\,\psi_n(-\vec{p},\vec{k}_n,\ldots,\vec{k}_{2n-2})
		}
		{\mathbb{R}e\,\psi_2(\vec{p})}
		\Bigg] \,,
		\\
		& \ldots\ldots \, .
	\end{split}
\end{equation}
Note that in our single-cut rule, the discontinuity of a cosmological correlator gets decomposed in terms of lower point discontinuities of the correlators $\mathcal{B}$, and also the objects $\widetilde{\mathcal{B}}$ which are not exactly lower-point correlators but are auxiliary objects that involve imaginary parts of the lower point wave-function coefficients. As we will discuss the appearance of this $\widetilde{\mathcal{B}}$ in the discontinuity relations is one of the significant points that we found in our analysis. They will also play crucial role when we will compare with the existing literature such as in \cite{Donath:2024utn}.

For cosmological correlators at the loop level, we need to be more careful with our notation and the apllication of the `$\text{Disc}_p$' operation. A $2$-site $1$-loop diagram $\mathcal{B}^{(2)}\left( \lbrace \mathbf{k}_L, \mathbf{k}_R \rbrace\right)$ will be obtained by performing integrations over the momenta running in the loop, and the integrand of the loop-integration will be denoted by $\mathbb{B}^{(2)}\left( \lbrace \mathbf{k}_L, \mathbf{k}_R \rbrace ;p_1, p_2 \right)$, where $p_1, \, p_2$ are the magnitude of the momentum labelling the two arms of the loop. Our discontinuity relation will involve the `$\text{Disc}_p$' operation acting on this $\mathbb{B}^{(2)}\left( \lbrace \mathbf{k}_L, \mathbf{k}_R \rbrace ;p_1, p_2 \right)$, which is slightly different from the discontinuity relation for the tree-level diagrams \footnote{As mentioned before, our `$\text{Disc}_p$' operation involves analytically continuing the internal energy $p \to -p$, and therefore we need to apply this on the integrand before performing the loop integration. On the other hand, the discontinuity used in \cite{Melville:2021lst} didn't change the internal energy and can be applied on the loop-integrated expression for the correlator.}. Also, we will be applying the $\text{Disc}_p$ operation successively by cutting all the loop arms one-by-one, which is in contrast to the simultaneous-cutting of the loop lines as mentioned in \cite{Melville:2021lst}. Our final result for the $2$-site $1$-loop diagrams will be obtained in eq.\eqref{finalresult 1 loop}.  
\section{Identities involving in-in propagators from unitarity and analyticity} \label{ininpropagatorfromunitarity}

In this section, we derive specific identities for any tree-level cosmological correlator that involves scalar operators inserted at a late-time slice at $\eta_0$. We use the bulk-to-bulk and bulk-to-boundary propagators within the in-in formalism. These identities demonstrate that a suitable linear combination of the bulk-to-bulk propagators, as seen in the in-in formalism, can be expressed in a factorized form using bulk-to-boundary propagators.

The bulk-to-bulk propagators connect two interaction vertices at times $\eta_1$ and $\eta_2$. In particular, the Feynman propagator $G_p^{++}(\eta_1, \eta_2)$ and its conjugate $G_p^{--}(\eta_1,\eta_2)$ exhibit explicit time ordering, meaning that the integrations over $\eta_1$ and $\eta_2$ are coupled. We will demonstrate that by starting from a fundamental argument of unitarity and considering the hermitian analyticity properties of the bulk-to-boundary propagators, we can construct a specific combination of the four in-in propagators that is free of time ordering. This resulting expression depends separately on $\eta_1$ and $\eta_2$, and can be represented as a product of two bulk-to-boundary propagators evaluated at their respective times. This provides a natural interpretation of the relevant bulk-to-bulk propagator as a``split" object, created from building blocks that are anchored to the boundary.

For clarity, we will limit our discussion to conformally coupled and massless scalars, corresponding to $\nu = \frac{1}{2}$ and $\nu = \frac{3}{2}$, respectively, and derive the corresponding splitting formulas explicitly. However, the argument applies to any half-integer $\nu$, and we will indicate the generalization along the way. The case of a generic $\nu$ will require a separate treatment, which we will defer to Appendix~\ref{genericmassscalar}.

\noindent \textbf{Hermitian analyticity:} We start by writing the Hermitian analyticity in terms of the bulk-to-boundary propagators corresponding to the wave-function picture, see  \cite{Goodhew:2020hob}, i.e.,
\begin{equation}
\label{defHA}
\Big( K^{\psi}_{-k^{*}}(\eta_0,\eta) \Big)^* = K^{\psi}_{k}(\eta_0,\eta) \, , 
\end{equation}
where the comparison on both sides involve the following operation of $k \to -k^*$. Next, our goal is to use the relation between the bulk-to-boundary propagators corresponding to the in-in correlators and the wave-function coefficients. This was written in eq.\eqref{bulkboundary WF and inin} before, but we write it here again for convenience
\begin{equation}\label{in-in prop}
\begin{split}
      K_k^+(\eta_0, \eta) =  K^{\psi}_k(\eta_0, \eta) \Big{\vert} f_k^+(\eta_0) \Big{\vert}^2\, , \quad 
      K_k^-(\eta_0, \eta) =  \Big( K^{\psi}_k(\eta_0, \eta) \Big)^* \Big{\vert} f_k^+(\eta_0) \Big{\vert}^2 \, .
\end{split}
\end{equation}
Using eq.\eqref{defHA} and eq.\eqref{in-in prop}, we can derive the following identity 
\begin{equation}\label{in-in prop1}
\begin{split}
K_{-k^*}^{-}(\eta_0, \eta) = K^+_{k}( \eta) \frac{\Big{\vert} f_{-k^*}^+(\eta_0) \Big{\vert}^2 }{\Big{\vert} f_{k}^+(\eta_0) \Big{\vert}^2 } \, .
\end{split}
\end{equation}
Since our primary focus is to consider conformally coupled or massless scalar fields, the analytic structure of cosmological correlators has no branch cut, rather they develop a branch point at the origin in the complex momentum space. Therefore, we can restrict to real momentum in the above relation and simplify it further to obtain 
\begin{equation}\label{in-in prop2}
K_{-k}^{-}(\eta_0, \eta) = K_{k}^+(\eta_0, \eta) \frac{P_{-k}(\eta_0)}{P_{k}(\eta_0)} \, .
\end{equation}
In general, the power spectrum $P_k(\eta_0)\propto k^{-2\nu}$\footnote{considering light mass scalars, $m^2 < \frac{9}{4} H^2$}, and, therefore, for half-integer $\nu$, eq.\eqref{in-in prop2} produces the following expression 
\begin{equation}\label{HA for Corr}
K_{-k}^{-}(\eta_0, \eta) = -K_{k}^+(\eta_0, \eta)\,. 
\end{equation}
From here we can proceed further and incorporate the hermitian analyticity condition imprinted in eq.\eqref{HA for Corr} in terms of bulk-to-boundary propagators, to the bulk-to-bulk propagator as well. In the expressions of the bulk-to-bulk propagators written in eq.\eqref{in in bulk to bulk prop}, using eq.\eqref{HA for Corr} we get the following 
\begin{equation}\label{HA for Corr 1}
\begin{split}
     G_{-p}^{\pm \pm}(\eta_1, \eta_2) =  - G_p^{\mp \mp}(\eta_1, \eta_2)\, , \quad 
     G_{-p}^{\pm \mp}(\eta_1, \eta_2) = - G_p^{\mp \pm}(\eta_1, \eta_2) \, .
\end{split}
\end{equation}
Note that the above set of relations between the bulk-to-bulk propagators are valid for $\nu=$ half-integer. 

\noindent \textbf{Unitarity:} To formalize the implications of unitarity, let's begin by discussing the perturbative statement of unitarity. We denote the interaction picture time-evolution operator as follows 
\begin{equation}\label{unitary evolution op}
\begin{split}
    & \mathcal{U}_I(\eta_0, -\infty) =\mathcal{T} \text{exp} \left(-i\int_{-\infty}^{\eta_0} d\eta \mathcal{H}_I(\eta) \right) 
    = \mathbb{I} + \delta \mathcal{U}_I^{(1)} + \delta \mathcal{U}_I^{(2)}  +\cdots \equiv \mathbb{I} + \delta \mathcal{U}_I \, ,
\end{split}
\end{equation}
where we have expanded the exponential in a series with respect to the coupling constant of the theory. Note that in the RHS of the above equation, we have suppressed the time argument $\eta_0$ in the evolution operator. In the linear order of the coupling constant, we define the evolution operator as follows
\begin{equation}
\begin{split}
    & \delta \mathcal{U}_I^{(1)} = - i \int_{-\infty}^{\eta_0} d\eta \Big( \mathcal{H}^I_L(\eta) + \mathcal{H}^I_R(\eta) \Big) \, ,
\end{split}
\end{equation}
where $\mathcal{H}^I_{L/R}$ is the interaction picture Hamiltonian. Since we are going to discuss a tree-level correlator, there must be two interaction vertices in the bulk, which is why we consider the Hamiltonian as a sum of two interactions localized in two vertices, namely the left and right vertices of the correlator. 
At the quadratic order in the coupling constant, the evolution operator involves ordering between two times at which the interactions are localized and denoted as follows  
\begin{equation}
\begin{split}
    & \delta \mathcal{U}_I^{(2)} = - \frac{1}{2}\int_{-\infty}^{\eta_0} d\eta_1  d\eta_2~ \bigg( \mathcal{H}_L^I(\eta_1) \mathcal{H}_R^I(\eta_2)\theta(\eta_1 - \eta_2) + \mathcal{H}_L^I(\eta_2) \mathcal{H}_R^I(\eta_1)\theta(\eta_2 - \eta_1) \bigg) + \big( L \leftrightarrow R \big) \, .
\end{split}
\end{equation}
By definition unitarity implies the following 
\begin{equation}
\begin{split}
    \text{Unitarity:}~~~ \mathcal{U}_I \mathcal{U}_I^\dagger = \mathbb{I} \implies & \left( \mathbb{I} + \delta \mathcal{U}_I \right)\left( \mathbb{I} + \delta \mathcal{U}_I^\dagger \right) = \mathbb{I} \, ,
    \\
    \implies & \delta \mathcal{U}_I + \delta \mathcal{U}_I^\dagger = - \delta \mathcal{U}_I \delta \mathcal{U}_I^\dagger \, ,
\end{split}
\end{equation}
where the second equality follows from eq.\eqref{unitary evolution op}.\\
If we restrict ourselves at the first order, the statement of unitarity boils down to the following equation 
\begin{equation}
   \delta \mathcal{U}_I^{(1)} = - {\delta \mathcal{U}_I^{(1)}}^\dagger
\end{equation}
which is essentially the statement of the hermiticity of the interaction Hamiltonian
\begin{equation}
    \mathcal{H}^I_L(\eta) = \left( \mathcal{H}_L^I(\eta) \right)^\dagger,~~ \mathcal{H}^I_R(\eta) = \left( \mathcal{H}_R^I(\eta) \right)^\dagger \, .
\end{equation}
Now, going to the quadratic order in the coupling constant, the implication of unitarity takes the following form
\begin{equation}\label{second ord unitarity}
    \delta \mathcal{U}_I^{(2)} + \delta {\mathcal{U}_I^{(2)}}^\dagger = - \left( \delta \mathcal{U}_L^{(1)} \delta {\mathcal{U}_R^{(1)}}^\dagger + \delta \mathcal{U}_R^{(1)} \delta {\mathcal{U}_L^{(1)}}^\dagger \right) \, .
\end{equation}
We now introduce a generic $n$-particle state (of particle type $\alpha$) in the free Hilbert space and the corresponding completeness relation by the following expressions
\begin{equation}
\begin{split}
    & n\text{-particles state} : ~~~ \vert \lbrace \vec{k} , \alpha \rbrace_n \rangle \equiv \vert \vec{k}_1, \alpha_1 ; \cdots ; \vec{k}_n, \alpha_n\rangle = \left( \prod_{i=1}^n f_k^{+} a_I^\dagger (\vec{k}_i ,\alpha_i) \right) \vert 0 \rangle\, ,
    \\
    & \text{Completeness relation} : ~~~ \sum_{m=0}^\infty \sum_{\lbrace \beta_1, \cdots, \beta_m \rbrace} \int \frac{d^3 \vec{p}_1}{(2\pi)^3 P_{p_1}} \cdots \frac{d^3 \vec{p}_m}{(2\pi)^3 P_{p_m}} \vert \lbrace \vec{p}, \beta \rbrace_m \rangle \langle \lbrace \vec{p}, \beta \rbrace_m \vert = \mathbb{I} \, ,
\end{split}
\end{equation}
where $a_I^\dagger$ is the standard particle creation operator in the interaction picture.

With the above definitions, we can write the  eq.\eqref{second ord unitarity} in the following form  by taking an expectation value with respect to the free vacuum of the theory and an $2n-2$-particles state
\begin{equation}\label{unitarity at order 2 with particle avg}
\begin{split}
    & \Big{\langle} \lbrace \vec{k}, \phi \rbrace_{2n-2} \Big{\vert} \delta \mathcal{U}_I^{(2)} \Big{\vert} 0 \Big{\rangle} + \Big{\langle} \lbrace \vec{k}, \phi \rbrace_{2n-2} \Big{\vert} \delta {\mathcal{U}_I^{(2)}}^\dagger \Big{\vert} 0 \Big{\rangle} 
    \\
    =& - \left( \prod_{j=1}^{m} \frac{1}{P_{p_j}(\eta_0)}\int \frac{d^3 \vec{p}_j}{(2\pi)^3} \right) \Big{\langle} \lbrace \vec{k}, \phi \rbrace_{2n-2} \Big{\vert} \delta \mathcal{U}_L^{(1)} \Big{\vert} \lbrace \vec{p}, \phi \rbrace_{m} \Big{\rangle} \Big{\langle} \lbrace \vec{p} , \phi \rbrace_{m} \Big{\vert} \delta {\mathcal{U}_R^{(1)}}^\dagger \Big{\vert} 0 \Big{\rangle}
    \\
    &  - \left( \prod_{j=1}^{m} \frac{1}{P_{p_j}(\eta_0)} \int \frac{d^3 \vec{p}_j}{(2\pi)^3} \right) \Big{\langle} \lbrace \vec{k}, \phi \rbrace_{2n-2} \Big{\vert} \delta {\mathcal{U}_R^{(1)}} \Big{\vert} \lbrace \vec{p} , \phi\rbrace_{n} \Big{\rangle} \Big{\langle} \lbrace \vec{p}, \phi {\rbrace}_{m} \Big{\vert}  \delta {\mathcal{U}_L^{(1)}}^\dagger  \Big{\vert} 0 \Big{\rangle} \, .
\end{split}
\end{equation}
Note that in the RHS we have sandwiched a completeness relation where the fields are inserted at the boundary time slice $\eta=\eta_0$. Since we are interested in the conformally coupled or massless polynomial interaction of the form $\lambda_{L/R} \phi^n$, a tree-level correlator will have $2n-2$ external lines, each corresponding to a boundary field. 

Now we will manipulate each term on the LHS and the RHS of eq.\eqref{unitarity at order 2 with particle avg} to obtain our desired result. We start with the first term in the LHS of  eq.\eqref{unitarity at order 2 with particle avg}, and wick contractions among the fields in all possible ways \footnote{To be precise, throughout this paper we will restrict ourselves to the correlators only at $s$-channel. However, the formalism developed in this paper works straight-forwardly for other channels as well.} yield the following expression 
\begin{equation} \label{LHSpart1}
\begin{split}
    & \Big{\langle} \lbrace \vec{k}, \phi \rbrace_{2n-2} \Big{\vert} \delta \mathcal{U}_I^{(2)} \Big{\vert} 0 \Big{\rangle} \\
    & = - \frac{\lambda_L \lambda_R}{2}  \int d\eta_1\int d\eta_2 \Bigg( \mathcal{K}^+_{\eta_0 \eta_1} \left(\lbrace \mathbf{k}_L \rbrace \right) \mathcal{K}^+_{\eta_0 \eta_2} \left(\lbrace \mathbf{k}_R \rbrace \right) +  \lbrace \mathbf{k}_R \rbrace \leftrightarrow \lbrace \mathbf{k}_L \rbrace  \Bigg) G^{++}_{p}(\eta_1, \eta_2) \, .
\end{split}
\end{equation}
where we used the notation 
\begin{equation}
   \int d\eta_r = \int_{-\infty}^{\eta_0} \frac{d\eta_r}{(H \eta_r)^4}
\end{equation}
Similarly, the second term in the LHS of  eq.\eqref{unitarity at order 2 with particle avg} can be analyzed in the same way and eventually can be written as the following
\begin{equation} \label{LHSpart2}
\begin{split}
    & \Big{\langle} \lbrace \vec{k}, \phi \rbrace_{2n-2} \Big{\vert} \delta{\mathcal{U}_I^{(2)}}^\dagger \Big{\vert} 0 \Big{\rangle} \\
    & = - \frac{\lambda_L \lambda_R}{2}  \int d\eta_1 \int d\eta_2 \Bigg( \mathcal{K}^+_{\eta_0 \eta_1} \left(\lbrace \mathbf{k}_L \rbrace \right) \mathcal{K}^+_{\eta_0 \eta_2} \left(\lbrace \mathbf{k}_R \rbrace \right) +  \lbrace \mathbf{k}_R \rbrace \leftrightarrow \lbrace \mathbf{k}_L \rbrace \Bigg)  G^{--}_{p}(\eta_1, \eta_2) \, .
\end{split}
\end{equation}
In writing eq.\eqref{LHSpart1} and eq.\eqref{LHSpart2} in a compact form, we have defined 
\begin{equation} \label{defmathcalK}
\begin{split}
\mathcal{K}_{\eta_0 \eta_1}^+(\lbrace \mathbf{k}_L \rbrace) \equiv & \prod_{i=1}^{n-1} K_{k_i}^+(\eta_0, \eta_1) = \prod_{i=1}^{n-1} f^{-}_{k_i}(\eta_0) f^{+}_{k_i}(\eta_1) \, , \\
\mathcal{K}_{\eta_0 \eta_2}^+(\lbrace \mathbf{k}_R \rbrace) \equiv & \prod_{i=n}^{2n-2} K_{k_i}^+(\eta_0, \eta_2) = \prod_{i=n}^{2n-2} f^{-}_{k_i}(\eta_0) f^{+}_{k_i}(\eta_2) \, , 
\end{split}
\end{equation}
such that $\mathcal{K}_{\eta_0 \eta_1}^+$ denotes products of bulk-to-boundary propagators (each denoted by $K_{k_i}^+(\eta_0, \eta_1)$) from one vertex (`$+$'-type) at a bulk-time $\eta_1$ to the late-time slice at $\eta_0$. Similarly, we can obtain $\mathcal{K}_{\eta_0 \eta_1}^-$ if the vertex is of `$-$'-type. It is important to distinguish between $\mathcal{K}_{\eta_0 \eta_1}^\pm$ and $K_{k_i}^\pm$, since we will be using these notations frequently going ahead. Also, the argument of $\mathcal{K}_{\eta_0 \eta_1}^\pm$ as $\lbrace \mathbf{k}_L \rbrace$ or $\lbrace \mathbf{k}_R \rbrace$, denotes all the external momenta at the left or right vertex, respectively. 

Adding eq.\eqref{LHSpart1} and eq.\eqref{LHSpart2}, we get 
\begin{equation}\label{preLHS}
\begin{split}
     \text{LHS of eq.}\eqref{unitarity at order 2 with particle avg}& \\
    = \Big{\langle} \lbrace \vec{k}, \phi \rbrace_{2n-2} \Big{\vert} \delta \mathcal{U}_I^{(2)} \Big{\vert} 0 \Big{\rangle}& +  \Big{\langle} \lbrace \vec{k}, \phi \rbrace_{2n-2} \Big{\vert} \delta {\mathcal{U}_I^{(2)}}^\dagger \Big{\vert} 0 \Big{\rangle} 
    \\
     = -\frac{\lambda_L \lambda_R}{2} \int d \eta_1 \int d\eta_2 & \Bigg( \mathcal{K}^+_{\eta_0 \eta_1} \left(\lbrace \mathbf{k}_L \rbrace \right) \mathcal{K}^+_{\eta_0 \eta_2} \left(\lbrace \mathbf{k}_R \rbrace \right) + \mathcal{K}^+_{\eta_0 \eta_1} \left(\lbrace \mathbf{k}_R \rbrace \right) \mathcal{K}^+_{\eta_0 \eta_2} \left(\lbrace \mathbf{k}_L \rbrace \right) \Bigg) \times 
     \\
     & \Bigg( G^{++}_{p}(\eta_1, \eta_2) + G^{--}_{p}(\eta_1, \eta_2) \Bigg)  .
\end{split}
\end{equation}
Using the hermitian analyticity of the bulk-to-bulk propagators, see eq.\eqref{HA for Corr 1}, which is valid for all half-integer $\nu$, and therefore, obviously for conformally coupled and massless scalar fields, we can re-write the above expression as follows
\begin{equation}\label{LHS}
	\begin{split}
		\text{LHS of eq.}\eqref{unitarity at order 2 with particle avg}& \\
		= -\frac{\lambda_L \lambda_R}{2} \int d \eta_1 \int d\eta_2 & \Bigg( \mathcal{K}^+_{\eta_0 \eta_1} \left(\lbrace \mathbf{k}_L \rbrace \right) \mathcal{K}^+_{\eta_0 \eta_2} \left(\lbrace \mathbf{k}_R \rbrace \right) + \mathcal{K}^+_{\eta_0 \eta_1} \left(\lbrace \mathbf{k}_R \rbrace \right) \mathcal{K}^+_{\eta_0 \eta_2} \left(\lbrace \mathbf{k}_L \rbrace \right) \Bigg) \times 
		\\
		& \Bigg( G^{++}_{p}(\eta_1, \eta_2) - G^{++}_{-p}(\eta_1, \eta_2) \Bigg)  \, .
	\end{split}
\end{equation}

Now we compute the RHS of the eq.\eqref{unitarity at order 2 with particle avg}. The first term can be processed by Wick contracting the $(2n-2)$ numbers of boundary fields at time $\eta_0$ with the same number of intermediate fields collectively denoted by $\lbrace \vec{p}, \phi \rbrace_{m}$, and then there will be self-contraction between a pair of the intermediate fields. Finally, one needs to apply the three momentum conservation imposed by $\delta^{(3)}(\vec{k}_i - \vec{p}_j)$ coming out of each contraction, which sets $\vec{p}_i = \vec{k}_i$, and we then obtain the following
\begin{equation}\label{2nd ord unitarity RHS 1st term}
\begin{split}
    & - \left( \prod_{j=1}^{m} \frac{1}{P_{p_j}(\eta_0)} \int \frac{d^3 \vec{p}_j}{(2\pi)^3} \right) \Big{\langle} \lbrace \vec{k}, \phi \rbrace_{n-1} \Big{\vert} \delta \mathcal{U}_L^{(1)} \Big{\vert} \lbrace \vec{p}, \phi \rbrace_{m} \Big{\rangle} \Big{\langle} \lbrace \vec{p} , \phi \rbrace_{m} \Big{\vert} \delta {\mathcal{U}_R^{(1)}}^\dagger \Big{\vert} 0 \Big{\rangle}
    \\
    & \, = -\lambda_L \lambda_R \int d\eta_1 \int d\eta_2 \Bigg( \prod_{i=1}^{n-1} f^{-}_{k_i}(\eta_0) f^{+}_{k_i}(\eta_1) \prod_{i=n}^{2n-2} f^{-}_{k_i}(\eta_0) f^{+}_{k_i}(\eta_2) \Bigg) \Big( f_p^-(\eta_1)f_p^+(\eta_2) \Big)
    \\
    &\, = -\lambda_L \lambda_R \int d\eta_1 \int d\eta_2~ \mathcal{K}^+_{\eta_0\eta_1}(\lbrace \mathbf{k}_L \rbrace) \mathcal{K}^+_{\eta_0\eta_2}(\lbrace \mathbf{k}_R \rbrace) \times G^{-+}_p(\eta_1, \eta_2) \, ,
\end{split}
\end{equation}
where to write the last equality, we used the definition of the bulk to boundary propagators in terms of the mode functions. Following similar arguments, the second term (which is just $L \leftrightarrow R$ of the above expression) in the RHS of  eq.\eqref{unitarity at order 2 with particle avg} turns out to have the following expression
\begin{equation}\label{2nd ord unitarity RHS 2nd term}
\begin{split}
    & - \left( \prod_{j=1}^{m} \frac{1}{P_{p_j}(\eta_0)} \int \frac{d^3 \vec{p}_j}{(2\pi)^3} \right) \Big{\langle} \lbrace \vec{k}, \phi \rbrace_{2n-2} \Big{\vert} \delta {\mathcal{U}_R^{(1)}} \Big{\vert} \lbrace \vec{p} , \phi\rbrace_{m} \Big{\rangle} \Big{\langle} \Big{\lbrace} \vec{p}, \phi \rbrace_{m} \Big{\vert} \delta {\mathcal{U}_L^{(1)}}^\dagger  \Big{\vert} 0 \Big{\rangle}
    \\
    \\
    & \, = -\lambda_L \lambda_R \int d\eta_1 \int d\eta_2 \Big( \mathcal{K}^+_{\eta_0\eta_2}(\lbrace \mathbf{k}_L \rbrace) \mathcal{K}^+_{\eta_0\eta_1}(\lbrace \mathbf{k}_R \rbrace) \Big) \Big( G^{+-}_p(\eta_1, \eta_2) \Big) \, .
\end{split}
\end{equation}
The RHS of eq.\eqref{unitarity at order 2 with particle avg}  can be now obtained by adding the two contributions written above, as follows
\begin{equation} \label{RHSa}
\begin{split}
  & \text{RHS of eq.} \eqref{unitarity at order 2 with particle avg}
    \\
    & = -\lambda_L \lambda_R \int d\eta_1 \int d\eta_2 \Bigg( \mathcal{K}^+_{\eta_0\eta_1}(\lbrace \mathbf{k}_L \rbrace) \mathcal{K}^+_{\eta_0\eta_2}(\lbrace \mathbf{k}_R \rbrace) G^{-+}_p(\eta_1, \eta_2) \\
    & \quad \quad + \mathcal{K}^+_{\eta_0\eta_2}(\lbrace \mathbf{k}_L \rbrace) \mathcal{K}^+_{\eta_0\eta_1}(\lbrace \mathbf{k}_R \rbrace) G^{+-}_p(\eta_1, \eta_2) \Bigg) \, .
\end{split}
\end{equation}
We should implement symmetrization under the exchange of the external legs at the left and the right vertices, which is explicit in eq.\eqref{LHS} \footnote{This could be physically justified since the two vertices in a tree-level exchange diagram should be indistinguishable and equivalent to the process.}. Under $(\eta_{1},\, k_{L}) \leftrightarrow (\eta_{2}, k_{R})$ eq.\eqref{RHSa} becomes  
\begin{equation}
\begin{split} \label{RHS1}
      \text{RHS of eq.} \eqref{unitarity at order 2 with particle avg} \quad \quad \quad&
    \\
     =
    -\frac{\lambda_L \lambda_R}{2} \int d \eta_1 \int d\eta_2& \Bigg( \mathcal{K}^+_{\eta_0 \eta_1} \left(\lbrace \mathbf{k}_L \rbrace \right) \mathcal{K}^+_{\eta_0 \eta_2} \left(\lbrace \mathbf{k}_R \rbrace \right) + \mathcal{K}^+_{\eta_0 \eta_1} \left(\lbrace \mathbf{k}_R \rbrace \right) \mathcal{K}^+_{\eta_0 \eta_2} \left(\lbrace \mathbf{k}_L \rbrace \right) \Bigg) \times 
     \\
     &  \Bigg( G^{+-}_{p}(\eta_1, \eta_2) + G^{-+}_{p}(\eta_1, \eta_2) \Bigg)  \, ,
     \\
\end{split}
\end{equation}
which upon using the hermitian analyticity eq.\eqref{HA for Corr 1} can be further manipulated as 
\begin{equation}
\begin{split} \label{RHS}
      \text{RHS of eq.} \eqref{unitarity at order 2 with particle avg} \quad \quad \quad&
     \\
     =
   -\frac{\lambda_L \lambda_R}{2} \int d \eta_1 \int d\eta_2& \Bigg( \mathcal{K}^+_{\eta_0 \eta_1} \left(\lbrace \mathbf{k}_L \rbrace \right) \mathcal{K}^+_{\eta_0 \eta_2} \left(\lbrace \mathbf{k}_R \rbrace \right) + \mathcal{K}^+_{\eta_0 \eta_1} \left(\lbrace \mathbf{k}_R \rbrace \right) \mathcal{K}^+_{\eta_0 \eta_2} \left(\lbrace \mathbf{k}_L \rbrace \right) \Bigg) \times 
     \\
    &  \Bigg( G^{+-}_{p}(\eta_1, \eta_2) - G^{+-}_{-p}(\eta_1, \eta_2) \Bigg) \, .
\end{split}
\end{equation}
 The above identity holds for any half-integer $\nu$ (which includes particularly the conformally coupled and massless scalars). However, for generic other values of $\nu$ we refer to the discussion in Appendix-\ref{genericmassscalar}.

By comparing the equations eq.\eqref{LHS} and eq.\eqref{RHS}, we obtain the following equality
\begin{equation}\label{propid1}
\begin{split}
    G^{++}_{p}(\eta_1, \eta_2) - G^{++}_{-p}(\eta_1, \eta_2) = G^{+-}_{p}(\eta_1, \eta_2) - G^{+-}_{-p}(\eta_1, \eta_2) \, .
\end{split}
\end{equation}
We could have compared eq.\eqref{LHS} and eq.\eqref{RHS}, both at the intermediate steps before using eq.\eqref{HA for Corr 1}, and obtain the following identity \footnote{Actually, this relation is known in the Schwinger-Keldysh literature, see eq.(2.38) in Kamenev's book. We thank Chandramouli Chowdhury for pointing this to us.}
\begin{equation}\label{propid1a}
\begin{split}
    G^{++}_{p}(\eta_1, \eta_2) + G^{--}_{p}(\eta_1, \eta_2) = G^{+-}_{p}(\eta_1, \eta_2) + G^{-+}_{p}(\eta_1, \eta_2) \, ,
\end{split}
\end{equation}
which upon using eq.\eqref{HA for Corr 1} and eq.\eqref{in in bulk to bulk prop} can be expressed solely in terms of the bulk-to-boundary propagators 
\begin{equation}\label{propid1 RHS most}
\begin{split}
    G^{++}_{p}(\eta_1, \eta_2) - G^{++}_{-p}(\eta_1, \eta_2) =& G^{+-}_{p}(\eta_1, \eta_2) - G^{+-}_{-p}(\eta_1, \eta_2) \\
    = &\frac{K_p^-(\eta_0, \eta_1) K_p^+(\eta_0, \eta_2) + K_p^+(\eta_0, \eta_1) K_p^-(\eta_0, \eta_2)}{P_p(\eta_0)}  \, ,
\end{split}
\end{equation}

Alternatively, we could have followed the same steps as mentioned above, but used the condition of Hermitian analyticity while going from the second line to the third line in eq.\eqref{LHS} and eq.\eqref{RHS} on the other possible terms, and, thereby, obtain a different set of identities as follows 
\begin{equation}\label{propid2}
\begin{split}
    G^{--}_{p}(\eta_1, \eta_2) - G^{--}_{-p}(\eta_1, \eta_2) &= G^{-+}_{p}(\eta_1, \eta_2) - G^{-+}_{-p}(\eta_1, \eta_2) \\ 
   & =\frac{K_p^-(\eta_0, \eta_1) K_p^+(\eta_0, \eta_2) + K_p^+(\eta_0, \eta_1) K_p^-(\eta_0, \eta_2)}{P_p(\eta_0)} \, .
 \end{split}
\end{equation}

Therefore, to summarise, using Hermitian analyticity written in terms of the bulk-to-boundary propagators and unitarity, we have derived the following identities for the bulk-to-bulk propagators
\begin{equation} \label{discG}
    G^{\pm \pm}_{p}(\eta_1, \eta_2) - G^{\pm \pm}_{-p}(\eta_1, \eta_2) = \frac{K_p^-(\eta_0, \eta_1) K_p^+(\eta_0, \eta_2) + K_p^+(\eta_0, \eta_1) K_p^-(\eta_0, \eta_2)}{P_p(\eta_0)}\, .
\end{equation}
We will frequently use these identities in the derivation of the single-cut rule in the next section for the in-in cosmological correlator. It is instructive to note that on the LHS of eq.\eqref{discG} above, the operation of subtracting the quantity from itself after flipping the sign of the magnitude of the exchange momentum $p$, is consistent with the definition of discontinuity given in eq.\eqref{defdisc}. 

A similar strategy and hence an equivalent relation as in eq.\eqref{propid1 RHS most} for the in-in bulk-to-bulk propagators has also been obtained in \cite{MarinMacedo:2025jco} focussing on conformally coupled and massless scalar cases. However, in Appendix \ref{genericmassscalar} we have obtained all these propagator identities, see eq.\eqref{discGgenericnu}, which are  valid for cases beyond conformally coupled and massless scalars (i.e., for generic $\nu$, as defined in eq.\eqref{nu def}, not restricted to half-integer values). Using them we have also argued how one can derive a discontinuity relation for such cases, as in eq.\eqref{discb2tgeneric} which is valid for generic $\mathcal{R}$ defined in eq.\eqref{defR}. For conformally coupled and massless scalars, $\mathcal{R} = -1$, and using that we can recover eq.\eqref{propid1 RHS most}. This analysis in Appendix \ref{genericmassscalar} based on the analytic properties of Hankel functions was not there in \cite{MarinMacedo:2025jco}. 


\section{Derivation of the single-cut discontinuity relations at tree-level} \label{confscalartreelevel}

In this section, we discuss the discontinuities of cosmological correlators. A practical motivation for concentrating on discontinuities is that they are typically simpler to evaluate than the full correlator. To proceed, we express the correlator using the in-in formalism, and then we use the single-cut relation to demonstrate that the discontinuity of a higher-point correlator gets decomposed into a sum of the product of two lower-point objects, namely the correlator $\mathcal{B}$ and the auxiliary counterparts $\widetilde{\mathcal{B}}$. Here we will primarily focus on the conformally coupled scalar with polynomial interaction and will explicitly derive the discontinuity relation for $2$-site and $r$-site correlators.

We consider a theory of scalar fields with the following action
\begin{equation}\label{HInt}
    S= \int d^3\vec{x} \int_{-\infty}^{\eta_0} d\eta \left(- \frac{1}{2} \partial_\mu \phi \partial^\mu \phi -\frac{m^2 \phi^2}{2} +\frac{\lambda}{n!} \phi^n \right)
\end{equation}
\subsection{single-cut rule for $2$-site correlator}\label{2-site correlator}

A $2$-site correlator has $2$ vertices, and following the in-in Feynman rules, there are four possible ways to label the vertices as $``+"$ or $``-"$, and eventually we sum up the contributions due to all possible labelling to calculate the full correlator. So the $2$-site correlator, $\mathcal{B}^{(2)}$ can be written as follows
\begin{equation}
\label{inin2site}
\begin{split}
    & \mathcal{B}^{(2)}\left( \lbrace \mathbf{k}_L, \mathbf{k}_R \rbrace ;p \right) \\ =  & (i\lambda)^2 \int^{(2)}  d\eta~\bigg[ \mathcal{K}_{\eta_0 \eta_1}^+(\lbrace \mathbf{k}_L \rbrace) G^{++}_{p}(\eta_1,\eta_2) \mathcal{K}_{\eta_0 \eta_2}^+(\lbrace \mathbf{k}_R \rbrace) - \mathcal{K}_{\eta_0 \eta_1}^+(\lbrace \mathbf{k}_L \rbrace) G^{+-}_{p}(\eta_1,\eta_2) \mathcal{K}_{\eta_0 \eta_2}^-(\lbrace \mathbf{k}_R \rbrace)
    \\
    & \quad \quad  - \mathcal{K}_{\eta_0 \eta_1}^-(\lbrace \mathbf{k}_L \rbrace) G^{-+}_{p}(\eta_1,\eta_2) \mathcal{K}_{\eta_0 \eta_2}^+(\lbrace \mathbf{k}_R \rbrace) + \mathcal{K}_{\eta_0 \eta_1}^-(\lbrace \mathbf{k}_L \rbrace) G^{--}_{p}(\eta_1,\eta_2) \mathcal{K}_{\eta_0 \eta_2}^-(\lbrace \mathbf{k}_R \rbrace) \bigg]\,,
\end{split}
\end{equation}
where $\mathcal{K}_{\eta_0 \eta_1}^\pm$ are defined in eq.\eqref{defmathcalK}.  
In the above expression, the first term on the RHS corresponds to the case where both vertices are of the $``+"$ type, leading to the Feynman propagator in the bulk. Likewise, the last term comes from choosing both vertices of the $``-"$ type. On the other hand, the second and third terms are due to cases where one of the vertices is of the $``+"$ type, and the other one is of $``-"$ type.

We now apply the single-cut discontinuity operation, denoted by `${\text {Disc}}_{p}$' and defined in eq.\eqref{defdisc}, on eq.\eqref{inin2site} of a $2$-site correlator. Then, using the propagator identities written in eq.\eqref{discG}, we can rewrite the RHS in the following way
\begin{equation}\label{B2 disc step1}
\begin{split}
   & {\text {Disc}}_{p} \mathcal{B}^{(2)}( \lbrace \mathbf{k}_L, \mathbf{k}_R \rbrace ;p) \equiv \mathcal{B}^{(2)}( \lbrace \mathbf{k}_L, \mathbf{k}_R \rbrace ;p)-\mathcal{B}^{(2)}( \lbrace \mathbf{k}_L, \mathbf{k}_R \rbrace ;-p) 
   \\
   & = (i\lambda)^2 \int^{(2)} d\eta~ \bigg( \mathcal{K}_{\eta_0 \eta_1}^+(\lbrace \mathbf{k}_L \rbrace) - \mathcal{K}_{\eta_0 \eta_1}^-(\lbrace \mathbf{k}_L \rbrace) \bigg) \times \\
   & \frac{\bigg(K_{p}^-(\eta_0, \eta_1) K_{p}^+(\eta_0, \eta_2) + K_{p}^+(\eta_0, \eta_1) K_{p}^-(\eta_0, \eta_2) \bigg)}{{P_{p}(\eta_0)}} \times 
    \bigg( \mathcal{K}_{\eta_0 \eta_2}^+(\lbrace \mathbf{k}_R \rbrace) - \mathcal{K}_{\eta_0 \eta_2}^-(\lbrace \mathbf{k}_R \rbrace) \bigg)\,,
\end{split}
\end{equation}
where we have used the following shorthand notation 
\begin{equation}\label{n time integrals}
\int^{(n)} d\eta = \prod_{j=1}^n \int_{-\infty}^{\eta_0}\frac{d\eta_j}{H^4\eta_j^4} \, , 
\end{equation}
which we will repeatedly use in the rest of the paper. 

It is obvious from eq.\eqref{B2 disc step1} that the expression on the RHS is not fully factorized with respect to the time integrals, since the term in the second line of the above equation still mixes $\eta_1$ and $\eta_2$. Therefore, the way out is to write the second line of eq.\eqref{B2 disc step1} in such a way that factorizes the $\eta_1$ and $\eta_2$ dependence. To facilitate this, we use the following relation 
\begin{equation}\label{bulk to bulk ab+cd type}
\begin{split}
    & K_{p}^-(\eta_0, \eta_1) K_{p}^+(\eta_0, \eta_2) + K_{p}^+(\eta_0, \eta_1) K_{p}^-(\eta_0, \eta_2) \\
    & =  \frac{1}{2} \Big( K_{p}^-(\eta_0, \eta_1)+ K_{p}^+(\eta_0, \eta_1) \Big)\Big( K_{p}^+(\eta_0, \eta_2) + K_{p}^-(\eta_0, \eta_2) \Big)
    \\
    & + \frac{1}{2} \Big( K_{p}^-(\eta_0, \eta_1)- K_{p}^+(\eta_0, \eta_1) \Big)\Big( K_{p}^+(\eta_0, \eta_2) - K_{p}^-(\eta_0, \eta_2) \Big) \, .
\end{split}
\end{equation}
Now substituting eq.\eqref{bulk to bulk ab+cd type} in eq.\eqref{B2 disc step1} we obtain the following expression
\begin{equation}\label{B2 disc step2}
\begin{split}
      {\text {Disc}}_{p} \mathcal{B}^{(2)}( \lbrace \mathbf{k}_L, \mathbf{k}_R \rbrace ;p)  =  {\text {Disc}}_{p} \mathcal{B}^{(2)}( \lbrace \mathbf{k}_L, \mathbf{k}_R \rbrace ;p)\Bigg{\vert}_{\text{term-1}}+  \,  {\text {Disc}}_{p} \mathcal{B}^{(2)}( \lbrace \mathbf{k}_L, \mathbf{k}_R \rbrace ;p)\Bigg{\vert}_{\text{term-2}} \, ,
\end{split}
\end{equation}
such that 
\begin{equation}\label{definition term1 from B2 disc step1}
\begin{split}
     &{\text {Disc}}_{p} \mathcal{B}^{(2)}( \lbrace \mathbf{k}_L, \mathbf{k}_R \rbrace ;p)\Bigg{\vert}_{\text{term-1}} \\
     & = \frac{(i\lambda)^2}{2 P_p(\eta_0)} \int^{(2)} d\eta~ \bigg( \mathcal{K}_{\eta_0 \eta_1}^+(\lbrace \mathbf{k}_L \rbrace) - \mathcal{K}_{\eta_0 \eta_1}^-(\lbrace \mathbf{k}_L \rbrace) \bigg) \Big( K_{p}^-(\eta_0, \eta_1)+ K_{p}^+(\eta_0, \eta_1) \Big) 
     \\
     & ~~~~~~~~~~~~~~~~~~~~~~ \Big( K_{p}^+(\eta_0, \eta_2) + K_{p}^-(\eta_0, \eta_2) \Big) \bigg( \mathcal{K}_{\eta_0 \eta_2}^+(\lbrace \mathbf{k}_R \rbrace) - \mathcal{K}_{\eta_0 \eta_2}^-(\lbrace \mathbf{k}_R \rbrace) \bigg) \, ,
\end{split}
\end{equation}
and 
\begin{equation}\label{definition term2 from B2 disc step1}
\begin{split}
     & {\text {Disc}}_{p} \mathcal{B}^{(2)}( \lbrace \mathbf{k}_L, \mathbf{k}_R \rbrace ;p)\Bigg{\vert}_{\text{term-2}} \\ 
     & = -\frac{(i\lambda)^2}{2 P_p(\eta_0)} \int^{(2)} d\eta~ \bigg( \mathcal{K}_{\eta_0 \eta_1}^+(\lbrace \mathbf{k}_L \rbrace) - \mathcal{K}_{\eta_0 \eta_1}^-(\lbrace \mathbf{k}_L \rbrace) \bigg) \Big( K_{p}^-(\eta_0, \eta_1)- K_{p}^+(\eta_0, \eta_1) \Big) 
     \\
     & ~~~~~~~~~~~~~~~~~~~~~~ \Big( K_{p}^-(\eta_0, \eta_2) - K_{p}^+(\eta_0, \eta_2) \Big) \bigg( \mathcal{K}_{\eta_0 \eta_2}^+(\lbrace \mathbf{k}_R \rbrace) - \mathcal{K}_{\eta_0 \eta_2}^-(\lbrace \mathbf{k}_R \rbrace) \bigg) \, .
\end{split}
\end{equation}
From eq.\eqref{B2 disc step2},\eqref{definition term1 from B2 disc step1} and 
\eqref{definition term2 from B2 disc step1} we notice that the time integrals over $\eta_1$ and $\eta_2$ are now independent of each other. Therefore, this will be useful going ahead to obtain the discontinuity relation for the $2$-site correlator written in terms of two lower-point contact correlators, respectively, at the left and right sites or vertices. 

Next, we note that in both the terms on the RHS of eq.\eqref{B2 disc step2} the double-integration can be split into products of two integrations over the variables $\eta_1$ and $\eta_2$. These integrations can now be performed independently, and the final result, after a few manipulations, can be written as a product of two $1$-site objects, which are essentially contact objects. 

Let us see this factorization into lower-point discontinuities in some more detail. The part in ${\text {Disc}}_{p} \mathcal{B}^{(2)}( \lbrace \mathbf{k}_L, \mathbf{k}_R \rbrace ;p) \vert_{\text{term-1}}$ in eq.\eqref{definition term1 from B2 disc step1} that involves only integration over $\eta_1$, can be manipulated as follows
\begin{equation}\label{2site Disc eta1 part}
\begin{split}
    & \text{$\eta_1$ integration in ${\text {Disc}}_{p} \mathcal{B}^{(2)}( \lbrace \mathbf{k}_L, \mathbf{k}_R \rbrace ; p ) \vert_{\text{term}-1}$ in eq.\eqref{definition term1 from B2 disc step1}} \\
    & = (i \lambda) \int d\eta_1 \bigg( \mathcal{K}_{\eta_0 \eta_1}^+(\lbrace \mathbf{k}_L \rbrace) - \mathcal{K}_{\eta_0 \eta_1}^-(\lbrace \mathbf{k}_L \rbrace) \bigg) \Big( K_{p}^-(\eta_0, \eta_1)+ K_{p}^+(\eta_0, \eta_1) \Big) 
    \\
    & =  (i \lambda) \int d\eta_1 \bigg( \mathcal{K}_{\eta_0 \eta_1}^+(\lbrace \mathbf{k}_L \rbrace) K_{p}^+ (\eta_0,\eta_1) -  \mathcal{K}_{\eta_0 \eta_1}^-(\lbrace \mathbf{k}_L \rbrace) K_{p}^- (\eta_0,\eta_1) 
    \\
    & ~~~~~~~~~~~~~~~~~~~~ -  \mathcal{K}_{\eta_0 \eta_1}^+(\lbrace \mathbf{k}_L \rbrace) K_{-p}^+ (\eta_0,\eta_1) + \mathcal{K}_{\eta_0 \eta_1}^-(\lbrace \mathbf{k}_L \rbrace) K_{-p}^- (\eta_0,\eta_1) \bigg)
    \\
    & = (i \lambda) \int d\eta_1 \bigg( \mathcal{K}_{\eta_0 \eta_1}^+(\lbrace \mathbf{k}_L\rbrace,p ) -  \mathcal{K}_{\eta_0 \eta_1}^-(\lbrace \mathbf{k}_L \rbrace,p ) -  \mathcal{K}_{\eta_0 \eta_1}^+(\lbrace \mathbf{k}_L\rbrace, -p )  + \mathcal{K}_{\eta_0 \eta_1}^-(\lbrace \mathbf{k}_L\rbrace , -p) \bigg)
    \\
    & = \mathcal{B}^{(1)}( \lbrace \mathbf{k}_{L} \rbrace ,p) - \mathcal{B}^{(1)}( \lbrace \mathbf{k}_{L} \rbrace ,-p)  ~~~~ \equiv ~~~~\text{Disc}_p \mathcal{B}^{(1)}(\lbrace \mathbf{k}_{L} \rbrace,p) \, .
\end{split}
\end{equation}
where we have introduced the following shorthand notation 
\begin{equation}\label{defmathcalKLKp}
	\mathcal{K}_{\eta_0 \eta_1}^{\pm} (\lbrace \mathbf{k}_L\rbrace, \pm p )  = \mathcal{K}_{\eta_0 \eta_1}^{\pm} (\lbrace \mathbf{k}_L \rbrace) K_{\pm p}^{\pm} (\eta_0,\eta_1)  \, .
\end{equation}
In deriving the third equality in eq.\eqref{2site Disc eta1 part}  we have used the fact that $K_p^+(\eta_0, \eta) = - K_{-p}^-(\eta_0, \eta)$ from eq.\eqref{HA for Corr}. Also, $\mathcal{B}^{(1)}( \lbrace \mathbf{k}_{L} \rbrace ,p)$ is the contact correlator or a $1$-site correlator, and the last line directly follows from the definition of the discontinuity for the contact correlator.

Now, from eq.\eqref{definition term1 from B2 disc step1} it is clear that the other part in ${\text {Disc}}_{p} \mathcal{B}^{(2)}( \lbrace \mathbf{k}_L, \mathbf{k}_R \rbrace ;p) \vert_{\text{term-1}}$ that only includes integration over $\eta_2$ is nothing but $\eta_1 \leftrightarrow \eta_2, \lbrace \mathbf{k}_L \rbrace \leftrightarrow \lbrace \mathbf{k}_R \rbrace$ of the above term we have computed in eq.\eqref{2site Disc eta1 part}. Hence, we can directly write down the result of this $\eta_2$ integration in eq.\eqref{definition term1 from B2 disc step1} as follows 
\begin{equation}\label{right most Disc 2site}
\begin{split}
    & \text{$\eta_2$ integration in ${\text {Disc}}_{p} \mathcal{B}^{(2)}( \lbrace \mathbf{k}_L, \mathbf{k}_R \rbrace ; p ) \vert_{\text{term}-1}$ in eq.\eqref{definition term1 from B2 disc step1}} \\
    & = (i \lambda) \int d\eta_2  \Big( K_{p}^+(\eta_0, \eta_2)+ K_{p}^-(\eta_0, \eta_2) \Big)  \bigg( \mathcal{K}_{\eta_0 \eta_2}^+(\lbrace \mathbf{k}_R \rbrace) - \mathcal{K}_{\eta_0 \eta_2}^-(\lbrace \mathbf{k}_R \rbrace) \bigg) 
    \\
    & = \mathcal{B}^{(1)}( \lbrace \mathbf{k}_{R} \rbrace ,p) - \mathcal{B}^{(1)}( \lbrace \mathbf{k}_{R} \rbrace ,-p)  ~~~~ \equiv ~~~~\text{Disc}_p \mathcal{B}^{(1)}( \lbrace \mathbf{k}_{R} \rbrace,p) \, .
\end{split}
\end{equation} 
Therefore, eq.\eqref{definition term1 from B2 disc step1} can be written as follows
\begin{equation} \label{FactorDisc term1}
    {\text {Disc}}_{p} \mathcal{B}^{(2)}(\lbrace \mathbf{k}_L, \mathbf{k}_R \rbrace ;p)\Bigg{\vert}_{\text{term-1}} = \frac{1}{2 P_p(\eta_0)} \text{Disc}_p \mathcal{B}^{(1)}( \lbrace \mathbf{k}_{L} \rbrace,p) \times  \text{Disc}_p \mathcal{B}^{(1)}( \lbrace \mathbf{k}_{R} \rbrace,p) \, .
\end{equation}

We now turn our focus to the last term on the RHS of eq.\eqref{B2 disc step2}, i.e., ${\text {Disc}}_{p} \mathcal{B}^{(2)}( \lbrace \mathbf{k}_L, \mathbf{k}_R \rbrace ;p)\vert_{\text{term-2}}$ with the expression given in eq.\eqref{definition term2 from B2 disc step1}. Although the manipulations for this second term will be the same as for the first term, which we computed above, the final result for this term reveals rather non-trivial and intriguing subtleties. To see this, we proceed to compute the $\eta_1$ integration part of term-$2$ as written in eq.\eqref{definition term2 from B2 disc step1} in the following way
\begin{equation} \label{2site Disc 1 eta1 part 1}
\begin{split}
     & \text{$\eta_1$ integration in ${\text {Disc}}_{p} \mathcal{B}^{(2)}( \lbrace \mathbf{k}_L, \mathbf{k}_R \rbrace ; p ) \vert_{\text{term}-2}$ in eq.\eqref{definition term2 from B2 disc step1}} \\
     & = (i \lambda) \int d\eta_1 \bigg( \mathcal{K}_{\eta_0 \eta_1}^+(\lbrace \mathbf{k}_L \rbrace) - \mathcal{K}_{\eta_0 \eta_1}^-(\lbrace \mathbf{k}_L \rbrace) \bigg) \Big( K_{p}^-(\eta_0, \eta_1) - K_{p}^+(\eta_0, \eta_1) \Big) 
    \\
    & =  (i \lambda) \int d \eta_1 \bigg( - \mathcal{K}_{\eta_0 \eta_1}^+(\lbrace \mathbf{k}_L \rbrace) K_{p}^+ (\eta_0,\eta_1) -  \mathcal{K}_{\eta_0 \eta_1}^-(\lbrace \mathbf{k}_L \rbrace) K_{p}^- (\eta_0,\eta_1) 
    \\
    & ~~~~~~~~~~~~~~~~~~~~~~~~ -  \mathcal{K}_{\eta_0 \eta_1}^+(\lbrace \mathbf{k}_L \rbrace) K_{-p}^+ (\eta_0,\eta_1) - \mathcal{K}_{\eta_0 \eta_1}^-(\lbrace \mathbf{k}_L \rbrace) K_{-p}^- (\eta_0,\eta_1) \bigg)
    \\
    & = -(i \lambda) \int d\eta_1 \bigg( \mathcal{K}_{\eta_0 \eta_1}^+(\lbrace \mathbf{k}_L \rbrace,p ) + \mathcal{K}_{\eta_0 \eta_1}^-(\lbrace \mathbf{k}_L\rbrace,p ) +  \mathcal{K}_{\eta_0 \eta_1}^+(\lbrace \mathbf{k}_L\rbrace, -p )  + \mathcal{K}_{\eta_0 \eta_1}^-(\lbrace \mathbf{k}_L\rbrace , -p) \bigg) \, .
\end{split}
\end{equation}
Surprisingly, we note that, unlike what we have seen in eq.\eqref{2site Disc eta1 part} for the first term, the combination of the bulk-to-boundary propagators in the final expression on the RHS above is not something that produces a lower-point correlator (a contact correlator as expected in this case). Nevertheless, to tackle them, we first rewrite them in terms of the bulk-to-boundary propagators corresponding to wave-function coefficients using the following relations, which can be derived from eq.\eqref{bulkboundary WF and inin} and eq.\eqref{defmathcalK}, 
\begin{equation} \label{relation mathcalK}
\begin{split}
    & \mathcal{K}_{\eta_0 \eta_1}^+(\lbrace \mathbf{k}_L\rbrace,p ) = \mathcal{P}_{\mathbf{k}_L}(\eta_0) P_{p}(\eta_0) \mathcal{K}_{\eta_0 \eta_1}^{\psi}(\lbrace \mathbf{k}_L \rbrace,p ) \, ,
    \\
    & \mathcal{K}_{\eta_0 \eta_1}^-(\lbrace \mathbf{k}_L\rbrace,p ) = \mathcal{P}_{\mathbf{k}_L}(\eta_0) P_{p}(\eta_0) \mathcal{K}_{\eta_0 \eta_1}^{\psi}(\lbrace \mathbf{k}_L \rbrace,p )^* \, ,
\end{split}
\end{equation}
where we have defined $ \mathcal{P}_{\mathbf{k}_L}(\eta_0) $ as 
\begin{equation}\label{PkL PkR}
\begin{split}
   & \mathcal{P}^{-1}_{\mathbf{k}_L}(\eta_0) = \prod_{i=1}^{n-1} \left( 2 \mathbb{R}e \psi_2(k_i) \right) \, , \quad \mathcal{P}^{-1}_{\mathbf{k}_R}(\eta_0) = \prod_{i=n}^{2n-2} \left( 2 \mathbb{R}e \psi_2(k_i) \right) \, ,
\end{split}
\end{equation}
at the left and right vertices denoted by $\mathbf{k}_L$ and $\mathbf{k}_R$ respectively in a $2$-site correlator. 

Then, using eq.\eqref{relation mathcalK} in the last expression on the RHS in eq.\eqref{2site Disc 1 eta1 part 1} we obtain the following expression for the $\eta_1$ integration part in eq.\eqref{definition term2 from B2 disc step1}
\begin{equation}\label{2site Disc 1 eta1 part 1a}
\begin{split}
    & \text{$\eta_1$ integration in ${\text {Disc}}_{p} \mathcal{B}^{(2)}( \lbrace \mathbf{k}_L, \mathbf{k}_R \rbrace ; p ) \vert_{\text{term}-2}$ in eq.\eqref{definition term2 from B2 disc step1}} \\
    & = (i \lambda) \int d\eta_1 \bigg( \mathcal{K}_{\eta_0 \eta_1}^+(\lbrace \mathbf{k}_L \rbrace) - \mathcal{K}_{\eta_0 \eta_1}^-(\lbrace \mathbf{k}_L \rbrace) \bigg) \Big( K_{p}^-(\eta_0, \eta_1) - K_{p}^+(\eta_0, \eta_1) \Big) \\
    & =  -(i \lambda) \int d\eta_1 \bigg( \frac{\mathcal{K}_{\eta_0 \eta_1}^{\psi}(\lbrace \mathbf{k}_L\rbrace,p ) +  \mathcal{K}_{\eta_0 \eta_1}^{\psi}(\lbrace \mathbf{k}_L\rbrace,p )^*}{\mathcal{P}^{-1}_{\mathbf{k}_L}(\eta_0) P^{-1}_{p}(\eta_0)} +  \frac{\mathcal{K}_{\eta_0 \eta_1}^{\psi}(\lbrace \mathbf{k}_L\rbrace, -p ) + \mathcal{K}_{\eta_0 \eta_1}^{\psi}(\lbrace \mathbf{k}_L\rbrace, -p )^*}{\mathcal{P}^{-1}_{\mathbf{k}_L}(\eta_0) P^{-1}_{-p}(\eta_0)} \bigg) 
    \\
    & = - \left( \frac{2i ~\mathbb{I}m ~\psi^{(1)} (\lbrace \mathbf{k}_L\rbrace,p )}{\mathcal{P}^{-1}_{\mathbf{k}_L}(\eta_0) P^{-1}_{p}(\eta_0)} + \frac{2i ~\mathbb{I}m ~\psi^{(1)} (\lbrace \mathbf{k}_L\rbrace, -p )}{\mathcal{P}^{-1}_{\mathbf{k}_L}(\eta_0) P^{-1}_{-p}(\eta_0)} \right)
    \\
    & = - \Big( \widetilde{\mathcal{B}}^{(1)}(\lbrace \mathbf{k}_{L} \rbrace ,p) + \widetilde{\mathcal{B}}^{(1)}( \lbrace \mathbf{k}_{L} \rbrace,-p) \Big) ~~~~ \equiv ~~~~ - ~ \widetilde{\text{Disc}}_{p} \,  \widetilde{\mathcal{B}}^{(1)}( \lbrace \mathbf{k}_{L} \rbrace,p) \, ,
\end{split}
\end{equation}
where we have used the operation $\widetilde{\text{Disc}}_{p}$ and the new  quantity $\widetilde{\mathcal{B}}^{(1)}( \lbrace \mathbf{k}_{L} \rbrace,p)$ as 
\begin{equation}\label{mathfrak B1}
   \widetilde{\mathcal{B}}^{(1)}( \lbrace \mathbf{k}_{L} \rbrace,p) = \frac{2i ~\mathbb{I}m ~\psi^{(1)} (\lbrace \mathbf{k}_L\rbrace,p )}{\mathcal{P}^{-1}_{\mathbf{k}_L}(\eta_0) P^{-1}_{p}(\eta_0)} \, .
\end{equation}
where $\psi^{(1)}$ in a contact wavefunction coefficient\footnote{Conventionally it is denoted by $\psi_n$ as an $n$-point wavefunction coefficient. However, throughout the paper we will use the superscript to denote the number of sites in the bulk.} and hence the superscript denotes the number of vertices in the bulk which is one at the contact level.\\
Alternatively, looking at eq.\eqref{2site Disc 1 eta1 part 1}, we can also define the auxiliary counterpart of the $1$-site correlator,
$\widetilde{\mathcal{B}}^{(1)}\left( \lbrace \mathbf{k}_L \rbrace , p \right)$,
in terms of the in-in propagators as follows:
\begin{equation}
\widetilde{\mathcal{B}}^{(1)}\left( \lbrace \mathbf{k}_L \rbrace ,p\right) = (i \lambda) \int d\eta_1 \bigg( \mathcal{K}_{\eta_0 \eta_1}^+(\lbrace \mathbf{k}_L \rbrace,p ) + \mathcal{K}_{\eta_0 \eta_1}^-(\lbrace \mathbf{k}_L\rbrace,p ) \bigg) \, .
\end{equation}

Similarly, the remaining part of the term-2, i.e., $\eta_2$ integration part in eq.\eqref{definition term2 from B2 disc step1}, can be obtained by $\eta_1 \leftrightarrow \eta_2, \lbrace \mathbf{k}_L \rbrace \leftrightarrow \lbrace \mathbf{k}_R \rbrace$ in eq.\eqref{2site Disc 1 eta1 part 1a} 
\begin{equation}\label{right most disc tilde 2site}
\begin{split}
    & \text{$\eta_2$ integration in ${\text {Disc}}_{p} \mathcal{B}^{(2)}( \lbrace \mathbf{k}_L, \mathbf{k}_R \rbrace ; p ) \vert_{\text{term}-2}$ in eq.\eqref{definition term2 from B2 disc step1}} \\
    & = (i \lambda) \int d\eta_2  \Big( K_{p}^-(\eta_0, \eta_2) - K_{p}^+(\eta_0, \eta_2) \Big) \bigg( \mathcal{K}_{\eta_0 \eta_2}^+(\lbrace \mathbf{k}_R \rbrace) - \mathcal{K}_{\eta_0 \eta_2}^-(\lbrace \mathbf{k}_R \rbrace) \bigg)
    \\
    & = - \Big( \widetilde{\mathcal{B}}^{(1)}( \lbrace \mathbf{k}_{R} \rbrace ,p) + \widetilde{\mathcal{B}}^{(1)}( \lbrace \mathbf{k}_{R} \rbrace ,-p) \Big) ~~~~ \equiv ~~~~ - ~ \widetilde{\text{Disc}}_{p} \widetilde{\mathcal{B}}^{(1)}( \lbrace \mathbf{k}_{R} \rbrace ,p) \, .
\end{split}
\end{equation}
Therefore eq.\eqref{definition term2 from B2 disc step1} can be written by multiplying eq.\eqref{2site Disc 1 eta1 part 1a} and eq.\eqref{right most disc tilde 2site} as follows
\begin{equation} \label{FactorDisc term2}
    {\text {Disc}}_{p} \mathcal{B}^{(2)}( \lbrace \mathbf{k}_L, \mathbf{k}_R \rbrace ;p)\Bigg{\vert}_{\text{term-2}} = - \frac{1}{2 P_p(\eta_0)} \widetilde{\text{Disc}}_{p} \widetilde{B}^{(1)}( \lbrace \mathbf{k}_{L} \rbrace ,p) \times \widetilde{\text{Disc}}_{p} \widetilde{B}^{(1)}( \lbrace \mathbf{k}_{R} \rbrace,p)
\end{equation}
Finally, combining eq.\eqref{FactorDisc term1} and eq.\eqref{FactorDisc term2}, we obtain the following expression for eq.\eqref{B2 disc step2}
\begin{equation}\label{2siteDiscFinal Result}
\begin{split}
     {\text {Disc}}_{p} \mathcal{B}^{(2)}( \lbrace \mathbf{k}_L, \mathbf{k}_R \rbrace ;p) = \frac{1}{2 P_p(\eta_0)} \Bigg(& \text{Disc}_p \mathcal{B}^{(1)}( \lbrace \mathbf{k}_{L} \rbrace ,p)  \times \text{Disc}_p \mathcal{B}^{(1)}(\lbrace \mathbf{k}_{R} \rbrace ,p) 
     \\
     & - \widetilde{\text{Disc}}_{p} \widetilde{\mathcal{B}}^{(1)}( \lbrace \mathbf{k}_{L} \rbrace ,p) \times \widetilde{\text{Disc}}_{p} \widetilde{\mathcal{B}}^{(1)}( \lbrace \mathbf{k}_{R} \rbrace ,p) \Bigg) \, .
\end{split}
\end{equation}
This is one of the main results in our paper demonstrating the single-cut discontinuity rule for a $2$-site correlator involving conformally coupled and massless scalar polynomial interaction. 
\begin{figure}[h]
    \centering
   \includegraphics[width=0.95\textwidth]{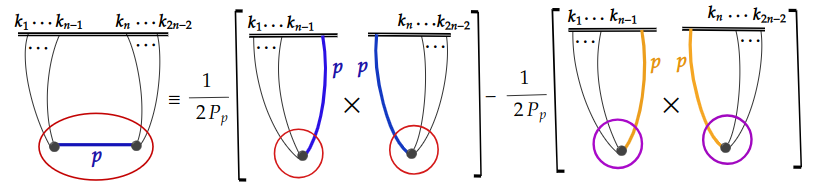}
    \caption{Diagrammatic representation of the single-cut rule of a $2$-site correlator as in eq.\eqref{2siteDiscFinal Result}. Red circle always denotes a correlator $\mathcal{B}$, whereas the purple circle denotes the auxiliary object $\widetilde{\mathcal{B}}$. The blue shaded line implies taking Disc with respect to the modulus of momentum corresponding to that line, and the yellow shaded line implies taking $\widetilde{\text{Disc}}$ with respect to the modulus of momentum corresponding to that line.}
    \label{single cut 2 site correlator}
\end{figure}

With the expression in eq.\eqref{2siteDiscFinal Result} at our disposal let us comment on certain non-trivial aspects of this single-cut discontinuity relation and highlight the novelties of our result. First and foremost, the second term on the RHS of eq.\eqref{2siteDiscFinal Result} involving $\widetilde{\text{Disc}}_{p} \widetilde{\mathcal{B}}^{(1)}( \lbrace \mathbf{k}_{L} \rbrace ,p)$ is very counter-intuitive and to the best of our knowledge has not been derived before. The first term on the RHS involves ${\mathcal{B}}^{(1)}$, which is the standard contact correlator. The origin of the product of two contact correlators obtained through cutting the only available internal line in a $2$-site correlator is understandable, since after cutting the internal line, if we take the cut line to the boundary, we readily obtain the lower-point contact correlator, see the first term on the RHS of Fig.-\ref{single cut 2 site correlator}. 
However, the second term on the RHS of eq.\eqref{2siteDiscFinal Result} is rather non-trivial because the object $\widetilde{\mathcal{B}}^{(1)}$ is something which does not resemble any contact correlator. We call them auxiliary objects. 

Secondly, we observe that for late–time IR-convergent, conformally coupled polynomial $\phi^n$ interactions with even $n$, only the first term on the RHS of eq.\eqref{2siteDiscFinal Result} contributes, whereas for odd $n$ the second term alone survives. 

There is also an interesting aspect to this in terms of leading vs sub-leading contributions to the cosmological correlators as the late-time limit is taken, i.e., $\eta_0 \to 0$. In the following, we substantiate this pattern by examining the leading late–time behaviour of the one-site correlator $\mathcal{B}^{(1)}(\lbrace \mathbf{k} \rbrace, p) $ and its auxiliary counterpart $\widetilde{\mathcal{B}}^{(1)}(\lbrace \mathbf{k} \rbrace, p)$. In particular, we show that the relative contribution of these two building blocks is controlled entirely by their leading $\eta_0$-scaling in the limit $\eta_0 \to 0$.
For IR-convergent conformally coupled interactions with $n \geq 4$ and even $n$, the leading late–time contribution of the contact correlator $\mathcal{B}^{(1)}$ scales as $\mathcal{O}(\eta_0^n)$. Consequently, the product of discontinuities of two such correlators behaves as
\begin{equation} \label{discBdiscB contribution1}
    \text{Disc}_p \mathcal{B}^{(1)}( \lbrace \mathbf{k}_{L} \rbrace ,p)  \times \text{Disc}_p \mathcal{B}^{(1)}(\lbrace \mathbf{k}_{R} \rbrace ,p)\Big{\vert}_{n=\text{even}}  \sim \mathcal{O}(\eta_0^{2n}) \, .
\end{equation}
By contrast, the auxiliary structure $\widetilde{\mathcal{B}}^{(1)}$  has a leading contribution at order $\mathcal{O}\left( \eta_0^{n+1} \right)$ implying
\begin{equation} \label{discBtdiscBt contribution1}
 \widetilde{\text{Disc}}_{p} \widetilde{\mathcal{B}}^{(1)}( \lbrace \mathbf{k}_{L} \rbrace ,p) \times \widetilde{\text{Disc}}_{p} \widetilde{\mathcal{B}}^{(1)}( \lbrace \mathbf{k}_{R} \rbrace ,p) \Big{\vert}_{n=\text{even}} \sim \mathcal{O}\left( \eta_0^{2n+2} \right)\, .
\end{equation}
Thus, in the late–time limit $\eta_0 \to 0$, the correlator contribution in eq.\eqref{discBdiscB contribution1} dominates for even $n$ over the auxiliary counterparts in eq.\eqref{discBtdiscBt contribution1}. 

Interestingly, for odd $n$, the situation reverses. In this case, the leading late–time behaviour of the contact correlator scales as $\mathcal{O}(\eta_0^{n+1})$, resulting
\begin{equation}\label{discBdiscB contribution2}
    \text{Disc}_p \mathcal{B}^{(1)}( \lbrace \mathbf{k}_{L} \rbrace ,p) \times \text{Disc}_p \mathcal{B}^{(1)}(\lbrace \mathbf{k}_{R} \rbrace ,p)\Big{\vert}_{n=\text{odd}}  \sim \mathcal{O}(\eta_0^{2n+2}) \, .
\end{equation}
Meanwhile, the auxiliary object now has a leading behaviour of order $\mathcal{O}\left( \eta_0^{n} \right)$ so that
\begin{equation}\label{discBtdiscBt contribution2}
 \widetilde{\text{Disc}}_{p} \widetilde{\mathcal{B}}^{(1)}( \lbrace \mathbf{k}_{L} \rbrace ,p) \times \widetilde{\text{Disc}}_{p} \widetilde{\mathcal{B}}^{(1)}( \lbrace \mathbf{k}_{R} \rbrace ,p) \Big{\vert}_{n=\text{odd}} \sim \mathcal{O}\left( \eta_0^{2n} \right) \, .
\end{equation}
Accordingly, in the $\eta_0 \to 0$ limit, the auxiliary terms in eq.\eqref{discBtdiscBt contribution2} provides the dominant contribution for odd $n$ over the contribution in eq.\eqref{discBdiscB contribution2}. 
Explicit computations for both even and odd conformally coupled polynomial interactions, presented in \S \ref{sec:Explicit checks}, make these scaling arguments fully manifest.

At this point it is worth highlighting that in \cite{MarinMacedo:2025jco} a relation equivalent to eq.\eqref{2siteDiscFinal Result} was obtained following a similar strategy. Consequently, the single-cut discontinuity relations at the tree-level have similarities. Additionally, they have also identified the contribution from the auxilary objects $\widetilde{\mathcal{B}}$'s, although in a different notation. However, we have been able to relate these auxiliary objects $\widetilde{\mathcal{B}}$'s to the imaginary parts of the contact wave-function coefficients, which is absent in \cite{MarinMacedo:2025jco}. Most importantly, the relative dominance between $\mathcal{B}$ and $\widetilde{\mathcal{B}}$ depending on whether we have $\phi^n$ interactions with $n = $ even or odd, which we discussed above, has been missed in \cite{MarinMacedo:2025jco}. As a consequence, \cite{MarinMacedo:2025jco} failed to capture the significant subtlety of the relative contribution of $\mathcal{B}$ and $\widetilde{\mathcal{B}}$'s to the discontinuity relation at late times when $\eta_0 \to 0^-$, which we discussed above in eq.\eqref{discBdiscB contribution1}, \eqref{discBtdiscBt contribution1}, \eqref{discBdiscB contribution2},  \eqref{discBtdiscBt contribution2}) for IR-convergent cases, i.e., $\phi^n$ with $n\ge 4$.

Finally, before we conclude this subsection, let us point out that the discontinuity relations for the $2$-site correlator derived in this subsection used technical machineries borrowing from the in-in formalism. However, one might wonder if we could have derived these discontinuity relations staying completely within the wave-function picture. This is very relevant since the discontinuity relations for the wave-function coefficients are already known in the literature and they are the building blocks for the cosmological correlators. In Appendix~\ref{2sitediscontinuityWVFN}, we show that one can indeed obtain the discontinuity relations, written in eq.\eqref{2site disc WVFN} from wave-function picture, same as in eq.\eqref{2siteDiscFinal Result}, for a $2$-site tree-level correlator focussing on conformally coupled scalar with polynomial interaction. Previously in \cite{Goodhew:2020hob} in eq.(2.69) using the cutting rules for the wave-function coefficients a similar cutting rule for the $2$-site correlator was obtained. However, the pieces involving the auxiliary objects $\widetilde{B}$ were missed, which is not a surprise since in \cite{Goodhew:2020hob} only such cases were considered where the imaginary parts of the wave-function coefficients were absent. On the contrary, in our discontinuity relations, eq.\eqref{2siteDiscFinal Result} or eq.\eqref{2site disc WVFN}, we capture the $\widetilde{B}$'s consistently, and hence they are valid in more general cases \footnote{Let us highlight that our derivation of the $2$-site discontinuity relation using the wave-function approach in Appendix \ref{2sitediscontinuityWVFN} is not discussed in \cite{MarinMacedo:2025jco}, as they only focussed on in-in method machineries.}.

\subsection{single-cut rule for r-site correlator}

Following the steps in the previous subsection, we now generalise the result obtained in the previous subsection and derive a single-cut rule for a tree-level $r$-site correlator considering $\phi^n$ interactions \footnote{The analysis in this section applies to cases with $r\ge2$.}. The tree-level correlator with $r$-sites will have $(r-1)$ internal lines where we will label the modulus of exchange momentum with $p_1, \, p_2, \cdots, p_{r-1}$. Our single-cut rule will be obtained by cutting the right-most internal line. Following the in-in formalism, such a correlator can be written as 
\begin{equation}\label{r site from in in}
\begin{split}
    & \mathcal{B}^{(r)}\left( \lbrace \mathbf{k}_L,...,\mathbf{k}_R \rbrace ; \lbrace p_1,..,p_{r-1} \rbrace \right) \\
    & = (i\lambda)^2 \int_{-\infty}^{\eta_0} \frac{d\eta_r}{(H \eta_r)^4} \int_{-\infty}^{\eta_0} \frac{d\eta_{r-1}}{(H \eta_{r-1})^4} \bigg[ \mathcal{V}_{(r-2)}^+ \mathcal{K}_{\eta_0 \eta_{r-1}}^+(\lbrace \mathbf{k}_{M_{r-2}} \rbrace) G^{++}_{p_{r-1}}(\eta_{r-1},\eta_r) \mathcal{K}_{\eta_0 \eta_r}^+(\lbrace \mathbf{k}_R \rbrace) 
    \\& \quad 
    -\mathcal{V}_{(r-2)}^+ \mathcal{K}_{\eta_0 \eta_{r-1}}^+(\lbrace \mathbf{k}_{M_{r-2}} \rbrace) G^{+-}_{p_{r-1}}(\eta_{r-1},\eta_r) \mathcal{K}_{\eta_0 \eta_r}^-(\lbrace \mathbf{k}_R \rbrace) 
    \\& \quad 
    -\mathcal{V}_{(r-2)}^- \mathcal{K}_{\eta_0 \eta_{r-1}}^-(\lbrace \mathbf{k}_{M_{r-2}} \rbrace) G^{-+}_{p_{r-1}}(\eta_{r-1},\eta_r) \mathcal{K}_{\eta_0 \eta_r}^+(\lbrace \mathbf{k}_ R\rbrace)
    \\& \quad 
    + \mathcal{V}_{(r-2)}^- \mathcal{K}_{\eta_0 \eta_{r-1}}^-(\lbrace \mathbf{k}_{M_{r-2}} \rbrace) G^{--}_{p_{r-1}}(\eta_{r-1},\eta_r) \mathcal{K}_{\eta_0 \eta_r}^-(\lbrace \mathbf{k}_R \rbrace) \bigg]\, ,
\end{split}
\end{equation}
where $\mathcal{K}_{\eta_0 \eta_{r}}^\pm$ has been defined in eq.\eqref{defmathcalK}, and the argument $\lbrace \mathbf{k}_{M_{j}} \rbrace$ in $\mathcal{K}_{\eta_0 \eta_{r}}^\pm(\lbrace \mathbf{k}_{M_{j}} \rbrace)$ denotes that the vertex at which the bulk-to-boundary propagators are anchored in the bulk, is in the middle - the $j$-th one - in a $r$-site correlator.  

Also, we have introduced a new object, namely $\mathcal{V}_{(r-2)}^\pm$ in eq.\eqref{r site from in in}, which captures the collection of other $(r-2)$ number of vertices from the $r$-site chain we are considering and hence is of order $\mathcal{O}\left((i \lambda)^{r-2}\right)$ in the coupling, which is consistent since $\mathcal{B}^{(r)}$ should be of order $\mathcal{O}\left((i \lambda)^{r}\right)$. Note that it consists of $(r-2)$ number of bulk-to-bulk propagators. In particular one can define $\mathcal{V}_{(r-2)}^{\pm}$ as follows 
\begin{equation} \label{defmathcalV}
\begin{split}
    & \mathcal{V}_{(r-2)}^{\sigma \in \pm} \left( \lbrace \eta_1, \cdots, \eta_{r-1} \rbrace \right) \\ 
    &=  (i \lambda)^{r-2} \int^{(r-2)} d\eta \sum_{\lbrace \sigma_i \rbrace_{i=1}^{r-2} \in \pm }  \Bigg[ \left( \prod_{j=1}^{r-2} \sigma_j \right) \left( \mathcal{K}^{\sigma_1}_{\eta_0 \eta_1}\left( \lbrace \mathbf{k}_L \rbrace \right) \prod_{j=2}^{r-2}  \mathcal{K}^{\sigma_{j}}_{\eta_0\eta_j} \left(\lbrace \mathbf{k}_{M_{j-1}} \rbrace \right)\right) \times 
    \\
    & \quad \quad \left( \prod_{j=1}^{r-3} G_{p_j}^{\sigma_j \sigma_{j+1}} ( \eta_j, \eta_{j+1} )\right) G_{p_{r-2}}^{\sigma_{r-2}\sigma} ( \eta_{r-2}, \eta_{r-1} ) \Bigg].
\end{split}
\end{equation}
Note that in the expressions above and in the following analysis, we have implicitly made a choice, that our single-cut discontinuity will be obtained by cutting the right-most internal line. The expressions such as in eq.\eqref{r site from in in} and eq.\eqref{defmathcalV} depend on that choice of convention. 

Now, we take the discontinuity of the $r$-site correlator with respect to the bulk-to-bulk energy $p_{r-1}$, and by using the propagator properties, we obtain the following 
\begin{equation}\label{step1 for Disc Brsite}
\begin{split}
    & \text{Disc}_{p_{r-1}} \mathcal{B}^{(r)}\left( \lbrace \mathbf{k}_L,...,\mathbf{k}_R \rbrace ; \lbrace p_1,..,p_{r-1} \rbrace \right)
    \\
    = & (i\lambda)^2 \int (d\eta_r)  \int (d\eta_{r-1}) \bigg( \mathcal{V}_{(r-2)}^+ \mathcal{K}_{\eta_0 \eta_{r-1}}^+(\lbrace \mathbf{k}_{M_{r-2}} \rbrace)  - \mathcal{V}_{(r-2)}^- \mathcal{K}_{\eta_0 \eta_{r-1}}^-(\lbrace \mathbf{k}_{M_{r-2}} \rbrace) \bigg)  \times 
   \\&
    ~~~~~~~~~~ \frac{\bigg(K_{p_{r-1}}^-(\eta_0, \eta_{r-1}) K_{p_{r-1}}^+(\eta_0, \eta_r) + K_{p_{r-1}}^+(\eta_0, \eta_{r-1}) K_{p_{r-1}}^-(\eta_0, \eta_r) \bigg)}{{P_{p_{r-1}}(\eta_0)}} \times
    \\&
    ~~~~~~~~~~ \bigg( \mathcal{K}_{\eta_0 \eta_r}^+(\lbrace \mathbf{k}_R \rbrace) - \mathcal{K}_{\eta_0 \eta_r}^-(\lbrace \mathbf{k}_R \rbrace) \bigg) \, , 
\end{split}
\end{equation}
where we have used a short-hand notation such that 
\begin{equation}\label{rth time integral}
\int (d\eta_r) = \int_{-\infty}^{\eta_0} \frac{d\eta_r}{(H \eta_r)^4} \, ,
\end{equation}
and we will be using it in the rest of this paper as well. 

We can write the term in the third line of eq.\eqref{step1 for Disc Brsite} in a similar way following eq.\eqref{bulk to bulk ab+cd type}
\begin{equation} \label{bulk to bulk ab+cd type 1}
\begin{split}
    & K_{p_{r-1}}^-(\eta_0, \eta_{r-1}) K_{p_{r-1}}^+(\eta_0, \eta_r) + K_{p_{r-1}}^+(\eta_0, \eta_{r-1}) K_{p_{r-1}}^-(\eta_0, \eta_r) 
    \\
    =&  \frac{1}{2} \Big( K_{p_{r-1}}^-(\eta_0, \eta_{r-1})+ K_{p_{r-1}}^+(\eta_0, \eta_{r-1}) \Big)\Big( K_{p_{r-1}}^+(\eta_0, \eta_r) + K_{p_{r-1}}^-(\eta_0, \eta_r) \Big)
    \\
    &  + \frac{1}{2} \Big( K_{p_{r-1}}^-(\eta_0, \eta_{r-1})- K_{p_{r-1}}^+(\eta_0, \eta_{r-1}) \Big)\Big( K_{p_{r-1}}^+(\eta_0, \eta_r) - K_{p_{r-1}}^-(\eta_0, \eta_r) \Big) \, ,
\end{split}
\end{equation}
which will help us dis-entangle the two time integrations over $\eta_r$ and $\eta_{r-1}$. 
Now substituting eq.\eqref{bulk to bulk ab+cd type 1} in eq.\eqref{step1 for Disc Brsite}, we therefore obtain the following 
\begin{equation}\label{Br disc step2}
\begin{split}
      {\text {Disc}}_{p_{r-1}} \mathcal{B}^{(r)}( \lbrace \mathbf{k}_L,..., \mathbf{k}_R \rbrace ; \lbrace p_1, ..., p_{r-1} \rbrace ) & = {\text {Disc}}_{p_{r-1}} \mathcal{B}^{(r)}( \lbrace \mathbf{k}_L,..., \mathbf{k}_R \rbrace ; \lbrace p_1, ..., p_{r-1} \rbrace )\Bigg{\vert}_{\text{term}-1}\\
     & + {\text {Disc}}_{p_{r-1}} \mathcal{B}^{(r)}( \lbrace \mathbf{k}_L,..., \mathbf{k}_R \rbrace ; \lbrace p_1, ..., p_{r-1} \rbrace )\Bigg{\vert}_{\text{term}-2} \, ,
\end{split}
\end{equation}
where the two terms on the RHS are given by 
\begin{equation} \label{discBrTerm1}
\begin{split}
    & {\text {Disc}}_{p_{r-1}} \mathcal{B}^{(r)}( \lbrace \mathbf{k}_L,..., \mathbf{k}_R \rbrace ; \lbrace p_1, ..., p_{r-1} \rbrace )\Bigg{\vert}_{\text{term}-1}
    \\
     & = \frac{(i\lambda)^2}{2 P_{p_{r-1}} (\eta_0)} \int (d\eta_r) \int (d\eta_{r-1}) \bigg( \mathcal{V}_{(r-2)}^+ \mathcal{K}_{\eta_0 \eta_{r-1}}^+(\lbrace \mathbf{k}_{M_{r-2}} \rbrace)  - \mathcal{V}_{(r-2)}^- \mathcal{K}_{\eta_0 \eta_{r-1}}^-(\lbrace \mathbf{k}_{M_{r-2}} \rbrace) \bigg) \times 
     \\
     & ~~~~~~~ \Big( K_{p_{r-1}}^-(\eta_0, \eta_{r-1})+ K_{p_{r-1}}^+(\eta_0, \eta_{r-1}) \Big) \times \Big( K_{p_{r-1}}^+(\eta_0, \eta_r) + K_{p_{r-1}}^-(\eta_0, \eta_r) \Big) \times \\
     & ~~~~~~~ \bigg( \mathcal{K}_{\eta_0 \eta_r}^+(\lbrace \mathbf{k}_R \rbrace) - \mathcal{K}_{\eta_0 \eta_r}^-(\lbrace \mathbf{k}_R \rbrace) \bigg) \, ,
\end{split}
\end{equation}
and
\begin{equation} \label{discBrTerm2}
\begin{split}
    & {\text {Disc}}_{p_{r-1}} \mathcal{B}^{(r)}( \lbrace \mathbf{k}_L,..., \mathbf{k}_R \rbrace ; \lbrace p_1, ..., p_{r-1} \rbrace )\Bigg{\vert}_{\text{term}-2}
    \\
    & = \frac{(i\lambda)^2}{2 P_{p_{r-1}}(\eta_0)} \int (d\eta_r) \int (d\eta_{r-1}) \bigg( \mathcal{V}_{(r-2)}^+ \mathcal{K}_{\eta_0 \eta_{r-1}}^+(\lbrace \mathbf{k}_{M_{r-2}} \rbrace)  - \mathcal{V}_{(r-2)}^- \mathcal{K}_{\eta_0 \eta_{r-1}}^-(\lbrace \mathbf{k}_{M_{r-2}} \rbrace) \bigg) \times 
     \\
     & ~~~~~~~ \Big( K_{p_{r-1}}^-(\eta_0, \eta_{r-1})- K_{p_{r-1}}^+(\eta_0, \eta_{r-1}) \Big) \times  \Big( K_{p_{r-1}}^+(\eta_0, \eta_r) - K_{p_{r-1}}^-(\eta_0, \eta_r) \Big) \times \\
     &  ~~~~~~~ \bigg( \mathcal{K}_{\eta_0 \eta_r}^+(\lbrace \mathbf{k}_R \rbrace) - \mathcal{K}_{\eta_0 \eta_r}^-(\lbrace \mathbf{k}_R \rbrace) \bigg) \, .
\end{split}
\end{equation}
Now that the integrations over $\eta_r$ and $\eta_{r-1}$ have become independent, we can easily perform the $\eta_r$ integral in both the above terms in eq.\eqref{discBrTerm1} and eq.\eqref{discBrTerm2}, in the similar way as shown in eq.\eqref{right most Disc 2site} and eq.\eqref{right most disc tilde 2site} respectively 
\begin{equation}\label{right most Disc r site}
\begin{split}
    & \text{$\eta_r$ integration in ${\text {Disc}}_{p_{r-1}} \mathcal{B}^{(r)}( \lbrace \mathbf{k}_L,..., \mathbf{k}_R \rbrace ; \lbrace p_1, ..., p_{r-1} \rbrace ) \vert_{\text{term}-1}$ in eq.\eqref{discBrTerm1}} \\
    & = (i \lambda) \int (d\eta_r)   \Big( K_{p_{r-1}}^+(\eta_0, \eta_r)+ K_{p_{r-1}}^-(\eta_0, \eta_r) \Big)  \bigg( \mathcal{K}_{\eta_0 \eta_r}^+(\lbrace \mathbf{k}_R \rbrace) - \mathcal{K}_{\eta_0 \eta_r}^-(\lbrace \mathbf{k}_R \rbrace) \bigg) 
    \\
    & = \mathcal{B}^{(1)}( \lbrace \mathbf{k}_{R} \rbrace ,p_{r-1}) - \mathcal{B}^{(1)}( \lbrace \mathbf{k}_{R} \rbrace ,-p_{r-1})  ~~~~ \equiv ~~~~\text{Disc}_{p_{r-1}} \mathcal{B}^{(1)}( \lbrace \mathbf{k}_{R} \rbrace, p_{r-1})\, ,
\end{split}
\end{equation}
and
\begin{equation}\label{right most disc tilde r site}
\begin{split}
    & \text{$\eta_r$ integration in ${\text {Disc}}_{p_{r-1}} \mathcal{B}^{(r)}( \lbrace \mathbf{k}_L,..., \mathbf{k}_R \rbrace ; \lbrace p_1, ..., p_{r-1} \rbrace ) \vert_{\text{term}-2}$ in eq.\eqref{discBrTerm2}} \\
    & =  - (i \lambda) \int (d\eta_r)   \Big( K_{p_{r-1}}^-(\eta_0, \eta_r) - K_{p_{r-1}}^+(\eta_0, \eta_r) \Big) \bigg( \mathcal{K}_{\eta_0 \eta_r}^+(\lbrace \mathbf{k}_R \rbrace) - \mathcal{K}_{\eta_0 \eta_r}^-(\lbrace \mathbf{k}_R \rbrace) \bigg)
    \\
    & =  \widetilde{\mathcal{B}}^{(1)}( \lbrace \mathbf{k}_{R} \rbrace ,p_{r-1}) + \widetilde{\mathcal{B}}^{(1)}( \lbrace \mathbf{k}_{R} \rbrace ,-p_{r-1})  ~~~~ \equiv ~~~~  ~ \widetilde{\text{Disc}}_{p_{r-1}} \widetilde{\mathcal{B}}^{(1)}( \lbrace \mathbf{k}_{R} \rbrace, p_{r-1}) \, .
\end{split}
\end{equation}
Next, we analyze the remaining parts in both term-1 and term-2. First, the remaining part of term-1 can be written as 
\begin{equation} \label{eta r-1 in term 1}
\begin{split}
    & \text{$\eta_{r-1}$ integration in ${\text {Disc}}_{p_{r-1}} \mathcal{B}^{(r)}( \lbrace \mathbf{k}_L,..., \mathbf{k}_R \rbrace ; \lbrace p_1, ..., p_{r-1} \rbrace )\Bigg{\vert}_{\text{term}-1}$} \\
    & =  (i \lambda) \int (d\eta_{r-1})~ \bigg( \mathcal{V}_{(r-2)}^+ \mathcal{K}_{\eta_0 \eta_{r-1}}^+(\lbrace \mathbf{k}_{M_{r-2}} \rbrace)  - \mathcal{V}_{(r-2)}^- \mathcal{K}_{\eta_0 \eta_{r-1}}^-(\lbrace \mathbf{k}_{M_{r-2}} \rbrace) \bigg) \times \\
    & \quad \quad \Big( K_{p_{r-1}}^-(\eta_0, \eta_{r-1})+ K_{p_{r-1}}^+(\eta_0, \eta_{r-1}) \Big)
    \\
    & = ~ \mathcal{B}^{(r-1)}\left(  \lbrace \mathbf{k}_L,...,\mathbf{k}_{M_{r-2}} \rbrace, p_{r-1} ; \lbrace p_1,..,p_{r-2} \rbrace \right) - \mathcal{B}^{(r-1)}\left(  \lbrace \mathbf{k}_L,...,\mathbf{k}_{M_{r-2}} \rbrace, - p_{r-1} ; \lbrace p_1,..,p_{r-2} \rbrace \right) 
    \\
    & \equiv  ~ \text{Disc}_{p_{r-1}} \mathcal{B}^{(r-1)}\left(  \lbrace \mathbf{k}_L,...,\mathbf{k}_{M_{r-2}} \rbrace, p_{r-1} ; \lbrace p_1,..,p_{r-2} \rbrace \right) \, .
\end{split}
\end{equation}
Combining both the parts, we now have the final expression for term-1 as follows 
\begin{equation}\label{r site term1}
\begin{split}
    & {\text {Disc}}_{p_{r-1}} \mathcal{B}^{(r)}( \lbrace \mathbf{k}_L,..., \mathbf{k}_R \rbrace ; \lbrace p_1, ..., p_{r-1} \rbrace )\Bigg{\vert}_{\text{term}-1} 
    \\
    & = \frac{1}{2 P_{p_{r-1}} (\eta_0)}  \text{Disc}_{p_{r-1}} \mathcal{B}^{(r-1)}\left(  \lbrace \mathbf{k}_L,...,\mathbf{k}_{M_{r-2}} \rbrace, p_{r-1} ; \lbrace p_1,..,p_{r-2} \rbrace \right) \times \text{Disc}_{p_{r-1}} \mathcal{B}^{(1)}( \lbrace \mathbf{k}_{R} \rbrace, p_{r-1}) \, .
\end{split}
\end{equation}

As we have seen before for the $2$-site correlator case in the previous subsection, the term-2 in eq.\eqref{discBrTerm2} for the $r$-site correlator is non-trivial to analyze. Unlike what we observed in eq.\eqref{eta r-1 in term 1}, the $\eta_{r-1}$ integration in eq.\eqref{discBrTerm2} does not manifest itself in terms of any lower point correlator (i.e., $\mathcal{B}^{(r-1)}$), which can be seen as follows 
\begin{equation}
\label{eta r-1 in term 2a}
\begin{split}
& \eta_{r-1}\,\text{integration in $\text{Disc}_{p_{r-1}}
\mathcal{B}^{(r)}\!\left(
\lbrace \mathbf{k}_L,\ldots,\mathbf{k}_R \rbrace \,;\,
\lbrace p_1,\ldots,p_{r-1} \rbrace
\right)\Bigg|_{\text{term-2}}$ in eq.\eqref{discBrTerm2}}
\\
&= (i\lambda)\int (d\eta_{r-1})\,
\bigg(
\mathcal{V}_{(r-2)}^+\,
\mathcal{K}_{\eta_0\eta_{r-1}}^+(\lbrace \mathbf{k}_{M_{r-2}} \rbrace)
-
\mathcal{V}_{(r-2)}^-\,
\mathcal{K}_{\eta_0\eta_{r-1}}^-(\lbrace \mathbf{k}_{M_{r-2}} \rbrace)
\bigg)
\times
\\
&\hspace{2cm}
\Big(
K_{p_{r-1}}^-(\eta_0,\eta_{r-1})
-
K_{p_{r-1}}^+(\eta_0,\eta_{r-1})
\Big)
\\
&= (i\lambda)\int (d\eta_{r-1})\,
\bigg[
\bigg(
-\mathcal{V}_{(r-2)}^+\,
\mathcal{K}_{\eta_0\eta_{r-1}}^+(\lbrace \mathbf{k}_{M_{r-2}} \rbrace)
K_{p_{r-1}}^+(\eta_0,\eta_{r-1})
\\
&\hspace{3cm}
-\mathcal{V}_{(r-2)}^-\,
\mathcal{K}_{\eta_0\eta_{r-1}}^-(\lbrace \mathbf{k}_{M_{r-2}} \rbrace)
K_{p_{r-1}}^-(\eta_0,\eta_{r-1})
\bigg)
\\
&\hspace{3cm}
+
\bigg(
-\mathcal{V}_{(r-2)}^+\,
\mathcal{K}_{\eta_0\eta_{r-1}}^+(\lbrace \mathbf{k}_{M_{r-2}} \rbrace)
K_{-p_{r-1}}^+(\eta_0,\eta_{r-1})
\\
&\hspace{3cm}
-\mathcal{V}_{(r-2)}^-\,
\mathcal{K}_{\eta_0\eta_{r-1}}^-(\lbrace \mathbf{k}_{M_{r-2}} \rbrace)
K_{-p_{r-1}}^-(\eta_0,\eta_{r-1})
\bigg)
\bigg] \, .
\end{split}
\end{equation} 
Since the final expression on the RHS in eq.\eqref{eta r-1 in term 2a} cannot be written in terms of lower point $\mathcal{B}^{(m)}$, we need to define what we call the auxiliary counterparts of the correlator $\widetilde{\mathcal{B}}^{(m)}$
\begin{equation}\label{rsiteauxinin}
\begin{split}
    &\widetilde{\mathcal{B}}^{(r)}\left( \lbrace \mathbf{k}_L,...,\mathbf{k}_R \rbrace ; \lbrace p_1,\ldots,p_{r-1} \rbrace \right) \\&= (i\lambda)^2 \int (d\eta_r) \int (d\eta_{r-1})~ \bigg[ \mathcal{V}_{(r-2)}^+ \mathcal{K}_{\eta_0 \eta_{r-1}}^+(\lbrace \mathbf{k}_{M_{r-2}} \rbrace) G^{++}_{p_{r-1}}(\eta_{r-1},\eta_r) \mathcal{K}_{\eta_0 \eta_r}^+(\lbrace \mathbf{k}_R \rbrace) 
    \\&
    +\mathcal{V}_{(r-2)}^+ \mathcal{K}_{\eta_0 \eta_{r-1}}^+(\lbrace \mathbf{k}_{M_{r-2}} \rbrace) G^{+-}_{p_{r-1}}(\eta_{r-1},\eta_r) \mathcal{K}_{\eta_0 \eta_r}^-(\lbrace \mathbf{k}_R \rbrace) 
    \\&
    -\mathcal{V}_{(r-2)}^- \mathcal{K}_{\eta_0 \eta_{r-1}}^-(\lbrace \mathbf{k}_{M_{r-2}} \rbrace) G^{-+}_{p_{r-1}}(\eta_{r-1},\eta_r) \mathcal{K}_{\eta_0 \eta_r}^+(\lbrace \mathbf{k}_ R\rbrace)
    \\&
    - \mathcal{V}_{(r-2)}^- \mathcal{K}_{\eta_0 \eta_{r-1}}^-(\lbrace \mathbf{k}_{M_{r-2}} \rbrace) G^{--}_{p_{r-1}}(\eta_{r-1},\eta_r) \mathcal{K}_{\eta_0 \eta_r}^-(\lbrace \mathbf{k}_R \rbrace) \bigg]\,,
\end{split}
\end{equation}
which can be further simplified as 
\begin{equation} \label{rsiteauxinin1}
\begin{split}
   \widetilde{\mathcal{B}}^{(r)}\left( \lbrace \mathbf{k}_L,\ldots,\mathbf{k}_R \rbrace ; \lbrace p_1,\ldots,p_{r-1} \rbrace \right) = (i\lambda) \int (d\eta_r) ~ \bigg[ \mathcal{V}_{(r-1)}^+ \mathcal{K}_{\eta_0 \eta_r}^+(\lbrace \mathbf{k}_R \rbrace) 
    +\mathcal{V}_{(r-1)}^- \mathcal{K}_{\eta_0 \eta_r}^-(\lbrace \mathbf{k}_R \rbrace) \bigg]\,.
\end{split}
\end{equation}
Since $\mathcal{V}_{(r-2)}^\pm \sim \mathcal{O}\left( (i\lambda)^{r-2} \right) $,  it is obvious from eq.\eqref{rsiteauxinin} and eq.\eqref{rsiteauxinin1} that $\widetilde{\mathcal{B}}^{(r)} \sim \mathcal{O}\left( (i\lambda)^r \right)$ in the coupling, which is consistent with what is expected.

Note that we had encountered similar obstacles in the previous subsection where we looked at $2$-site correlator discontinuity, see eq.\eqref{2site Disc 1 eta1 part 1a}. However, here for the $r$-site case, we define the auxiliary objects more generally which reduces to eq.\eqref{2site Disc 1 eta1 part 1a} for $r=2$. In eq.\eqref{rsiteauxinin} the auxiliary counterparts of cosmological correlators are defined using in-in propagators. However, like we did for the $2$-site case before in eq.\eqref{2site Disc 1 eta1 part 1a}, the auxiliary counterparts can be expressed in terms of the wave-function coefficients as follows
\begin{equation}
\begin{split}
& \widetilde{\mathcal{B}}^{(1)}(k_1,\ldots,k_{n-1},p)
= \frac{2i\,\mathbb{I}m\,\psi_n(\vec{k}_1,\ldots,\vec{k}_{n-1},\vec{p})}
{\left(\prod_{i=1}^{n-1} 2\,\mathbb{R}e\,\psi_2(\vec{k}_i)\right)
\,2\,\mathbb{R}e\,\psi_2(\vec{p})} \, ,
\\[6pt]
& \widetilde{\mathcal{B}}^{(2)}(k_1,\ldots,k_{2n-2};p)
= \frac{2i}{\prod_{i=1}^{2n-2} 2\,\mathbb{R}e\,\psi_2(\vec{k}_i)}
\Bigg[
\mathbb{I}m\,\psi_{2n-2}(\vec{k}_1,\ldots,\vec{k}_{2n-2};\vec{p})
\\
& \qquad \qquad \qquad \qquad + \frac{
\mathbb{I}m\,\psi_n(\vec{k}_1,\ldots,\vec{k}_{n-1},\vec{p})\,
\mathbb{R}e\,\psi_n(-\vec{p},\vec{k}_n,\ldots,\vec{k}_{2n-2})
}
{\mathbb{R}e\,\psi_2(\vec{p})}
\Bigg] \,,
\\
& \ldots\ldots \, .
\end{split}
\end{equation}

Now, using eq.\eqref{rsiteauxinin} in eq.\eqref{eta r-1 in term 2a} we obtain 
\begin{equation}
\label{rsitecutterm2a}
\begin{split}
& \eta_{r-1}\,\text{integration in $\text{Disc}_{p_{r-1}}
\mathcal{B}^{(r)}\!\left(
\lbrace \mathbf{k}_L,\ldots,\mathbf{k}_R \rbrace \,;\,
\lbrace p_1,\ldots,p_{r-1} \rbrace
\right)\Bigg|_{\text{term-2}}$ in eq.\eqref{discBrTerm2}}
\\
&= -
\bigg(
\widetilde{\mathcal{B}}^{(r-1)}\!\left(
\lbrace \mathbf{k}_L,\ldots,\mathbf{k}_{M_{r-2}} \rbrace,
p_{r-1} \,;\,
\lbrace p_1,\ldots,p_{r-2} \rbrace
\right)
\\
&\hspace{1.5cm}
+
\widetilde{\mathcal{B}}^{(r-1)}\!\left(
\lbrace \mathbf{k}_L,\ldots,\mathbf{k}_{M_{r-2}} \rbrace,
- p_{r-1} \,;\,
\lbrace p_1,\ldots,p_{r-2} \rbrace
\right)
\bigg)
\\
&=
-\,\widetilde{\text{Disc}}_{p_{r-1}}\,
\widetilde{\mathcal{B}}^{(r-1)}\!\left(
\lbrace \mathbf{k}_L,\ldots,\mathbf{k}_{M_{r-2}} \rbrace,
p_{r-1} \,;\,
\lbrace p_1,\ldots,p_{r-2} \rbrace
\right)\,,
\end{split}
\end{equation}

Combining eq.\eqref{right most disc tilde r site} and eq.\eqref{rsitecutterm2a} we finally obtain
\begin{equation}\label{r site term2}
\begin{split}
    & {\text {Disc}}_{p_{r-1}} \mathcal{B}^{(r)}( \lbrace \mathbf{k}_L,..., \mathbf{k}_R \rbrace ; \lbrace p_1, ..., p_{r-1} \rbrace )\Bigg{\vert}_{\text{term}-2}
    \\
     & =  - \frac{1}{2 P_{p_{r-1}} (\eta_0)} \widetilde{\text{Disc}}_{p_{r-1}} \widetilde{\mathcal{B}}^{(r-1)}\left(  \lbrace \mathbf{k}_L,...,\mathbf{k}_{M_{r-2}} \rbrace, p_{r-1} ; \lbrace p_1,..,p_{r-2} \rbrace \right) \times \widetilde{\text{Disc}}_{p_{r-1}} \widetilde{\mathcal{B}}^{(1)}( \lbrace \mathbf{k}_{R} \rbrace, p_{r-1}) \, .
\end{split}
\end{equation}

\begin{figure}[h]
    \centering
   \includegraphics[width=1.00\textwidth]{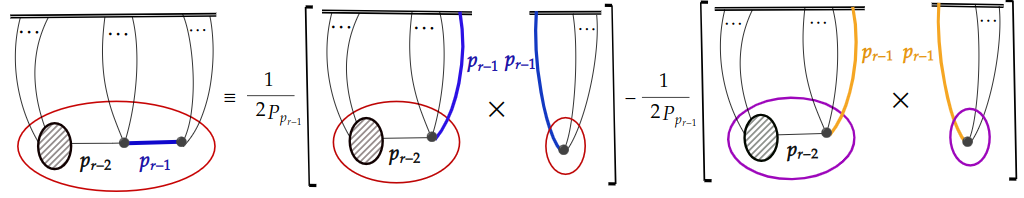}
    \caption{Diagrammatic representation of the single-cut rule of a $r$-site correlator as in eq.\eqref{r site correlator}. Red circle denotes correlator $\mathcal{B}$ , the purple circle denotes the auxiliary object $\widetilde{\mathcal{B}}$ and the patterned blob represents $\mathcal{V}^\sigma_{(r-2)}$. The first term in the RHS denotes the term $\text{Disc}_{p_{r-1}}\mathcal{B}^{(r-1)}\text{Disc}_{p_{r-1}}\mathcal{B}^{(1)}$ and the second term describes the term $\widetilde{\text{Disc}}_{p_{r-1}}\widetilde{\mathcal{B}}^{(r-1)}\widetilde{\text{Disc}}_{p_{r-1}}\widetilde{\mathcal{B}}^{(1)}$ as given in eq.\eqref{r site correlator} .}
    \label{single-cut r-site correlator}
\end{figure}

Finally, adding eq.\eqref{r site term1} and eq.\eqref{r site term2} we get the following discontinuity relation of a r-site correlator
\begin{equation}\label{r site correlator}
\begin{split}
    & {\text {Disc}}_{p_{r-1}} \mathcal{B}^{(r)}( \lbrace \mathbf{k}_L,..., \mathbf{k}_R \rbrace ; \lbrace p_1, ..., p_{r-1} \rbrace )
    \\
    & = \frac{1}{2 P_{p_{r-1}} (\eta_0)} \bigg[ \text{Disc}_{p_{r-1}} \mathcal{B}^{(r-1)}\left(  \lbrace \mathbf{k}_L,...,\mathbf{k}_{M_{r-2}} \rbrace, p_{r-1} ; \lbrace p_1,..,p_{r-2} \rbrace \right)  \times  \text{Disc}_{p_{r-1}} \mathcal{B}^{(1)}( \lbrace \mathbf{k}_{R} \rbrace, p_{r-1})\\
    & ~~ - \widetilde{\text{Disc}}_{p_{r-1}} \widetilde{\mathcal{B}}^{(r-1)}\left(  \lbrace \mathbf{k}_L,...,\mathbf{k}_{M_{r-2}} \rbrace, p_{r-1} ; \lbrace p_1,..,p_{r-2} \rbrace \right)  \times  \widetilde{\text{Disc}}_{p_{r-1}} \widetilde{\mathcal{B}}^{(1)}( \lbrace \mathbf{k}_{R} \rbrace, p_{r-1}) \bigg] \, ,
\end{split}
\end{equation}
which is one of the main results we derive in this paper. This relation can be pictorially presented as in Fig.\ref {single-cut r-site correlator}. 

In appendix-\ref{3 site factorization} we discuss the derivation of discontinuity relation for a  tree-level $3$-site correlator in detail, which might help one to understand the abstract calculations involving notational complexities performed in this section in the simplest non-trivial example beyond the $2$-site case.

We conclude this section by making some comments on the single-cut discontinuity relation in the context of the flat space cosmological correlators. It is easy to see that in the flat space, the bulk to bulk propagators satisfy the same identity as we have obtained in eq.\eqref{discG} for the half-integer $\nu$. Therefore, it is trivially expected that the single-cut discontinuity relation we have derived in this section will work through and we have discussed this exhaustively in Appendix-\ref{DiscFlatCorrelator}. 

\section{Generalizations to other cases} \label{sec:generalization_cases}
In this section, we generalise the derivation of the single-cut discontinuity relation to other cases involving massless scalar fields with interactions of the type $\chi^n$, interactions involving the derivative of the scalar field, and to loop diagrams beyond the tree-level. 
\subsection{Massless scalar} \label{massless scalar}
It is rather straightforward to argue that the derivation of the single-cut discontinuity relation, which so far has focused on conformally coupled scalar fields, also applies to massless interacting scalar fields. The crucial reason supporting this is to note the following property of the power spectrum for massless scalar fields, i.e., 
\begin{equation} \label{PowerSpectrum_masslessscalar}
	P_p(\eta_0) = \Big{\vert}f_p^+(\eta_0)\Big{\vert}^2 = \frac{H^2}{2p^3}\big(1 + p^2\eta_0^2\big) \,, ~~ \text{and hence}~~	P_{-p}(\eta_0) = -P_p(\eta_0)\,,
\end{equation}
which follows from its definition and the corresponding mode expansion, eq.\eqref{mode functions},
\begin{equation}
	\begin{split}
		f^+_p(\eta) = \frac{H}{\sqrt{2p^3}}(1 - i p\eta)e^{i p \eta} \, , \quad 
		f_p^-(\eta) = \frac{H}{\sqrt{2p^3}}(1 + i p\eta)e^{-i p \eta} \, .
	\end{split}
\end{equation}
Once eq.\eqref{PowerSpectrum_masslessscalar} is established, one can check that the calculations performed in \S \ref{ininpropagatorfromunitarity} and \S \ref{confscalartreelevel} also apply rather straightforwardly to massless interacting scalar fields. In other words, all the identities involving in-in bulk-to-boundary and bulk-to-bulk propagators, and therefore the basic relations significant for the derivation of the single-cut discontinuity relation, hold for massless scalars as well. Consequently, the cutting rule for the two-site massless correlator with a $\chi^n $ bulk interaction takes the same form as before. Hence, the discontinuity relation for $2$-site correlator with massless scalar fields is given by the following
\begin{equation} \label{2site massless}
	\begin{split}
		{\text {Disc}}_{p} \mathcal{B}_{\chi^n}^{(2)}( \lbrace \mathbf{k}_L, \mathbf{k}_R \rbrace ;p) = \frac{1}{2 P_p(\eta_0)} \Bigg(& \text{Disc}_p \mathcal{B}_{\chi^n}^{(1)}( \lbrace \mathbf{k}_{L} \rbrace ,p) \times \text{Disc}_p \mathcal{B}_{\chi^n}^{(1)}(\lbrace \mathbf{k}_{R} \rbrace ,p) 
		\\
		& - \widetilde{\text{Disc}}_{p} \widetilde{\mathcal{B}}_{\chi^n}^{(1)}( \lbrace \mathbf{k}_{L} \rbrace ,p) \times \widetilde{\text{Disc}}_{p} \widetilde{\mathcal{B}}_{\chi^n}^{(1)}( \lbrace \mathbf{k}_{R} \rbrace ,p) \Bigg) \, ,
	\end{split}
\end{equation}
where $\mathcal{B}_{\chi^n}^{(2)}$ denotes the $2$-site  correlator for a massless theory with $\chi^n$ interaction and $\mathcal{B}_{\chi^n}^{(1)}$ denotes the $n$-point contact correlator for the massless theory.

For IR-finite interactions involving massless scalars, the wave-function coefficients are purely real \cite{Stefanyszyn:2023qov, Cabass_2023}. In this case, the second term on the RHS of eq.\eqref{2site massless} does not contribute, and the result is entirely determined by the first term. This agrees with the result obtained in \cite{Goodhew:2020hob}. For IR-divergent interactions, however, the wave-function coefficients of massless scalars acquire imaginary contributions. In this case, both terms on the right-hand side of eq.\eqref{2site massless} contribute.

\subsection{Derivative interaction} \label{derivative interaction}
Here in this subsection, we will derive the single-cut discontinuity rule for a derivative interaction with the following interaction term in the Lagrangian
\begin{equation}
	\mathcal{L}_{\text{int}} = \lambda ~ \phi^{n-1} \partial_{\eta}^{(s)}{\phi} \, ,
\end{equation}
where the superscript $s$ indicates that the derivative is taken $s$ times with respect to conformal time $\eta$. 
For convenience, we treat the field  $\partial_{\eta}^{(s)}{\phi}$ as if it propagates through the bulk. Therefore, in this case, the in-in bulk-to-bulk propagators take the following form
\begin{equation}
	\begin{split}
		G^{++}_{p}(\eta_1, \eta_2) &= \frac{1}{P_p(\eta_0)} \bigg[ \partial_{\eta_1}^{(s)} K^{-}_{p}(\eta_0, \eta_1) ~ \partial_{\eta_2}^{(s)}  K^{+}_{p}(\eta_0, \eta_2) ~ \theta(\eta_1 - \eta_2) +(\eta_1 \leftrightarrow \eta_2) \bigg] \, ,\\
		G^{--}_{p}(\eta_1, \eta_2) &= \frac{1}{P_p(\eta_0)} \bigg[ \partial_{\eta_1} ^{(s)}  K^{+}_{p}(\eta_0, \eta_1) \, \partial_{\eta_2}^{(s)}  K^{-}_{p}(\eta_0, \eta_2) ~ \theta(\eta_1 - \eta_2)+(\eta_1 \leftrightarrow \eta_2) \bigg] \, , \\
		G^{+-}_{p}(\eta_1, \eta_2) &= \frac{\partial_{\eta_1}^{(s)}  K^{+}_{p}(\eta_0, \eta_1) ~ \partial_{\eta_2}^{(s)}  K^{-}_{p}(\eta_0, \eta_2)}{P_p(\eta_0)} \, ,
		\\G^{-+}_{p}(\eta_1, \eta_2) &= \frac{\partial_{\eta_1}^{(s)}  K^{-}_{p}(\eta_0, \eta_1) ~\partial_{\eta_2}^{(s)}  K^{+}_{p}(\eta_0, \eta_2)}{P_p(\eta_0)} \, .   
	\end{split}    
\end{equation}
The $2$-site correlator then has the following form by definition of the in-in formalism
\begin{equation}
	\label{ininderivative}
	\begin{split}
		&\mathcal{B}_{\partial_ \eta}^{(2)}\left( \lbrace \mathbf{k}_L, \mathbf{k}_R \rbrace ;p \right)  = (i\lambda)^2 \int^{(2)}  d\eta~\bigg[ \mathcal{K}_{\eta_0 \eta_1}^+(\lbrace \mathbf{k}_L \rbrace) G^{++}_{p}(\eta_1,\eta_2) \mathcal{K}_{\eta_0 \eta_2}^+(\lbrace \mathbf{k}_R \rbrace) \\
		&  \quad \quad - \mathcal{K}_{\eta_0 \eta_1}^+(\lbrace \mathbf{k}_L \rbrace) G^{+-}_{p}(\eta_1,\eta_2) \mathcal{K}_{\eta_0 \eta_2}^-(\lbrace \mathbf{k}_R \rbrace)
		- \mathcal{K}_{\eta_0 \eta_1}^-(\lbrace \mathbf{k}_L \rbrace) G^{-+}_{p}(\eta_1,\eta_2) \mathcal{K}_{\eta_0 \eta_2}^+(\lbrace \mathbf{k}_R \rbrace) \\
		& \quad \quad  + \mathcal{K}_{\eta_0 \eta_1}^-(\lbrace \mathbf{k}_L \rbrace) G^{--}_{p}(\eta_1,\eta_2) \mathcal{K}_{\eta_0 \eta_2}^-(\lbrace \mathbf{k}_R \rbrace) \bigg]\,.
	\end{split}
\end{equation}
where to distinguish from the many other correlators studied in this paper, we are using the subscript $\partial_\eta$ to remind the readers that we are considering a derivative interaction.\\
Now we take the discontinuity of the above expression in eq.\eqref{ininderivative} with respect to the bulk-to-bulk energy (denoted by $p$) followed by manipulations using the in-in propagator identities written in eq.\eqref{discG}, to get the following
\begin{equation}\label{B2_deleta step1}
	\begin{split}
		&{\text {Disc}}_{p} \mathcal{B}_{\partial_\eta}^{(2)}( \lbrace \mathbf{k}_L, \mathbf{k}_R \rbrace ;p) = (i\lambda)^2 \int^{(2)} d\eta~  \bigg( \mathcal{K}_{\eta_0 \eta_1}^+(\lbrace \mathbf{k}_L \rbrace) - \mathcal{K}_{\eta_0 \eta_1}^-(\lbrace \mathbf{k}_L \rbrace) \bigg) \times 
		\\& \quad \quad \quad \quad  \frac{\bigg(\partial_{\eta_1}^{(s)}  K_{p}^-(\eta_0, \eta_1)~ \partial_{\eta_2}^{(s)}  K_{p}^+(\eta_0, \eta_2) + \partial_{\eta_1}^{(s)}  K_{p}^+(\eta_0, \eta_1)~ \partial_{\eta_2}^{(s)}  K_{p}^-(\eta_0, \eta_2) \bigg)}{{P_{p}(\eta_0)}} \times \\
		& \quad \quad \quad \quad  \bigg( \mathcal{K}_{\eta_0 \eta_2}^+(\lbrace \mathbf{k}_R \rbrace) - \mathcal{K}_{\eta_0 \eta_2}^-(\lbrace \mathbf{k}_R \rbrace) \bigg)\,.
	\end{split}
\end{equation}
The above expression is analogous to the one in eq.\eqref{B2 disc step1} except that the mode function associated with the bulk-to-bulk propagators is swapped with its time derivatives. Therefore, the subsequent derivation follows the same pattern as the non-derivative case done in \S \ref{2-site correlator}. Therefore, following the same steps, we obtain the single-cut rule for $2$-site correlator with derivative interaction
\begin{equation}\label{Disc for derivative}
	\begin{split}
		{\text {Disc}}_{p} \mathcal{B}_{\partial_\eta}^{(2)}( \lbrace \mathbf{k}_L, \mathbf{k}_R \rbrace ;p) = \frac{1}{2 P_p(\eta_0)} \Bigg( & \text{Disc}_p \mathcal{B}_{\partial_\eta}^{(1)}( \lbrace \mathbf{k}_{L} \rbrace ,p)  \text{Disc}_p \mathcal{B}_{\partial_\eta}^{(1)}(\lbrace \mathbf{k}_{R} \rbrace ,p) 
		\\
		& - \widetilde{\text{Disc}}_{p} \widetilde{\mathcal{B}}_{\partial_\eta}^{(1)}( \lbrace \mathbf{k}_{L} \rbrace ,p) \widetilde{\text{Disc}}_{p} \widetilde{\mathcal{B}}_{\partial \eta}^{(1)}( \lbrace \mathbf{k}_{R} \rbrace ,p) \Bigg) \, .
	\end{split}
\end{equation}
Note that here both the contact correlators $\mathcal{B}_{\partial_\eta}^{(1)}$ and the one site auxiliary object we have defined in $\widetilde{\mathcal{B}}_{\partial_\eta}^{(1)}$ must contain a boundary field $\partial_{\eta}^{(s)}{\phi} $ along with $(n-1)$- number of $\phi(\eta_0)$.

\subsection{Loop diagrams} \label{loopdiagrams}
In this subsection, considering $\lambda\phi^n$ interaction, we derive the discontinuity relation involving loop diagrams for the $2$-site correlator, $\mathcal{B}^{(2)}$. As we discussed in \S\ref{basicsetup}, we first define the object which appears as the the integrand in the loop integration over the momenta running over the loop as follows
\begin{equation} \label{defmathbbB}
\mathcal{B}^{(2)}\left( \lbrace \mathbf{k}_L, \mathbf{k}_R \rbrace\right) = \int_{\vec{p_1},\vec{p_2}} \mathbb{B}^{(2)}\left( \lbrace \mathbf{k}_L, \mathbf{k}_R \rbrace ;p_1,p_2\right) \, .
\end{equation}
Note that $\mathcal{B}^{(2)}\left( \lbrace \mathbf{k}_L, \mathbf{k}_R \rbrace\right)$ is the $2$-site $1$-loop correlator, and $\mathbb{B}^{(2)}\left( \lbrace \mathbf{k}_L, \mathbf{k}_R \rbrace ;p_1,p_2\right)$ is the integrand in the loop integration.  Also the two momenta flowing in the two sides of the loop are being denoted by $\vec{p}_1$ and $\vec{p}_2$ and the subscript $\vec{p}_1, \vec{p}_2$ in the integral in eq.\eqref{defmathbbB} implies the loop integrals over $3$-momenta.

Using the bulk-to-boundary and bulk-to-bulk propagators in the in-in formalism, for a $2$-site $1$-loop diagram one can write  
\begin{equation}
	\label{inin2site1loop}
	\begin{split}
		& \mathbb{B}^{(2)}\left( \lbrace \mathbf{k}_L, \mathbf{k}_R \rbrace ;p_1,p_2\right) \\
		& = (i\lambda)^2   \int^{(2)}  d\eta \bigg[ \mathcal{K}_{\eta_0 \eta_1}^+(\lbrace \mathbf{k}_L \rbrace) G^{++}_{p_1}(\eta_1,\eta_2) G^{++}_{p_2}(\eta_1,\eta_2) \mathcal{K}_{\eta_0 \eta_2}^+(\lbrace \mathbf{k}_R \rbrace) \\
		& \quad \quad \quad \quad - \mathcal{K}_{\eta_0 \eta_1}^+(\lbrace \mathbf{k}_L \rbrace) G^{+-}_{p_1}(\eta_1,\eta_2) G^{+-}_{p_2}(\eta_1,\eta_2)\mathcal{K}_{\eta_0 \eta_2}^-(\lbrace \mathbf{k}_R \rbrace)
		\\
		& \quad \quad \quad \quad  - \mathcal{K}_{\eta_0 \eta_1}^-(\lbrace \mathbf{k}_L \rbrace) G^{-+}_{p_1}(\eta_1,\eta_2) G^{-+}_{p_2}(\eta_1,\eta_2) \mathcal{K}_{\eta_0 \eta_2}^+(\lbrace \mathbf{k}_R \rbrace) \\
		& \quad \quad \quad \quad + \mathcal{K}_{\eta_0 \eta_1}^-(\lbrace \mathbf{k}_L \rbrace) G^{--}_{p_1}(\eta_1,\eta_2) G^{--}_{p_2}(\eta_1,\eta_2) \mathcal{K}_{\eta_0 \eta_2}^-(\lbrace \mathbf{k}_R \rbrace) \bigg] \, 
	\end{split}
\end{equation}
where by $\lbrace \mathbf{k}_L, \mathbf{k}_R \rbrace$ we denote the collection of external momenta $\lbrace \mathbf{k}_L \rbrace \equiv \lbrace \vec{k}_1, \cdots, \vec{k}_{n-2}\rbrace$ and $\lbrace \mathbf{k}_R \rbrace \equiv \lbrace \vec{k}_{n-1}, \cdots, \vec{k}_{2n-4}\rbrace$. Substituting this in eq.\eqref{defmathbbB} will produce the full correlator for us.

As we will see, to write the discontinuity relation we will have to cut both the sides of the loop, resulting in taking successive discontinuities with respect to $p_1$ and $p_2$\footnote{One could have thought of defining an operation $\text{Disc}_{p_1,p_2}$ to take the discontinuity with respect to both $p_1$ and $p_2$ simultaneously as was done, e.g., in eq.(3.11) in \cite{Melville:2021lst}, which we have previously mentioned as the \textit{simultaneous-cut}. This is different from what we have employed, namely the the \textit{successive-cuts}, where we have applied the discontinuity operation successively on all the sides of the loops.}. Let us first  compute the discontinuity of the above correlator with respect to $p_2$
\begin{equation}\label{loop disc step1}
	\begin{split}
		& {\text {Disc}}_{p_2} \mathbb{B}^{(2)}( \lbrace \mathbf{k}_L, \mathbf{k}_R \rbrace ;p_1,p_2) \\
		&=  (i\lambda)^2 \int^{(2)} d\eta  \Bigg[ \mathcal{K}_{\eta_0 \eta_1}^+(\lbrace \mathbf{k}_L \rbrace) \Big(  G^{++}_{p_1}(\eta_1,\eta_2) \mathcal{K}_{\eta_0 \eta_2}^+(\lbrace \mathbf{k}_R \rbrace) - G^{+-}_{p_1} (\eta_1,\eta_2) \mathcal{K}_{\eta_0 \eta_2}^-(\lbrace \mathbf{k}_R \rbrace) \Big)   \\
		& ~~~~~~~~~~~~ -\mathcal{K}_{\eta_0 \eta_1}^-(\lbrace \mathbf{k}_L \rbrace) \Big(  G^{-+}_{p_1}(\eta_1,\eta_2) \mathcal{K}_{\eta_0 \eta_2}^+(\lbrace \mathbf{k}_R \rbrace)- G^{--}_{p_1} (\eta_1,\eta_2)  \mathcal{K}_{\eta_0 \eta_2}^-(\lbrace \mathbf{k}_R \rbrace)\Big)\Bigg]  \times  
		\\& ~~~~~~~~~~~~~
		\frac{\bigg(K_{p_2}^-(\eta_0, \eta_1) K_{p_2}^+(\eta_0, \eta_2) + K_{p_2}^+(\eta_0, \eta_1) K_{p_2}^-(\eta_0, \eta_2) \bigg)}{{P_{p_2}(\eta_0)}} \, .
	\end{split}
\end{equation}
Again, applying the discontinuity with respect to $p_1$ we get the following
\begin{equation}\label{discp1}
	\begin{split}
		&  {\text {Disc}}_{p_1} \Big[  {\text {Disc}}_{p_2} \mathbb{B}^{(2)}( \lbrace \mathbf{k}_L, \mathbf{k}_R \rbrace ;p_1,p_2) \Big] \\
		& =  \frac{(i\lambda)^2}{{{P_{p_1}(\eta_0)}}{{P_{p_2}(\eta_0)}}}  \int^{(2)} d\eta~    \bigg( \mathcal{K}_{\eta_0 \eta_1}^+(\lbrace \mathbf{k}_L \rbrace) -  \mathcal{K}_{\eta_0 \eta_1}^-(\lbrace \mathbf{k}_L \rbrace)\bigg) \times  \\
		& \bigg( \mathcal{K}_{\eta_0 \eta_2}^+(\lbrace \mathbf{k}_R \rbrace) -  \mathcal{K}_{\eta_0 \eta_2}^-(\lbrace \mathbf{k}_R \rbrace)\bigg) \times \bigg(K_{p_1}^-(\eta_0, \eta_1) K_{p_1}^+(\eta_0, \eta_2) + K_{p_1}^+(\eta_0, \eta_1) K_{p_1}^-(\eta_0, \eta_2) \bigg) \times \\
		& \bigg(K_{p_2}^-(\eta_0, \eta_1) K_{p_2}^+(\eta_0, \eta_2) + K_{p_2}^+(\eta_0, \eta_1) K_{p_2}^-(\eta_0, \eta_2)\bigg) \, .
	\end{split}  
\end{equation}
To proceed further, we need to rewrite the following pieces in the above equation as follows 
\begin{equation}\label{ab+cd type1}
	\begin{split}
		&K_{p_1}^-(\eta_0, \eta_1) K_{p_1}^+(\eta_0, \eta_2) + K_{p_1}^+(\eta_0, \eta_1) K_{p_1}^-(\eta_0, \eta_2) \\
		& =  \frac{1}{2} \Big( K_{p_1}^-(\eta_0, \eta_1)+ K_{p_1}^+(\eta_0, \eta_1) \Big)\Big( K_{p_1}^+(\eta_0, \eta_2) + K_{p_1}^-(\eta_0, \eta_2) \Big)
		\\
		& ~ + \frac{1}{2} \Big( K_{p_1}^-(\eta_0, \eta_1)- K_{p_1}^+(\eta_0, \eta_1) \Big)\Big( K_{p_1}^+(\eta_0, \eta_2) - K_{p_1}^-(\eta_0, \eta_2) \Big) \, ,
	\end{split}
\end{equation}
and, similarly,
\begin{equation}\label{ab+cd type2}
	\begin{split}
		& K_{p_2}^-(\eta_0, \eta_1) K_{p_2}^+(\eta_0, \eta_2) + K_{p_2}^+(\eta_0, \eta_1) K_{p_2}^-(\eta_0, \eta_2) \\
		& =  \frac{1}{2} \Big( K_{p_2}^-(\eta_0, \eta_1)+ K_{p_2}^+(\eta_0, \eta_1) \Big)\Big( K_{p_2}^+(\eta_0, \eta_2) + K_{p_2}^-(\eta_0, \eta_2) \Big)
		\\
		& ~ + \frac{1}{2} \Big( K_{p_2}^-(\eta_0, \eta_1)- K_{p_2}^+(\eta_0, \eta_1) \Big)\Big( K_{p_2}^+(\eta_0, \eta_2) - K_{p_2}^-(\eta_0, \eta_2) \Big) \, .
	\end{split}
\end{equation}
Next we substitute  eq.\eqref{ab+cd type1} and eq.\eqref{ab+cd type2} into eq.\eqref{discp1}. The resulting expression produces four terms which we simplify individually. After some tedious but straightforward manipulations which we defer to Appendix \ref{loopdetails} we obtain the final result as follows 
\begin{equation} \label{finalresult 1 loop}
	\begin{split}
		& {\text {Disc}}_{p_1} \Big[  {\text {Disc}}_{p_2} \mathbb{B}^{(2)}( \lbrace \mathbf{k}_L, \mathbf{k}_R \rbrace ;p_1,p_2) \Big] 
		= \frac{1}{4 {{P_{p_1}(\eta_0)}}{{P_{p_2}(\eta_0)}}}  \times \\
		& \Bigg[  \bigg({\text {Disc}}_{p_1}{\text {Disc}}_{p_2}\mathcal{B}^{(1)}(\lbrace \mathbf{k}_{L} \rbrace,p_1,p_2)\times (L \to R)  + \widetilde{{\text {Disc}}}_{p_1} \widetilde{{\text{Disc}}}_{p_2}\mathcal{B}^{(1)}(\lbrace \mathbf{k}_{L} \rbrace,p_1,p_2)\times (L \to R)  \bigg)\\
		& - \bigg({\text {Disc}}_{p_1} \widetilde{{\text {Disc}}}_{p_2}\widetilde{\mathcal{B}}^{(1)}(\lbrace \mathbf{k}_{L} \rbrace,p_1,p_2)\times (L \to R) +\widetilde{{\text {Disc}}}_{p_1}{\text {Disc}}_{p_2} \widetilde{\mathcal{B}}^{(1)}(\lbrace \mathbf{k}_{L} \rbrace,p_1,p_2)\times (L \to R) \bigg)\Bigg] \, . 
	\end{split}
\end{equation}
It is important to note that in eq.\eqref{finalresult 1 loop} on the RHS we have the integrand of the loop integration for the full correlator, however, on the RHS we get the contact correlator $\mathcal{B}^{(1)}$ or the contact auxiliary counterpart $\widetilde{\mathcal{B}}^{(1)}$. 
 
\begin{figure}[h]
	\centering
	\includegraphics[width=1.00\textwidth]{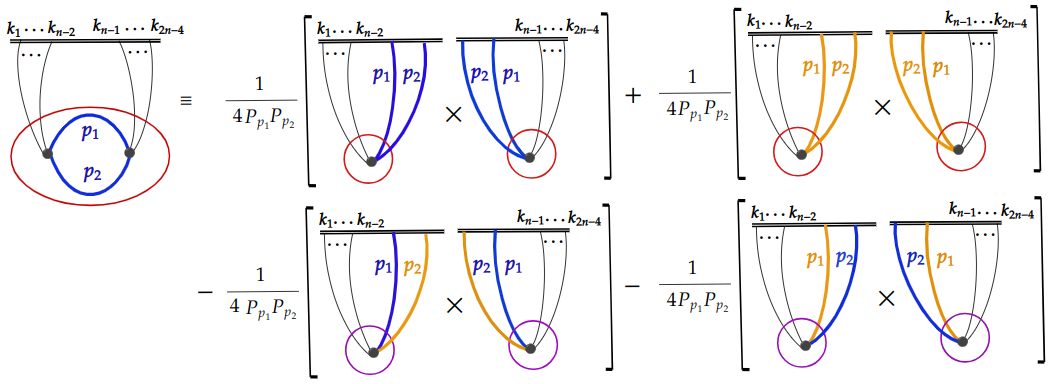}
	\caption{Diagrammatic representation of single-cut rule for the loop integrand of a 1-loop $2$-site correlator. Each blue shaded line corresponds to taking Disc with respect to the modulus of momentum of that line. On the other hand, each yellow line corresponds to taking $\widetilde{\text{Disc}}$ with respect to the modulus of the momentum of that line.}
	\label{1-loop correlator}
\end{figure}

In Fig.\ref{1-loop correlator} we represent eq.\eqref{finalresult 1 loop} diagrammatically. 

Motivated by specific structures in the derivation for the $1$-loop discontinuity result, we conjecture the following cutting rule for the $2$-site correlator at arbitrary $\ell$-loop order
\begin{equation} \label{finalresult l loop}
	\begin{split}
		&\mathrm{Disc}_{p_{\ell+1}}\cdots \mathrm{Disc}_{p_1}\,
		\mathbb{B}^{(2)}\!\left(\{\mathbf{k}_L,\mathbf{k}_R\};p_1,\dots,p_{\ell+1}\right) =
		\frac{1}{2^{\ell+1} \left(\displaystyle\prod_{j=1}^{\ell+1}P_{p_j}(\eta_0)\right)} \times \\
		& \Bigg[
		\sum_{\substack{S\subseteq\{1,\dots,\ell+1\} \\ |S|\ \mathrm{even}}}
		\left(
		\mathcal{D}_S\,
		\mathcal{B}^{(1)}(\{\mathbf{k}_L\};p_1,\dots,p_{\ell+1})
		\right) \times 
		\left(
		\mathcal{D}_S\,
		\mathcal{B}^{(1)}(\{\mathbf{k}_R\};p_1,\dots,p_{\ell+1})
		\right)
		\\
		&\qquad -
		\sum_{\substack{S\subseteq\{1,\dots,\ell+1\} \\ |S|\ \mathrm{odd}}}
		\left(
		\mathcal{D}_S\,
		\widetilde{\mathcal{B}}^{(1)}(\{\mathbf{k}_L\};p_1,\dots,p_{\ell+1})
		\right) \times 
		\left(
		\mathcal{D}_S\,
		\widetilde{\mathcal{B}}^{(1)}(\{\mathbf{k}_R\};p_1,\dots,p_{\ell+1})
		\right)
		\Bigg]\,,
	\end{split}
\end{equation}
where the operator $\mathcal{D}_S$ (with the same subset $S$ as in the sums above) is defined by
\begin{equation*}
	\mathcal{D}_S
	\;\equiv\;
	\left(\prod_{i\in S}\widetilde{\mathrm{Disc}}_{p_i}\right) \times 
	\left(\prod_{i\notin S}\mathrm{Disc}_{p_i}\right)\, .
\end{equation*}
Here, $|S|$ is just the number of elements in $S$. Each subset $S$ determines which internal momenta get a $\widetilde{\mathrm{Disc}}$: if $|S|$ is even, $\mathcal{D}_S$ acts on $\mathcal{B}^{(1)}$; if odd, it acts on $\widetilde{\mathcal{B}}^{(1)}$.

It is worth noting that in our discontinuity relation for loops we have used successive-cut of the loop arms, which is different from the simultaneous-cut employed in earlier studies, although at the level of wave-function coefficients, such as \cite{Melville:2021lst} or at the correlator level in \cite{Donath:2024utn} \footnote{For the loop diagrams \cite{MarinMacedo:2025jco} employed the same simultaneous-cut for the discontinuity but we
have adopted the successive single-cut method which are different at the level of operations.}. In our opinion, the successive single cut in the case of loop diagrams are more advantageous than the simultaneous cut. One advantage is that, if we take successive cuts, the RHS in eq.\eqref{finalresult 1 loop} and eq.\eqref{finalresult l loop} can be obtained solely in terms of the contact level objects (both $\mathcal{B}$ and $\widetilde{\mathcal{B}}$) which is relatively easier to compute. However, for simultaneous cut, since one needs to take all possible combinations of the cut lines and hence need to consider tree-level diagrams (for one loop at the LHS). Now the situation becomes complicated if you start increasing the number of loops. In that case, we believe our eq.\eqref{finalresult l loop} is more elegant.


\section{Explicit checks} \label{sec:Explicit checks}
In this section, we present several explicit examples with various interactions that demonstrate the validity and generality of the single-cut rules or the discontinuity relations derived formally in the previous section. 



For concreteness, in this section we focus on conformally coupled scalar fields with IR convergent polynomial self-interactions of the type $\phi^n$. This allows us to make analytic progress while still capturing the structural features that are essential for the single-cut construction. In particular, to illustrate the interplay between the two types of building blocks, $\mathcal{B}^{(1)}$ and $\widetilde{\mathcal{B}}^{(1)}$ we analyse both odd and even polynomial interactions (i.e., $\phi^n$ with $n=$ even and $n=$ odd) in the simplest non-trivial setting of a $2$-site correlator. These examples provide a clear demonstration of how the two building blocks enter differently depending on the odd and even polynomial interaction. 

Additionally, we are also able to highlight how the leading and sub-leading pieces in the correlator play an essential role in making the discontinuity relation work as we take the late-time limit in the $\eta_0 \to 0^-$ perturbative approximation. In this context, we also identify a shortcoming of the recently proposed cutting rule in \cite{Donath:2024utn}, providing explicit examples that clarify the situations where the formula is effective and where it fails. We also establish that our result is universally true in every example we have examined. 

Furthermore, we extend the analysis to correlators with more than two sites. This allows us to substantiate a vital claim made in one of our previous sections: for odd interactions (i.e., $\phi^n$ with $n=$ odd) with more than two sites, both the terms on the RHS in eq.\eqref{2siteDiscFinal Result}: $\text{Disc}_p\mathcal{B}^{(1)}\text{Disc}_p\mathcal{B}^{(1)}$ and $\widetilde{\text{Disc}_p}\widetilde{\mathcal{B}}^{(1)} \widetilde{\text{Disc}_p}\widetilde{\mathcal{B}}^{(1)}$, contribute at the same order in the late-time expansion. Consequently, both must be included on equal footing to correctly capture the leading behaviour as $\eta_0 \to 0$. The explicit examples we work through confirm this expectation and illustrate how the single-cut rule naturally incorporates both types of contributions when required.
\subsection{$2$-site correlator for conformally coupled interaction}

\subsubsection{$2$-site correlator with $\phi^5$} \label{2site tree phi5}
We first compute a $2$-site correlator for a conformally coupled $\phi^5$ interaction using the in-in formalism. We have done the analytical computation in Mathematica, and the notebook can be found in the following link, see the file \texttt{`Nonderivative in dS.nb'} available at - 
\href{https://github.com/DKaran98/Single-cut-disc}{(link)} \footnote{The file can be accessed at \texttt{`github.com/DKaran98/Single-cut-disc'}.}. The explicit expression of the correlator takes the following form 
\begin{equation}
	\begin{split}
		& \mathcal{B}^{(2)}_{\phi^5}(\lbrace \mathbf{k}_L, \mathbf{k}_R \rbrace; p) = \left(- \frac{\lambda^2 H^{10}\eta_0^8}{64 k_1...k_8} \right) \left( \frac{k_L^2 + 3 k_L k_R + k_R^2 + 2 p(k_L + k_R) + p^2}{(p+k_L)^2(p+k_R)^2(k_L+k_R)^3} \right) \, ,
		\end{split}
\end{equation}
where $k_L = k_1+k_2+k_3+k_4$, and $k_R = k_5+k_6+k_7+k_8$. 
Then by taking a discontinuity with respect to the bulk-to-bulk exchange energy $p$ we get the following expression
\begin{equation}\label{Disc 2site LHS for phi5}
	\text{Disc}_p \mathcal{B}^{(2)}_{\phi^5}(\lbrace \mathbf{k}_L, \mathbf{k}_R \rbrace; p) = \left( \frac{\lambda^2 H^{10} \eta_0^8}{16 k_1 k_2...k_8} \right) \left(  \frac{k_L k_R p}{(k_L^2 - p^2)^2 (k_R^2 - p^2)^2} \right) \, .
\end{equation}
To compare this with the RHS of eq.\eqref{2siteDiscFinal Result} we need to know the expression of $\widetilde{\mathcal{B}}(\lbrace \mathbf{k}_L \rbrace,p)$ and $\widetilde{\mathcal{B}}(\lbrace \mathbf{k}_R \rbrace,p)$
\begin{equation}
	\begin{split}
		&  \widetilde{\mathcal{B}}^{(1)}_{\phi^5}(\lbrace \mathbf{k}_L \rbrace, p) = \frac{2i ~\mathbb{I}m ~\psi^{(1)} (\lbrace \mathbf{k}_L\rbrace,p )}{\mathcal{P}^{-1}_{\mathbf{k}_L}(\eta_0) P^{-1}_{p}(\eta_0)} = \frac{2i ~\mathbb{I}m ~\psi^{(1)} (\lbrace \mathbf{k}_L\rbrace,p )}{\Big( P_{k_1}(\eta_0) P_{k_2}(\eta_0) P_{k_3}(\eta_0) P_{k_4}(\eta_0) P_{p}(\eta_0) \Big)^{-1}} \, ,
	\end{split}
\end{equation}
where $P_{k}(\eta_0)^{-1} = \frac{2k}{H^2 \eta_0^2}$, etc. 

For a $\phi^5$ theory, a contact wave-function has the following expression at the leading order in the late time $\eta_0$
\begin{equation}
	\psi^{(1)}_{\phi^5} (\lbrace \mathbf{k}_L\rbrace,p ) = \frac{i\lambda}{H^4\eta_0^5} \frac{1}{(k_L+p)^2} \implies \mathbb{I}m ~\psi^{(1)} (\lbrace \mathbf{k}_L\rbrace,p )=  \frac{\lambda}{H^4\eta_0^5} \frac{1}{(k_L+p)^2}\, .
\end{equation}
Therefore, we can obtain 
\begin{equation}
	\begin{split}
		& \widetilde{\mathcal{B}}^{(1)}_{\phi^5}(\lbrace \mathbf{k}_L \rbrace, p) = i \left( \frac{\lambda H^6 \eta_0^5}{16  k_1 k_2 k_3 k_4} \right) \frac{1}{p (k_L+p)^2} \, ,
		\\
		\implies & ~ \widetilde{\text{Disc}}_{p} \widetilde{\mathcal{B}}^{(1)}_{\phi^5}(\lbrace \mathbf{k}_L \rbrace, p)  = \left( \frac{-i \lambda H^6 \eta_0^5}{4 k_1 k_2 k_3 k_4} \right) \frac{k_L}{(k_L^2-p^2)^2} \, .
	\end{split}
\end{equation}
Then the RHS of eq.\eqref{2siteDiscFinal Result} gives the following
\begin{equation}\label{Disc 2site RHS for phi5}
	\begin{split}
		& - \frac{1}{2 P_p(\eta_0)} \widetilde{\text{Disc}}_{p} \widetilde{\mathcal{B}}^{(1)}_{\phi^5}(\lbrace \mathbf{k}_L \rbrace, p) \widetilde{\text{Disc}}_{p} \widetilde{\mathcal{B}}^{(1)}_{\phi^5}(\lbrace \mathbf{k}_R \rbrace, p) 
		 = \left( \frac{\lambda^2 H^{10} \eta_0^8}{16 k_1 k_2...k_8} \right) \left(  \frac{k_L \, k_R \, p}{(k_L^2 - p^2)^2 (k_R^2 - p^2)^2} \right) \, .
	\end{split}
\end{equation}
Comparing eq.\eqref{Disc 2site RHS for phi5} with eq.\eqref{Disc 2site LHS for phi5}, we see that the discontinuity relation holds for $2$-sites tree level exchange correlator with a conformally coupled $\phi^{n}$ interaction with $n=odd$, what we have chosen to be $n=5$.
\begin{figure}[h]
	\centering
	\includegraphics[width=0.65\textwidth]{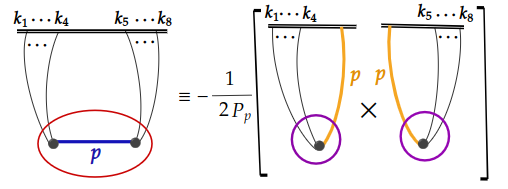 }
	\caption{Diagrammatic representations of the discontinuity relation of a $2$-site correlator with $\phi^5$ interaction in de Sitter. LHS signifies the eq.\eqref{Disc 2site LHS for phi5} and the RHS denotes eq.\eqref{Disc 2site RHS for phi5}.}
	\label{single-cut 2-site correlator explicit check}
\end{figure}

\subsubsection{$2$-site correlator with $\phi^6$}

We now consider an even polynomial interaction. Following the standard in-in prescription, a $2$-site correlator with a conformally coupled $\phi^6$ interaction takes the following expression 
\begin{equation}
	\begin{split}
		& \mathcal{B}^{(2)}_{\phi^6}(\lbrace \mathbf{k}_L, \mathbf{k}_R \rbrace; p) \\ 
		& = \left( \frac{\lambda^2 H^{14}\eta_0^{10}}{128 k_1 ... k_{10}} \right) \left( \frac{ \substack{(k_L+k_R)^5 + 3(k_L^4 + 5 k_L^3 k_R + 10 k_L^2 k_R^2 + 5 k_L k_R^3 + k_R^4)p \\ + 9 (k_L+k_R)(k_L^2 + 4 k_L k_R  + k_R^2) p^2 + (19  k_L^2 + 50 k_L k_R + 19 k_R^2)  p^3 + 18 (k_L + k_R) p^4 + 6 p^5} }{p (k_L + k_R)^5 (k_L + p)^3  (k_R + p)^3} \right) \, ,
	\end{split}
\end{equation}
where $k_L = k_1+k_2+k_3+k_4+k_5$, and $k_R = k_6+k_7+k_8+k_9+k_{10}$. By taking the discontinuity with respect to the bulk-to-bulk exchange energy $p$ the following expression can be obtained
\begin{equation}\label{Disc 2site LHS for phi6}
	\begin{split}
		&  \text{Disc}_p \mathcal{B}^{(2)}_{\phi^6}(\lbrace \mathbf{k}_L, \mathbf{k}_R \rbrace; p) = \left( \frac{\lambda^2 H^{14} \eta_0^{10}}{64 k_1 k_2...k_{10}} \right) \left(  \frac{k_L k_R (k_L^2 + 3 p^2)(k_R^2 + 3p^2)}{p(k_L^2 - p^2)^3 (k_R^2 - p^2)^3} \right) \, .
	\end{split}
\end{equation}
To compare this with the RHS of the discontinuity eq.\eqref{2siteDiscFinal Result} we need to know the expression of $\mathcal{B}^{(1)}_{\phi^6}(\lbrace \mathbf{k}_L \rbrace,p)$ and $\mathcal{B}^{(1)}_{\phi^6}(\lbrace \mathbf{k}_R \rbrace,p)$ which can be computed using the standard in-in formalism. One can show that it takes the following form 
\begin{equation}
	\begin{split}
		& \mathcal{B}^{(1)}_{\phi^6}(\lbrace \mathbf{k}_L \rbrace,p) = \left( \frac{- \lambda H^8 \eta_0^6}{16 k_1 k_2 ... k_5} \right) \frac{1}{p(k_L+p)^3}
		\\
		\implies & \text{Disc}_p \mathcal{B}^{(1)}_{\phi^6}(\lbrace \mathbf{k}_L \rbrace,p) = \left( \frac{- \lambda H^8 \eta_0^6}{8 k_1 k_2 ... k_5} \right) \frac{k_L( k_L^2 + 3 p^2)}{p (k_L^2  -p^2)^3}
	\end{split}
\end{equation}
As a result the RHS of the discontinuity equation takes the following form 
\begin{equation}\label{Disc 2site RHS for phi6}
	\begin{split}
		& \frac{1}{2 P_p(\eta_0)} \text{Disc}_p \mathcal{B}^{(1)}_{\phi^6}(\lbrace \mathbf{k}_L \rbrace, p) \text{Disc}_p \mathcal{B}^{(1)}_{\phi^6}(\lbrace \mathbf{k}_R \rbrace, p)  \\
		& = ~ \left( \frac{\lambda^2 H^{14} \eta_0^{10}}{64 k_1 k_2...k_{10}} \right) \left( \frac{k_L ( k_L^2 + 3 p^2) k_R ( k_R^2 + 3 p^2) }{p (k_L^2 - p^2)^3 (k_R^2 - p^2)^3} \right)
	\end{split}
\end{equation}
By comparing eq.\eqref{Disc 2site RHS for phi6} with eq.\eqref{Disc 2site LHS for phi6} we see that the both side agrees showing that the discontinuity relation holds for a $2$-sites tree level exchange with conformally coupled $\phi^{n}$ interaction with $n=even$. Here we have chosen $n$ to be 6.\\
\begin{figure}[h]
	\centering
	\includegraphics[width=0.65\textwidth]{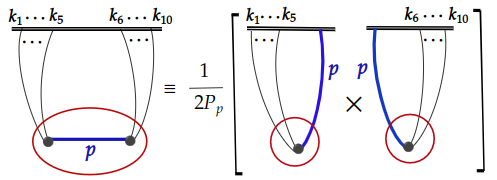 }
	\caption{Diagrammatic representations of the discontinuity relation of a $2$-site correlator with $\phi^6$ interaction in de Sitter. LHS signifies the eq.\eqref{Disc 2site LHS for phi6} and the RHS denotes eq.\eqref{Disc 2site RHS for phi6}.}
	\label{single-cut 2-site correlator phi6}
\end{figure}

\subsection{$3$-site correlator for conformally coupled interaction}

\subsubsection{$3$-site correlator with $\phi^5$}

We now analyze a three-site correlator generated by a $\phi^5$ interaction. This case is particularly instructive because both types of discontinuities—the correlator discontinuities $\mathcal{B}$ and the auxiliary discontinuities $\widetilde{\mathcal{B}}$ —contribute simultaneously. 

The analytic computation is rather involved, so as an additional consistency check we have verified the single-cut rule numerically by fixing representative values for the external energies in Mathematica. The notebook can be found in the following link, see the file \texttt{`Nonderivative in dS.nb'} available at - 
\href{https://github.com/DKaran98/Single-cut-disc}{(link)} \footnote{The file can be accessed at \texttt{`github.com/DKaran98/Single-cut-disc'}.}. 

It can be seen from  the mathematica file \texttt{`Nonderivative in dS.nb'} that in this case, the leading $\eta_0$ contribution from $\mathcal{B}^{(2)}$ comes at $\mathcal{O}(\eta_0^{8})$ and the same happens for $\mathcal{B}^{(1)}$ at order $\mathcal{O}(\eta_0^{6})$. Therefore, the first term in the RHS of the eq.\eqref{r site correlator} with $r=3$ becomes 
\begin{equation}\label{B term in RHS for phi5 3site}
	\text{Disc}_{p_2} \mathcal{B}^{(2)}_{\phi^5}( \lbrace \mathbf{k}_{L}, \mathbf{k}_M \rbrace , p_2; p_1) \text{Disc}_{p_2} \mathcal{B}^{(1)}_{\phi^5}( \lbrace \mathbf{k}_{R} \rbrace ,p_2) \sim \mathcal{O}(\eta_0^{14}) \, ,
\end{equation}
where $\lbrace \mathbf{k}_{L} \rbrace = \lbrace k_1,k_2,k_3,k_4 \rbrace$, $\lbrace \mathbf{k}_{M} \rbrace = \lbrace k_5,k_6,k_7 \rbrace$ and $\lbrace \mathbf{k}_{R} \rbrace = \lbrace k_8,k_9,k_{10},k_{11} \rbrace$. 

And for the other piece in the eq.\eqref{r site correlator} the leading order in $\eta_0$ contribution also comes at $\mathcal{O}(\eta_0^{14})$
\begin{equation}\label{Bt term in RHS for phi5 3site}
	\widetilde{\text{Disc}}_{p_2} \widetilde{\mathcal{B}}^{(2)}_{\phi^5}( \lbrace \mathbf{k}_{L}, \mathbf{k}_M \rbrace , p_2; p_1) \widetilde{\text{Disc}}_{p_2} \widetilde{\mathcal{B}}^{(1)}_{\phi^5}( \lbrace \mathbf{k}_{R} \rbrace ,p_2) \sim \mathcal{O}(\eta_0^{14}) \, .
\end{equation}

\begin{figure}[h]
	\centering
	\includegraphics[width=0.95\textwidth]{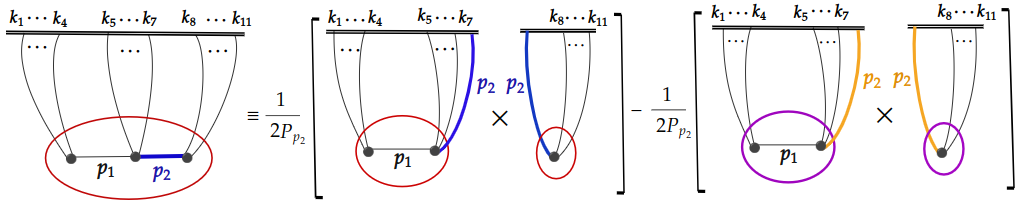 }
	\caption{Diagrammatic representations of the single-cut discontinuity relation of a 3-site correlator with $\phi^5$ interaction. The first term in the RHS signifies the eq.\eqref{B term in RHS for phi5 3site} and the second term denotes eq.\eqref{Bt term in RHS for phi5 3site}} 
	\label{single-cut 3-site correlator phi5}
\end{figure}

Therefore, we need to take both the contributions into the account since they come at the same order in $\eta_0$.

\subsubsection{$3$-site correlator with $\phi^6$}

We now analyze a three-site correlator generated by a $\phi^6$ interaction. In this case, only the correlator discontinuities $\mathcal{B}$ contribute which is expected since we are now  dealing with an $n=even$ interaction.
The analytic computation is more involved than the previous case, so as a consistency check we have verified the single-cut rule numerically by fixing representative values for the external energies. For details please check the mathematica notebook \texttt{``Nonderivative in dS.nb"}.
\begin{figure}[h]
	\centering
	\includegraphics[width=0.75\textwidth]{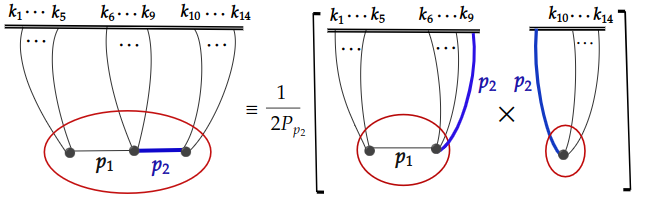 }
	\caption{Diagrammatic representations of a single-cut 3-site correlator with $\phi^6$ interaction. The RHS represents the equation eq.\eqref{B term in RHS of of 3site phi6} since it is leading in $\eta_0$.}
	\label{single-cut 3-site correlator phi6}
\end{figure}
In this case, we have found that the leading $\eta_0$ contribution from $\mathcal{B}^{(2)}$ comes at $\mathcal{O}(\eta_0^{10})$ and from the contact correlator $\mathcal{B}^{(1)}$ the leading $\eta_0$ contribution comes at $\mathcal{O}(\eta_0^6)$. As a result the first term in eq.\eqref{r site correlator} with $r=3$ becomes
\begin{equation}\label{B term in RHS of of 3site phi6}
	\text{Disc}_{p_2} \mathcal{B}^{(2)}_{\phi^6}( \lbrace \mathbf{k}_{L}, \mathbf{k}_M \rbrace , p_2; p_1) \text{Disc}_{p_2} \mathcal{B}^{(1)}_{\phi^6}( \lbrace \mathbf{k}_{R} \rbrace ,p_2) \sim \mathcal{O}(\eta_0^{16}) \, ,
\end{equation}
where $\lbrace \mathbf{k}_{L} \rbrace = \lbrace k_1,k_2,k_3,k_4,k_5 \rbrace$, $\lbrace \mathbf{k}_{M} \rbrace = \lbrace k_6,k_7,k_8,k_9 \rbrace$ and $\lbrace \mathbf{k}_{R} \rbrace = \lbrace k_{10},k_{11},k_{12},k_{13},k_{14} \rbrace$.

While for the other piece in the eq.\eqref{3 site correlator} the leading $\eta_0$ contribution from $\widetilde{\mathcal{B}}^{(2)}$ comes at $\mathcal{O}(\eta_0^{11})$ and from $\widetilde{\mathcal{B}}^{(1)}$ the leading $\eta_0$ contribution appears at $\mathcal{O}(\eta_0^{7})$. Therefore, the other piece in eq.\eqref{r site correlator} becomes sub-leading in $\eta_0$
\begin{equation}
	\widetilde{\text{Disc}}_{p_2} \widetilde{\mathcal{B}}^{(2)}_{\phi^6}( \lbrace \mathbf{k}_{L}, \mathbf{k}_M \rbrace , p_2; p_1) \widetilde{\text{Disc}}_{p_2} \widetilde{\mathcal{B}}^{(1)}_{\phi^6}( \lbrace \mathbf{k}_{R} \rbrace ,p_2) \sim \mathcal{O}(\eta_0^{18}) \, .
\end{equation}
As a result, as long as we are concerned with the leading $\eta_0$ behaviour in the limit $\eta_0 \to 0$, it is enough to consider the product of the correlator discontinuity ignoring the sub-leading contribution due to the discontinuity of the auxiliary object. In the mathematica notebook \texttt{``Nonderivative in dS.nb"}, it has been shown that the result is in a perfect reflection of the above argument. 

\subsection{$2$-site correlator for conformally coupled derivative interaction}

\subsubsection{Conformally coupled $\phi^4\partial_\eta \phi$}

We are now interested in a $2$-site correlator where the at the boundary $\eta=\eta_0$, we have inserted the fields $\phi(\eta_0)$ (but not its derivative) and in the bulk we let the field $\partial_\eta \phi$ propagate.\\
From the standard in-in prescription one can easily compute the 8-point correlator which takes the following form 
\begin{equation}
	\begin{split}
		& \mathcal{B}^{(2)}_{\phi^4 \partial_\eta \phi}(\lbrace \mathbf{k}_L, \mathbf{k}_R \rbrace; p) =  \left( \frac{\lambda^2 H^{10} \eta_0^{8}}{128 k_1 k_2...k_{8}} \right) \left( \frac{ \substack{ 2 p^5 + 4 p^4 (k_L + k_R) - 2 p k_L k_R (k_L + k_R)^2 - k_L k_R (k_L + k_R)^3 \\ + 2 p^3 (k_L^2 + 3 k_L k_R + k_R^2)}}{p (p + k_L)^2  (p + k_R)^2  (k_L + k_R)^3}  \right) \, ,
	\end{split}
\end{equation}
where $k_L = k_1+k_2+k_3+k_4$, and $k_R = k_5+k_6+k_7+k_8$.
		By taking its discontinuity with respect to the bulk-to-bulk energy we obtain the following simplified form which is essentially the LHS of the eq.\eqref{Disc for derivative}
\begin{equation}\label{Disc 2site LHS for phi4phidot}
	\begin{split}
		&  \text{Disc}_p \mathcal{B}^{(2)}_{\phi^4 \partial_\eta \phi}(\lbrace \mathbf{k}_L, \mathbf{k}_R \rbrace; p) = \left( \frac{\lambda^2 H^{10} \eta_0^{8}}{64 k_1 k_2...k_{8}} \right) \left( \frac{ k_L k_R (k_L^2 + p^2)(k_R^2 + p^2)}{p (k_L^2 - p^2)^2 (k_R^2 - p^2)^2} \right) \, .
	\end{split}
\end{equation}
To compare this with the RHS of the discontinuity eq.\eqref{Disc for derivative} we need to know the expression of $\mathcal{B}^{(1)}_{\phi^4 \partial_\eta \phi}(\lbrace \mathbf{k}_L \rbrace,p)$ and $\mathcal{B}^{(1)}_{\phi^4 \partial_\eta \phi}(\lbrace \mathbf{k}_R \rbrace,p)$ which can be computed using the standard in-in formalism. One can show that it takes the following form 
\begin{equation}
	\begin{split}
		& \mathcal{B}^{(1)}_{\phi^4 \partial_\eta \phi}(\lbrace \mathbf{k}_L \rbrace,p) = \left( \frac{ \lambda H^6 \eta_0^5}{16 k_1...k_4} \right) \left( \frac{k_L}{p(k_L+p)^2} \right)
		\\
		\implies & \text{Disc}_p \mathcal{B}^{(1)}_{\phi^4 \partial_\eta \phi}(\lbrace \mathbf{k}_L \rbrace,p) = \left( \frac{ \lambda H^6 \eta_0^5}{8 k_1... k_4} \right) \left( \frac{k_L (k_L^2 + p^2)}{p(k_L^2-p^2)^2} \right) \, .
	\end{split}
\end{equation}
\begin{figure}[h]
	\centering
	\includegraphics[width=0.65\textwidth]{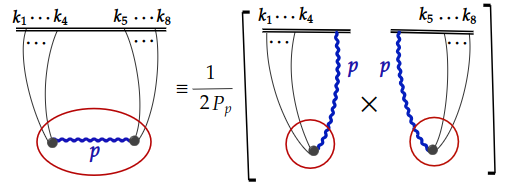}
	\caption{Diagrammatic representations of the discontinuity relation of a $2$-site correlator with $\phi^4 \partial_{\eta} \phi$ interaction. Wavy line denotes the field $\partial_{\eta}\phi$. The LHS denotes the eq.\eqref{Disc 2site LHS for phi4phidot} and the RHS signifies the eq.\eqref{Disc 2site RHS for phi4phidot}.}
	\label{single-cut 2-site correlator phi4phidot}
\end{figure}
Therefore the leading $\eta_0$ contribution in the RHS of the eq.\eqref{Disc for derivative} becomes the following
\begin{equation}\label{Disc 2site RHS for phi4phidot}
	\begin{split}
		&\frac{1}{2 P_p(\eta_0)} \text{Disc}_p \mathcal{B}^{(1)}_{\phi^4 \partial_\eta \phi} (\lbrace \mathbf{k}_L \rbrace, p) \text{Disc}_p \mathcal{B}^{(1)}_{\phi^4 \partial_\eta \phi}(\lbrace \mathbf{k}_R \rbrace, p) \\
	&	=  \left( \frac{\lambda^2 H^{10} \eta_0^{8}}{64 k_1 k_2...k_8} \right) \left(  \frac{ k_L k_R (k_L^2 + p^2)(k_R^2 + p^2)}{p (k_L^2 - p^2)^2 (k_R^2 - p^2)^2} \right) \, ,
	\end{split}
\end{equation}
which agrees with the LHS of the discontinuity equation that we have obtained in eq.\eqref{Disc 2site LHS for phi4phidot}.

\subsubsection{Conformally coupled $\phi^5\partial_\eta \phi$}

Let's now consider an interaction $\phi^5\partial_\eta \phi$ involving conformally coupled fields and similar to the case formerly studied we let the field $\partial_\eta \phi$ propagate in the bulk. 
From the standard in-in prescription one can easily compute the 8-point correlator which takes the following form 
\begin{equation}
	\begin{split}
		& \mathcal{B}^{(2)}_{\phi^5 \partial_\eta \phi}(\lbrace \mathbf{k}_L, \mathbf{k}_R \rbrace; p)
		= \left( \frac{\lambda^2 H^{14} \eta_0^{10}}{128 k_1...k_{10}} \right) \times \mathcal{N}\, ,
	\end{split}
\end{equation}
such that
\begin{equation}
	\begin{split}
\mathcal{N}  = \left(  \frac{\substack{k_L k_R (k_L + k_R)^2 (k_L^2 + 3 k_L k_R + k_R^2) - (k_L + k_R)^3  (k_L^2 - 
				k_L k_R + k_R^2)p -(3 k_L^4 \\ + 14 k_L^3 k_R  + 28 k_L^2 k_R^2 + 14 k_L k_R^3 + 3 k_R^4) p^2  -9  (k_L + k_R)  (k_L^2 + 4 k_L k_R + k_R^2) p^3\\-(19 k_L^2 + 50 k_L k_R + 19 k_R^2)p^4 - 18 (k_L + k_R) p^5 - 6 p^6}}{(k_L + k_R)^5  (k_L + p)^3  (k_R + p)^3}  \right) \, ,
	\end{split}
\end{equation}
where $k_L = k_1+k_2+k_3+k_4+k_5$, and $k_R = k_6+k_7+k_8 + k_9 + k_{10}$. 

By taking discontinuity with respect to the bulk-to-bulk energy $p$ we obtain the following expression that corresponds to the LHS of the eq.\eqref{Disc for derivative}
\begin{equation}\label{Disc 2site LHS for phi5phidot}
	\begin{split}
		&  \text{Disc}_p \mathcal{B}^{(2)}_{\phi^5 \partial_\eta \phi}(\lbrace \mathbf{k}_L, \mathbf{k}_R \rbrace; p) = \left( \frac{\lambda^2 H^{14}\eta_0^{10}}{16 k_1...k_{10}} \right) \left(  \frac{ p k_L k_R (k_L^2 + p^2)(k_R^2 + p^2)}{(k_L^2 - p^2)^3 (k_R^2 - p^2)^3} \right) \, .
	\end{split}
\end{equation}
To compare this with the RHS of the discontinuity eq.\eqref{Disc for derivative} we need to know the expression of $\widetilde{\mathcal{B}}^{(1)}_{\phi^5 \partial_\eta \phi}(\lbrace \mathbf{k}_L \rbrace,p)$ and $\widetilde{\mathcal{B}}^{(1)}_{\phi^5 \partial_\eta \phi}(\lbrace \mathbf{k}_R \rbrace,p)$ which can be computed using the standard in-in formalism. One can show that it takes the following form 
\begin{equation}\label{Disc 2site LHS for phi5phidot1}
	\begin{split}
		& \widetilde{\mathcal{B}}^{(1)}_{\phi^5 \partial_\eta \phi}(\lbrace \mathbf{k}_L \rbrace,p) = i \left( \frac{ \lambda H^8 \eta_0^6}{32 k_1...k_5} \right) \left( \frac{k_L-p}{p(k_L+p)^3} \right)
		\\
		\implies & \widetilde{\text{Disc}}_p \widetilde{\mathcal{B}}^{(1)}_{\phi^5 \partial_\eta \phi}(\lbrace \mathbf{k}_L \rbrace,p) = -i \left( \frac{ \lambda H^8 \eta_0^6}{4 k_1... k_5} \right) \left( \frac{k_L(k_L^2 + p^2)}{(k_L^2-p^2)^3} \right) \, .
	\end{split}
\end{equation}
\begin{figure}[h]
	\centering
	\includegraphics[width=0.65\textwidth]{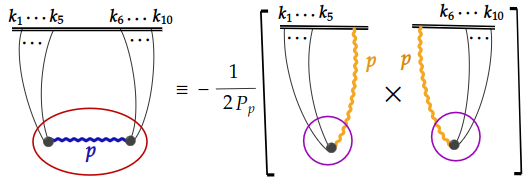}
	\caption{Diagrammatic representations of the discontinuity relation of a $2$-site correlator with $\phi^5 \partial_{\eta} \phi$ interaction. The LHS denotes the eq.\eqref{Disc 2site LHS for phi5phidot} and the RHS signifies the eq.\eqref{Disc 2site RHS for phi5phidot}.}
	\label{single-cut 2-site correlator phi5phidot}
\end{figure}

Therefore the leading $\eta_0$ piece in the RHS of the eq.\eqref{Disc for derivative} becomes the following
\begin{equation}\label{Disc 2site RHS for phi5phidot}
	\begin{split}
		& -\frac{1}{2 P_p(\eta_0)} \widetilde{\text{Disc}}_p \widetilde{\mathcal{B}}^{(1)}_{\phi^5 \partial_\eta \phi} (\lbrace \mathbf{k}_L \rbrace, p) \widetilde{\text{Disc}}_p \widetilde{\mathcal{B}}^{(1)}_{\phi^5 \partial_\eta \phi}(\lbrace \mathbf{k}_R \rbrace, p) \\
		&  = \left( \frac{\lambda^2 H^{14}\eta_0^{10}}{16 k_1...k_{10}} \right) \left(  \frac{ p k_L k_R (k_L^2 + p^2)(k_R^2 + p^2)}{(k_L^2 - p^2)^3 (k_R^2 - p^2)^3} \right) \, ,
	\end{split}
\end{equation}
which agrees with the LHS of the discontinuity equation that we have obtained in eq.\eqref{Disc 2site LHS for phi5phidot}.

All these calculations are given in the mathematica notebook \texttt{``Derivative in dS.nb"} available at - 
\href{https://github.com/DKaran98/Single-cut-disc}{(link)} \footnote{The file can be accessed at \texttt{`github.com/DKaran98/Single-cut-disc'}.}.

\subsection{$1$-loop correlator for conformally coupled interaction }
One should note that for a tree level diagram it is enough to take discontinuity with respect to any one bulk-to-bulk energy to separate out the two adjacent vertices. However, in case of the loop one does not have this freedom. For example, in our current case, to separate out the two adjacent vertices one must take discontinuities with respect to modulus of both the bulk-to-bulk momenta $\vec{p}_1$ and $\vec{p}_2$.
In the following we present two examples of the derivative interaction. Depending on the nature of the interaction, whether it is odd or even polynomial, the discontinuities of the correlator or the auxiliary part will contribute.
To find the detailed computation see the mathematica notebook \texttt{``1 loop cut in dS.nb"} available at - 
\href{https://github.com/DKaran98/Single-cut-disc}{(link)} \footnote{The file can be accessed at \texttt{`github.com/DKaran98/Single-cut-disc'}.}.

\subsubsection{$2$-site $1$-loop correlator with $\phi^4$}

With a $\phi^4$ conformally coupled interaction we study a 1-loop, 4 point correlator. Using standard in-in formalism one can find the integrand of the loop integration as follows
\begin{equation}
	\begin{split}
		& \mathbb{B}^{(2)}_{\phi^4\text{-loop}}(\lbrace \mathbf{k}_L, \mathbf{k}_R \rbrace; \lbrace p_1; p_2 \rbrace) =\left(  \frac{\lambda^2 H^4 \eta_0^4}{16 k_1 k_2 k_3 k_4} \right) \left( \frac{k_L + k_R + p_1 + p_2}{p_1 p_2(k_L+k_R)(k_L+p_1+p_2)(k_R+p_1+p_2)}  \right) \, .
	\end{split}
\end{equation}
So in principle, to obtain the 2-site correlator we indeed have to perform two three momenta integrations over $\vec{p}_1$ and $\vec{p}_2$. 
Taking discontinuities with respect to the magnitude of both the bulk-to-bulk momenta successively we obtain the following 
\begin{equation}
	\begin{split}
		& \text{Disc}_{p_1} \text{Disc}_{p_2} \mathbb{B}^{(2)}_{\phi^4\text{-loop}}(\lbrace \mathbf{k}_L, \mathbf{k}_R \rbrace; \lbrace p_1; p_2 \rbrace) \\
		& = \left( \frac{\lambda^2 H^4 \eta_0^4}{4 k_1 k_2 k_3 k_4} \right) \left( \frac{(k_L^2 - p_1^2)(k_R^2 - p_1^2) - (k_L^2 + k_R^2 - 6 p_1^2)p_2^2 + p_2^4} {p_1 p_2 (p_+^2 - k_L^2)(p_+^2 - k_R^2)(p_-^2 - k_L^2)(p_-^2 - k_R^2)} \right) \, ,
	\end{split}
\end{equation}
where $k_L = k_1 + k_2,~~k_R = k_3 + k_4, \,  p_+ = p_1+p_2, \, p_- = p_1 - p_2$, and this is the LHS of the discontinuity equation.

To compute the RHS we just need the information of a one site correlator, to be precise, a 4 point contact correlator in this case. In $\lambda \phi^4$ theory, a 4 point contact correlator is given by the following 
\begin{equation}
	\begin{split}
		\mathcal{B}^{(1)}_{\phi^4} (k_1, k_2, p_1, p_2) = \left( \frac{\lambda (H \eta_0)^4 }{8 k_1 k_2} \right) \frac{1}{p_1 p_2 (k_1 + k_2 + p_1 + p_2)} \, .
	\end{split}
\end{equation}
Using this, one can find the following quantities required to compute the RHS of the Disc equation given by eq.\eqref{finalresult 1 loop}\footnote{Note that since this is an even interaction, only the contributions from the correlator $\mathcal{B}$ will only contribute.}
\begin{equation}\label{phi41loopUS}
	\begin{split}
		&  \text{Disc}_{p_1} \text{Disc}_{p_2} \mathcal{B}^{(1)}_{\phi^4} ( \lbrace \mathbf{k}_L \rbrace, p_1, p_2) = \left( \frac{\lambda H^4 \eta_0^4}{2 k_1 k_2} \right) \left(\frac{ k_L( k_L^2 - p_1^2 - p_2^2)}{p_1 p_2 \left( p_+^2 - k_L^2 \right) \left( p_-^2 - k_L^2 \right) } \right) \, ,
		\\
		&  \text{Disc}_{p_1} \text{Disc}_{p_2} \mathcal{B}^{(1)}_{\phi^4} ( \lbrace \mathbf{k}_R \rbrace, p_1, p_2) = \left( \frac{\lambda H^4 \eta_0^4}{2 k_3 k_4} \right) \left( \frac{k_R ( k_R^2 - p_1^2 - p_2^2) }{p_1 p_2 \left( p_+^2 - k_R^2 \right) \left( p_-^2 - k_R^2 \right) } \right) \, ,
		\\
		& \widetilde{\text{Disc}}_{p_1} \widetilde{\text{Disc}}_{p_2} \mathcal{B}^{(1)}_{\phi^4} ( \lbrace \mathbf{k}_L \rbrace, p_1, p_2) = \left( \frac{\lambda H^4 \eta_0^4}{2 k_1 k_2} \right) \left(\frac{ k_L}{ \left( p_+^2 - k_L^2 \right) \left( p_-^2 - k_L^2 \right) } \right) \, ,
		\\
		& \widetilde{\text{Disc}}_{p_1} \widetilde{\text{Disc}}_{p_2} \mathcal{B}^{(1)}_{\phi^4} ( \lbrace \mathbf{k}_R \rbrace, p_1, p_2) = \left( \frac{\lambda H^4 \eta_0^4}{2 k_3 k_4} \right) \left(\frac{ k_R}{ \left( p_+^2 - k_R^2 \right) \left( p_-^2 - k_R^2 \right) } \right) \, .
	\end{split}
\end{equation}
Substituting them in the RHS of the Disc equation it is straight forward to verify that it correctly reproduces what we obtained in the LHS.

\subsubsection{$2$-site $1$-loop correlator with $\phi^5$} \label{2site 1loop phi5}

Using a $\phi^5$ conformally coupled interaction we study a 1-loop, 6 point correlator. Using standard in-in formalism one can find the integrand of the loop integration as follows
\begin{equation}
	\begin{split}
		& \mathbb{B}^{(2)}_{\phi^5 \text{-loop}}(\lbrace \mathbf{k}_L, \mathbf{k}_R \rbrace; \lbrace p_1; p_2 \rbrace) \\
		& =\left( - \frac{\lambda^2 H^8 \eta_0^6}{32 k_1 k_2 k_3 k_4 k_5 k_6} \right) \left(\frac{ (p_1+p_2)( k_L^2 + 3 k_L k_R + 2 k_L( p_1 + p_2) + (k_R + p_1+p_2)^2)}{p_1 p_2(k_L+k_R)^3 (k_L+p_1+p_2)^2 (k_R+p_1+p_2)^2}  \right) \, .
	\end{split}
\end{equation}
Now one can take the discontinuity with respect to both $p_1$ and $p_2$ successively by using the following definition. The result is slightly cumbersome and obtained in the mathematica notebook \texttt{``1 loop cut in dS.nb"} available at - 
\href{https://github.com/DKaran98/Single-cut-disc}{(link)} \footnote{The file can be accessed at \texttt{`github.com/DKaran98/Single-cut-disc'}.}. 

The next step is to check how to obtain the RHS of the Disc. equation and hence we need to know just one information that is the imaginary part of a $5$-point contact wave-function coefficient.
\begin{equation}
	\begin{split}
		\widetilde{\mathcal{B}}^{(1)}_{\phi^5} (k_1, k_2, k_3, p_1, p_2)  = & \frac{2i ~\mathbb{I}m \psi^{(1)}\left( k_1, k_2, k_3, p_1, p_2\right)}{P_{k_1}^{-1}(\eta_0) P_{k_2}^{-1}(\eta_0) P_{k_3}^{-1}(\eta_0) P_{p_1}^{-1}(\eta_0)P_{p_1}^{-1}(\eta_0)}
		\\
		= & ~ \left( \frac{i \lambda H^6 \eta_0^5}{16 k_1 k_2 k_3} \right) \left( \frac{1}{p_1 p_2 (p_1 + p_2 + k_L)^2} \right) \, .
	\end{split}
\end{equation}
From the above expression one can find the following expressions which constitute the RHS of the Disc equation given by eq.\eqref{finalresult 1 loop} \footnote{Since this is an odd interaction, only the contributions from the auxiliary object $\widetilde{\mathcal{B}}$ will only contribute.}
\begin{equation} \label{Dp1Dtp2Btphi5}
	\begin{split}
		&  \text{Disc}_{p_1} \widetilde{\text{Disc}}_{p_2} \widetilde{\mathcal{B}}^{(1)}_{\phi^5} ( \lbrace \mathbf{k}_L \rbrace, p_1, p_2) = \left( \frac{- i \lambda H^6 \eta_0^5}{2 k_1 k_2 k_3}\right) \left( \frac{k_L \left( k_L^4 - 3 p_1^4 + 2 p_1^2 p_2^2 + p_2^4 + 2 k_L^2 p_- p_+ \right)}{p_1 \left( p_+^2 - k_L^2 \right)^2 \left( p_-^2 - k_L^2 \right)^2}\right)\, ,
		\\
		&  \text{Disc}_{p_1} \widetilde{\text{Disc}}_{p_2} \widetilde{\mathcal{B}}^{(1)}_{\phi^5} ( \lbrace \mathbf{k}_R \rbrace, p_1, p_2) = \left( \frac{- i \lambda H^6 \eta_0^5}{2 k_4 k_5 k_6}\right) \left( \frac{k_R \left( k_R^4 - 3 p_1^4 + 2 p_1^2 p_2^2 + p_2^4 + 2 k_R^2 p_- p_+ \right)}{p_1 \left( p_+^2 - k_R^2 \right)^2 \left( p_-^2 - k_R^2 \right)^2}\right)\, ,
		\\
		& \widetilde{\text{Disc}}_{p_1} {\text{Disc}}_{p_2} \widetilde{\mathcal{B}}^{(1)}_{\phi^5} ( \lbrace \mathbf{k}_L \rbrace, p_1, p_2) = \left( \frac{- i \lambda H^6 \eta_0^5}{2 k_1 k_2 k_3}\right) \left( \frac{k_L \left( (k_L^2 - p_1^2)^2 + 2 (k_L^2 + p_1^2) p_2^2 - 3 p_2^4\right)}{p_2 \left( p_+^2 - k_L^2 \right)^2 \left( p_-^2 - k_L^2 \right)^2}\right)\, ,
		\\
		& \widetilde{\text{Disc}}_{p_1} {\text{Disc}}_{p_2} \widetilde{\mathcal{B}}^{(1)}_{\phi^5} ( \lbrace \mathbf{k}_R \rbrace, p_1, p_2) = \left( \frac{- i \lambda H^6 \eta_0^5}{2 k_4 k_5 k_6}\right) \left( \frac{k_R \left( (k_R^2 - p_1^2)^2 + 2 (k_R^2 + p_1^2) p_2^2 - 3 p_2^4\right)}{p_2 \left( p_+^2 - k_R^2 \right)^2 \left( p_-^2 - k_R^2 \right)^2}\right)\, .
	\end{split}
\end{equation}
Substituting them in the RHS of the Disc equation it is straight forward to verify that it correctly reproduces what we obtained in the LHS.

To summarise this section, we have provided various explicit checks above in support of our discontinuity relations written explicitly in terms of the cosmological correlators, which are summarised in eq.\eqref{2siteDiscFinal Result} and eq.\eqref{r site correlator} for tree-level correlators, and in \S\ref{derivative interaction} and \S\ref{loopdiagrams} for derivative intercations and loop diagrams respectively. 

Before we conclude this section let us comment on how our results compare with recent results available in the literature. In particular, in \cite{Donath:2024utn}, cutting rules for a $2$-site correlator, even with loop-diagrams was written down, using the Veltman's largest time equation and the in-out formalism developed directly in the context of cosmological correlator. Although we followed a different method in this paper, it is certainly instructive to compare both the results. 

Our attempts to understand the cutting rule written in eq.(5.32) in \cite{Donath:2024utn} yield the following - we find that for conformally coupled $\phi^4$ theory both at the tree-level and at $1$-loop level, eq.(5.32) in \cite{Donath:2024utn} is consistent, However, for $\phi^5$ interactions it is incomplete in a sense that the RHS of eq.(5.32) in \cite{Donath:2024utn} fails to capture the leading $\eta_0$ contribution and hence does not agree with the LHS for odd polynomial interaction, both at the tree and at the loop level. The details of this check are given in Appendix \ref{subsec:comparing in-out}. On the other hand, in previous subsections, namely in \S\ref{2site tree phi5} and \S\ref{2site 1loop phi5} we have checked that our formulae eq.\eqref{2siteDiscFinal Result} and eq.\eqref{finalresult 1 loop} work for conformally coupled scalars with $\phi^5$ interaction, both at the tree level and at the $1$-loop level respectively. In this sense, we believe that our discontinuity relations are improvements over the existing results in the literature \footnote{The comparison to the discontinuity relation in \cite{Donath:2024utn} (see the discussions above and in the Appendix \ref{subsec:comparing in-out}) is not present in \cite{MarinMacedo:2025jco}.}.


\section{Partial energy singularity: a corollary derived from discontinuity relation}\label{partial energy}

In this section we argue that starting from the single-cut discontinuity relation that we derived in \S\ref{confscalartreelevel}, one can gain further insight about the analytic structure of cosmological correlators. Particularly, our aim will be to look at what is known as the partial energy singularity of the cosmological correlators. 

It is known that the cosmological wave-function coefficients exhibit two types of singularities: total energy singularities which occur when the sum of external energies or absolute values of external momenta vanishes, and partial energy singularities arising when the total energy flowing into a interaction vertex vanishes \cite{Baumann:2020dch, Baumann:2021fxj}. Furthermore, the residue of the total energy singularity of a wave-function coefficient is known to give us the corresponding flat-space scattering amplitude \cite{Raju:2012zr, Baumann:2020dch, Baumann:2021fxj}. The same statement holds for the total energy singularity of cosmological correlators \cite{Goodhew:2020hob}. While making this statement, the choice of initial conditions as the Bunch-Davies one has been an important assumption, since otherwise one might encounter other singularities such as the folded singularities \footnote{Configurations where two or more momenta become collinear.}. Consequently, cosmological correlators also inherit these two types of singularities from the wave-function coefficients \cite{Goodhew:2020hob}. In the following, we analyze the behavior of cosmological correlators at a partial energy pole. 

Our strategy is to start from our single-cut discontinuity relation that we have derived for the cosmological correlators and take the limit $E_L \to 0$, where $E_L = k_1 + k_2 + p$ denotes the partial energy that flows into the left most vertex in a $r$-site correlator $\mathcal{B}^{(r)}$ considering cubic interaction in the bulk. To start with, we only consider the $2$-site correlator $\mathcal{B}^{(2)}(k_1,k_2,k_3,k_4;p)$ and the corresponding discntinuity relation was derived in eq.\eqref{2siteDiscFinal Result}, which we write here again for convenience
\begin{equation}\label{2siteDiscSec7}
	\begin{split}
		{\text {Disc}}_{p} \mathcal{B}^{(2)}( \lbrace \mathbf{k}_L, \mathbf{k}_R \rbrace ;p) = \frac{1}{2 P_p(\eta_0)} \Bigg(& \text{Disc}_p \mathcal{B}^{(1)}( \lbrace \mathbf{k}_{L} \rbrace ,p)  \times \text{Disc}_p \mathcal{B}^{(1)}(\lbrace \mathbf{k}_{R} \rbrace ,p) 
		\\
		& - \widetilde{\text{Disc}}_{p} \widetilde{\mathcal{B}}^{(1)}( \lbrace \mathbf{k}_{L} \rbrace ,p) \times \widetilde{\text{Disc}}_{p} \widetilde{\mathcal{B}}^{(1)}( \lbrace \mathbf{k}_{R} \rbrace ,p) \Bigg) \, .
	\end{split}
\end{equation}
Recalling the definition of ${\text {Disc}}_{p}$ and $ \widetilde{\text{Disc}}_{p}$ from eq.\eqref{defdisc} and eq.\eqref{defdisct}, we can now take the $E_L \to 0$ limit on the LHS and RHS of eq.\eqref{2siteDiscSec7} independently and then equate them to obtain our desired result. For the LHS we get 
\begin{equation}
	\label{paritialsingourLHS}
	\begin{split}
		\lim_{E_L \to 0} {\text {Disc}}_{p} \mathcal{B}^{(2)}( \lbrace \mathbf{k}_L, \mathbf{k}_R \rbrace ;p) & = \lim_{E_L \to 0} \left[\mathcal{B}^{(2)}(k_1,k_2,k_3,k_4;p)
		- \mathcal{B}^{(2)}(k_1,k_2,k_3,k_4;-p) \right]\\
		& 
		= \lim_{E_L \to 0} \mathcal{B}^{(2)}(k_1,k_2,k_3,k_4;p)\,,
	\end{split}
\end{equation} 
where we have used that the exchange correlator does not have a folded singularity due to our choice of Bunch-Davies initial condition. Also, note that $\mathcal{B}^{(2)}( \lbrace \mathbf{k}_L, \mathbf{k}_R \rbrace ;p)$ developes a pole as $E_L \to 0$, however, $ \mathcal{B}^{(2)}(k_1,k_2,k_3,k_4;-p)$ remains finite, as the latter has a pole at $E_L - 2p \to 0$. So in the strict $E_L \to 0$ limit, the second term, i.e., $\lim_{E_L \to 0} \mathcal{B}^{(2)}(k_1,k_2,k_3,k_4;-p)$ can be neglected and the last equality is justified. 

Similarly, for the RHS of eq.\eqref{2siteDiscSec7} we obtain
\begin{equation}
	\label{paritialsingourRHS}
	\begin{split}
		& \lim_{E_L \to 0} \bigg[ \frac{\text{Disc}_p \mathcal{B}^{(1)}( \lbrace \mathbf{k}_{L} \rbrace ,p)  ~ \text{Disc}_p \mathcal{B}^{(1)}(\lbrace \mathbf{k}_{R} \rbrace ,p) 
			- \widetilde{\text{Disc}}_{p} \widetilde{\mathcal{B}}^{(1)}( \lbrace \mathbf{k}_{L} \rbrace ,p)~\widetilde{\text{Disc}}_{p} \widetilde{\mathcal{B}}^{(1)}( \lbrace \mathbf{k}_{R} \rbrace ,p)}{2 P_p(\eta_0)}\bigg] \\ 
		&= \frac{1}{2P_p(\eta_0)} \lim_{E_L \to 0}\bigg[ \mathcal{B}^{(1)}(k_1,k_2,p)
		\left( \mathcal{B}^{(1)}(k_3,k_4,p) - \mathcal{B}^{(1)}(k_3,k_4,-p) \right)\\
		& \hspace{3cm}-\widetilde{\mathcal{B}}^{(1)}(k_1,k_2,p)
		\left( \widetilde{\mathcal{B}}^{(1)}(k_3,k_4,p) + \widetilde{\mathcal{B}}^{(1)}(k_3,k_4,-p) \right)\bigg]\,, 
	\end{split}
\end{equation}
in which we have excluded singularities in the contact correlator, $\lim_{E_L \to 0} \mathcal{B}^{(2)}(k_1,k_2,k_3,k_4;-p)$ and in the auxiliary object, $\lim_{E_L \to 0} \widetilde{\mathcal{B}}^{(2)}(k_1,k_2,k_3,k_4;-p)$ because of the reason mentioned above. 

Now, compairing eq.\eqref{paritialsingourLHS} and eq.\eqref{paritialsingourRHS} we get 
\begin{equation}
	\label{paritialsingour}
	\begin{split}
		\lim_{E_L \to 0} \mathcal{B}^{(2)}(k_1,k_2,k_3,k_4;p) =& \frac{1}{2P_p(\eta_0)} \lim_{E_L \to 0}\bigg[ \mathcal{B}^{(1)}(k_1,k_2,p)
		\left( \mathcal{B}^{(1)}(k_3,k_4,p) - \mathcal{B}^{(1)}(k_3,k_4,-p) \right)\\
		& -\widetilde{\mathcal{B}}^{(1)}(k_1,k_2,p)
		\left( \widetilde{\mathcal{B}}^{(1)}(k_3,k_4,p) + \widetilde{\mathcal{B}}^{(1)}(k_3,k_4,-p) \right)\bigg]\,.
	\end{split}
\end{equation} 
Eq.\eqref{paritialsingour} shows that the $4$-point $2$-site scalar correlator $\mathcal{B}^{(2)}(k_1,k_2,k_3,k_4;p)$ can be decomposed into a sum of two factorized contributions composed of lower-point data at the partial energy pole. It is worth highlighting that in the above relation the auxiliary lower-point objects $\widetilde{\mathcal{B}}^{(1)}$ are involved \footnote{This application of the cutting rules to obtain the decomposition of the residue at the partial energy singularity was not addressed in \cite{MarinMacedo:2025jco}.}. 

Eq.\eqref{paritialsingour} can also be generalized to higher-point correlators (involving higher sites in the bulk) as well. Using eq.\eqref{r site correlator}, we propose the following decomposition rule for any $r$-site correlator at a partial-energy singularity
\begin{equation} \label{El zero residue r site final}
	\begin{split}
		&\lim_{E_L \to 0} \mathcal{B}^{(r)}( \lbrace \mathbf{k}_L,..., \mathbf{k}_R \rbrace ; \lbrace p_1, ..., p_{r-1} \rbrace )\\&
		= \frac{1}{2 P_{p_{r-1}} (\eta_0)} \bigg[ \mathcal{B}^{(r-1)}\left(  \lbrace \mathbf{k}_L,...,\mathbf{k}_{M_{r-2}} \rbrace, p_{r-1} ; \lbrace p_1,..,p_{r-2} \rbrace \right) \text{Disc}_{p_{r-1}} \mathcal{B}^{(1)}( \lbrace \mathbf{k}_{R} \rbrace, p_{r-1})\\&\quad-\widetilde{\mathcal{B}}^{(r-1)}\left(  \lbrace \mathbf{k}_L,...,\mathbf{k}_{M_{r-2}} \rbrace, p_{r-1} ; \lbrace p_1,..,p_{r-2} \rbrace \right) \widetilde{\text{Disc}}_{p_{r-1}} \widetilde{\mathcal{B}}^{(1)}( \lbrace \mathbf{k}_{R} \rbrace, p_{r-1})\bigg]\,,
	\end{split}
\end{equation}
where, $E_L = k_L+\ldots+k_{M_{r-1}}+p_{r-1}$.

With our derived result eq.\eqref{paritialsingour} at hand, let us now verify it in a particular example. We consider a $2$-site correlator of conformally coupled scalars with $\phi^3$ interaction. The $2$-site correlator for this case is given in \cite{arkanihamed2015cosmologicalcolliderphysics}:
\begin{equation}
	\label{arkanimaldacena}
	\begin{split}
		\mathcal{B}^{(2)}(k_1,k_2,k_3,k_4;p) = \frac{\eta_0^4 (\lambda H)^2}{16\,k_1 k_2 k_3 k_4\,p} 
		\Bigg[&\text{Li}_2\Big(\frac{k_L-p}{k_L+k_R}\Big) + \text{Li}_2\Big(\frac{k_R-p}{k_L+k_R}\Big) \\
		&+ \log\Big(\frac{k_L+p}{k_L+k_R}\Big)\log\Big(\frac{k_R+p}{k_L+k_R}\Big) + \frac{\pi^2}{3} \Bigg]\,,
	\end{split}
\end{equation}
where $k_L = k_1 + k_2$ and $k_R = k_3 + k_4$. Substituting eq.\eqref{arkanimaldacena} in the LHS of eq.\eqref{paritialsingour} we get
\begin{equation}
	\label{partialenergypolecc4}
	\text{LHS} = \lim_{E_L\rightarrow0} \mathcal{B}^{(2)}(k_1,k_2,k_3,k_4;p) = \frac{H^2\eta_0^4 \lambda^2}{16\,k_1 k_2 k_3 k_4\,p} 
	\Bigg[ \log(k_L+p) \, \log\Big(\frac{k_R+p}{k_R-p}\Big) \Bigg]\,.
\end{equation}

Next, to evaluate the RHS of eq.\eqref{paritialsingour}, we require the corresponding contact correlator $\mathcal{B}^{(1)}$ and the auxiliary object $\widetilde{\mathcal{B}}^{(1)}$.  
We first consider the contact wave-function coefficient which is IR-divergent and given by
\begin{equation}
	\label{phi3def}
	\psi^{(1)}_{\phi^3} (\lbrace \mathbf{k}_L\rbrace,p ) = \frac{i\lambda}{H^4\eta_0^3}\log\big((k_L+p)\eta_0\big) + \frac{\lambda\pi}{2H^4\eta_0^3}\,.
\end{equation}
From this, we obtain
\begin{equation}
	\mathcal{B}^{(1)}(k_1,k_2,p) = \frac{\lambda \pi H^2 \eta_0^3}{8\,k_1 k_2 p}\,,
\end{equation}
and, using the definition in eq.\eqref{mathfrak B1},
\begin{equation}
	\widetilde{\mathcal{B}}^{(1)}(k_1,k_2,p) = \frac{i\,H^2 \eta_0^3 \lambda}{4 k_1 k_2 p} \, \log\big((k_1+k_2+p)\eta_0\big)\,.
\end{equation}

The RHS of eq.\eqref{paritialsingour} can now be expressed as
\begin{equation}
	\label{factorization4pt}
	\begin{split}
		\text{RHS} 
		&= \lim_{E_L \rightarrow 0} \Bigg[ \frac{H^2 \eta_0^4 \lambda^2 \pi^2}{32\, k_1 k_2 k_3 k_4 p} 
		+ \frac{H^2 \eta_0^4 \lambda^2}{16\, k_1 k_2 k_3 k_4 p} \, \log((k_1+k_2+p)\eta_0) 
		\log\Big(\frac{k_3+k_4+p}{k_3+k_4-p}\Big) \Bigg] \\
		&= \lim_{E_L \rightarrow 0} \frac{H^2 \eta_0^4 \lambda^2}{16\, k_1 k_2 k_3 k_4 p} \, 
		\log(k_L+p) \, \log\Big(\frac{k_R+p}{k_R-p}\Big)\,.
	\end{split}
\end{equation}
Comparing eqs.~\eqref{partialenergypolecc4} and~\eqref{factorization4pt}, we see a perfact match between the LHS and RHS of eq.\eqref{paritialsingour}, furthermore, confirming that the $4$-point $2$-site correlator can be decomposed as eq.\eqref{paritialsingour} at the partial-energy pole.

It is worth noting that this conformally coupled $\phi^3$ theory is an example where both $\mathcal{B}^{(1)}$ and $\widetilde{\mathcal{B}}^{(1)}$ appear at the same order in the leading late-time, i.e. at  $\mathcal{O}(\eta_0^3)$. Also, this decomposition of the residue at partial energy singularity serves as a consistency of our discontinuity relation derived previously. Most importantly, we must take note of the significance of the auxiliary objects $\widetilde{\mathcal{B}}^{(1)}$ which are not only present and play a crucial role in the discontinuity relation and also in the decomposition of the partial energy residues for the cosmological correlators.
\section{Conclusions and outlook} \label{conclusions}

In this section, we will conclude by discussing the implications of our main results and highlighting the novelties thereof. Following that, we will also comment on some possible future directions. 

First, we briefly summarise the technical results in this paper and their significance:
\begin{itemize}
	\item In eq.\eqref{discG} we have obtained the discontinuity of the bulk-to-bulk propagators in the in-in picture using Hermitian analyticity written in terms of the bulk-to-boundary propagators (see eq.\eqref{HA for Corr}) and unitarity valid for conformally coupled and massless scalar fields in $4$-dimensional de Sitter space-time. 
	
	\item In eq.\eqref{2siteDiscFinal Result} we have derived the single-cut discontinuity relation for a tree-level correlator with $2$-sites or vertices, working with conformally coupled scalar fields with polynomial $\phi^n$ type interactions. The discontinuity of the higher point correlator (written on the LHS of eq.\eqref{2siteDiscFinal Result}) gets decomposed into discontinuities of the lower-point correlator (denoted by $\mathcal{B}^{(1)}( \lbrace \mathbf{k}_L;p)$) and lower point auxiliary objects (denoted by $\widetilde{\mathcal{B}}^{(1)}( \lbrace \mathbf{k}_{L} \rbrace ,p)$). These auxiliary objects are not exactly cosmological correlators but can be written in terms of imaginary pieces of the lower-point wave-function coefficients (see eq.\eqref{mathfrak B1}). Also, on the RHS of eq.\eqref{2siteDiscFinal Result}, we have obtained two types of discontinuity operation - the `$\text{Disc}_p$' operator acts on the standard correlator $\mathcal{B}^{(1)}( \lbrace \mathbf{k}_L;p)$, and the other operator `$\widetilde{\text{Disc}}_p$' acts on the auxiliary objects $\widetilde{\mathcal{B}}^{(1)}( \lbrace \mathbf{k}_{L} \rbrace ,p)$. These operators are defined in eq.\eqref{defdisc} and eq.\eqref{defdisct} respectively.
	
	\item In eq.\eqref{r site correlator} we have generalised our single-cut discontinuity relation to a cosmological correlator with generic $r$-sites at tree-level, again focussing on conformally coupled and massless scalar field theory with $\phi^n$ interactions.   
	
	\item That the auxiliary objects $\widetilde{\mathcal{B}}^{(m)}$ appear in the discontinuity relations for the cosmological correlators is one of the most significant findings of our paper. We have checked that for conformally coupled scalar interactions $\phi^n$, which are IR-convergent, i.e., $n \ge 4$, depending on whether $n$ is even or odd, one or the other term on the RHS of eq.\eqref{2siteDiscFinal Result} contributes to the leading late-time limit. More specifically, for $n=$ even, the product of discontinuities of the lower point correlator $\mathcal{B}^{(1)}$ dominates as $\eta_0 \to 0^-$ over the other term involving $\widetilde{\mathcal{B}}^{(1)}$'s. On the contrary, for $n=$ odd, the product of $\widetilde{\text{Disc}}_p$'s acting on the lower-point auxiliary counterparts $\widetilde{\mathcal{B}}^{(1)}$ is the dominant contribution that matches exactly with the discontinuity of the $2$-site correlator on the LHS of eq.\eqref{2siteDiscFinal Result}. 
	
	We have made similar observations for tree-level diagrams with more than $2$-sites as well, where for $\phi^n$ interactions with $n=\text{even}$, the product of discontinuities of the lower point correlators $\mathcal{B}$ dominates at the late time limit, $\eta_0 \to 0^-$. However, for $n=\text{odd}$, contributions from both the lower point correlators and the auxiliary objects contribute. To be precise, e.g., following eq.\eqref{n odd B piece 3site} and \eqref{n odd Bt piece 3site} we see that for a $3$-site correlator, with $n=\text{odd}$ the product of discontinuity of the lower point correlators and the auxiliary objects scale at the same order in $\eta_0$ at the late time limit. Hence, both of them must be taken into account.
	
	Why the discontinuity relation works in this peculiar way in the leading late-time limit, depending on whether $n$ is even or odd in a $\phi^n$ theory, is an interesting question. Still, at this point, we lack a deeper understanding of this. 

    \item In Appendix \ref {2sitediscontinuityWVFN}, we derive the discontinuity relation for the $2$-site tree-level correlator, staying entirely within the wave-function picture, starting from the cutting rules already known for the wave-function coefficients. In the final result eq.\eqref{2site disc WVFN} we see that the auxiliary counterparts of the correlator, i.e., $\widetilde{\mathcal{B}}^{(1)}$'s are also present. This derivation provides further justification for the discontinuity relation obtained using the in-in picture. However, as far as extending this wave-function picture-based derivation to more general $r$-site correlator or loop diagrams is concerned, we think our derivation based on the in-in picture is more controlled and less technically involved. 
	
	\item In \S\ref{sec:generalization_cases}, we have generalized the discontinuity relation to massless scalars, cases with derivative interactions, and also to loop diagrams. In eq.\eqref{finalresult 1 loop} we have derived the discontinuity relation for $2$-site $1$-loop correlators with $\phi^n$ interactions. Based on an intuitive understanding of the structures, in eq.\eqref{finalresult l loop} we have conjectured a discontinuity relation for a $2$-site correlator but at the arbitrary $\ell$-loop order. 
	
	\item Using our discontinuity relation, we have also gained insight about the analytic pole-structure of cosmological correlators. In eq.\eqref{paritialsingour} we have derived the partial energy singularity of a $2$-site correlator. This equation demonstrates that in the limit when the total energy flowing in an internal site or vertex (i.e., the sum of the magnitude of individual momentum in the external legs and the internal line attached to that vertex) vanishes, the correlator develops a simple pole, and the residue gets decomposed into lower point objects and the discontinuity of them.
	
	\item As a check of principle, we have derived and checked the discontinuity relation for the flat space cosmological correlator at the tree-level in Appendix \ref{DiscFlatCorrelator}. It highlights the differences that one notices when we derive the discontinuity relation for de Sitter.  
	
	\item At various relevant places throughout the bulk of this paper, we have highlighted the similarities and differences of our analysis with \cite{MarinMacedo:2025jco}, which we learned about while our project was nearing completion. To summarise, in terms of the basic idea and the strategy, such as the in-in propagator identities and the tree-level discontinuity relations, there are similarities. However, our treatment of the loop diagrams is different. Furthermore, at the level of executing the strategy and various final interpretations of the discontinuity relations, our work provides further refinements and physical understanding. 
	
\end{itemize}

Next, we discuss some future directions that, in our opinion, will be worth pursuing. 
\begin{itemize}
	
	\item We have seen that the discontinuity relation for cosmological correlators gets decomposed into a sum of products of lower-point data. However, do we call them a factorization of the discontinuity relation? There might be a better basis of variables where the discontinuity relation for the cosmological correlators gets factorized in the true sense of its meaning. Another related interesting direction is to reformulate the cutting rule using alternative kinematic variables, such as spinor-helicity or twistor variables. Recent work has shown that cosmological (A)dS correlators, especially those involving spin, can be written in twistor space with remarkable simplification \cite{Baumann:2025twist,  David:2019dSspinor}. Such a reformulation could expose hidden simplicity, making kinematic constraints more manifest, and offer a more intuitive understanding of analytic structures, potentially including loop corrections. The factorisation aspect may also manifest when we use these alternative kinematic variables.
	
	\item In a recent work, it has been argued that cosmological correlators can be computed from flat-space Feynman diagrams. In several cases, this approach becomes easier than the existing in-in or wave-function approaches. This method is known as ``dressing rules" \cite{Chowdhury:2025ohm, Chowdhury:2025nnk}. Although an algorithm to determine the dressing rules for a particular cosmological correlator was developed, a basic understanding of why this fantastic idea of dressing Feynman rules yields the cosmological correlators remained a mystery. It would have been great if the dressing rules could be derived from fundamental principles, such as the optical theorem and cutting rules. We have been able to reproduce the dressing rules using the single-cut discontinuity relation, which has been formulated in this paper based on inputs from principles such as unitarity and hermitian analyticity. We will report this in detail in an upcoming work \cite{DKKK}. 
	
	\item In this work, we restricted our attention to scalar operators. A natural next step is to extend the single-cut discontinuity framework to spinning fields. The so-called cosmological bootstrap has demonstrated that wave-function coefficients with spin‑$1$ and spin‑$2$ exchanges can be systematically built using symmetry and factorization methods from simpler scalar ``seed" solutions \cite{Baumann:2020dch}. Similarly, by embedding spinning correlators into our cutting-rule framework, we can clarify how unitarity and analytic-structure constraints structure interactions involving spins.
	 
	
	\item Recent work on spinning ``unparticle" exchange at tree level \cite{Pimentel:2025rds} shows that unparticle propagators have enhanced symmetries, causing cosmological correlators to follow differential equations. An interesting question is whether a similar single-cut rule can be established for these propagators.
	
	
	\item In \cite{Chakraborty:2023los}, the authors introduced a framework for cosmological correlators in quantum gravity, noting the lack of local gauge-invariant observables, which sets it apart from conventional QFT. They observed that cosmological correlators in the Bunch-Davies vacuum do not exhibit increased symmetry beyond the solutions of the Wheeler–DeWitt equation. In follow-up work \cite{Chakraborty:2025izq}, they examined correlators in a general excited state as a perturbation of the Bunch-Davies vacuum. It will be interesting to explore if the states in \cite{Chakraborty:2023los} can be formulated within the single-cut discontinuity relation and how unitarity appears in this framework. 
	
	
	\item An interesting direction is to extend our cutting rule to cosmological correlators computed in non‑standard initial states related to the Bunch–Davies vacuum by a Bogoliubov transformation. Previous studies have shown that such vacua lead to additional analytic features (folded or collinear singularities) in late-time correlators \cite{Shukla:2016bnu, Jain:2022uja, Chopping:2024bog, Jain:2022uja, Ghosh:2024aqd, Ansari:2024pgq, Ghosh:2025pxn}. Investigating the applicability of cutting rules in this context would test the robustness of unitarity constraints and partial-energy factorization. It could provide new insights into the analytic structure of correlators beyond the standard vacuum. 
	
	
	\item Although the single-cut discontinuity relations we have derived in this paper are valid for any order in the perturbation theory, we want to make it clear that to derive them, we have imposed the unitarity constraint, which is perturbative in principle. It would be interesting to derive a set of single-cut discontinuity relations solely from the non-perturbative setting discussed in \cite{Loparco:2023rug, Hogervorst:2021uvp, SalehiVaziri:2024joi}.
\end{itemize}


\section*{Acknowledgements}

We would like to thank Chandramouli Chowdhury and Suvrat Raju for their useful comments on our manuscript. We are especially grateful to Chandramouli Chowdhuri for sharing notes with us and providing clarifications on several related topics through private communications. We appreciate the various stimulating conversations we had with Chandramouli Chowdhury, R. Loganayagam, Suvrat Raju, and Ashoke Sen. We are also thankful to Sourish Banerjee, Jyotirmoy Bhattacharya, Sabyasachi Chakraborty, Sanmay Ganguly, Sachin Jain, Apratim Kaviraj, Suman Kundu, Akashdeep Roy, and Ashish Shukla for valuable discussions. SD, DK, and BK would like to thank the organizers of the Indian Strings Meeting 2025 for the hospitality. NK would like to acknowledge the warm hospitality from ICTS, Bengaluru, BITS Goa, RKMVERI Belur, and SINP Kolkata during an academic visit while this work was in progress. DK and BK are supported by the Department of Atomic Energy, Government of India, under project no. RTI4001. NK acknowledges support from a MATRICS research grant (MTR/2022/000794) from the Anusandhan National Research Foundation (ANRF) and Science and Engineering Research Board (SERB), India. 

\appendix
\section{Notation and convention} \label{notation_convention}
In this appendix we summarise various short-hand notations that we defined in the bulk of this paper. These definitions have been repeatedly used in the technical calculations throughout the paper. \\

\noindent $\bullet$ For a polynomial interaction of type $\phi^n$ we have used the following shorthand notation:\\
\begin{equation}
\begin{split}
   & \text{Left most vertex}: \prod_{i=1}^{n-1} \left( 2 \mathbb{R}e \psi_2(k_i) \right) =  \mathcal{P}^{-1}_{\mathbf{k}_L}(\eta_0)~~~~~~~~~~~~~~~~~~~~~~~~~~~~~~~~(\text{defined in eq.}~\eqref{PkL PkR})
   \\
   & \text{Right most vertex}: \prod_{i=(r-1)(n-2)+2}^{r(n-2)+2} \left( 2 \mathbb{R}e \psi_2(k_i) \right) =  \mathcal{P}^{-1}_{\mathbf{k}_R}(\eta_0)~~~~~~~~~~~~~~~~~(\text{defined in eq.}~\eqref{PkL PkR})
   \\
   & \text{Any $j$-th vertex in the middle}: \prod_{i=j(n-2)+2}^{(j+1)(n-2)+1} \left( 2 \mathbb{R}e \psi_2(k_i) \right) =  \mathcal{P}^{-1}_{\mathbf{k}_{M_j}}(\eta_0)~~~(\text{defined in eq.}~\eqref{PkM})
\end{split}
\end{equation}
where $r$ is the number of vertices in the bulk.\\
$\bullet$ For the 2 point wave-function coefficient we use (e.g. see eq.\eqref{cosmo-cor-wvfn-def})
\begin{equation}
\begin{split}
   & \mathbb{R}e \psi_2(k_i) = \psi_2(k_i) = \frac{k_i}{H^2\eta_0^2} = \left( 2 P_{k_i}(\eta_0) \right)^{-1}
   \\
   & \text{Power spectrum}: P_k(\eta_0) = \frac{1}{2 \mathbb{R}e \psi_2(k)} 
\end{split}
\end{equation}
$\bullet$ For the objects at $r$-site we will use superscript $r$ 
\begin{equation}
\begin{split}
    & \text{Correlator}: \mathcal{B}^{(r)}\left(\lbrace k_i \rbrace;\lbrace p_i \rbrace\right)~~~~~~~~~~~~~~~~~~~~~~~(\text{see e.g. eq.}\eqref{corr 1 2 def})
    \\
    &  \text{Auxiliary object}: \widetilde{\mathcal{B}}^{(r)}\left(\lbrace k_i\rbrace;\lbrace p_i \rbrace\right)~~~~~~~~~~~~~~~(\text{see e.g. eq.}\eqref{corrt 1 2 def})
    \\
    & \text{Wavefunction coefficient}: \psi^{(r)}\left(\lbrace k_i\rbrace;\lbrace p_i \rbrace\right)~~~~~(\text{see e.g. eq.}\eqref{mathfrak B1})
\end{split}
\end{equation}
$\bullet$ For the time integral we use
\begin{equation}
\begin{split}
    & \left( \prod_{j=1}^n \int_{-\infty}^{\eta_0}\frac{d\eta_j}{H^4\eta_j^4} \right) = \int^{(n)} d\eta~~~~(\text{defined in eq.}\eqref{n time integrals})
    \\
    & \int (d\eta_r) = \int_{-\infty}^{\eta_0} \frac{d\eta_r}{(H \eta_r)^4}~~~~~~~~~~~~(\text{defined in eq.}\eqref{rth time integral})
\end{split}
\end{equation}
$\bullet$ For the product of bulk-to-boundary propagators for correlator we will use 
\begin{equation}
\begin{split}
    & \text{Left most vertex}: \prod_{i=1}^{n-1} K_{k_i}^+(\eta_0, \eta') = \mathcal{K}_{\eta_0 \eta'}^+(\lbrace \mathbf{k}_L \rbrace)~~~~~~~~~~~~~~~~~~~~~~~~~~~~~~~~(\text{defined in eq.}~\eqref{defmathcalK})
    \\
    & \text{Right most vertex}: \prod_{i=(r-1)(n-2)+2}^{r(n-2)+2} K_{k_i}^+(\eta_0, \eta') = \mathcal{K}_{\eta_0 \eta'}^+(\lbrace \mathbf{k}_R \rbrace)~~~~~~~~~~~~~~~~~(\text{defined in eq.}~\eqref{defmathcalK})
    \\
    & \text{Any $j$-th vertex in the middle}: \prod_{i=j(n-2)+2}^{(j+1)(n-2)+1} K_{k_i}^+(\eta_0, \eta') = \mathcal{K}_{\eta_0 \eta'}^+ (\lbrace \mathbf{k}_{M_j} \rbrace)~~(\text{used in eq.}~\eqref{3 site from inin})
\end{split}
\end{equation}
$\bullet$ Conversion between bulk-to-boundary propagators for correlator and that of the wave-function 
\begin{equation}
\begin{split}
    \prod_{i=1}^{n-1} K_{k_i}^+(\eta_0, \eta_i) K_{p_j}^+(\eta_0,\eta_i) = \frac{1}{\mathcal{P}^{-1}_{\mathbf{k}_L}(\eta_0) P_{p_j}^{-1}(\eta_0)}  \mathcal{K}_{\eta_0 \eta_i}^{\psi}(\lbrace \mathbf{k}_L,p_j\rbrace)~~~~~~~~~~~(\text{defined in eq.}~\eqref{relation mathcalK}) 
\end{split}
\end{equation}
where in the RHS, the superscript $\psi$ stands for the wave-function coefficient.

\section{Scalars with generic masses: analytic continuation, discontinuity relation} \label{genericmassscalar}
In the bulk of the paper we have focused on the special cases with half-integer $\nu$, in particular the conformally coupled and massless scalars. In this appendix we discuss the analytic continuation to the magnitude of the 3-momentum, discontinuity operation and the discontinuity relation for a $2$-site correlator with the generic $\nu$ case.

From eq.\eqref{mode f+-}, the mode function for a free scalar of mass $m$ in $dS_4$ is given as 
\begin{equation}
	\label{massivemodes}
	\begin{split}
		f_{k}^-(\eta) &= \frac{-i\sqrt{\pi}}{2} H\,e^{\frac{i \pi}{2}\left(\nu+\frac{1}{2} \right)}
		(-\eta)^{\frac{3}{2}} 
		\mathbb{H}^{(1)}_{\nu}(-k\eta)\,,
		\\f_{k}^+(\eta) &= \frac{i\sqrt{\pi}}{2} H\,e^{\frac{-i \pi}{2}\left(\nu+\frac{1}{2} \right)}
		(-\eta)^{\frac{3}{2}} 
		\mathbb{H}^{(2)}_{\nu}(-k\eta)\,,
	\end{split}
\end{equation}
where, $\nu$ is given by
\begin{equation} \label{nu def}
\nu = \sqrt{\frac{9}{4} - \frac{m^{2}}{H^{2}} } \, .
\end{equation}
Using the expressions mode functions, we get the power spectrum as following
\begin{equation}
	\label{psgeneralm}
	P_k(\eta_0) = \lim_{\eta_0\rightarrow 0}|f_k^-(\eta_0)|^2
	= H\,\eta_0^{3/2} \Bigg[
	\frac{\sqrt{\pi}\,(-k\eta_0)^{\nu}}{2^{1+\nu}\Gamma(\nu+1)}
	\left( 1 + i\cot(\pi\nu) \right)
	- 
	\frac{i(-k\eta_0)^{-\nu}}{\sqrt{\pi}\,2^{1-\nu}\Gamma(\nu)}
	\Bigg] + \cdots \, .
\end{equation}
For light mass fields where $m^2 < \frac{9}{4}H^2$, $\nu$ is real and positive. In this case, the first term in brackets of eq.\eqref{psgeneralm} approaches zero faster than the second
and can be neglected. So the power spectrum becomes
\begin{equation}
	\label{pslight}
	P_k(\eta_0)
	= \frac{H^2}{\pi^2 2^{2(\nu - 1)} \Gamma(\nu)^2}
	(-\eta_0)^{3 - 2\nu}\,k^{-2\nu}\,,
	\qquad  \text{for}\quad m^2 < \tfrac{9}{4}H^2 \, .
\end{equation}
The second case is when fields are massive and
\(m^2 > \frac{9}{4}H^2\), so that \(\nu\) becomes complex. In this case both of the two
terms inside the brackets of eq.\eqref{psgeneralm} are of the same order.
The resulting power spectrum oscillates while decaying as \(\eta_0^3\).

Now, let us focus on convergence of any bulk-time integration for general massive field. Using the asymptotic form of Hankel functions in \cite{DLMF}, we can write following from eq.\eqref{massivemodes}
\begin{equation} \label{asympmodefn}
	\begin{split}
		&f_{k}^-(\eta)\sim \frac{\,H\eta\,e^{-ik\eta}}{\sqrt{2k}}\,, \qquad 
		f_{k}^+(\eta)\sim \frac{\,H\eta\,e^{ik\eta}}{\sqrt{2k}}\,, \qquad \text{as} ~ \eta\rightarrow-\infty \, .
	\end{split}
\end{equation}
Since the in-in propagators are constructed from the mode functions, we can write their asymptotic behaviours readily using eq,\eqref{asympmodefn}. Let us write down the asymptotic early time behavior of in-in bulk-to-boundary propagators, defined in eq.\eqref{in in bulk-bndry prop}, 
\begin{equation}
	\label{asympK}
	\begin{split}
		&\lim_{\eta\rightarrow-\infty}K_k^{+}(\eta_0, \eta) = \lim_{\eta\rightarrow-\infty}\bigg(f_{k}^-(\eta_0) f_{k}^+(\eta)\bigg)\sim \eta\,e^{ik\eta} \, ,\\
		&
		\lim_{\eta\rightarrow-\infty}K_k^{-}(\eta_0, \eta) = \lim_{\eta\rightarrow-\infty}\bigg(f_{k}^+(\eta_0) f_{k}^-(\eta)\bigg)\sim \eta\,e^{-ik\eta} \, .
	\end{split}
\end{equation}
This shows oscillatory behavior at early time ($\eta\rightarrow-\infty$) and this should be regularized by some damping factor to ensure the convergence of bulk time integral running from $-\infty$ to 0. Usually this is done by rotating the conformal time corrdinate $\eta\rightarrow \eta(1\pm i\epsilon')$ (with $\epsilon'\rightarrow0^+$). We can achieve the same without rotating $\eta$-coordinate but instead analytically continuing momenta, $k\rightarrow (k\pm i\epsilon)$ (with $\epsilon\rightarrow0^+$). From the asymptotic expressions of in-in bulk-to-boundary propagators in eq.\eqref{asympK}, we realize that the convergence of bulk time integral requires that we analytically continue the momenta of $K_k^{+}$ to the lower-half complex k-plane. Similarly, the momenta of $K_k^{-}$ needs to be analytically continued to the upper-half complex k-plane. For convergence of bulk-time integral, we will define the Hankel functions as function of the following momenta
\begin{equation}
\label{hankelconvergence}
	\begin{split}
		&\mathbb{H}^{(1)}_{\nu}(-k\eta)\rightarrow   \mathbb{H}^{(1)}_{\nu}\left(-k^- \eta \right)\quad;\, \text{such that} ~~ k^- = (k-i\epsilon) \, ,~ \text{and}\\
		& \mathbb{H}^{(2)}_{\nu}(-k\eta)\rightarrow   \mathbb{H}^{(2)}_{\nu}\left(-k^+ \eta \right)\quad;\, \text{such that} ~~  k^+ = (k+i\epsilon)\,.
	\end{split}
\end{equation}
Now, we focus on analytically continuing the mode functions from the positive $k$-axis to the negative $k$-axis\footnote{We will analytically continue only the magnitude of the cut-line (any internal line) momentum to the negative $k$-axis.}
. This continuation is subtle because the Hankel functions $\mathbb{H}^{(1)}_{\nu}(-k\eta)$ and $\mathbb{H}^{(2)}_{\nu}(-k\eta)$ for generic $\nu$ (not half-integer) have a branch cut along the negative real axis, $k \in (-\infty,0)$, in the complex $k$-plane.

To maintain convergence of the Hankel functions throughout the analytic continuation, we must specify whether the continuation approaches the branch cut from above or below. Using eq.~\eqref{hankelconvergence}, we find that $\mathbb{H}^{(2)}_{\nu}$ should be continued from $k \to -k$ from below, while $\mathbb{H}^{(1)}_{\nu}$ should be continued from above\footnote{The conjugation in the analytic continuation implies
\[
k^+ \rightarrow e^{i\pi}(k^+)^* \implies (k+i\epsilon) \rightarrow (-k+i\epsilon),
\]
indicating continuation in the upper half of the complex $k$-plane. Similarly, the analytic continuation of $k^-$ occurs in the lower half of the complex $k$-plane.}.
\begin{equation}
	\begin{split}
		&\mathbb{H}^{(1)}_{\nu}(-k^+\eta)\xrightarrow{k^+\rightarrow e^{i\pi}(k^+)^*}  
		\mathbb{H}^{(1)}_{\nu}((k^+)^*\eta)\,,\qquad
		\mathbb{H}^{(2)}_{\nu}(-k^-\eta)\xrightarrow{k^-\rightarrow e^{-i\pi}(k^-)^*}   
		\mathbb{H}^{(2)}_{\nu}((k^-)^*\eta)\,.
	\end{split}
\end{equation}
Now, we can use the following properties of Hankel functions \cite{DLMF}
\begin{equation}
	\label{hankelid}
	\begin{split}
		&\mathbb{H}^{(1)}_{\nu}(z\,e^{i\pi}) = -e^{-\nu\pi i}\, \mathbb{H}^{(2)}_{\nu}(z)\,,\qquad 
		\mathbb{H}^{(2)}_{\nu}(z\,e^{-i\pi}) = -e^{\nu\pi i}\, \mathbb{H}^{(1)}_{\nu}(z)\,.
	\end{split}
\end{equation}
Using the relations written above the following property of bulk-to-boundary propagator in wave-function picture can be found
\begin{equation}
	\label{bbrelationw}
	\begin{split}
		& \left( K^{\psi}_{-k^*}(\eta_0,\eta) \right)^*= K_k^{\psi}(\eta_0,\eta)\,,\qquad \text{or, equivalently,} \qquad 
		K^{\psi}_{-k^*}(\eta_0,\eta) = \left( K_k^{\psi}(\eta_0,\eta) \right)^*\,,
	\end{split}
\end{equation}
which can be verified using the expression of bulk-to-boundary propgator in eq. eq.\eqref{wvfnbulkbndry} and the property eq.\eqref{hankelid}. The relation in eq.\eqref{bbrelationw} was mentioned as Hermitian analyticity in \cite{Goodhew:2020hob}, which we have also used in eq.\eqref{defHA}. 

Furthermore, using the relation between bulk-to-boundary propagators in the wave-function picture and the in-in picture, eq.\eqref{bulkboundary WF and inin}, we can write eq.\eqref{bbrelationw} in terms of the in-in bulk-to-boundary propagators
\begin{equation}
	\label{bbrelationinin}
	\begin{split}
		&K_{-k^*}^{\mp}(\eta_0, \eta) = K_{k}^\pm (\eta_0, \eta) \frac{P_{-k^*}(\eta_0)}{P_{k}(\eta_0)} \, ,
	\end{split}
\end{equation}
where $P_k(\eta_0)$ is given in eq.~\eqref{psgeneralm}. From the expression of $P_k(\eta_0)$, we observe that a branch point can occur at $k=0$ for certain values of $\nu$. In all such cases, the multivaluedness of $P_{-k^*}(\eta_0)$ can be resolved by analytically continuing from below. Eq.~\eqref{bbrelationinin} can then be interpreted as the statement of Hermitian analyticity in terms of the in-in bulk-to-boundary propagators, corresponding to eq.~\eqref{in-in prop2} where we restrict to real momenta.

Now, we can substitute eq.\eqref{bbrelationinin} in the expressions of the bulk-to-bulk propagators written in the in-in picture in eq.\eqref{in in bulk to bulk prop} and obtain 
\begin{equation}
	\label{cutbbiningeneralm}
	\begin{split}
		&G^{++}_{-p^*}(\eta_1,\eta_2) = G^{--}_{p}(\eta_1,\eta_2)/\,\mathcal{R}\quad,\quad
		G^{+-}_{-p^*}(\eta_1,\eta_2) = G^{-+}_{p}(\eta_1,\eta_2)/\,\mathcal{R}\,,\\&
		G^{-+}_{-p^*}(\eta_1,\eta_2) = G^{+-}_{p}(\eta_1,\eta_2)/\,\mathcal{R}\quad,\quad
		G^{--}_{-p^*}(\eta_1,\eta_2) = G^{++}_{p}(\eta_1,\eta_2)/\,\mathcal{R}\, ,
	\end{split}
\end{equation}
where, $\mathcal{R} $ is given by
\begin{equation}
	\label{defR}
	\mathcal{R} = \frac{P_{p}(\eta_0)}{P_{-p^*}(\eta_0)} \,.
\end{equation}
Note that eq.\eqref{cutbbiningeneralm} is the statement of Hermitian analyticity written in terms of bulk-to-bulk in-in propagators and is the generalisation of eq.\eqref{HA for Corr 1} for generic $\nu$.

Following the same steps as in \S\ref{ininpropagatorfromunitarity} we can argue that the following relations can be derived
\begin{equation} \label{discGgenericnu}
    G^{\pm \pm}_{p}(\eta_1, \eta_2) +  \mathcal{R}\, G^{\pm \pm}_{-p^*}(\eta_1, \eta_2) = \frac{K_p^-(\eta_0, \eta_1) K_p^+(\eta_0, \eta_2) + K_p^+(\eta_0, \eta_1) K_p^-(\eta_0, \eta_2)}{P_p(\eta_0)}\, ,
\end{equation}
which are the statements of Hermitian analyticity and unitarity written in terms of bulk-to-bulk in-in propagators and are the generalisation of eq.\eqref{discG} which has a restricted validity for half-integer $\nu$.  

Applying these modifications to the derivation of the cutting rule for the $2$-site correlator, we obtain the following result, which holds for scalars with arbitrary generic $\nu$
\begin{equation}
	\label{discb2tgeneric}
	\begin{split}
		& \mathcal{B}^{(2)}( \lbrace \mathbf{k}_L, \mathbf{k}_R \rbrace ;p) +\,\mathcal{R}\,\mathcal{B}^{(2)}( \lbrace \mathbf{k}_L, \mathbf{k}_R \rbrace ;-p^*) =  \\&\frac{1}{2\,P_{p}(\eta_0)}\bigg[ \bigg(\mathcal{B}^{(1)}( \lbrace \mathbf{k}_L \rbrace,p)+\,\mathcal{R}\,\mathcal{B}^{(1)}( \lbrace \mathbf{k}_L \rbrace,-p^*)\bigg) \bigg(\mathcal{B}^{(1)}( \lbrace \mathbf{k}_R \rbrace,p)+\,\mathcal{R}\,\mathcal{B}^{(1)}( \lbrace \mathbf{k}_R \rbrace,-p^*)\bigg)  \\
		& \qquad -  \bigg(\widetilde{\mathcal{B}}^{(1)}( \lbrace \mathbf{k}_{L} \rbrace,p)-\,\mathcal{R}\,\widetilde{\mathcal{B}}^{(1)}( \lbrace \mathbf{k}_{L} \rbrace,-p^*)\bigg)\bigg(\widetilde{\mathcal{B}}^{(1)}( \lbrace \mathbf{k}_{R} \rbrace,p)-\,\mathcal{R}\,\widetilde{\mathcal{B}}^{(1)}( \lbrace \mathbf{k}_{R} \rbrace,-p^*)\bigg)\bigg]\, .
	\end{split}
\end{equation}
Note that this is the discontinuity relation involving a $2$-site cosmological correlator in a theory of scalar fields but with generic masses. 

If we restrict ourselves to scalar fields with integer valued $\nu$ (that includes conformally coupled and massless scalars), using eq.\eqref{pslight}, eq.\eqref{defR}, we get $\mathcal{R} = -1$. Substituting $\mathcal{R} = -1$, the LHS of eq.\eqref{discb2tgeneric} reduces to
\begin{equation}
\label{discgenericR-1}
	\mathcal{B}^{(2)}( \lbrace \mathbf{k}_L, \mathbf{k}_R \rbrace ;p) -\,\mathcal{B}^{(2)}( \lbrace \mathbf{k}_L, \mathbf{k}_R \rbrace ;-p^*)\, .
\end{equation}
By redefining the complex momentum $p$ as $p = i\, q$ (with $q \in \mathbb{R}$) \footnote{This treatment has also been done for interpreting cosmological optical theorem of wave-function coefficients in \cite{Goodhew:2020hob}.}, we can write
\begin{equation}
\label{discgenericB2R-1}
	\begin{split}
		& \mathcal{B}^{(2)}( \lbrace \mathbf{k}_L, \mathbf{k}_R \rbrace ;p) -\,\mathcal{B}^{(2)}( \lbrace \mathbf{k}_L, \mathbf{k}_R \rbrace ;-p^*) \\ 
		& = \lim_{\epsilon\to 0^+}   \mathcal{B}^{(2)}( \lbrace \mathbf{k}_L, \mathbf{k}_R \rbrace ;i\, q-\epsilon) -\,\mathcal{B}^{(2)}( \lbrace \mathbf{k}_L, \mathbf{k}_R \rbrace ;i\,q+\epsilon)\Big]  = \text{Disc}\!\left[   \mathcal{B}^{(2)}( \lbrace \mathbf{k}_L, \mathbf{k}_R \rbrace ;i \,q)\right] .
	\end{split}
\end{equation}
This shows that, for scalar fields with half-integer $\nu$, our discontinuity operation for correlators with complex momenta reduces to the standard notion of discontinuity in complex analysis.

For scalar fields with half-integer $\nu$ (including the massless and conformally coupled cases), the Hankel functions $\mathbb{H}^{(1)}_{\nu}(-k\eta)$ and $\mathbb{H}^{(2)}_{\nu}(-k\eta)$ for  $\nu = n + \tfrac12$ (with $n \in \mathbb{Z}_{\ge 0}$) are defined on the principal branch \cite{DLMF} as follows,
\begin{equation}
\begin{split}
\mathbb{H}^{(1)}_{n+\frac12}(z)
&= \sqrt{\frac{2}{\pi z}}\,
e^{iz}
\sum_{k=0}^{n}
\frac{i^{\,(k-n-1)}\,(n+k)!}{k!\,(n-k)!\,(2z)^k},
\\
\mathbb{H}^{(2)}_{n+\frac12}(z)
&= \sqrt{\frac{2}{\pi z}}\,
e^{-iz}
\sum_{k=0}^{n}
\frac{(-i)^{\,(k-n-1)}\,(n+k)!}{k!\,(n-k)!\,(2z)^k}\,.
\end{split}
\end{equation}
From these series expansions and eq.~\eqref{in in bulk-bndry prop}, we see that the in-in bulk-to-boundary propagators have a pole only at $k=0$ for scalar fields with half-integer $\nu$\footnote{For half-integer $\nu$, $\mathbb{H}^{(1,2)}_{\nu}(-k\eta)$ include a factor of $\sqrt{k}$, which is multivalued along the negative $k$-axis. But, in-in propagators are built from products of two mode functions, making them single-valued on the negative $k$-axis.}
. This eliminates any ambiguity in analytically continuing the in-in propagators to the negative $k$ axis. In such cases, the correlator can be continued to the negative $k$-axis without introducing $i\epsilon$ shifts, which allows us to write eq.~\eqref{discgenericR-1} as follows.
\begin{equation}
	\text{LHS of eq.\eqref{discb2tgeneric}} = \mathcal{B}^{(2)}( \lbrace \mathbf{k}_L, \mathbf{k}_R \rbrace ;p) -\,\mathcal{B}^{(2)}( \lbrace \mathbf{k}_L, \mathbf{k}_R \rbrace ;-p)\, .
\end{equation}
According, to our definition in eq.\eqref{defdisc}, this becomes 
\[
	\text{Disc}_p\,\mathcal{B}^{(2)}( \lbrace \mathbf{k}_L, \mathbf{k}_R \rbrace ;p) = \mathcal{B}^{(2)}( \lbrace \mathbf{k}_L, \mathbf{k}_R \rbrace ;p) -\,\mathcal{B}^{(2)}( \lbrace \mathbf{k}_L, \mathbf{k}_R \rbrace ;-p)\, .
\]
 One might worry that subtleties could still arise for half-integer $\nu$ in IR-divergent settings. However, analytically continuing the internal energy to negative values does not introduce any multivaluedness, even when IR divergence is present \footnote{For example, analytically continuing the internal energy to negative values sends $\log(k_1+k_2+p_1)\to \log(k_1+k_2-p_1)$, where $p_1 = |\vec{k}_1+\vec{k}_2|$. This remains single valued because the argument of the logarithm stays positive by the triangle inequality.} in the half-integer $\nu$ cases.

The preceding set of arguments also justifies removing the complex conjugation on the momenta and writing the cutting rule for the correlator as in eq.\eqref{2siteDiscFinal Result}. It also suggests that our cutting rule for the exchange correlator in eq.\eqref{2siteDiscFinal Result} remains valid in IR-divergent scenarios, provided $\nu$ is a half-integer.

The situation is more subtle for fields with non–half-integer $\nu$, where we could not find any closed-form expression for the bulk time integral. In these cases, one must follow our analytic continuation prescription, explicitly approaching the negative-$k$ axis from below to fix the branch structure of $\mathcal{R}$. We expect eq.~\eqref{discb2tgeneric} to hold in this setting, although it has not been explicitly tested.

\section{Discontinuity of $2$-site correlator using wave-function cutting rules} \label{2sitediscontinuityWVFN}
In this appendix, we show that the discontinuity relation that we have obtained from the in-in picture in \S\ref{2-site correlator}, can also be derived from the wave-function formalism. In particular, the strategy we will follow is that we will first write a $2$-site tree-level correlator in terms of the wave-function coefficients. In the next step, we will use the definition of the discontinuity operation and then by using the cutting-rules derived in \cite{Goodhew:2021oqg} at the wave-function level we will derive the discontinuity relation for the correlator.

For convenience, we will restrict ourselves to the following polynomial self-interaction of conformally coupled scalar field written in eq.\eqref{HInt}.
Although our single-cut rule perfectly works for the IR divergent interaction, in the current context to avoid the technical complications, we will restrict to $n \geq 4$ which describes the IR-convergent interaction. In this theory, a $2$-site correlator can be expressed in terms of the wave-function coefficients as follows
\begin{equation}\label{2site corr def in terms of WF}
\begin{split}
	\mathcal{B}^{(2)}(\lbrace \mathbf{k}_L, \mathbf{k}_R \rbrace; p) = \frac{2\left( \mathbb{R}e \psi^{(2)}(\lbrace \mathbf{k}_L, \mathbf{k}_R \rbrace; p) + \frac{\mathbb{R}e \psi^{(1)}(\lbrace \mathbf{k}_L\rbrace, p) ~ \mathbb{R}e \psi^{(1)}(\lbrace \mathbf{k}_R \rbrace, p)}{\mathbb{R}e \psi_2(p)} \right)}{\mathcal{P}^{-1}_{\mathbf{k}_L}(\eta_0) \mathcal{P}^{-1}_{\mathbf{k}_R}(\eta_0)} \, ,
\end{split}
\end{equation}
where $p$ is the bulk-to-bulk momenta and $\lbrace \mathbf{k}_L \rbrace = \lbrace k_1, \cdots, k_{n-1} \rbrace$, $\lbrace \mathbf{k}_R \rbrace = \lbrace k_n, \cdots, k_{2n-2} \rbrace$. 
Applying the discontinuity operation eq.\eqref{defdisc} in the $2$-site correlator, by definition we obtain the following expression that we need to expand in terms of the wave-function coefficients
\begin{equation}
\begin{split}
	\text{Disc}_p  \mathcal{B}^{(2)}(\lbrace \mathbf{k}_L, \mathbf{k}_R \rbrace; p)  =  \mathcal{B}^{(2)}(\lbrace \mathbf{k}_L, \mathbf{k}_R \rbrace; p)  -  \mathcal{B}^{(2)}(\lbrace \mathbf{k}_L, \mathbf{k}_R \rbrace; -p) \, . 
\end{split}
\end{equation}
Employing eq.\eqref{2site corr def in terms of WF} we can write the discontinuity of the $2$-site correlator as follows 
\begin{equation}\label{2site corr disc in terms of WF}
\begin{split}
	 & \text{Disc}_p  \mathcal{B}^{(2)}(\lbrace k_i \rbrace; p) =\frac{2}{\mathcal{P}^{-1}_{\mathbf{k}_L}(\eta_0) \mathcal{P}^{-1}_{\mathbf{k}_R}(\eta_0)} \Bigg(  \mathbb{R}e \psi^{(2)}(\lbrace \mathbf{k}_L, \mathbf{k}_R \rbrace; p) - \mathbb{R}e \psi^{(2)}(\lbrace \mathbf{k}_L, \mathbf{k}_R \rbrace; -p) 
	\\
	& + \frac{\mathbb{R}e \psi^{(1)}(\lbrace \mathbf{k}_L\rbrace, p) ~ \mathbb{R}e \psi^{(1)}(\lbrace \mathbf{k}_R \rbrace, p)}{\mathbb{R}e \psi_2(p)} + \frac{\mathbb{R}e \psi^{(1)}(\lbrace \mathbf{k}_L\rbrace, -p) ~ \mathbb{R}e \psi^{(1)}(\lbrace \mathbf{k}_R \rbrace, -p)}{\mathbb{R}e \psi_2(p)}\Bigg) \, .
\end{split}
\end{equation}
At this point we have to make use of the cutting rules \cite{Goodhew:2020hob,Goodhew:2021oqg,Goodhew:2023bcu} for the wave-function coefficients. Note that the terms in the first line on the RHS of the above equation can be written as follows
\begin{equation}\label{real part of 2site Disc WF}
\begin{split}
	\mathbb{R}e \Bigg( \psi^{(2)}(\lbrace \mathbf{k}_L, \mathbf{k}_R \rbrace; p) - \psi^{(2)}(\lbrace \mathbf{k}_L, \mathbf{k}_R \rbrace; -p) \Bigg) =  \mathbb{R}e \Bigg( \text{Disc}_p \psi^{(2)}(\lbrace \mathbf{k}_L, \mathbf{k}_R \rbrace; p) \Bigg) \, .
\end{split}
\end{equation}
Cutting rule for a tree level wave-function coefficient suggests that the discontinuity of the exchange diagram can be thought of as a product of the discontinuity of two contact level wave-function coefficients as follows\footnote{Our convention of the Disc. is slightly different from the convention used in \cite{Goodhew:2020hob}. However, for a conformally coupled and real scalar polynomial interaction the final result of the discontinuity of a wave-function coefficient turns out to be same in both the conventions.}
\begin{equation}\label{2site WF cut}
\begin{split}
	\text{Disc}_{p} \psi^{(2)}\left( \lbrace \mathbf{k}_L, \mathbf{k}_R \rbrace; p\right) = - \frac{1}{2 \mathbb{R}e \psi_2(p)} \text{Disc}_{p} \psi^{(1)}\left(\lbrace \mathbf{k}_L \rbrace, p\right) \text{Disc}_{p} \psi^{(1)}\left(\lbrace \mathbf{k}_R \rbrace, p \right) \, .
\end{split}
\end{equation}
For a contact correlator, from the definition of discontinuity we obtain the following
\begin{equation}\label{1site WF Disc}
\begin{split}
	&\text{Disc}_{p} \psi^{(1)}\left(\lbrace \mathbf{k}_L\rbrace, p\right)  = \psi^{(1)}(\lbrace \mathbf{k}_L \rbrace, p)- \psi^{(1)}(\lbrace \mathbf{k}_L \rbrace, -p)  
	\\
	=& \Big( \mathbb{R}e\psi^{(1)}(\lbrace \mathbf{k}_L \rbrace, p) - \mathbb{R}e\psi^{(1)}(\lbrace \mathbf{k}_L \rbrace, -p) \Big) + \Big( i \mathbb{I}m \psi^{(1)} (\lbrace \mathbf{k}_L \rbrace, p) - i \mathbb{I}m \psi^{(1)} (\lbrace \mathbf{k}_L \rbrace, -p) \Big)  \, .
\end{split}
\end{equation}
Substituting eq.\eqref{1site WF Disc} and eq.\eqref{2site WF cut} in eq.\eqref{real part of 2site Disc WF}, after some simple algebraic simplification, yields the following expression 
\begin{equation}
\begin{split}
	& \mathbb{R}e \Bigg( \psi^{(2)}(\lbrace \mathbf{k}_L, \mathbf{k}_R \rbrace; p) - \psi^{(2)}(\lbrace \mathbf{k}_L, \mathbf{k}_R \rbrace; -p) \Bigg) = - \frac{1}{2 \mathbb{R}e \psi_2(p)} \times 
	\\
	& \Big[ \Big( \mathbb{R}e\psi^{(1)}(\lbrace \mathbf{k}_L\rbrace, p) - \mathbb{R}e\psi^{(1)}(\lbrace \mathbf{k}_L\rbrace, -p) \Big) \Big( \mathbb{R}e \psi^{(1)}(\lbrace \mathbf{k}_R\rbrace, p) - \mathbb{R}e \psi^{(1)}(\lbrace \mathbf{k}_R\rbrace, -p) \Big) +
	\\
	&\Big( i \, \mathbb{I}m \psi^{(1)} (\lbrace \mathbf{k}_L\rbrace, p) -  i  \, \mathbb{I}m \psi^{(1)} (\lbrace \mathbf{k}_L\rbrace, -p) \Big)  \Big( i \,  \mathbb{I}m \psi^{(1)} (\lbrace \mathbf{k}_R\rbrace, p) -  i \,  \mathbb{I}m \psi^{(1)} (\lbrace \mathbf{k}_R\rbrace, -p) \Big)\Big] \, .
\end{split}
\end{equation}
With the above expression, the RHS of the eq.\eqref{2site corr disc in terms of WF} cab be simplified and put in the following form 
\begin{equation}\label{2site cut from WF prior to final}
\begin{split}
	\text{Disc}_p  \mathcal{B}^{(2)}(\lbrace k_i \rbrace; p) = &\frac{2}{\mathcal{P}^{-1}_{\mathbf{k}_L}(\eta_0) \mathcal{P}^{-1}_{\mathbf{k}_R}(\eta_0) P_p^{-1}(\eta_0)} \times \\
	& \Bigg( \Big( \mathbb{R}e\psi^{(1)}(\lbrace \mathbf{k}_L\rbrace, p) + \mathbb{R}e\psi^{(1)}(\lbrace \mathbf{k}_L\rbrace, -p) \Big) \Big( \lbrace \mathbf{k}_L \rbrace  \to \lbrace \mathbf{k}_R \rbrace \Big)
	\\
	& - \Big( i\,\mathbb{I}m \psi^{(1)} (\lbrace \mathbf{k}_L\rbrace, p) -  i \,\mathbb{I}m \psi^{(1)} (\lbrace \mathbf{k}_L\rbrace, -p) \Big) \Big( \lbrace \mathbf{k}_L \rbrace  \to \lbrace \mathbf{k}_R \rbrace \Big) \Bigg).
\end{split}
\end{equation}
At this level, it is worth highlting that in the RHS of the above equation, the first line inside the parenthesis will contribute for $n=$ even interaction only. This happens because a contact wave-function coefficient is completely real only for conformally coupled even interaction. On the other hand, the second line inside the parenthesis will contribute for $n=$ odd since a contact wave-function is completely imaginary for conformally coupled odd interaction.

The above expression can be further written in terms of the contact correlator $\mathcal{B}^{(1)}(\lbrace \mathbf{k} \rbrace, p)$ and the auxiliary object $\widetilde{\mathcal{B}}^{(1)}(\lbrace \mathbf{k}\rbrace,p)$ we have encountered before. Note that, a contact correlator can be expressed in terms of contact wave-function coefficient by the following relation 
\begin{equation}\label{def B1}
\begin{split}
	\mathcal{B}^{(1)} (\lbrace \mathbf{k}_L \rbrace, p)= \frac{2 ~\mathbb{R}e \psi^{(1)}(\lbrace \mathbf{k}_L \rbrace, p)}{\mathcal{P}^{-1}_{\mathbf{k}_L}(\eta_0) P_p^{-1}(\eta_0)} \, ,
\end{split}
\end{equation}
and recall that we defined 
\begin{equation}\label{def B1 tilde}
\widetilde{\mathcal{B}}^{(1)} (\lbrace \mathbf{k}_L \rbrace, p)= \frac{2i ~\mathbb{I}m \psi^{(1)}(\lbrace \mathbf{k}_L \rbrace, p)}{\mathcal{P}^{-1}_{\mathbf{k}_L}(\eta_0) P_p^{-1}(\eta_0)} \, .
\end{equation}
By using eq.\eqref{def B1} and eq.\eqref{def B1 tilde} we can finally rewrite the eq.\eqref{2site cut from WF prior to final} as follows
\begin{equation} \label{2site disc WVFN}
\begin{split}
	{\text {Disc}}_{p} \mathcal{B}^{(2)}( \lbrace \mathbf{k}_L, \mathbf{k}_R \rbrace ;p) = & \frac{1}{2 P_p(\eta_0)} \Bigg( \text{Disc}_p \mathcal{B}^{(1)}( \lbrace \mathbf{k}_{L} \rbrace ,p) \times \text{Disc}_p \mathcal{B}^{(1)}(\lbrace \mathbf{k}_{R} \rbrace ,p) 
	\\
	& - \widetilde{\text{Disc}}_{p} \widetilde{\mathcal{B}}^{(1)}( \lbrace \mathbf{k}_{L} \rbrace ,p) \times \widetilde{\text{Disc}}_{p} \widetilde{\mathcal{B}}^{(1)}( \lbrace \mathbf{k}_{R} \rbrace ,p) \Bigg) \, ,
\end{split}
\end{equation}
which is essentially the single-cut rule we have derived in eq.\eqref{2siteDiscFinal Result} starting directly from the in-in perspective of $2$-site correlator.

\section{A single-cut discontinuity relation for $3$-site tree-level correlator}\label{3 site factorization}
Here we will derive the single-cut rule for 3-site correlator for the polynomial interaction written in eq.\eqref{HInt}.\\
A 3-site correlator contains 3 vertices, so there are $8$ possible ways to label the vertices as $``+"$ or $``-"$. Therefore the full  correlator $\mathcal{B}^{(3)}$ is the sum of the contributions from each of these configurations, which is expressed as follows

\begin{equation}\label{3 site from inin}
	\begin{split}
		& \mathcal{B}^{(3)} \left( \lbrace \mathbf{k}_L, \mathbf{k}_M , \mathbf{k}_R \rbrace ;p_1;p_2 \right) \\ 
		& = (i\lambda)^3 \int^{(3)} d\eta \Bigg[ \Bigg( \mathcal{K}_{\eta_0 \eta_1}^+(\lbrace \mathbf{k}_L \rbrace) G^{++}_{p_1}(\eta_1,\eta_2) \mathcal{K}_{\eta_0 \eta_2}^+(\lbrace \mathbf{k}_M \rbrace) G^{++}_{p_2}(\eta_2,\eta_3) \mathcal{K}_{\eta_0 \eta_3}^+(\lbrace \mathbf{k}_R \rbrace)
		\\&  - \mathcal{K}_{\eta_0 \eta_1}^+(\lbrace \mathbf{k}_L \rbrace) G^{++}_{p_1}(\eta_1,\eta_2) \mathcal{K}_{\eta_0 \eta_2}^+(\lbrace \mathbf{k}_M \rbrace) G^{+-}_{p_2}(\eta_2,\eta_3) \mathcal{K}_{\eta_0 \eta_3}^-(\lbrace \mathbf{k}_R \rbrace) 
		\\& - \mathcal{K}_{\eta_0 \eta_1}^+(\lbrace \mathbf{k}_L \rbrace) G^{+-}_{p_1}(\eta_1,\eta_2) \mathcal{K}_{\eta_0 \eta_2}^-(\lbrace \mathbf{k}_M \rbrace) G^{-+}_{p_2}(\eta_2,\eta_3) \mathcal{K}_{\eta_0 \eta_3}^+(\lbrace \mathbf{k}_R \rbrace) 
		\\& + \mathcal{K}_{\eta_0 \eta_1}^+(\lbrace \mathbf{k}_L \rbrace) G^{+-}_{p_1}(\eta_1,\eta_2) \mathcal{K}_{\eta_0 \eta_2}^-(\lbrace \mathbf{k}_M \rbrace) G^{--}_{p_2}(\eta_2,\eta_3) \mathcal{K}_{\eta_0 \eta_3}^-(\lbrace \mathbf{k}_R \rbrace) \Bigg)
		\\&
		-  \Bigg( \mathcal{K}_{\eta_0 \eta_1}^-(\lbrace \mathbf{k}_L \rbrace) G^{-+}_{p_1}(\eta_1,\eta_2) \mathcal{K}_{\eta_0 \eta_2}^+(\lbrace \mathbf{k}_M \rbrace) G^{++}_{p_2}(\eta_2,\eta_3) \mathcal{K}_{\eta_0 \eta_3}^+(\lbrace \mathbf{k}_R \rbrace) 
		\\&
		- \mathcal{K}_{\eta_0 \eta_1}^-(\lbrace \mathbf{k}_L \rbrace) G^{-+}_{p_1}(\eta_1,\eta_2) \mathcal{K}_{\eta_0 \eta_2}^+(\lbrace \mathbf{k}_M \rbrace) G^{+-}_{p_2}(\eta_2,\eta_3) \mathcal{K}_{\eta_0 \eta_1}^-(\lbrace \mathbf{k}_R \rbrace)
		\\&
		- \mathcal{K}_{\eta_0 \eta_1}^-(\lbrace \mathbf{k}_L \rbrace) G^{--}_{p_1}(\eta_1,\eta_2) \mathcal{K}_{\eta_0 \eta_2}^-(\lbrace \mathbf{k}_M \rbrace) G^{-+}_{p_2}(\eta_2,\eta_3) \mathcal{K}_{\eta_0 \eta_3}^+(\lbrace \mathbf{k}_R \rbrace) 
		\\&
		+ \mathcal{K}_{\eta_0 \eta_1}^-(\lbrace \mathbf{k}_L \rbrace) G^{--}_{p_1}(\eta_1,\eta_2)\mathcal{K}_{\eta_0 \eta_2}^-(\lbrace \mathbf{k}_M \rbrace) G^{--}_{p_2}(\eta_2,\eta_3) \mathcal{K}_{\eta_0 \eta_3}^-(\lbrace \mathbf{k}_R \rbrace) \Bigg) \Bigg]
	\end{split}
\end{equation}
Note that, $\lbrace \mathbf{k}_M \rbrace$ describes the collection of the external energies at the internal vertex of the diagram. And $\lbrace \mathbf{k}_L \rbrace$, $\lbrace \mathbf{k}_R \rbrace$ as usual denote the collection of external energies at the left and right most vertex.\\
Applying discontinuity with respect to $p_2$ and following the propagator identities eq.\eqref{discG} we can write
\begin{equation}\label{discb3p2}
	\begin{split}
		& \text{Disc}_{p_2} \mathcal{B}^{(3)} \left( \lbrace \mathbf{k}_L, \mathbf{k}_M , \mathbf{k}_R \rbrace ;p_1;p_2 \right)= (i\lambda)^3 \int^{(3)} d \eta ~\times 
		\\&
		\bigg( \mathcal{K}_{\eta_0 \eta_1}^+(\lbrace \mathbf{k}_L \rbrace) G^{++}_{p_1}(\eta_1,\eta_2) \mathcal{K}_{\eta_0 \eta_2}^+(\lbrace \mathbf{k}_M \rbrace)
		- \mathcal{K}_{\eta_0 \eta_1}^+(\lbrace \mathbf{k}_L \rbrace) G^{+-}_{p_1}(\eta_1,\eta_2) \mathcal{K}_{\eta_0 \eta_2}^-(\lbrace \mathbf{k}_M  \rbrace) 
		\\& -  \mathcal{K}_{\eta_0 \eta_1}^-(\lbrace \mathbf{k}_L \rbrace) G^{-+}_{p_1}(\eta_1,\eta_2) \mathcal{K}_{\eta_0 \eta_2}^+(\lbrace \mathbf{k}_M \rbrace) + \mathcal{K}_{\eta_0 \eta_1}^-(\lbrace \mathbf{k}_L \rbrace) G^{--}_{p_1}(\eta_1,\eta_2) \mathcal{K}_{\eta_0 \eta_2}^-(\lbrace \mathbf{k}_M \rbrace) \bigg) 
		\\& \bigg( \frac{K_{p_2}^-(\eta_0, \eta_2) K_{p_2}^+(\eta_0, \eta_3) + K_{p_2}^+(\eta_0, \eta_2) K_{p_2}^-(\eta_0, \eta_3) }{{P_{p_2}(\eta_0)}}\bigg) \bigg( \mathcal{K}_{\eta_0 \eta_3}^+(\lbrace \mathbf{k}_R \rbrace) - \mathcal{K}_{\eta_0 \eta_3}^-(\lbrace \mathbf{k}_R \rbrace) \bigg) \, .
	\end{split}
\end{equation}
The left-most term in the last line can again be written in the same form as given in eq.\eqref{bulk to bulk ab+cd type} and then we obtain the following two terms that we have to analyze further 
\begin{equation}\label{definition term 1 disc 3site}
	\begin{split}
		& {\text {Disc}}_{p_2} \mathcal{B}^{(3)}(\lbrace \mathbf{k}_L, \mathbf{k}_M , \mathbf{k}_R \rbrace ;p_1;p_2)\Bigg{\vert}_{\text{term-1}}= \frac{(i\lambda)^3}{2 P_{p_2}(\eta_0)} \int^{(3)} d \eta ~ \times 
		\\
		&\bigg( \mathcal{K}_{\eta_0 \eta_1}^+(\lbrace \mathbf{k}_L \rbrace) G^{++}_{p_1}(\eta_1,\eta_2) \mathcal{K}_{\eta_0 \eta_2}^+(\lbrace \mathbf{k}_M \rbrace)
		- \mathcal{K}_{\eta_0 \eta_1}^+(\lbrace \mathbf{k}_L \rbrace) G^{+-}_{p_1}(\eta_1,\eta_2) \mathcal{K}_{\eta_0 \eta_2}^-(\lbrace \mathbf{k}_M  \rbrace) 
		\\& -  \mathcal{K}_{\eta_0 \eta_1}^-(\lbrace \mathbf{k}_L \rbrace) G^{-+}_{p_1}(\eta_1,\eta_2) \mathcal{K}_{\eta_0 \eta_2}^+(\lbrace \mathbf{k}_M \rbrace) + \mathcal{K}_{\eta_0 \eta_1}^-(\lbrace \mathbf{k}_L \rbrace) G^{--}_{p_1}(\eta_1,\eta_2) \mathcal{K}_{\eta_0 \eta_2}^-(\lbrace \mathbf{k}_M \rbrace) \bigg) 
		\\
		&  \Big( K_{p_2}^-(\eta_0, \eta_2)+ K_{p_2}^+(\eta_0, \eta_2) \Big)  \Big( K_{p_2}^+(\eta_0, \eta_3) + K_{p_2}^-(\eta_0, \eta_3) \Big) \bigg( \mathcal{K}_{\eta_0 \eta_3}^+(\lbrace \mathbf{k}_R \rbrace) - \mathcal{K}_{\eta_0 \eta_3}^-(\lbrace \mathbf{k}_R \rbrace) \bigg) \, ,
	\end{split}
\end{equation}
\begin{equation}\label{definition term 2 disc 3site}
	\begin{split}
		& {\text {Disc}}_{p_2} \mathcal{B}^{(3)}( \lbrace \mathbf{k}_L, \mathbf{k}_M , \mathbf{k}_R \rbrace ;p_1;p_2)\Bigg{\vert}_{\text{term-2}} =\frac{(i\lambda)^3}{2 P_{p_2}(\eta_0)} \int^{(3)} d\eta  ~ \times
		\\
		& \bigg( \mathcal{K}_{\eta_0 \eta_1}^+(\lbrace \mathbf{k}_L \rbrace) G^{++}_{p_1}(\eta_1,\eta_2) \mathcal{K}_{\eta_0 \eta_2}^+(\lbrace \mathbf{k}_M \rbrace)
		- \mathcal{K}_{\eta_0 \eta_1}^+(\lbrace \mathbf{k}_L \rbrace) G^{+-}_{p_1}(\eta_1,\eta_2) \mathcal{K}_{\eta_0 \eta_2}^-(\lbrace \mathbf{k}_M  \rbrace) 
		\\&
		 -  \mathcal{K}_{\eta_0 \eta_1}^-(\lbrace \mathbf{k}_L \rbrace) G^{-+}_{p_1}(\eta_1,\eta_2) \mathcal{K}_{\eta_0 \eta_2}^+(\lbrace \mathbf{k}_M \rbrace) + \mathcal{K}_{\eta_0 \eta_1}^-(\lbrace \mathbf{k}_L \rbrace) G^{--}_{p_1}(\eta_1,\eta_2) \mathcal{K}_{\eta_0 \eta_2}^-(\lbrace \mathbf{k}_M \rbrace) \bigg) 
		\\
		& \Big( K_{p_2}^-(\eta_0, \eta_2) - K_{p_2}^+(\eta_0, \eta_2) \Big)  \Big( K_{p_2}^+(\eta_0, \eta_3) - K_{p_2}^-(\eta_0, \eta_3) \Big) \bigg( \mathcal{K}_{\eta_0 \eta_3}^+(\lbrace \mathbf{k}_R \rbrace) - \mathcal{K}_{\eta_0 \eta_3}^-(\lbrace \mathbf{k}_R \rbrace) \bigg) \, .
	\end{split}
\end{equation}
Now we will analyze the term-1 in the following. Note that the $\eta_3$ integral in eq.\eqref{definition term 1 disc 3site} can be handled in the similar way as given in eq.\eqref{right most Disc 2site}
\begin{equation}
	\begin{split}
		&{\text {Disc}}_{p_2} \mathcal{B}^{(3)}( \lbrace \mathbf{k}_L, \mathbf{k}_R \rbrace ;p_1;p_2)\Bigg{\vert}_{\text{term-1}}^{\eta_3-\text{part}}  \\
		&= (i \lambda) \int d\eta_3   \Big( K_{p_2}^+(\eta_0, \eta_3)+ K_{p_2}^-(\eta_0, \eta_3) \Big)  \bigg( \mathcal{K}_{\eta_0 \eta_3}^+(\lbrace \mathbf{k}_R \rbrace) - \mathcal{K}_{\eta_0 \eta_3}^-(\lbrace \mathbf{k}_R \rbrace) \bigg) 
		\\
		& = \mathcal{B}^{(1)}( \lbrace \mathbf{k}_{R} \rbrace ,p_2) - \mathcal{B}^{(1)}( \lbrace \mathbf{k}_{R} \rbrace ,-p_2)  ~~~~ \equiv ~~~~\text{Disc}_{p_2} \mathcal{B}^{(1)}(\lbrace \mathbf{k}_{R} \rbrace,p_2)
	\end{split}
\end{equation}
Next, the part involving $\eta_1$ and $\eta_2$ in eq.\eqref{definition term 1 disc 3site} can be written as follows just by using the property of the bulk-to-boundary propagator $K^+_p(\eta_0, \eta_i) = - K^{-}_{-p}(\eta_0, \eta_i)$ for $p_2$ and so on
\begin{equation}
	\begin{split}
		& {\text {Disc}}_{p_2} \mathcal{B}^{(3)}( \lbrace \mathbf{k}_L, \mathbf{k}_M , \mathbf{k}_R \rbrace ;p_1;p_2)\Bigg{\vert}_{\text{term-1}}^{\eta_1\eta_2-\text{part}} = (i\lambda)^2 \int^{(2)}  d\eta ~ \times
		\\
		& \Bigg[ \bigg( \mathcal{K}_{\eta_0 \eta_1}^+(\lbrace \mathbf{k}_L \rbrace) G^{++}_{p_1}(\eta_1,\eta_2) \mathcal{K}_{\eta_0 \eta_2}^+(\lbrace \mathbf{k}_M\rbrace,p_2 )
		- \mathcal{K}_{\eta_0 \eta_1}^+(\lbrace \mathbf{k}_L \rbrace) G^{+-}_{p_1}(\eta_1,\eta_2) \mathcal{K}_{\eta_0 \eta_2}^-(\lbrace \mathbf{k}_M\rbrace,p_2) 
		\\&
		-  \mathcal{K}_{\eta_0 \eta_1}^-(\lbrace \mathbf{k}_L \rbrace) G^{-+}_{p_1}(\eta_1,\eta_2) \mathcal{K}_{\eta_0 \eta_2}^+(\lbrace \mathbf{k}_M \rbrace ,p_2) + \mathcal{K}_{\eta_0 \eta_1}^-(\lbrace \mathbf{k}_L \rbrace) G^{--}_{p_1}(\eta_1,\eta_2) \mathcal{K}_{\eta_0 \eta_2}^-(\lbrace \mathbf{k}_M\rbrace,p_2) \bigg) 
		\\
		& - \bigg(  \mathcal{K}_{\eta_0 \eta_1}^+(\lbrace \mathbf{k}_L \rbrace) G^{++}_{p_1}(\eta_1,\eta_2) \mathcal{K}_{\eta_0 \eta_2}^+(\lbrace \mathbf{k}_M \rbrace,- p_2 )
		- \mathcal{K}_{\eta_0 \eta_1}^+(\lbrace \mathbf{k}_L \rbrace) G^{+-}_{p_1}(\eta_1,\eta_2) \mathcal{K}_{\eta_0 \eta_2}^-(\lbrace \mathbf{k}_M\rbrace,- p_2 ) 
		\\&
		-  \mathcal{K}_{\eta_0 \eta_1}^-(\lbrace \mathbf{k}_L \rbrace) G^{-+}_{p_1}(\eta_1,\eta_2) \mathcal{K}_{\eta_0 \eta_2}^+(\lbrace \mathbf{k}_M \rbrace, -p_2) + \mathcal{K}_{\eta_0 \eta_1}^-(\lbrace \mathbf{k}_L \rbrace) G^{--}_{p_1}(\eta_1,\eta_2) \mathcal{K}_{\eta_0 \eta_2}^-(\lbrace \mathbf{k}_M \rbrace, -p_2)    \bigg) \Bigg]
		\\
		= &  \mathcal{B}^{(2)}( \lbrace \mathbf{k}_{L}, \mathbf{k}_M \rbrace, p_2; p_1) - \mathcal{B}^{(2)}( \lbrace \mathbf{k}_{L},\mathbf{k}_M \rbrace , -p_2; p_1)\equiv \text{Disc}_{p_2} \mathcal{B}^{(2)}( \lbrace \mathbf{k}_{L}, \mathbf{k}_M \rbrace , p_2; p_1)  \, .
	\end{split}
\end{equation}
The last equality is obtained from the definition of a $2$-site tree level correlator.

Therefore, the term-1 from the eq.\eqref{definition term 1 disc 3site} can finally be written as follows
\begin{equation}
	{\text {Disc}}_{p_2} \mathcal{B}^{(3)}( \lbrace \mathbf{k}_L, \mathbf{k}_M , \mathbf{k}_R \rbrace ;p_1;p_2)\Bigg{\vert}_{\text{term-1}} = \frac{1}{2 P_{p_2}(\eta_0)} \text{Disc}_{p_2} \mathcal{B}^{(2)}( \lbrace \mathbf{k}_{L}, \mathbf{k}_M \rbrace, p_2; p_1) \text{Disc}_{p_2} \mathcal{B}^{(1)}(\lbrace \mathbf{k}_{R} \rbrace ,p_2)
\end{equation}
Now we move to the term-2. Note that the $\eta_3$ integral in eq.\eqref{definition term 2 disc 3site} can be written in the same token as eq.\eqref{right most disc tilde 2site}
\begin{equation}\label{3site disc term 2 eta3}
	\begin{split}
		& {\text {Disc}}_{p_2} \mathcal{B}^{(3)}( \lbrace \mathbf{k}_L, \mathbf{k}_M , \mathbf{k}_R \rbrace ;p_1;p_2)\Bigg{\vert}_{\text{term-2}}^{\eta_3-\text{part}} 
		\\
		= &  (i \lambda) \int d\eta_3   \Big( K_{p_2}^+(\eta_0, \eta_3) - K_{p_2}^-(\eta_0, \eta_3) \Big) \bigg( \mathcal{K}_{\eta_0 \eta_3}^+(\lbrace \mathbf{k}_R \rbrace) - \mathcal{K}_{\eta_0 \eta_3}^-(\lbrace \mathbf{k}_R \rbrace) \bigg)
		\\
		= & ~\Big( \widetilde{\mathcal{B}}^{(1)}( \lbrace \mathbf{k}_{R} \rbrace ,p_2) + \widetilde{\mathcal{B}}^{(1)}( \lbrace \mathbf{k}_{R} \rbrace ,-p_2) \Big) ~~~~ \equiv ~~~~ \widetilde{\text{Disc}_{p_2}} \widetilde{\mathcal{B}}^{(1)}(\lbrace \mathbf{k}_{R} \rbrace ,p_2)
	\end{split}
\end{equation}
where $\widetilde{\mathcal{B}}$ was defined earlier in eq.\eqref{mathfrak B1}.\\

Now we analyze the  remaining part of the term-2 in eq.\eqref{definition term 2 disc 3site} involving $\eta_1$ and $\eta_2$ integral 
\begin{equation}\label{disc 3site term 2 eta1eta2}
	\begin{split}
		& {\text {Disc}}_{p_2} \mathcal{B}^{(3)}(\lbrace \mathbf{k}_L, \mathbf{k}_M , \mathbf{k}_R \rbrace ;p_1;p_2)\Bigg{\vert}_{\text{term-2}}^{\eta_1\eta_2-\text{part}} = \frac{(i\lambda)^2}{2 P_{p_2}(\eta_0)} \int^{(2)} d\eta ~ \times 
		\\
		&   \bigg( \mathcal{K}_{\eta_0 \eta_1}^+(\lbrace \mathbf{k}_L \rbrace) G^{++}_{p_1}(\eta_1,\eta_2) \mathcal{K}_{\eta_0 \eta_2}^+(\lbrace \mathbf{k}_M \rbrace)
		- \mathcal{K}_{\eta_0 \eta_1}^+(\lbrace \mathbf{k}_L \rbrace) G^{+-}_{p_1}(\eta_1,\eta_2) \mathcal{K}_{\eta_0 \eta_2}^-(\lbrace \mathbf{k}_M  \rbrace) 
		\\&
		-  \mathcal{K}_{\eta_0 \eta_1}^-(\lbrace \mathbf{k}_L \rbrace) G^{-+}_{p_1}(\eta_1,\eta_2) \mathcal{K}_{\eta_0 \eta_2}^+(\lbrace \mathbf{k}_M \rbrace) + \mathcal{K}_{\eta_0 \eta_1}^-(\lbrace \mathbf{k}_L \rbrace) G^{--}_{p_1}(\eta_1,\eta_2) \mathcal{K}_{\eta_0 \eta_2}^-(\lbrace \mathbf{k}_M \rbrace) \bigg) 
		\\
		& \Big( K_{p_2}^-(\eta_0, \eta_2) - K_{p_2}^+(\eta_0, \eta_2) \Big)  
		\\
		=& \frac{-(i\lambda)^2}{2 P_{p_2}(\eta_0)} \int^{(2)} d\eta  ~ \times \\
		& \Bigg[ \bigg( \mathcal{K}_{\eta_0 \eta_1}^+(\lbrace \mathbf{k}_L \rbrace) G^{++}_{p_1}(\eta_1,\eta_2) \mathcal{K}_{\eta_0 \eta_2}^+(\lbrace \mathbf{k}_M \rbrace,p_2)
		+ \mathcal{K}_{\eta_0 \eta_1}^+(\lbrace \mathbf{k}_L \rbrace) G^{+-}_{p_1}(\eta_1,\eta_2) \mathcal{K}_{\eta_0 \eta_2}^-(\lbrace \mathbf{k}_M \rbrace,p_2) 
		\\&- \mathcal{K}_{\eta_0 \eta_1}^-(\lbrace \mathbf{k}_L \rbrace) G^{-+}_{p_1}(\eta_1,\eta_2) \mathcal{K}_{\eta_0 \eta_2}^+(\lbrace \mathbf{k}_M \rbrace,p_2) - \mathcal{K}_{\eta_0 \eta_1}^-(\lbrace \mathbf{k}_L \rbrace) G^{--}_{p_1}(\eta_1,\eta_2) \mathcal{K}_{\eta_0 \eta_2}^-(\lbrace \mathbf{k}_M\rbrace,p_2 ) \bigg)
		\\
		& + \bigg( p_2 \to - p_2 \bigg)  \Bigg] \, .
	\end{split}
\end{equation}
 We now analyze the first two lines. But note that the combinations of the four pieces in both pair of lines is in such a way that we can't write them in terms of a correlator. In particular the signs prevent us from writing them as in-in correlator. Therefore, one can try to write it in terms of the wave-function coefficients both at the contact and the exchange level.

In eq.\eqref{bulkbulkreln} we have seen that the bulk-to-bulk propagators for in-in correlator are related to that of the wave-function coefficient in the following way
\begin{equation}
	\begin{split}
		& G_{p_1}^{++} (\eta_1, \eta_2) = G_{p_1}^\psi(\eta_1, \eta_2) + P_{p_1}(\eta_0) K^\psi_{p_1}(\eta_0,\eta_1) K^\psi_{p_1}(\eta_0,\eta_2)  \, ,
		\\
		& G_{p_1}^{--} (\eta_1, \eta_2) = G_{p_1}^\psi(\eta_1, \eta_2)^* + P_{p_1}(\eta_0) K^\psi_{p_1}(\eta_0,\eta_1)^* K^\psi_{p_1}(\eta_0,\eta_2)^* \, .
	\end{split}
\end{equation}
Using these relations, let's first analyze the following set of terms appeared in the first two lines of eq.\eqref{disc 3site term 2 eta1eta2}
\begin{equation}
	\begin{split}
		& (i\lambda)^2 \int^{(2)} d\eta ~\bigg( -\mathcal{K}_{\eta_0 \eta_1}^+(\lbrace \mathbf{k}_L \rbrace) G^{++}_{p_1}(\eta_1,\eta_2) \mathcal{K}_{\eta_0 \eta_2}^+(\lbrace \mathbf{k}_M \rbrace,p_2) 
		\\ & + \mathcal{K}_{\eta_0 \eta_1}^-(\lbrace \mathbf{k}_L \rbrace) G^{--}_{p_1}(\eta_1,\eta_2) \mathcal{K}_{\eta_0 \eta_2}^-(\lbrace \mathbf{k}_M\rbrace,p_2 ) - \mathcal{K}_{\eta_0 \eta_1}^+(\lbrace \mathbf{k}_L \rbrace) G^{+-}_{p_1}(\eta_1,\eta_2) \mathcal{K}_{\eta_0 \eta_2}^-(\lbrace \mathbf{k}_M \rbrace,p_2)  
		\\& + \mathcal{K}_{\eta_0 \eta_1}^-(\lbrace \mathbf{k}_L \rbrace) G^{-+}_{p_1}(\eta_1,\eta_2) \mathcal{K}_{\eta_0 \eta_2}^+(\lbrace \mathbf{k}_M \rbrace,p_2)  \bigg) \, ,
	\end{split}
\end{equation}
where the first two terms will produce the following expression 
\begin{equation}\label{PkM}
	\begin{split}
		& -(i\lambda)^2 \int^{(2)} d\eta \bigg( \mathcal{K}_{\eta_0 \eta_1}^+(\lbrace \mathbf{k}_L \rbrace) G^{++}_{p_1}(\eta_1,\eta_2) \mathcal{K}_{\eta_0 \eta_2}^+(\lbrace \mathbf{k}_M \rbrace,p_2)  
		\\& - \mathcal{K}_{\eta_0 \eta_1}^-(\lbrace \mathbf{k}_L \rbrace) G^{--}_{p_1}(\eta_1,\eta_2) \mathcal{K}_{\eta_0 \eta_2}^-(\lbrace \mathbf{k}_M\rbrace,p_2 ) \bigg)
		\\
		= & - \mathcal{P}_{ \mathbf{k}_L }(\eta_0) \mathcal{P}_{ \mathbf{k}_M }(\eta_0)  P_{p_2}(\eta_0) \Big( 2 i ~\mathbb{I}m \psi^{(2)} ( \lbrace \mathbf{k}_{L}, \mathbf{k}_M \rbrace , p_2; p_1) \\
		&+ \psi^{(1)}(\lbrace \mathbf{k}_L \rbrace, p_1) \psi^{(1)}(\lbrace \mathbf{k}_M \rbrace, p_1, p_2) - \psi^{(1)}(\lbrace \mathbf{k}_L \rbrace, p_1)^* \psi^{(1)}(\lbrace \mathbf{k}_M \rbrace, p_1, p_2)^* \Big) \, .
	\end{split}
\end{equation}
And the other two terms can be written as follows
\begin{equation}
	\begin{split}
		& - (i\lambda)^2 \int^{(2)} d\eta \bigg( \mathcal{K}_{\eta_0 \eta_1}^+(\lbrace \mathbf{k}_L \rbrace) G^{+-}_{p_1}(\eta_1,\eta_2) \mathcal{K}_{\eta_0 \eta_2}^-(\lbrace \mathbf{k}_M \rbrace,p_2)    
		\\& - \mathcal{K}_{\eta_0 \eta_1}^-(\lbrace \mathbf{k}_L \rbrace) G^{-+}_{p_1}(\eta_1,\eta_2) \mathcal{K}_{\eta_0 \eta_2}^+(\lbrace \mathbf{k}_M \rbrace,p_2)  \bigg)
		\\
		= & \mathcal{P}_{ \mathbf{k}_L }(\eta_0) \mathcal{P}_{ \mathbf{k}_M }(\eta_0)  P_{p_1}(\eta_0) P_{p_2}(\eta_0) \Big(  \psi^{(1)}(\lbrace \mathbf{k}_L \rbrace, p_1) \psi^{(1)}(\lbrace \mathbf{k}_M \rbrace, p_1, p_2)^*   
		\\& - \psi^{(1)}(\lbrace \mathbf{k}_L \rbrace, p_1)^* \psi^{(1)}(\lbrace \mathbf{k}_M \rbrace, p_1, p_2) \Big) \, .
	\end{split}
\end{equation}
Combining the above two expressions we obtain the following expression which in particular corresponds to the first two lines of eq.\eqref{disc 3site term 2 eta1eta2}
\begin{equation}\label{definition of 2 site auxiliary}
	\begin{split}
		 &\frac{-2 i \Bigg( \mathbb{I}m \psi^{(2)} ( \lbrace \mathbf{k}_{L}, \mathbf{k}_M \rbrace , p_2; p_1)  + \frac{\mathbb{R}e \psi^{(1)}(\lbrace \mathbf{k}_L\rbrace,p_1) \times \mathbb{I}m \psi^{(1)}(\lbrace \mathbf{k}_M\rbrace,p_1,p_2)}{\left( 2 P_{p_1}(\eta_0) \right)^{-1}} \Bigg)}{\left( \mathcal{P}_{ \mathbf{k}_L }(\eta_0) \mathcal{P}_{ \mathbf{k}_M }(\eta_0) P_{p_2}(\eta_0) \right)^{-1}} 
		\\
		 &\equiv - \widetilde{\mathcal{B}}^{(2)} \left( \lbrace \mathbf{k}_{L}, \mathbf{k}_M \rbrace , p_2; p_1 \right) \, ,
	\end{split}
\end{equation}
where we defined an auxiliary object at $2$-sites $\widetilde{\mathcal{B}}^{(2)} \left( \lbrace \mathbf{k}_{L}, \mathbf{k}_M \rbrace , p_2; p_1 \right)$.
One can also write an expression for $\widetilde{\mathcal{B}}^{(3)}$ as follows\footnote{At three site, we don't need this expression. However, to write a single-cut rule for a 4-site tree level correlator, $\widetilde{\mathcal{B}}^{(3)}$ will be necessary.}
\begin{equation}
	\begin{split}
		& \widetilde{\mathcal{B}}^{(3)} \left( \lbrace \mathbf{k}_{L}, \mathbf{k}_{M_1}, \mathbf{k}_{M_2} \rbrace , p_3; p_1 ; p_2 \right) = \frac{2 i}{\left( \mathcal{P}_{ \mathbf{k}_L }(\eta_0) \mathcal{P}_{ \mathbf{k}_{M_1} }(\eta_0) \mathcal{P}_{ \mathbf{k}_{M_2} }(\eta_0) P_{p_3}(\eta_0) \right)^{-1}} \times \\
		& \Bigg( \mathbb{I}m \psi^{(3)} \left( \lbrace \mathbf{k}_{L}, \mathbf{k}_{M_1}, \mathbf{k}_{M_2} \rbrace , p_3; p_1; p_2 \right) + \frac{\mathbb{R}e \psi^{(2)}(\lbrace \mathbf{k}_L, \mathbf{k}_{M_1}\rbrace,p_2; p_1) \times \mathbb{I}m \psi^{(1)}(\lbrace \mathbf{k}_{M_2}\rbrace,p_1,p_2)}{\mathbb{R}e\psi_2(p_2)}  
		\\
		& + \frac{\mathbb{R}e \psi^{(1)}(\lbrace \mathbf{k}_L \rbrace, p_1) \times \mathbb{I}m \psi^{(2)}(\lbrace (\mathbf{k}_{M_1}, p_1), (\lbrace \mathbf{k}_{M_2}, p_3)\rbrace; p_2)}{\mathbb{R}e\psi_2(p_1)}  
		\\
		&  + \frac{\mathbb{R}e \psi^{(1)}(\lbrace \mathbf{k}_L \rbrace, p_1) \times \mathbb{R}e \psi^{(1)}(\lbrace \mathbf{k}_{M_1}\rbrace, p_1, p_2) \times \mathbb{I}m\psi^{(1)}(\lbrace \mathbf{k}_{M_2} \rbrace, p_2, p_3)}{\mathbb{R}e \psi_2(p_1)~ \mathbb{R}e \psi_2(p_2) }\Bigg) \, .
	\end{split}
\end{equation}

Notice that the auxiliary object, $\widetilde{\mathcal{B}}^{(2)}$ we have defined, has a structure very similar to a $2$-site correlator except the $\mathbb{I}m$ part. With this, we finally obtain the following expression that corresponds to eq.\eqref{disc 3site term 2 eta1eta2}
\begin{equation}\label{3 site disc term 2 eta1eta2}
	\begin{split}
		&{\text {Disc}}_{p_2} \mathcal{B}^{(3)}( \lbrace \mathbf{k}_L, \mathbf{k}_M , \mathbf{k}_R \rbrace ;p_1;p_2)\Bigg{\vert}_{\text{term-2}}^{\eta_1\eta_2-\text{part}} \\
		& = - \widetilde{\mathcal{B}}^{(2)} (  \lbrace \mathbf{k}_{L}, \mathbf{k}_M \rbrace , p_2; p_1 ) - \widetilde{\mathcal{B}}^{(2)} \left( \lbrace \mathbf{k}_{L}, \mathbf{k}_M \rbrace, - p_2; p_1 \right) \equiv - \widetilde{\text{Disc}_{p_2}} \widetilde{\mathcal{B}}^{(2)} \left( \lbrace \mathbf{k}_{L}, \mathbf{k}_M \rbrace , p_2; p_1 \right) \, .
	\end{split}
\end{equation}
Therefore, the term-2 from eq.\eqref{definition term 2 disc 3site} can finally be written by combining eq.\eqref{3site disc term 2 eta3} and eq.\eqref{3 site disc term 2 eta1eta2} as follows
\begin{equation}
	\begin{split}
	&{\text {Disc}}_{p_2} \mathcal{B}^{(3)}( \lbrace \mathbf{k}_L, \mathbf{k}_M , \mathbf{k}_R \rbrace ;p_1;p_2)\Bigg{\vert}_{\text{term-2}} = - \frac{\widetilde{\text{Disc}_{p_2}} \widetilde{\mathcal{B}}^{(2)}( \lbrace \mathbf{k}_{L}, \mathbf{k}_M \rbrace , p_2; p_1) \widetilde{\text{Disc}_{p_2}} \widetilde{\mathcal{B}}^{(1)}( \lbrace \mathbf{k}_{R} \rbrace,p_2)}{2 P_{p_2}(\eta_0)} \, .
	\end{split}
\end{equation}
So the single-cut rule of a 3-site correlator can be expressed in the following form 
\begin{equation}\label{3 site correlator}
	\begin{split}
		{\text {Disc}}_{p_2} \mathcal{B}^{(3)}( \lbrace \mathbf{k}_L, \mathbf{k}_M , \mathbf{k}_R \rbrace ;p_1;p_2) =&  \frac{1}{2 P_{p_2}(\eta_0)} \Big( \text{Disc}_{p_2} \mathcal{B}^{(2)}( \lbrace \mathbf{k}_{L}, \mathbf{k}_M \rbrace , p_2; p_1) \text{Disc}_{p_2} \mathcal{B}^{(1)}( \lbrace \mathbf{k}_{R} \rbrace ,p_2) 
		\\
		& - \widetilde{\text{Disc}_{p_2}} \widetilde{\mathcal{B}}^{(2)}( \lbrace \mathbf{k}_{L}, \mathbf{k}_M \rbrace , p_2; p_1) \widetilde{\text{Disc}_{p_2}} \widetilde{\mathcal{B}}^{(1)}( \lbrace \mathbf{k}_{R} \rbrace ,p_2)\Bigg) \, .
	\end{split}
\end{equation}

\begin{figure}[h]
	\centering
	\includegraphics[width=1.05\textwidth]{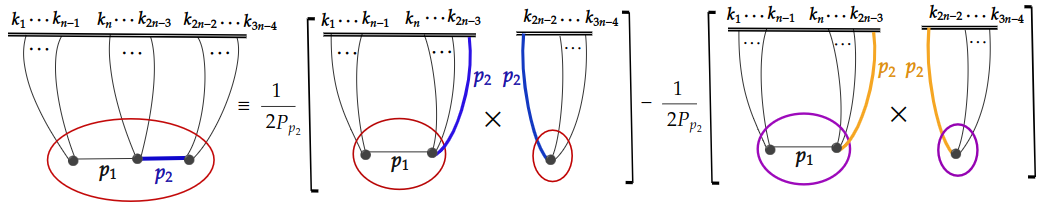}
	\caption{Diagrammatic representation of the single-cut rule of a $3$-site correlator as in eq.\eqref{3 site correlator}. Red circle denotes correlator $\mathcal{B}$ and the purple circle denotes the auxiliary object $\widetilde{\mathcal{B}}$. The blue shaded line implies taking Disc with respect to the energy corresponding to that line and the yellow shaded line implies taking $\widetilde{\text{Disc}}$ with respect to the energy corresponding to that line.}
	\label{singlecut 3 site}
\end{figure}
It turns out that for a conformally convergent polynomial interaction with $even~n$, the first piece only contributes since the second piece becomes sub-leading in $\eta_0$. However, for $odd~n$, both the terms contribute to the leading order in $\eta_0$. In the following we argue how this could be seen for a polynomial interaction $\phi^n$.\\

For a $2$-site correlator $\mathcal{B}^{(2)}$ with $\phi^n$ conformally coupled IR convergent interaction with $n=even$, the leading $\eta_0$ piece scales as $\mathcal{O}\left(\eta_0^{2n-2}\right)$ and for a contact correlator $\mathcal{B}^{(1)}$ the leading $\eta_0$ piece scales as $\mathcal{O}\left(\eta_0^n \right)$. Therefore the product of their discontinuities becomes at $\mathcal{O}\left(\eta_0^{3n-2} \right)$. As a result the first term in the parenthesis of the above expression has the following leading $\eta_0$ behaviour at $\eta_0 \to 0$ limit
\begin{equation}
	\text{Disc}_{p_2} \mathcal{B}^{(2)}( \lbrace \mathbf{k}_{L}, \mathbf{k}_M \rbrace , p_2; p_1) \text{Disc}_{p_2} \mathcal{B}^{(1)}( \lbrace \mathbf{k}_{R} \rbrace ,p_2)\Bigg{\vert}_{n=\text{even}}  \sim \mathcal{O} \left( \eta_0^{3n-2} \right).
\end{equation}
Turning to the auxiliary contributions, the leading real part of the one-site wave-function coefficient $\mathbb{R}e\psi^{(1)}$ and imaginary parts of the one–site wave-function coefficient $\mathbb{I}m \psi^{(1)}$ behave as $\mathcal{O}\left( \eta_0^{-n} \right)$ and $\mathcal{O} \left( \eta_0^{-n+1} \right)$ respectively, the latter being the next–to–leading real piece. From eq.\eqref{definition of 2 site auxiliary} the leading non–vanishing auxiliary contribution enters at $\mathcal{O} \left( \eta_0^{-2n+3} \right)$ which is consistent with $\mathbb{I}m\psi^{(2)}\sim \mathcal{O}\left( \eta_0^{-2n+3}\right)$ as it should be. This results in the leading $\eta_0$ contribution of $\widetilde{\mathcal{B}}^{(2)}$ contributing at $\mathcal{O}\left( \eta_0^{2n-1}\right)$. Next note that the leading $\eta_0$ behaviour of the one site auxiliary object $\widetilde{\mathcal{B}}^{(1)}$ scales as $\mathcal{O}\left( \eta_0^{n+1}\right)$. Therefore the second term in the RHS of eq.\eqref{3 site correlator} exhibits the following leading $\eta_0$ behaviour
\begin{equation}
	\widetilde{\text{Disc}_{p_2}} \widetilde{\mathcal{B}}^{(2)}( \lbrace \mathbf{k}_{L}, \mathbf{k}_M \rbrace , p_2; p_1) \widetilde{\text{Disc}_{p_2}} \widetilde{\mathcal{B}}^{(1)}( \lbrace \mathbf{k}_{R} \rbrace ,p_2)\Bigg{\vert}_{n=\text{even}} \sim \mathcal{O}\left( \eta_0^{3n} \right) \, .
\end{equation}
Since this is parametrically subleading relative to the first term $\mathcal{O}(\eta_0^{3n-2})$, we conclude that for even $n$ the correlator discontinuity alone captures the full leading late–time behaviour.

We now present a similar argument for $n=odd$ case. For odd interaction $2$-site correlator $\mathcal{B}^{(2)}$, contributes at $\mathcal{O}\left(\eta_0^{2n-2}\right)$ whereas the contact correlator $\mathcal{B}^{(1)}$ now behaves as $\mathcal{O}\left(\eta_0^{n+1} \right)$. So the product of the correlator discontinuities
\begin{equation}\label{n odd B piece 3site}
	\text{Disc}_{p_2} \mathcal{B}^{(2)}( \lbrace \mathbf{k}_{L}, \mathbf{k}_M \rbrace , p_2; p_1) \text{Disc}_{p_2} \mathcal{B}^{(1)}( \lbrace \mathbf{k}_{R} \rbrace ,p_2)\Bigg{\vert}_{n=\text{odd}}  \sim \mathcal{O} \left( \eta_0^{3n-1} \right) \, .
\end{equation}
On the other hand, For the wave-function coefficients, the hierarchy is reversed relative to the even $n$ case, meaning that $\mathbb{R}e \psi^{(1)}\sim \mathcal{O}\left( \eta_0^{-n+1} \right)$ and $\mathbb{I}m\psi^{(1)}\sim \mathcal{O} \left( \eta_0^{-n} \right)$, with the imaginary part now providing the leading contribution. As before, the combination entering in the RHS of equation (\ref{definition of 2 site auxiliary}) scales as $\mathcal{O} \left( \eta_0^{-2n+3} \right)$. The other term i.e. the imaginary part of the two site wave-function coefficient $\mathbb{I}m \psi^{(2)}$ contributes at $\mathcal{O}\left( \eta_0^{-2n+3}\right)$ as it was formerly noted. This makes the leading $\eta_0$ contribution of the $2$-site auxiliary piece $\widetilde{\mathcal{B}}^{(2)}$ appearing at $\mathcal{O}\left( \eta_0^{2n-1}\right)$. Next note that the one site auxiliary object $\widetilde{\mathcal{B}}^{(1)}$ contributes at $\mathcal{O}\left( \eta_0^{n}\right)$. Therefore the second term in the RHS of eq.\eqref{3 site correlator} exhibits the following leading $\eta_0$ behaviour
\begin{equation}\label{n odd Bt piece 3site}
	\widetilde{\text{Disc}_{p_2}} \widetilde{\mathcal{B}}^{(2)}( \lbrace \mathbf{k}_{L}, \mathbf{k}_M \rbrace , p_2; p_1) \widetilde{\text{Disc}_{p_2}} \widetilde{\mathcal{B}}^{(1)}( \lbrace \mathbf{k}_{R} \rbrace ,p_2)\Bigg{\vert}_{n=\text{odd}} \sim \mathcal{O}\left( \eta_0^{3n-1} \right) \, .
\end{equation}
Thus, for odd $n$, both terms in (\ref{3 site correlator}) contribute at the same order in $\eta_0$. In this case, retaining only the discontinuity term would miss part of the leading late–time structure, and the auxiliary contributions must be included.

\section{Details of deriving the discontinuity relation for loops} \label{loopdetails}
In this appendix we provide additional details which we skipped in \S\ref{loopdiagrams} while deriving the cutting rule for loop diagrams. Substituting  eq.\eqref{ab+cd type1} and eq.\eqref{ab+cd type2} into eq.\eqref{discp1} produces several terms, which we now analyze individually.
\begin{equation}\label{loop term1}
	\begin{split}
		& {\text {Disc}}_{p_1} \Big[  {\text {Disc}}_{p_2} \mathbb{B}^{(2)}( \lbrace \mathbf{k}_L, \mathbf{k}_R \rbrace ;p_1,p_2) \Big] \Bigg{\vert}_{\text{term 1}}
		= \frac{(i\lambda)^2}{4 {{P_{p_1}(\eta_0)}}{{P_{p_2}(\eta_0)}}}   \times \int^{(2)} d\eta ~ \times \\ 
		& \bigg( \mathcal{K}_{\eta_0 \eta_1}^+(\lbrace \mathbf{k}_L \rbrace) -  \mathcal{K}_{\eta_0 \eta_1}^-(\lbrace \mathbf{k}_L \rbrace)\bigg) \Big( K_{p_1}^-(\eta_0, \eta_1)+ K_{p_1}^+(\eta_0, \eta_1) \Big) \Big( K_{p_2}^-(\eta_0, \eta_1)+ K_{p_2}^+(\eta_0, \eta_1) \Big)  \\
		& \bigg( \mathcal{K}_{\eta_0 \eta_2}^+(\lbrace \mathbf{k}_R \rbrace) -  \mathcal{K}_{\eta_0 \eta_2}^-(\lbrace \mathbf{k}_R \rbrace)\bigg) \Big( K_{p_1}^+(\eta_0, \eta_2) + K_{p_1}^-(\eta_0, \eta_2) \Big) \Big( K_{p_2}^+(\eta_0, \eta_2) + K_{p_2}^-(\eta_0, \eta_2) \Big) \, ,
	\end{split}   
\end{equation}
\begin{equation}\label{loop term2}
	\begin{split}
		&  {\text {Disc}}_{p_1} \Big[  {\text {Disc}}_{p_2} \mathbb{B}^{(2)}( \lbrace \mathbf{k}_L, \mathbf{k}_R \rbrace ;p_1,p_2) \Big] \Bigg{\vert}_{\text{term 2}}
		=  \frac{(i\lambda)^2}{4 {{P_{p_1}(\eta_0)}}{{P_{p_2}(\eta_0)}}}   \times \int^{(2)} d\eta ~ \times\\
		& \bigg( \mathcal{K}_{\eta_0 \eta_1}^+(\lbrace \mathbf{k}_L \rbrace) -  \mathcal{K}_{\eta_0 \eta_1}^-(\lbrace \mathbf{k}_L \rbrace)\bigg) \Big( K_{p_1}^-(\eta_0, \eta_1)+ K_{p_1}^+(\eta_0, \eta_1) \Big)\Big( K_{p_2}^-(\eta_0, \eta_1)- K_{p_2}^+(\eta_0, \eta_1) \Big)
		\\
		& \bigg( \mathcal{K}_{\eta_0 \eta_1}^+(\lbrace \mathbf{k}_R \rbrace) -  \mathcal{K}_{\eta_0 \eta_1}^-(\lbrace \mathbf{k}_R \rbrace)\bigg)  \Big( K_{p_2}^+(\eta_0, \eta_2) - K_{p_2}^-(\eta_0, \eta_2) \Big)\Big( K_{p_1}^+(\eta_0, \eta_2) + K_{p_1}^-(\eta_0, \eta_2) \Big) \, ,
	\end{split}
\end{equation}
\begin{equation}\label{loop term3}
	\begin{split}
		&  {\text {Disc}}_{p_1} \Big[  {\text {Disc}}_{p_2} \mathbb{B}^{(2)}( \lbrace \mathbf{k}_L, \mathbf{k}_R \rbrace ;p_1,p_2) \Big] \Bigg{\vert}_{\text{term 3}}
		= \frac{(i\lambda)^2}{4 {{P_{p_1}(\eta_0)}}{{P_{p_2}(\eta_0)}}}   \times \int^{(2)} d\eta ~ \times\\
		& \bigg( \mathcal{K}_{\eta_0 \eta_1}^+(\lbrace \mathbf{k}_L \rbrace) -  \mathcal{K}_{\eta_0 \eta_1}^-(\lbrace \mathbf{k}_L \rbrace)\bigg) \Big( K_{p_1}^-(\eta_0, \eta_1)- K_{p_1}^+(\eta_0, \eta_1) \Big)\Big( K_{p_2}^-(\eta_0, \eta_1)+K_{p_2}^+(\eta_0, \eta_1) \Big)
		\\
		& \bigg( \mathcal{K}_{\eta_0 \eta_1}^+(\lbrace \mathbf{k}_R \rbrace) -  \mathcal{K}_{\eta_0 \eta_1}^-(\lbrace \mathbf{k}_R \rbrace)\bigg)  \Big( K_{p_2}^+(\eta_0, \eta_2) +K_{p_2}^-(\eta_0, \eta_2) \Big)\Big( K_{p_1}^+(\eta_0, \eta_2) - K_{p_1}^-(\eta_0, \eta_2) \Big) \, ,
	\end{split}
\end{equation}
\begin{equation}\label{loop term4}
	\begin{split}
		&  {\text {Disc}}_{p_1} \Big[  {\text {Disc}}_{p_2} \mathbb{B}^{(2)}( \lbrace \mathbf{k}_L, \mathbf{k}_R \rbrace ;p_1,p_2) \Big] \Bigg{\vert}_{\text{term 4}}
		= \frac{(i\lambda)^2}{4 {{P_{p_1}(\eta_0)}}{{P_{p_2}(\eta_0)}}}   \times \int^{(2)} d\eta ~ \times\\
		& \bigg( \mathcal{K}_{\eta_0 \eta_1}^+(\lbrace \mathbf{k}_L \rbrace) -  \mathcal{K}_{\eta_0 \eta_1}^-(\lbrace \mathbf{k}_L \rbrace)\bigg) \Big( K_{p_1}^-(\eta_0, \eta_1)- K_{p_1}^+(\eta_0, \eta_1) \Big)\Big( K_{p_2}^-(\eta_0, \eta_1)-K_{p_2}^+(\eta_0, \eta_1) \Big)
		\\
		& \bigg( \mathcal{K}_{\eta_0 \eta_1}^+(\lbrace \mathbf{k}_R \rbrace) -  \mathcal{K}_{\eta_0 \eta_1}^-(\lbrace \mathbf{k}_R \rbrace)\bigg)  \Big( K_{p_2}^+(\eta_0, \eta_2) -K_{p_2}^-(\eta_0, \eta_2) \Big)\Big( K_{p_1}^+(\eta_0, \eta_2) - K_{p_1}^-(\eta_0, \eta_2) \Big) \, .
	\end{split}
\end{equation}
Now we simplify a part of the first term in eq.\eqref{loop term1} as follows
\begin{equation}\label{term1 eta1 part}
	\begin{split}
		& (i \lambda) \int d\eta_1  \bigg( \mathcal{K}_{\eta_0 \eta_1}^+(\lbrace \mathbf{k}_L \rbrace) -  \mathcal{K}_{\eta_0 \eta_1}^-(\lbrace \mathbf{k}_L \rbrace)\bigg) \Big( K_{p_1}^-(\eta_0, \eta_1)+ K_{p_1}^+(\eta_0, \eta_1) \Big) \Big( K_{p_2}^-(\eta_0, \eta_1)+ K_{p_2}^+(\eta_0, \eta_1) \Big) 
		\\
		& =  (i \lambda) \int d\eta_1 \bigg( -\mathcal{K}_{\eta_0 \eta_1}^+(\lbrace \mathbf{k}_L \rbrace,-p_1) +\mathcal{K}_{\eta_0 \eta_1}^+(\lbrace \mathbf{k}_L \rbrace,p_1) -\mathcal{K}_{\eta_0 \eta_1}^-(\lbrace \mathbf{k}_L \rbrace,p_1)  +\mathcal{K}_{\eta_0 \eta_1}^-(\lbrace \mathbf{k}_L \rbrace,-p_1)\bigg)\\
		& ~~~~~~~~~~~~~~~~~~~~~~~~~~~~~~~~~~~~~~~~~~~~~~~~~~~~~~~~~~~~~~\times \Big( K_{p_2}^-(\eta_0, \eta_1)+ K_{p_2}^+(\eta_0, \eta_1) \Big)  \\
		& = (i \lambda) \int \eta_1 \bigg( \mathcal{K}_{\eta_0 \eta_1}^+(\lbrace \mathbf{k}_L\rbrace,-p_1,-p_2 ) -  \mathcal{K}_{\eta_0 \eta_1}^+(\lbrace \mathbf{k}_L \rbrace,-p_1,p_2 ) -  \mathcal{K}_{\eta_0 \eta_1}^+(\lbrace \mathbf{k}_L\rbrace, p_1,-p_2 )  \\
		& ~~~~~~~~~~ + \mathcal{K}_{\eta_0 \eta_1}^+(\lbrace \mathbf{k}_L\rbrace, p_1,p_2 ) - \mathcal{K}_{\eta_0 \eta_1}^-(\lbrace \mathbf{k}_L\rbrace , p_1,p_2)+\mathcal{K}_{\eta_0 \eta_1}^-(\lbrace \mathbf{k}_L\rbrace , p_1,-p_2) \\
		& ~~~~~~~~~~ + \mathcal{K}_{\eta_0 \eta_1}^-(\lbrace \mathbf{k}_L\rbrace , -p_1,p_2)-\mathcal{K}_{\eta_0 \eta_1}^-(\lbrace \mathbf{k}_L\rbrace , -p_1,-p_2) \bigg)
		\\
		& = \mathcal{B}^{(1)}( \lbrace \mathbf{k}_{L} \rbrace ,-p_1,-p_2) - \mathcal{B}^{(1)}( \lbrace \mathbf{k}_{L} \rbrace ,-p_1,p_2) -\mathcal{B}^{(1)}( \lbrace \mathbf{k}_{L} \rbrace ,p_1,-p_2)+\mathcal{B}^{(1)}( \lbrace \mathbf{k}_{L} \rbrace ,p_1,p_2) \, .
	\end{split}
\end{equation}
Here the last line simply follows from the definition of the contact correlator. 
The remaining contribution of eq.\eqref{loop term1} can be obtained from the above result by using $\eta_1 \leftrightarrow \eta_2, \lbrace \mathbf{k}_L \rbrace \leftrightarrow \lbrace \mathbf{k}_R \rbrace$. Therefore,
\begin{equation}
	\begin{split}
		& (i \lambda) \int d\eta_2  \bigg( \mathcal{K}_{\eta_0 \eta_2}^+(\lbrace \mathbf{k}_R \rbrace) -  \mathcal{K}_{\eta_0 \eta_2}^-(\lbrace \mathbf{k}_R \rbrace)\bigg) \Big( K_{p_1}^+(\eta_0, \eta_2) + K_{p_1}^-(\eta_0, \eta_2) \Big) \Big( K_{p_2}^+(\eta_0, \eta_2) + K_{p_2}^-(\eta_0, \eta_2) \Big) \\
		& = \mathcal{B}^{(1)}( \lbrace \mathbf{k}_{R} \rbrace ,-p_1,-p_2) - \mathcal{B}^{(1)}( \lbrace \mathbf{k}_{R} \rbrace ,-p_1,p_2) -\mathcal{B}^{(1)}( \lbrace \mathbf{k}_{R} \rbrace ,p_1,-p_2)+\mathcal{B}^{(1)}( \lbrace \mathbf{k}_{R} \rbrace ,p_1,p_2) \, .
	\end{split}
\end{equation}
Therefore, eq.\eqref{loop term1} can be written as follows
\begin{equation}\label{loop term1 final}
	\begin{split}
		& {\text {Disc}}_{p_1} \Big[  {\text {Disc}}_{p_2} \mathbb{B}^{(2)}( \lbrace \mathbf{k}_L, \mathbf{k}_R \rbrace ;p_1,p_2) \Big] \Bigg{\vert}_{\text{term 1}}\\
		&  = \frac{\sum_{(\sigma_1, \sigma_2) \in ( \pm , \pm)} (\sigma_1 \sigma_2) \mathcal{B}^{(1)} (\lbrace \mathbf{k}_{L} \rbrace,\sigma_1 p_1, \sigma_2 p_2) \sum_{(\sigma_1, \sigma_2) \in ( \pm , \pm)} (\sigma_1 \sigma_2) \mathcal{B}^{(1)} (\lbrace \mathbf{k}_{R} \rbrace,\sigma_1 p_1, \sigma_2 p_2)}{4 {{P_{p_1}(\eta_0)}}{{P_{p_2}(\eta_0)}}} \\
		&= \frac{{\text {Disc}}_{p_1}{\text {Disc}}_{p_2}\mathcal{B}^{(1)}(\lbrace \mathbf{k}_{L} \rbrace,p_1,p_2)\times {\text {Disc}}_{p_1}  {\text {Disc}}_{p_2} \mathcal{B}^{(1)}(\lbrace \mathbf{k}_{R} \rbrace,p_1,p_2)}{4\,P_{p_1}(\eta_0)\,P_{p_2}(\eta_0)} \, .
	\end{split}
\end{equation}
Similarly, eq.\eqref{loop term2} can be written as follows
\begin{equation}\label{loop term2 final}
	\begin{split}
		& {\text {Disc}}_{p_1} \Big[  {\text {Disc}}_{p_2} \mathbb{B}^{(2)}( \lbrace \mathbf{k}_L, \mathbf{k}_R \rbrace ;p_1,p_2) \Big] \Bigg{\vert}_{\text{term 2}}\\
		& =-  \frac{\sum_{(\sigma_1, \sigma_2) \in ( \pm , \pm)}  \sigma_1 \widetilde{\mathcal{B}}^{(1)} (\lbrace \mathbf{k}_{L} \rbrace,\sigma_1 p_1, \sigma_2 p_2) \sum_{(\sigma_1, \sigma_2) \in ( \pm , \pm)} \sigma_1 \widetilde{\mathcal{B}}^{(1)} (\lbrace \mathbf{k}_{R} \rbrace,\sigma_1 p_1, \sigma_2 p_2)}{4 {{P_{p_1}(\eta_0)}}{{P_{p_2}(\eta_0)}}} \\
		&= - \frac{{\text {Disc}}_{p_1} \widetilde{{\text {Disc}}}_{p_2}\widetilde{\mathcal{B}}^{(1)}(\lbrace \mathbf{k}_{L} \rbrace,p_1,p_2)\times {\text {Disc}}_{p_1}  \widetilde{{\text {Disc}}}_{p_2} \mathcal{B}^{(1)}(\lbrace \mathbf{k}_{R} \rbrace,p_1,p_2)}{4\,P_{p_1}(\eta_0)\,P_{p_2}(\eta_0)} \, .
	\end{split}
\end{equation}
Similarly, eq.\eqref{loop term3} can be written as follows
\begin{equation}\label{loop term3 final}
	\begin{split}
		& {\text {Disc}}_{p_1} \Big[  {\text {Disc}}_{p_2} \mathbb{B}^{(2)}( \lbrace \mathbf{k}_L, \mathbf{k}_R \rbrace ;p_1,p_2) \Big] \Bigg{\vert}_{\text{term 3}}\\
		& =- \frac{\sum_{(\sigma_1, \sigma_2) \in ( \pm , \pm)}  \sigma_2 \widetilde{\mathcal{B}}^{(1)} (\lbrace \mathbf{k}_{L} \rbrace,\sigma_1 p_1, \sigma_2 p_2) \sum_{(\sigma_1, \sigma_2) \in ( \pm , \pm)} \sigma_2 \widetilde{\mathcal{B}}^{(1)} (\lbrace \mathbf{k}_{R} \rbrace,\sigma_1 p_1, \sigma_2 p_2)}{4 {{P_{p_1}(\eta_0)}}{{P_{p_2}(\eta_0)}}} \\
		&=- \frac{\widetilde{{\text {Disc}}}_{p_1}{\text {Disc}}_{p_2} \widetilde{\mathcal{B}}^{(1)}(\lbrace \mathbf{k}_{L} \rbrace,p_1,p_2)\times   \widetilde{{\text {Disc}}}_{p_1} {\text {Disc}}_{p_2} \mathcal{B}^{(1)}(\lbrace \mathbf{k}_{R} \rbrace,p_1,p_2)}{4\,P_{p_1}(\eta_0)\,P_{p_2}(\eta_0)} \, .
	\end{split}
\end{equation}
Similarly, eq.\eqref{loop term4} can be written as follows
\begin{equation}\label{loop term4 final}
	\begin{split}
		& {\text {Disc}}_{p_1} \Big[  {\text {Disc}}_{p_2} \mathbb{B}^{(2)}( \lbrace \mathbf{k}_L, \mathbf{k}_R \rbrace ;p_1,p_2) \Big] \Bigg{\vert}_{\text{term 4}}\\
		& = \frac{\sum_{(\sigma_1, \sigma_2) \in ( \pm , \pm)}  \mathcal{B}^{(1)} (\lbrace \mathbf{k}_{L} \rbrace,\sigma_1 p_1, \sigma_2 p_2)\sum_{(\sigma_1, \sigma_2) \in ( \pm , \pm)} \mathcal{B}^{(1)} (\lbrace \mathbf{k}_{R} \rbrace,\sigma_1 p_1, \sigma_2 p_2)}{4 {{P_{p_1}(\eta_0)}}{{P_{p_2}(\eta_0)}}} \\
		&= \frac{\widetilde{{\text {Disc}}}_{p_1} \widetilde{{\text{Disc}}}_{p_2}\mathcal{B}^{(1)}(\lbrace \mathbf{k}_{L} \rbrace,p_1,p_2)\times \widetilde{{\text {Disc}}}_{p_1} \widetilde{{\text {Disc}}}_{p_2} \mathcal{B}^{(1)}(\lbrace \mathbf{k}_{R} \rbrace,p_1,p_2)}{4\,P_{p_1}(\eta_0)\,P_{p_2}(\eta_0)} \, .
	\end{split}
\end{equation}
Combining equations eq.\eqref{loop term1 final}- eq.\eqref{loop term4 final} yields the following result
\begin{equation}\label{1loopfinal}
	\begin{split}
		& {\text {Disc}}_{p_1} \Big[  {\text {Disc}}_{p_2} \mathbb{B}^{(2)}( \lbrace \mathbf{k}_L, \mathbf{k}_R \rbrace ;p_1,p_2) \Big] 
		=  \frac{1}{4 {{P_{p_1}(\eta_0)}}{{P_{p_2}(\eta_0)}}} \times  \\
		&\Bigg[  \bigg({\text {Disc}}_{p_1}{\text {Disc}}_{p_2}\mathcal{B}^{(1)}(\lbrace \mathbf{k}_{L} \rbrace,p_1,p_2)\times (L \to R )  + \widetilde{{\text {Disc}}}_{p_1} \widetilde{{\text{Disc}}}_{p_2}\mathcal{B}^{(1)}(\lbrace \mathbf{k}_{L} \rbrace,p_1,p_2)\times (L \to R ) \bigg)\\
		& - \bigg({\text {Disc}}_{p_1} \widetilde{{\text {Disc}}}_{p_2}\widetilde{\mathcal{B}}^{(1)}(\lbrace \mathbf{k}_{L} \rbrace,p_1,p_2)\times (L \to R ) +\widetilde{{\text {Disc}}}_{p_1}{\text {Disc}}_{p_2} \widetilde{\mathcal{B}}^{(1)}(\lbrace \mathbf{k}_{L} \rbrace,p_1,p_2)\times (L \to R ) \bigg) \Bigg] \, .
	\end{split}
\end{equation}
\begin{figure}[h]
	\centering
	\includegraphics[width=1.00\textwidth]{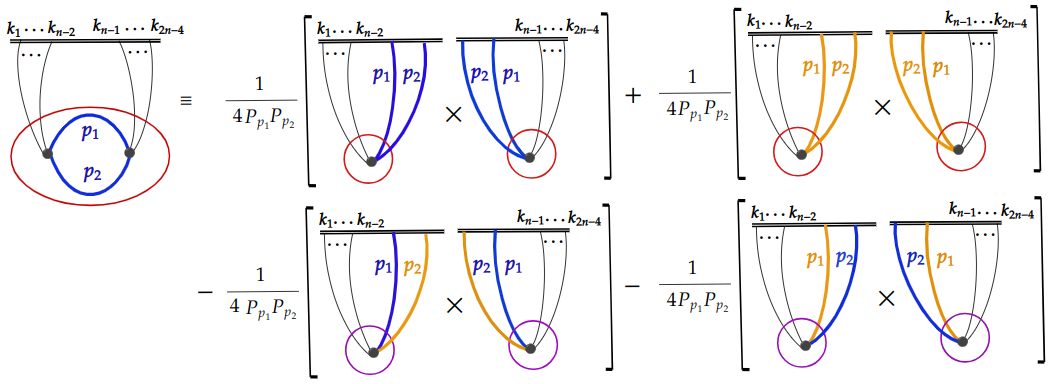}
	\caption{Diagrammatic representation of the single-cut rule the 1-loop $2$-site correlator.}
	\label{single-cut 2-site correlator}
\end{figure}
Motivated by the 1-loop cutting rule result, we propose the following cutting rule for the two-site correlator at arbitrary $\ell$-loop order.
\begin{equation}
	\begin{split}
		&\mathrm{Disc}_{p_{\ell+1}}\cdots \mathrm{Disc}_{p_1}\,
		\mathbb{B}^{(2)}\!\left(\{\mathbf{k}_L,\mathbf{k}_R\};p_1,\dots,p_{\ell+1}\right)
		=\frac{ \mathcal{F}_1}{2^{\ell+1} \left(\displaystyle\prod_{j=1}^{\ell+1}P_{p_j}(\eta_0)\right)} \, ,
	\end{split}
\end{equation}
such that 
\begin{equation}
\begin{split}
 &\mathcal{F}_1 = \sum_{\substack{S\subseteq\{1,\dots,\ell+1\} \\ |S|\ \mathrm{even}}} \left(
		\mathcal{D}_S\,
		\mathcal{B}^{(1)}(\{\mathbf{k}_L\};p_1,\dots,p_{\ell+1})
		\right)
		\left(
		\mathcal{D}_S\,
		\mathcal{B}^{(1)}(\{\mathbf{k}_R\};p_1,\dots,p_{\ell+1})
		\right)
		\\
		&\qquad-
		\sum_{\substack{S\subseteq\{1,\dots,\ell+1\} \\ |S|\ \mathrm{odd}}}
		\left(
		\mathcal{D}_S\,
		\widetilde{\mathcal{B}}^{(1)}(\{\mathbf{k}_L\};p_1,\dots,p_{\ell+1})
		\right)
		\left(
		\mathcal{D}_S\,
		\widetilde{\mathcal{B}}^{(1)}(\{\mathbf{k}_R\};p_1,\dots,p_{\ell+1})
		\right) \,,
\end{split}
\end{equation}
where the operator $\mathcal{D}_S$ (with the same subset $S$ as in the sums above) is defined by
\begin{equation*}
	\mathcal{D}_S
	\;\equiv\;
	\left(\prod_{i\in S}\widetilde{\mathrm{Disc}}_{p_i}\right)
	\left(\prod_{i\notin S}\mathrm{Disc}_{p_i}\right)\, .
\end{equation*}
Here, $|S|$ is just the number of elements in $S$. Each subset $S$ determines which internal momenta get a $\widetilde{\mathrm{Disc}}$: if $|S|$ is even, $\mathcal{D}_S$ acts on $\mathcal{B}^{(1)}$; if odd, it acts on $\widetilde{\mathcal{B}}^{(1)}$.

\section{Comparing our cutting rules with the existing literature and improvements} \label{subsec:comparing in-out}

 Recently, in \cite{Donath:2024utn}, the authors have derived cutting rules for a two sites correlator, even with loop-diagrams, and expressed them directly in the context of cosmological correlator. They have considered $(L+1)$-loop and $2$-site correlator using the following conformally coupled polynomial interaction 
\begin{equation} \label{inout interaction}
	H_{\text{int}} = \int_{\vec{x}} \frac{1}{(m+L+1)!} \phi^{m+L+1} + \int_{\vec{x}} \frac{1}{(n-m+L+1)!} \phi^{n-m+L+1} \, .
\end{equation}
For a $n$-point $(L+1)$-loop $2$-site correlator they obtain the following formula for the cutting rule\footnote{Note that we have rewritten the expression using the notation we have followed throughout in this paper. At tree level, since there is no loop integration, the LHS should have the full correlator $\mathcal{B}^{(2)}$ instead of $\mathbb{B}^{(2)}$ which denotes the loop integrand.} 
\begin{equation}\label{Pajer 2 site cut}
	\begin{split}
		&   \mathbb{B}^{(2)} \left( \lbrace \mathbf{k}_L, \mathbf{k}_R \rbrace ; \lbrace p_1;p_2;...;p_{L+1} \rbrace \right) + (-1)^n \mathbb{B}^{(2)} \left( \lbrace - \mathbf{k}_L, -\mathbf{k}_R \rbrace ; \lbrace p_1;p_2;...;p_{L+1} \rbrace \right)
		\\
		& = 2 \left( \frac{\mathcal{B}^{(1),\text{cut}} \left( \lbrace \mathbf{k}_L \rbrace, \lbrace p_i \rbrace_{i=1}^{L+1} \right)  \mathcal{B}^{(1), \text{cut}} \left( \lbrace \mathbf{k}_R \rbrace , \lbrace p_i \rbrace_{i=1}^{L+1} \right)}{\prod_{i=1}^{L+1} P_{p_i}( \eta_0)} \right) \, ,
	\end{split}
\end{equation}
where, $n$ is the total number of external lines and according to our notation $\lbrace \mathbf{k}_L \rbrace$ is the collection of external lines labelled by $\lbrace k_1, k_2,..., k_m \rbrace$ whereas $\lbrace \mathbf{k}_R \rbrace = \lbrace k_{m+1}, k_{m+2},...,k_{n} \rbrace$. 
$\mathcal{B}^{(1),\text{cut}} \left( \lbrace \mathbf{k} \rbrace, \lbrace p_i \rbrace_{i=1}^{L+1} \right)$ is defined by the following expression
\begin{equation}\label{Bc cut Pajer}
	\begin{split}
		\mathcal{B}^{(1),\text{cut}} \left( \lbrace \mathbf{k} \rbrace, \lbrace p_i \rbrace_{i=1}^{L+1} \right)  = \frac{1}{2} \Bigg( \mathcal{B}^{(1)} \left( \lbrace \mathbf{k} \rbrace, \lbrace p_i \rbrace_{i=1}^{L+1} \right) + (-1)^{L+1}  \mathcal{B}^{(1)} \left( \lbrace \mathbf{k} \rbrace, \lbrace - p_i \rbrace_{i=1}^{L+1} \right) \Bigg) \, .
	\end{split}
\end{equation}
Comparing with our $2$-site discontinuity relation derived in eq.\eqref{2siteDiscFinal Result} we find that at the tree-level (which corresponds to $L=0$) eq.\eqref{Pajer 2 site cut} gives consistent result for cases with $\phi^n$ interactions with $n=$ even (i.e., in eq.\eqref{inout interaction}, when $(m+L+1)~\text{and}~(n-m+L+1)$ are both even integers). However, for odd polynomial interactions ($n=$ odd with $\phi^n$ interactions), the RHS of the eq.\eqref{Pajer 2 site cut} fails to capture the leading $\eta_0$ contribution as $\eta_0 \to 0^-$, where $\eta_0$ is the late time-slice in de Sitter. In other words, for the leading $\eta_0$ behaviour the equality stops to hold. We have also checked that the situation is same at $1$-loop (which corresponds to $L=1$). In the following discussion we explicitly show this observation for both tree and $1$-loop cases using odd and even polynomial interaction. 
\begin{itemize}
\item \textbf{For $\phi^4$ theory at tree level:} We consider a conformally coupled $\phi^4$ theory\footnote{This corresponds to $m=3$ and $n=6$.}. Following the in-in formalism one can first derive a $2$-site correlator and then substitute it in the LHS of the eq.\eqref{Pajer 2 site cut} that yields the following expression
\begin{equation}\label{Pajer cut phi4 tree LHS}
	\begin{split}
		& \mathcal{B}^{(2)}_{\phi^4} \left( \lbrace \mathbf{k}_L, \mathbf{k}_R \rbrace ; p \right) + (-1)^6 \mathcal{B}^{(2)}_{\phi^4} \left( \lbrace - \mathbf{k}_L, -\mathbf{k}_R \rbrace ; p \right) = \left( \frac{\lambda^2 (H \eta_0)^6}{16 k_1...k_6} \right) \left( \frac{k_L k_R}{p (k_L^2 - p^2)(k_R^2  - p^2)} \right) \, ,
	\end{split}
\end{equation}
where $k_L=k_1+k_2+k_3$, and $k_R=k_4+k_5+k_6$. 
To obtain the RHS of eq.\eqref{Pajer 2 site cut} one need to know a contact correlator in this theory which can easily be computed from the standard in-in formalism. Then substituting it in eq.\eqref{Bc cut Pajer} one can obtain the following expression
\begin{equation}
	\begin{split}
	 \mathcal{B}^{(1),\text{cut}}_{\phi^4} \left( \lbrace \mathbf{k}_L \rbrace, p \right)  & = \left( \frac{\lambda (H \eta_0)^4}{8 k_1 k_2 k_3}\right) \frac{k_L}{p( k_L^2 - p^2)} \, , \\ 
	 \mathcal{B}^{(1),\text{cut}}_{\phi^4} \left( \lbrace \mathbf{k}_R \rbrace, p \right)  & = \left( \frac{\lambda (H \eta_0)^4}{8 k_4 k_5 k_6}\right) \frac{k_R}{p( k_R^2 - p^2)} \, .
	\end{split}
\end{equation}
Using the above expressions in the RHS of eq.\eqref{Pajer 2 site cut}, one can get the following expression
\begin{equation}
	\begin{split}
		&  \frac{ 2 ~ \mathcal{B}^{(1),\text{cut}}_{\phi^4} \left( \lbrace \mathbf{k}_L \rbrace, p \right)  \mathcal{B}^{(1), \text{cut}}_{\phi^4} \left( \lbrace \mathbf{k}_R \rbrace , p \right)}{P_{p}( \eta_0)}  =  \left( \frac{\lambda^2 (H \eta_0)^6}{16 k_1...k_6} \right) \left( \frac{k_L k_R}{p (k_L^2 - p^2)(k_R^2  - p^2)} \right) \, ,
	\end{split}
\end{equation}
which is in a perfect agreement with the eq.\eqref{Pajer cut phi4 tree LHS}.
\item \textbf{For $\phi^5$ theory at tree level:} Let us now consider an odd polynomial interaction in the form of conformally coupled $\phi^5$ interaction\footnote{It corresponds to $m=4$ and $n=8$.}. Again by standard in-in computation we can find the correlator which then yields the LHS of the eq.\eqref{Pajer 2 site cut} as follows 
\begin{equation}\label{Pajer cut phi5 tree LHS}
	\begin{split}
		& \mathcal{B}^{(2)}_{\phi^5} \left( \lbrace \mathbf{k}_L, \mathbf{k}_R \rbrace ; p \right) + (-1)^8 \mathcal{B}^{(2)}_{\phi^5} \left( \lbrace - \mathbf{k}_L, -\mathbf{k}_R \rbrace ; p \right) = \left( \frac{\lambda^2H^{10} \eta_0^8}{16 k_1...k_8} \right) \left( \frac{k_L k_R p }{(k_L^2 - p^2)^2(k_R^2  - p^2)^2} \right) \, .
	\end{split}
\end{equation}
where $k_L=k_1+k_2+k_3$, and $k_R=k_4+k_5+k_6$. 
To find the RHS, one needs to find a contact correlator in this theory that can be done using the in-in formalism. Substituting this in the RHS of eq.\eqref{Bc cut Pajer} we obtain the following 
\begin{equation}
	\begin{split}
		& \mathcal{B}^{(1),\text{cut}}_{\phi^5} \left( \lbrace \mathbf{k}_L \rbrace, p \right)  = \left( \frac{\lambda (H \eta_0)^6}{16 k_1 k_2 k_3 k_4}\right) \frac{k_L}{p( k_L^2 - p^2)} \, , \\
		& \mathcal{B}^{(1),\text{cut}}_{\phi^5} \left( \lbrace \mathbf{k}_R \rbrace, p \right)  = \left( \frac{\lambda (H \eta_0)^6}{16 k_5 k_6 k_7 k_8}\right) \frac{k_R}{p( k_R^2 - p^2)} \, .
	\end{split}
\end{equation}
Substituting this in the RHS of eq.\eqref{Pajer 2 site cut}, we obtain the following expression
\begin{equation}
	\begin{split}
		&  \frac{2 ~\mathcal{B}^{(1),\text{cut}}_{\phi^5} \left( \lbrace \mathbf{k}_L \rbrace, p \right)  \mathcal{B}^{(1), \text{cut}}_{\phi^5} \left( \lbrace \mathbf{k}_R \rbrace , p \right)}{P_{p}( \eta_0)}  = \left( \frac{\lambda^2 (H \eta_0)^{10}}{64 k_1...k_8} \right) \left( \frac{k_L k_R}{p (k_L^2 - p^2)(k_R^2  - p^2)} \right) \, ,
	\end{split}
\end{equation}
which does not agree with eq.\eqref{Pajer cut phi5 tree LHS}. 
In fact, from the RHS what we have obtained is the subleading contribution to the correlator that can be verified by looking at the order of $\eta_0$ and the order of the singularities.
\item \textbf{For $\phi^4$ theory at loop level:} We consider a $\phi^4$ theory\footnote{This corresponds to $m=2, n=4$.} in which we first find the integrand of the loop integration of a $2$-site 1-loop correlator from the in-in formalism and then using it in the LHS of eq.\eqref{Pajer 2 site cut} we get the following expression
\begin{equation}\label{Pajer 2 site phi4 loop LHS}
	\begin{split}
		& \mathbb{B}^{(2)}_{\phi^4-\text{loop}} \left( \lbrace \mathbf{k}_L, \mathbf{k}_R \rbrace ; \lbrace p_1,p_2 \rbrace \right) + (-1)^4 \mathbb{B}^{(2)}_{\phi^4-\text{loop}} \left( \lbrace - \mathbf{k}_L, -\mathbf{k}_R \rbrace ; \lbrace p_1, p_2 \rbrace \right) 
		\\
		& = \left( \frac{\lambda^2 (H \eta_0)^4}{8 k_1...k_4} \right) \left( \frac{k_L k_R}{p_1 p_2 (k_L^2 - p_+^2)(k_R^2  - p_+^2)} \right) \, ,
	\end{split}
\end{equation}
where we define $k_L=k_1+k_2, k_R=k_3+k_4$ and $p_+=p_1+p_2$. 
To find the RHS of eq.\eqref{Pajer 2 site cut}, we need the following building blocks which are essentially contact level correlator, what we call $1$-site correlator 
\begin{equation}
	\begin{split}
		& \mathcal{B}^{(1),\text{cut}}_{\phi^4} \left( \lbrace \mathbf{k}_L \rbrace, p_1,p_2 \right)  = \left( \frac{\lambda (H \eta_0)^4}{8 k_1 k_2}\right) \frac{k_L}{p_1 p_2 ( k_L^2 - p_+^2)} \, ,
		\\
		& \mathcal{B}^{(1),\text{cut}}_{\phi^4} \left( \lbrace \mathbf{k}_R \rbrace, p_1,p_2 \right)  = \left( \frac{\lambda (H \eta_0)^4}{8 k_3 k_4}\right) \frac{k_R}{p_1 p_2 ( k_R^2 - p_+^2)} \, .
	\end{split}
\end{equation}
Substituting these expressions in the RHS of eq.\eqref{Pajer 2 site cut}, one can see the perfect agreement with the LHS computed in eq.\eqref{Pajer 2 site phi4 loop LHS}
\begin{equation}
	\begin{split}
		&  2 \left(  \frac{\mathcal{B}^{(1),\text{cut}}_{\phi^4} \left( \lbrace \mathbf{k}_L \rbrace, p_1, p_2  \right)  \mathcal{B}^{(1), \text{cut}}_{\phi^4} \left( \lbrace \mathbf{k}_R \rbrace , p_1 p_2  \right)}{P_{p_1}( \eta_0) P_{p_2}( \eta_0)} \right) \\
		& =  \left( \frac{\lambda^2 (H \eta_0)^6}{8 k_1...k_4} \right) \left( \frac{k_L k_R}{p_1 p_2  \left(k_L^2 - p_+^2\right) \left(k_R^2  - p_+^2\right)} \right) \, .
	\end{split}
\end{equation}
\item \textbf{For $\phi^5$ theory at loop level:} We now consider an odd polynomial interaction in the form of $\phi^5$ theory at $1$-loop\footnote{It corresponds to $m=3$ and $n=6$}. From the in-in formalism one can easily compute the integrand of the loop integration of the correlator. Then the LHS of the eq.\eqref{Pajer 2 site cut} yields the expression as follows
\begin{equation}\label{Pajer 2 site phi5 loop LHS}
	\begin{split}
		& \mathbb{B}^{(2)}_{\phi^5} \left( \lbrace \mathbf{k}_L, \mathbf{k}_R \rbrace ; \lbrace p_1, p_2 \rbrace \right) + (-1)^6 \mathbb{B}^{(2)}_{\phi^5} \left( \lbrace - \mathbf{k}_L, -\mathbf{k}_R \rbrace ; \lbrace p_1, p_2 \rbrace \right) 
		\\
		& = \left( \frac{\lambda^2H^{8}\eta_0^6}{8 k_1...k_6} \right) \left( \frac{k_L k_R  p_+^2}{p_1 p_2 \left(k_L^2 - p_+^2\right)^2 \left(k_R^2  - p_+^2\right)^2} \right) \, ,
	\end{split}
\end{equation}
where $k_L=k_1+k_2+k_3$ and $k_R=k_4+k_5+k_6$. To find the RHS of eq.\eqref{Pajer 2 site cut}, we need to make use of the following expression 
\begin{equation}
	\begin{split}
		& \mathcal{B}^{(1),\text{cut}}_{\phi^5} \left( \lbrace \mathbf{k}_L \rbrace, p_1,p_2 \right)  = \left( \frac{\lambda (H \eta_0)^6}{16 k_1 k_2 k_3}\right) \frac{k_L}{p_1 p_2 \left( k_L^2 - p_+^2\right)} \, ,
		\\
		& \mathcal{B}^{(1),\text{cut}}_{\phi^5} \left( \lbrace \mathbf{k}_R \rbrace, p_1,p_2 \right)  = \left( \frac{\lambda (H \eta_0)^6}{16 k_4 k_5 k_6 }\right) \frac{k_R}{p_1 p_2 \left( k_R^2 - p_+^2\right)} \, .
	\end{split}
\end{equation}
Substituting this in the RHS of eq.\eqref{Pajer 2 site cut} we obtain the following expression
\begin{equation}
	\begin{split}
		&  2 \left(  \frac{\mathcal{B}^{(1),\text{cut}}_{\phi^5} \left( \lbrace \mathbf{k}_L \rbrace, p_1, p_2  \right)  \mathcal{B}^{(1), \text{cut}}_{\phi^5} \left( \lbrace \mathbf{k}_R \rbrace , p_1 p_2  \right)}{P_{p_1}( \eta_0) P_{p_2}( \eta_0)}  \right)
		\\
		= & \left( \frac{\lambda^2 (H \eta_0)^8}{32 k_1...k_6} \right) \left( \frac{k_L k_R}{p_1 p_2  \left(k_L^2 - p_+^2 \right) \left(k_R^2  - p_+^2\right)} \right) \, .
	\end{split}
\end{equation}
It is easy to see that it does not match with the LHS computed in the eq.\eqref{Pajer 2 site phi5 loop LHS}.
\end{itemize}

\section{Discontinuity of flat space cosmological correlator} \label{DiscFlatCorrelator}
In this appendix we derive the discontinuity of the in-in correlator in flat Minkowski space-time. In case of flat-space the mode functions are  simple as compared to the de Sitter case. 
We assume that the fields are inserted at the time slice $t_0=0$. We will treat this time slice as future boundary analogous to the future boundary of de Sitter. We consider massless $\phi^n$ theory with the following action
\begin{equation}
	S=\int d^3\vec{x} \int_{-\infty}^{t_0=0} dt \left(- \frac{1}{2} \partial_\mu \phi \partial^\mu \phi +\frac{\lambda}{n!} \phi^n  \right) \, .
\end{equation}
The solution derived from the equation of motion for the given action yields the mode functions, which are expressed as
\begin{equation}
	f_k^-(t) = \frac{e^{-ikt}}{\sqrt{2k}} ,~~ f_k^+(t) = \frac{ e^{ikt}}{\sqrt{2k}} \, .
\end{equation}
It is manifest that the above mode functions closely resemble the conformally coupled scalar field upto an overall factor of $(-H\eta)$. However, the energy dependence of the mode function is identical to the conformally coupled case which is why the cutting rules we have derived in this paper must be valid for the flat space correlator as well.

A striking difference between the flat space and de Sitter case is rooted in the contribution from the auxiliary objects $\tilde{\mathcal{B}}^{(r)}$. Recall that, in flat space, a contact wave-function coefficient is real. Therefore, the auxiliary object at one site $\tilde{\mathcal{B}}^{(1)}$ vanishes in flat space. As a result, the discontinuity of a correlator can be obtained only from the discontinuity of lower site correlators. In particular, our single-cut rule in the context of the flat space correlator takes the following simple form 
\begin{equation}\label{r site single cut flat}
	\begin{split}
		& {\text {Disc}}_{p_{r-1}} \mathcal{B}^{(r)}( \lbrace \mathbf{k}_L,..., \mathbf{k}_R \rbrace ; \lbrace p_1, ..., p_{r-1} \rbrace )=
		\\
		& \frac{1}{2 P_{p_{r-1}} (\eta_0)}  \text{Disc}_{p_{r-1}} \mathcal{B}^{(r-1)}\left(  \lbrace \mathbf{k}_L,...,\mathbf{k}_{M_{r-2}} \rbrace, p_{r-1} ; \lbrace p_1,..,p_{r-2} \rbrace \right) \text{Disc}_{p_{r-1}} \mathcal{B}^{(1)}( \lbrace \mathbf{k}_{R} \rbrace, p_{r-1}) \, ,
	\end{split}
\end{equation}
where, there is no contribution from the auxiliary objects $\widetilde{\mathcal{B}}$. In other words, the other contribution we got in the eq.\eqref{r site correlator} in the de Sitter context for conformally coupled and massless scalars does not participate in the flat space part.

In the following, we will work through few explicit examples of the single-cut rule for a $2$-site and $3$-site correlator using $\phi^3$ interaction. Since in flat space, the singularity structures of a correlator is same for any polynomial massless interaction of type $\phi^n$, it suffices to work with the $\phi^3$ interaction. \\

\noindent \textbf{$2$-site correlator:} In flat space, a $2$-site correlator can be computed using the standard in-in formalism. It has the following expression
\begin{equation}
	\begin{split}
		& \mathcal{B}^{(2)}(\lbrace \mathbf{k}_L, \mathbf{k}_R \rbrace; p) = \left(\frac{\lambda^2 }{8 k_1...k_4} \right) \left( \frac{p+k_L+k_R}{p(p+k_L)(p+k_R)(k_L+k_R)} \right) \, ,
	\end{split}
\end{equation}
where $k_L = k_1+k_2$, and $k_R = k_3+k_4$. 
Taking discontinuity with respect to the bulk-to-bulk energy ($p$) yields the following expression 
\begin{equation}\label{Disc 2site LHS flat}
	\text{Disc}_p \mathcal{B}^{(2)}(\lbrace \mathbf{k}_L, \mathbf{k}_R \rbrace; p) = \left(\frac{\lambda^2 }{4 k_1...k_4} \right) \left(  \frac{k_L k_R }{p(k_L^2 - p^2) (k_R^2 - p^2)} \right).
\end{equation}
To compare this with the RHS of the discontinuity eq.\eqref{r site single cut flat} for $r=2$, we need to evaluate a $1$-site correlator which is quite straight forward from the in-in prescription. Therefore, we need to use the following building blocks to find the product of discontinuity of two $1$-site correlator
\begin{equation}
	\begin{split}
		&  {\mathcal{B}}^{(1)}(\lbrace \mathbf{k}_L \rbrace, p) = \left( \frac{\lambda}{4 k_1 k_2} \right) \left( \frac{1}{p(k_L+p)} \right) \, .
	\end{split}
\end{equation}
Then the discontinuity of the above 1 site correlator with respect to $p$ becomes the following
\begin{equation}
	\begin{split}
		{\text{Disc}_p} {\mathcal{B}}^{(1)} (\lbrace \mathbf{k}_L \rbrace, p)  = \left( \frac{\lambda}{2 k_1 k_2} \right) \frac{k_L}{p (k_L^2-p^2)} \, .
	\end{split}
\end{equation}
\begin{figure}[h]
	\centering
	\includegraphics[width=0.65\textwidth]{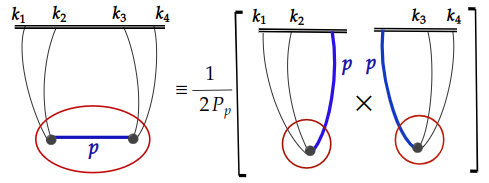 }
	\caption{Diagrammatic representations of a $2$-site correlator with $\phi^3$ interaction }
	\label{single-cut 2-site correlator flat}
\end{figure}
Then the RHS of the discontinuity eq.\eqref{r site single cut flat} for $r=2$ gives the following
\begin{equation}\label{Disc 2site RHS flat}
	\begin{split}
		\frac{1}{2 P_p} {\text{Disc}_p} {\mathcal{B}}^{(1)}(\lbrace \mathbf{k}_L \rbrace, p) {\text{Disc}_p} {\mathcal{B}}^{(1)}(\lbrace \mathbf{k}_R \rbrace, p) =& \left( \frac{\lambda^2} {4 k_1...k_4} \right) \frac{k_L k_R}{p (k_L^2 - p^2) (k_R^2 - p^2)} \, .
	\end{split}
\end{equation}
Comparing eq.\eqref{Disc 2site RHS flat} with eq.\eqref{Disc 2site LHS flat}, we see a perfect agreement. \\

\noindent \textbf{3-site correlator:} We now consider a 3-site correlator that can be computed using the standard in-in formalism. It has a slightly lengthy expression written ss follows
\begin{equation}
	\begin{split}
		& \mathcal{B}^{(3)}(\lbrace \mathbf{k}_L, \mathbf{k}_M, \mathbf{k}_R \rbrace; p_1;p_2) = \frac{\lambda^3 }{16 k_1...k_5}  \times \mathcal{F} \, ,
	\end{split}
\end{equation}
where $k_L = k_1+k_2, \, k_M=k_3$ and $k_R = k_4+k_5$ and 
\begin{equation}
\mathcal{F} = \frac{\substack{(k_M+k_R+p_1)(k_M+k_R+p_2)(k_M+p_1+p_2)+k_L(k_M + k_R + p_1 + p_2)(k_L + 2 k_M + k_R + p_1 + p_2)}}{\substack{p_1 p_2 (k_L + k_M + k_R) (k_L + p_1)  (k_M + k_R + p_1) (k_L + k_M + p_2)  (k_R + p_2)  (k_M + p_1 + p_2)}}  \, .
\end{equation} 
Taking discontinuity with respect to one of the the bulk-to-bulk energy $p_2$ yields the following expression 
\begin{equation}\label{Disc 3site LHS flat}
	\text{Disc}_{p_2} \mathcal{B}^{(3)}(\lbrace \mathbf{k}_L, \mathbf{k}_M, \mathbf{k}_R \rbrace; p_1;p_2) =  \frac{\lambda^3}{8 k_1...k_5}  \times \text{Disc}_{p_2} \mathcal{F} \, ,
\end{equation}
such that 
\begin{equation}
\text{Disc}_{p_2} \mathcal{F}  = \frac{k_R \Big( (k_L+k_M)(k_M+p_1)(k_L+k_M+p_1)-k_M p_2^2 \Big)}{p_1 p_2 (k_L+p_1) \left((k_L+k_M)^2 - p_2^2\right) \left( (k_M+p_1)^2 -p_2^2 \right)(k_R^2 - p_2^2)} \, .
\end{equation}
To compare this with the RHS of the eq.\eqref{r site single cut flat} for $r=3$, we need to evaluate a $1$-site and a $2$-site correlator which is quite straight forward from the in-in prescription. Therefore, we need to use the following building blocks to find the product of discontinuity of the $2$-site and a $1$-site correlator
\begin{figure}[h]
	\centering
	\includegraphics[width=0.75\textwidth]{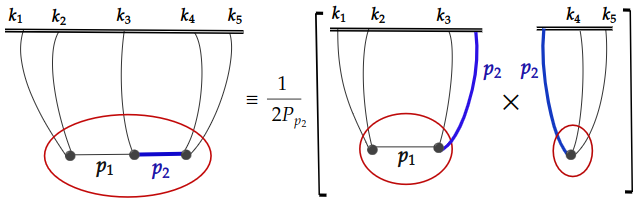 }
	\caption{Diagrammatic representations of a 3-site correlator with $\phi^3$ interaction }
	\label{single-cut 3-site correlator flat}
\end{figure}
\begin{equation}
	\begin{split}
		& \mathcal{B}^{(2)}(\lbrace \mathbf{k}_L, \mathbf{k}_M \rbrace, p_2 ; p_1) =  \frac{\lambda^2 }{8 k_1 k_2 k_3}  \, \frac{p_1+k_L+k_M+p_2}{p_1 p_2 (p_1+k_L)(p_1+k_M+p_2)(k_L+k_M+p_2)}  \, ,
		\\
		&  {\mathcal{B}}^{(1)}(\lbrace \mathbf{k}_R \rbrace, p_2) = \frac{\lambda}{4 k_4 k_5} \,  \frac{1}{p_2(k_R+p_2)} \, .
	\end{split}
\end{equation}
Now the discontinuity of both the correlators with respect to $p_2$ has the following expression 
\begin{equation}
	\begin{split}
		& \text{Disc}_{p_2}\mathcal{B}^{(2)}(\lbrace \mathbf{k}_L, \mathbf{k}_M \rbrace, p_2 ; p_1) =  \frac{\lambda^2}{4 k_1 k_2 k_3} \,  \frac{(k_L+k_M)(k_M+p_1)(k_L+k_M+p_1)-k_M p_2^2 }{p_1 p_2 (k_L+p_1) \left((k_L+k_M)^2 - p_2^2\right) \left( (k_M+p_1)^2 -p_2^2 \right)} \, ,
		\\
		& \text{Disc}_{p_2}{\mathcal{B}}^{(1)}(\lbrace \mathbf{k}_R \rbrace, p_2) = \frac{\lambda}{2 k_4 k_5} \,  \frac{k_R}{p_2(k_R^2-p_2^2)} \, .
	\end{split}
\end{equation}
Now substituting the above two expressions in the RHS of the eq.\eqref{r site single cut flat} it is simple to see that we indeed get the same expression as computed in eq.\eqref{Disc 3site LHS flat}.

\bibliographystyle{JHEP}

\bibliography{cuttingrules.bib}

\providecommand{\href}[2]{#2}\begingroup\raggedright\begin{thebibliography}{10}

\bibitem{Schwinger:1960qe}
J.S.~Schwinger, \emph{{Brownian motion of a quantum oscillator}},
  \href{https://doi.org/10.1063/1.1703727}{\emph{J. Math. Phys.} {\bfseries 2}
  (1961) 407}.

\bibitem{Keldysh:1964ud}
L.V.~Keldysh, \emph{{Diagram Technique for Nonequilibrium Processes}},
  \href{https://doi.org/10.1142/9789811279461_0007}{\emph{Sov. Phys. JETP}
  {\bfseries 20} (1965) 1018}.

\bibitem{Weinberg:2005vy}
S.~Weinberg, \emph{{Quantum contributions to cosmological correlations}},
  \href{https://doi.org/10.1103/PhysRevD.72.043514}{\emph{Phys. Rev. D}
  {\bfseries 72} (2005) 043514}
  [\href{https://arxiv.org/abs/hep-th/0506236}{{\ttfamily hep-th/0506236}}].

\bibitem{Maldacena:2002vr}
J.M.~Maldacena, \emph{{Non-Gaussian features of primordial fluctuations in
  single field inflationary models}},
  \href{https://doi.org/10.1088/1126-6708/2003/05/013}{\emph{JHEP} {\bfseries
  05} (2003) 013} [\href{https://arxiv.org/abs/astro-ph/0210603}{{\ttfamily
  astro-ph/0210603}}].

\bibitem{Anninos:2014lwa}
D.~Anninos, T.~Anous, D.Z.~Freedman and G.~Konstantinidis, \emph{{Late-time
  Structure of the Bunch-Davies De Sitter Wavefunction}},
  \href{https://doi.org/10.1088/1475-7516/2015/11/048}{\emph{JCAP} {\bfseries
  11} (2015) 048} [\href{https://arxiv.org/abs/1406.5490}{{\ttfamily
  1406.5490}}].

\bibitem{Mata:2012bx}
I.~Mata, S.~Raju and S.~Trivedi, \emph{{CMB from CFT}},
  \href{https://doi.org/10.1007/JHEP07(2013)015}{\emph{JHEP} {\bfseries 07}
  (2013) 015} [\href{https://arxiv.org/abs/1211.5482}{{\ttfamily 1211.5482}}].

\bibitem{Ghosh:2014kba}
A.~Ghosh, N.~Kundu, S.~Raju and S.P.~Trivedi, \emph{{Conformal Invariance and
  the Four Point Scalar Correlator in Slow-Roll Inflation}},
  \href{https://doi.org/10.1007/JHEP07(2014)011}{\emph{JHEP} {\bfseries 07}
  (2014) 011} [\href{https://arxiv.org/abs/1401.1426}{{\ttfamily 1401.1426}}].

\bibitem{Kundu:2014gxa}
N.~Kundu, A.~Shukla and S.P.~Trivedi, \emph{{Constraints from Conformal
  Symmetry on the Three Point Scalar Correlator in Inflation}},
  \href{https://doi.org/10.1007/JHEP04(2015)061}{\emph{JHEP} {\bfseries 04}
  (2015) 061} [\href{https://arxiv.org/abs/1410.2606}{{\ttfamily 1410.2606}}].

\bibitem{Kundu:2015xta}
N.~Kundu, A.~Shukla and S.P.~Trivedi, \emph{{Ward Identities for Scale and
  Special Conformal Transformations in Inflation}},
  \href{https://doi.org/10.1007/JHEP01(2016)046}{\emph{JHEP} {\bfseries 01}
  (2016) 046} [\href{https://arxiv.org/abs/1507.06017}{{\ttfamily
  1507.06017}}].

\bibitem{Creminelli:2011mw}
P.~Creminelli, \emph{{Conformal invariance of scalar perturbations in
  inflation}}, \href{https://doi.org/10.1103/PhysRevD.85.041302}{\emph{Phys.
  Rev. D} {\bfseries 85} (2012) 041302}
  [\href{https://arxiv.org/abs/1108.0874}{{\ttfamily 1108.0874}}].

\bibitem{Liu:1998ty}
H.~Liu and A.A.~Tseytlin, \emph{{On four point functions in the CFT / AdS
  correspondence}},
  \href{https://doi.org/10.1103/PhysRevD.59.086002}{\emph{Phys. Rev. D}
  {\bfseries 59} (1999) 086002}
  [\href{https://arxiv.org/abs/hep-th/9807097}{{\ttfamily hep-th/9807097}}].

\bibitem{Dey:2025kci}
P.~Dey, Z.~Huang and A.~Lipstein, \emph{{de Sitter locality from conformal
  field theory}},  \href{https://arxiv.org/abs/2508.15627}{{\ttfamily
  2508.15627}}.

\bibitem{Arkani-Hamed:2018kmz}
N.~Arkani-Hamed, D.~Baumann, H.~Lee and G.L.~Pimentel, \emph{{The Cosmological
  Bootstrap: Inflationary Correlators from Symmetries and Singularities}},
  \href{https://doi.org/10.1007/JHEP04(2020)105}{\emph{JHEP} {\bfseries 04}
  (2020) 105} [\href{https://arxiv.org/abs/1811.00024}{{\ttfamily
  1811.00024}}].

\bibitem{Baumann:2019oyu}
D.~Baumann, C.~Duaso~Pueyo, A.~Joyce, H.~Lee and G.L.~Pimentel, \emph{{The
  cosmological bootstrap: weight-shifting operators and scalar seeds}},
  \href{https://doi.org/10.1007/JHEP12(2020)204}{\emph{JHEP} {\bfseries 12}
  (2020) 204} [\href{https://arxiv.org/abs/1910.14051}{{\ttfamily
  1910.14051}}].

\bibitem{Baumann:2020dch}
D.~Baumann, C.~Duaso~Pueyo, A.~Joyce, H.~Lee and G.L.~Pimentel, \emph{{The
  Cosmological Bootstrap: Spinning Correlators from Symmetries and
  Factorization}},
  \href{https://doi.org/10.21468/SciPostPhys.11.3.071}{\emph{SciPost Phys.}
  {\bfseries 11} (2021) 071}
  [\href{https://arxiv.org/abs/2005.04234}{{\ttfamily 2005.04234}}].

\bibitem{Ansari:2025nsf}
A.~Ansari, A.~Banerjee, S.~Jain and S.~Lalsodagar, \emph{{Bootstrapping time
  non-locality in LSS perturbation theory}},
  \href{https://doi.org/10.1088/1475-7516/2025/11/084}{\emph{JCAP} {\bfseries
  11} (2025) 084} [\href{https://arxiv.org/abs/2504.01078}{{\ttfamily
  2504.01078}}].

\bibitem{Goodhew:2020hob}
H.~Goodhew, S.~Jazayeri and E.~Pajer, \emph{{The Cosmological Optical
  Theorem}}, \href{https://doi.org/10.1088/1475-7516/2021/04/021}{\emph{JCAP}
  {\bfseries 04} (2021) 021}
  [\href{https://arxiv.org/abs/2009.02898}{{\ttfamily 2009.02898}}].

\bibitem{Cespedes:2020xqq}
S.~C{\'e}spedes, A.-C.~Davis and S.~Melville, \emph{{On the time evolution of
  cosmological correlators}},
  \href{https://doi.org/10.1007/JHEP02(2021)012}{\emph{JHEP} {\bfseries 02}
  (2021) 012} [\href{https://arxiv.org/abs/2009.07874}{{\ttfamily
  2009.07874}}].

\bibitem{Baumann:2021fxj}
D.~Baumann, W.-M.~Chen, C.~Duaso~Pueyo, A.~Joyce, H.~Lee and G.L.~Pimentel,
  \emph{{Linking the singularities of cosmological correlators}},
  \href{https://doi.org/10.1007/JHEP09(2022)010}{\emph{JHEP} {\bfseries 09}
  (2022) 010} [\href{https://arxiv.org/abs/2106.05294}{{\ttfamily
  2106.05294}}].

\bibitem{Benincasa:2022omn}
P.~Benincasa, \emph{{Wavefunctionals/S-matrix techniques in de Sitter}},
  \href{https://doi.org/10.22323/1.406.0358}{\emph{PoS} {\bfseries CORFU2021}
  (2022) 358} [\href{https://arxiv.org/abs/2203.16378}{{\ttfamily
  2203.16378}}].

\bibitem{Jazayeri:2021fvk}
S.~Jazayeri, E.~Pajer and D.~Stefanyszyn, \emph{{From locality and unitarity to
  cosmological correlators}},
  \href{https://doi.org/10.1007/JHEP10(2021)065}{\emph{JHEP} {\bfseries 10}
  (2021) 065} [\href{https://arxiv.org/abs/2103.08649}{{\ttfamily
  2103.08649}}].

\bibitem{Melville:2021lst}
S.~Melville and E.~Pajer, \emph{{Cosmological Cutting Rules}},
  \href{https://doi.org/10.1007/JHEP05(2021)249}{\emph{JHEP} {\bfseries 05}
  (2021) 249} [\href{https://arxiv.org/abs/2103.09832}{{\ttfamily
  2103.09832}}].

\bibitem{Goodhew:2021oqg}
H.~Goodhew, S.~Jazayeri, M.H.G.~Lee and E.~Pajer, \emph{{Cutting cosmological
  correlators}},
  \href{https://doi.org/10.1088/1475-7516/2021/08/003}{\emph{JCAP} {\bfseries
  08} (2021) 003} [\href{https://arxiv.org/abs/2104.06587}{{\ttfamily
  2104.06587}}].

\bibitem{Bonifacio:2021azc}
J.~Bonifacio, E.~Pajer and D.-G.~Wang, \emph{{From amplitudes to contact
  cosmological correlators}},
  \href{https://doi.org/10.1007/JHEP10(2021)001}{\emph{JHEP} {\bfseries 10}
  (2021) 001} [\href{https://arxiv.org/abs/2106.15468}{{\ttfamily
  2106.15468}}].

\bibitem{Meltzer:2021zin}
D.~Meltzer, \emph{{The inflationary wavefunction from analyticity and
  factorization}},
  \href{https://doi.org/10.1088/1475-7516/2021/12/018}{\emph{JCAP} {\bfseries
  12} (2021) 018} [\href{https://arxiv.org/abs/2107.10266}{{\ttfamily
  2107.10266}}].

\bibitem{Hogervorst:2021uvp}
M.~Hogervorst, J.~Penedones and K.S.~Vaziri, \emph{{Towards the
  non-perturbative cosmological bootstrap}},
  \href{https://doi.org/10.1007/JHEP02(2023)162}{\emph{JHEP} {\bfseries 02}
  (2023) 162} [\href{https://arxiv.org/abs/2107.13871}{{\ttfamily
  2107.13871}}].

\bibitem{DiPietro:2021sjt}
L.~Di~Pietro, V.~Gorbenko and S.~Komatsu, \emph{{Analyticity and unitarity for
  cosmological correlators}},
  \href{https://doi.org/10.1007/JHEP03(2022)023}{\emph{JHEP} {\bfseries 03}
  (2022) 023} [\href{https://arxiv.org/abs/2108.01695}{{\ttfamily
  2108.01695}}].

\bibitem{Hillman:2021bnk}
A.~Hillman and E.~Pajer, \emph{{A differential representation of cosmological
  wavefunctions}}, \href{https://doi.org/10.1007/JHEP04(2022)012}{\emph{JHEP}
  {\bfseries 04} (2022) 012}
  [\href{https://arxiv.org/abs/2112.01619}{{\ttfamily 2112.01619}}].

\bibitem{Baumann:2022jpr}
D.~Baumann, D.~Green, A.~Joyce, E.~Pajer, G.L.~Pimentel, C.~Sleight et~al.,
  \emph{{Snowmass White Paper: The Cosmological Bootstrap}},
  \href{https://doi.org/10.21468/SciPostPhysCommRep.1}{\emph{SciPost Phys.
  Comm. Rep.} {\bfseries 2024} (2024) 1}
  [\href{https://arxiv.org/abs/2203.08121}{{\ttfamily 2203.08121}}].

\bibitem{Salcedo:2022aal}
S.A.~Salcedo, M.H.G.~Lee, S.~Melville and E.~Pajer, \emph{{The Analytic
  Wavefunction}}, \href{https://doi.org/10.1007/JHEP06(2023)020}{\emph{JHEP}
  {\bfseries 06} (2023) 020}
  [\href{https://arxiv.org/abs/2212.08009}{{\ttfamily 2212.08009}}].

\bibitem{AguiSalcedo:2023nds}
S.~Agui~Salcedo and S.~Melville, \emph{{The cosmological tree theorem}},
  \href{https://doi.org/10.1007/JHEP12(2023)076}{\emph{JHEP} {\bfseries 12}
  (2023) 076} [\href{https://arxiv.org/abs/2308.00680}{{\ttfamily
  2308.00680}}].

\bibitem{Stefanyszyn:2023qov}
D.~Stefanyszyn, X.~Tong and Y.~Zhu, \emph{{Cosmological correlators through the
  looking glass: reality, parity, and factorisation}},
  \href{https://doi.org/10.1007/JHEP05(2024)196}{\emph{JHEP} {\bfseries 05}
  (2024) 196} [\href{https://arxiv.org/abs/2309.07769}{{\ttfamily
  2309.07769}}].

\bibitem{Lee:2023kno}
M.H.G.~Lee, \emph{{From amplitudes to analytic wavefunctions}},
  \href{https://doi.org/10.1007/JHEP03(2024)058}{\emph{JHEP} {\bfseries 03}
  (2024) 058} [\href{https://arxiv.org/abs/2310.01525}{{\ttfamily
  2310.01525}}].

\bibitem{Chowdhury:2023arc}
C.~Chowdhury, A.~Lipstein, J.~Mei, I.~Sachs and P.~Vanhove, \emph{{The subtle
  simplicity of cosmological correlators}},
  \href{https://doi.org/10.1007/JHEP03(2025)007}{\emph{JHEP} {\bfseries 03}
  (2025) 007} [\href{https://arxiv.org/abs/2312.13803}{{\ttfamily
  2312.13803}}].

\bibitem{Stefanyszyn:2024msm}
D.~Stefanyszyn, X.~Tong and Y.~Zhu, \emph{{There and Back Again: Mapping and
  Factorizing Cosmological Observables}},
  \href{https://doi.org/10.1103/PhysRevLett.133.221501}{\emph{Phys. Rev. Lett.}
  {\bfseries 133} (2024) 221501}
  [\href{https://arxiv.org/abs/2406.00099}{{\ttfamily 2406.00099}}].

\bibitem{Liu:2024xyi}
H.~Liu, Z.~Qin and Z.-Z.~Xianyu, \emph{{Dispersive bootstrap of massive
  inflation correlators}},
  \href{https://doi.org/10.1007/JHEP02(2025)101}{\emph{JHEP} {\bfseries 02}
  (2025) 101} [\href{https://arxiv.org/abs/2407.12299}{{\ttfamily
  2407.12299}}].

\bibitem{Goodhew:2024eup}
H.~Goodhew, A.~Thavanesan and A.C.~Wall, \emph{{The Cosmological CPT Theorem}},
   \href{https://arxiv.org/abs/2408.17406}{{\ttfamily 2408.17406}}.

\bibitem{Lee:2024sks}
M.H.G.~Lee, E.~Pajer, M.~Giroux, H.S.~Hannesdottir, S.~Mizera and
  C.~Pasiecznik, \emph{{Records from the S-Matrix Marathon: A Timeless History
  of Time}},  9, 2024 [\href{https://arxiv.org/abs/2410.00227}{{\ttfamily
  2410.00227}}].

\bibitem{Chowdhury:2025ohm}
C.~Chowdhury, A.~Lipstein, J.~Marshall, J.~Mei and I.~Sachs,
  \emph{{Cosmological Dressing Rules}},
  \href{https://arxiv.org/abs/2503.10598}{{\ttfamily 2503.10598}}.

\bibitem{Cespedes:2025dnq}
S.~Cespedes and S.~Jazayeri, \emph{{The massive flat space limit of
  cosmological correlators}},
  \href{https://doi.org/10.1007/JHEP07(2025)032}{\emph{JHEP} {\bfseries 07}
  (2025) 032} [\href{https://arxiv.org/abs/2501.02119}{{\ttfamily
  2501.02119}}].

\bibitem{Melville:2024ove}
S.~Melville and G.L.~Pimentel, \emph{{A de Sitter S-matrix from amputated
  cosmological correlators}},
  \href{https://doi.org/10.1007/JHEP08(2024)211}{\emph{JHEP} {\bfseries 08}
  (2024) 211} [\href{https://arxiv.org/abs/2404.05712}{{\ttfamily
  2404.05712}}].

\bibitem{Arkani-Hamed:2023kig}
N.~Arkani-Hamed, D.~Baumann, A.~Hillman, A.~Joyce, H.~Lee and G.L.~Pimentel,
  \emph{{Differential equations for cosmological correlators}},
  \href{https://doi.org/10.1007/JHEP09(2025)009}{\emph{JHEP} {\bfseries 09}
  (2025) 009} [\href{https://arxiv.org/abs/2312.05303}{{\ttfamily
  2312.05303}}].

\bibitem{Werth:2024mjg}
D.~Werth, \emph{{Spectral representation of cosmological correlators}},
  \href{https://doi.org/10.1007/JHEP12(2024)017}{\emph{JHEP} {\bfseries 12}
  (2024) 017} [\href{https://arxiv.org/abs/2409.02072}{{\ttfamily
  2409.02072}}].

\bibitem{Anninos:2024fty}
D.~Anninos, T.~Anous and A.~Rios~Fukelman, \emph{{De Sitter at all loops: the
  story of the Schwinger model}},
  \href{https://doi.org/10.1007/JHEP08(2024)155}{\emph{JHEP} {\bfseries 08}
  (2024) 155} [\href{https://arxiv.org/abs/2403.16166}{{\ttfamily
  2403.16166}}].

\bibitem{Dey:2024zjx}
I.~Dey, K.K.~Nanda, A.~Roy and S.P.~Trivedi, \emph{{Aspects of dS/CFT
  holography}}, \href{https://doi.org/10.1007/JHEP05(2025)168}{\emph{JHEP}
  {\bfseries 05} (2025) 168}
  [\href{https://arxiv.org/abs/2407.02417}{{\ttfamily 2407.02417}}].

\bibitem{Nanda:2023wne}
K.K.~Nanda, S.K.~Sake and S.P.~Trivedi, \emph{{JT gravity in de Sitter space
  and the problem of time}},
  \href{https://doi.org/10.1007/JHEP02(2024)145}{\emph{JHEP} {\bfseries 02}
  (2024) 145} [\href{https://arxiv.org/abs/2307.15900}{{\ttfamily
  2307.15900}}].

\bibitem{Moitra:2022glw}
U.~Moitra, S.K.~Sake and S.P.~Trivedi, \emph{{Aspects of Jackiw-Teitelboim
  gravity in Anti-de Sitter and de Sitter spacetime}},
  \href{https://doi.org/10.1007/JHEP06(2022)138}{\emph{JHEP} {\bfseries 06}
  (2022) 138} [\href{https://arxiv.org/abs/2202.03130}{{\ttfamily
  2202.03130}}].

\bibitem{Afshordi:2017ihr}
N.~Afshordi, E.~Gould and K.~Skenderis, \emph{{Constraining holographic
  cosmology using Planck data}},
  \href{https://doi.org/10.1103/PhysRevD.95.123505}{\emph{Phys. Rev. D}
  {\bfseries 95} (2017) 123505}
  [\href{https://arxiv.org/abs/1703.05385}{{\ttfamily 1703.05385}}].

\bibitem{Antoniadis:1996dj}
I.~Antoniadis, P.O.~Mazur and E.~Mottola, \emph{{Conformal invariance and
  cosmic background radiation}},
  \href{https://doi.org/10.1103/PhysRevLett.79.14}{\emph{Phys. Rev. Lett.}
  {\bfseries 79} (1997) 14}
  [\href{https://arxiv.org/abs/astro-ph/9611208}{{\ttfamily
  astro-ph/9611208}}].

\bibitem{Larsen:2002et}
F.~Larsen, J.P.~van~der Schaar and R.G.~Leigh, \emph{{De Sitter holography and
  the cosmic microwave background}},
  \href{https://doi.org/10.1088/1126-6708/2002/04/047}{\emph{JHEP} {\bfseries
  04} (2002) 047} [\href{https://arxiv.org/abs/hep-th/0202127}{{\ttfamily
  hep-th/0202127}}].

\bibitem{McFadden:2010vh}
P.~McFadden and K.~Skenderis, \emph{{Holographic Non-Gaussianity}},
  \href{https://doi.org/10.1088/1475-7516/2011/05/013}{\emph{JCAP} {\bfseries
  05} (2011) 013} [\href{https://arxiv.org/abs/1011.0452}{{\ttfamily
  1011.0452}}].

\bibitem{Britto:2004ap}
R.~Britto, F.~Cachazo and B.~Feng, \emph{{New recursion relations for tree
  amplitudes of gluons}},
  \href{https://doi.org/10.1016/j.nuclphysb.2005.02.030}{\emph{Nucl. Phys. B}
  {\bfseries 715} (2005) 499}
  [\href{https://arxiv.org/abs/hep-th/0412308}{{\ttfamily hep-th/0412308}}].

\bibitem{Britto:2005fq}
R.~Britto, F.~Cachazo, B.~Feng and E.~Witten, \emph{{Direct proof of tree-level
  recursion relation in Yang-Mills theory}},
  \href{https://doi.org/10.1103/PhysRevLett.94.181602}{\emph{Phys. Rev. Lett.}
  {\bfseries 94} (2005) 181602}
  [\href{https://arxiv.org/abs/hep-th/0501052}{{\ttfamily hep-th/0501052}}].

\bibitem{Eden:1966dnq}
R.J.~Eden, P.V.~Landshoff, D.I.~Olive and J.C.~Polkinghorne, \emph{{The
  analytic S-matrix}}, Cambridge Univ. Press, Cambridge (1966).

\bibitem{Benincasa:2013faa}
P.~Benincasa, \emph{{New structures in scattering amplitudes: a review}},
  \href{https://doi.org/10.1142/S0217751X14300051}{\emph{Int. J. Mod. Phys. A}
  {\bfseries 29} (2014) 1430005}
  [\href{https://arxiv.org/abs/1312.5583}{{\ttfamily 1312.5583}}].

\bibitem{Taylor:2017sph}
T.R.~Taylor, \emph{{A Course in Amplitudes}},
  \href{https://doi.org/10.1016/j.physrep.2017.05.002}{\emph{Phys. Rept.}
  {\bfseries 691} (2017) 1} [\href{https://arxiv.org/abs/1703.05670}{{\ttfamily
  1703.05670}}].

\bibitem{Correia:2020xtr}
M.~Correia, A.~Sever and A.~Zhiboedov, \emph{{An analytical toolkit for the
  S-matrix bootstrap}},
  \href{https://doi.org/10.1007/JHEP03(2021)013}{\emph{JHEP} {\bfseries 03}
  (2021) 013} [\href{https://arxiv.org/abs/2006.08221}{{\ttfamily
  2006.08221}}].

\bibitem{Donath:2024utn}
Y.~Donath and E.~Pajer, \emph{{The in-out formalism for in-in correlators}},
  \href{https://doi.org/10.1007/JHEP07(2024)064}{\emph{JHEP} {\bfseries 07}
  (2024) 064} [\href{https://arxiv.org/abs/2402.05999}{{\ttfamily
  2402.05999}}].

\bibitem{Ema:2024hkj}
Y.~Ema and K.~Mukaida, \emph{{Cutting rule for in-in correlators and
  cosmological collider}},
  \href{https://doi.org/10.1007/JHEP12(2024)194}{\emph{JHEP} {\bfseries 12}
  (2024) 194} [\href{https://arxiv.org/abs/2409.07521}{{\ttfamily
  2409.07521}}].

\bibitem{Anninos_2015}
D.~Anninos, T.~Anous, D.Z.~Freedman and G.~Konstantinidis, \emph{Late-time
  structure of the bunch-davies de sitter wavefunction},
  \href{https://doi.org/10.1088/1475-7516/2015/11/048}{\emph{Journal of
  Cosmology and Astroparticle Physics} {\bfseries 2015} (2015) 048–048}.

\bibitem{Maldacena:2011nz}
J.M.~Maldacena and G.L.~Pimentel, \emph{{On graviton non-Gaussianities during
  inflation}}, \href{https://doi.org/10.1007/JHEP09(2011)045}{\emph{JHEP}
  {\bfseries 09} (2011) 045} [\href{https://arxiv.org/abs/1104.2846}{{\ttfamily
  1104.2846}}].

\bibitem{Raju:2012zr}
S.~Raju, \emph{{New Recursion Relations and a Flat Space Limit for AdS/CFT
  Correlators}}, \href{https://doi.org/10.1103/PhysRevD.85.126009}{\emph{Phys.
  Rev. D} {\bfseries 85} (2012) 126009}
  [\href{https://arxiv.org/abs/1201.6449}{{\ttfamily 1201.6449}}].

\bibitem{MarinMacedo:2025jco}
G.~Mar{\'\i}n~Mac{\^e}do, \emph{{Constraints from Unitarity: Cosmological
  Cutting Rules Using the Schwinger-Keldysh Path Integral Formalism and
  Celestial Strings}},  Master's thesis, U. Chile, Santiago, 2025.

\bibitem{Akhmedov:2013vka}
E.T.~Akhmedov, \emph{{Lecture notes on interacting quantum fields in de Sitter
  space}}, \href{https://doi.org/10.1142/S0218271814300018}{\emph{Int. J. Mod.
  Phys. D} {\bfseries 23} (2014) 1430001}
  [\href{https://arxiv.org/abs/1309.2557}{{\ttfamily 1309.2557}}].

\bibitem{Spradlin:2001pw}
M.~Spradlin, A.~Strominger and A.~Volovich, \emph{{Les Houches lectures on de
  Sitter space}},  in \emph{{Les Houches Summer School: Session 76: Euro Summer
  School on Unity of Fundamental Physics: Gravity, Gauge Theory and Strings}},
  pp.~423--453, 10, 2001
  [\href{https://arxiv.org/abs/hep-th/0110007}{{\ttfamily hep-th/0110007}}].

\bibitem{Galante:2023uyf}
D.A.~Galante, \emph{{Modave lectures on de Sitter space {\&} holography}},
  \href{https://doi.org/10.22323/1.435.0003}{\emph{PoS} {\bfseries Modave2022}
  (2023) 003} [\href{https://arxiv.org/abs/2306.10141}{{\ttfamily
  2306.10141}}].

\bibitem{Cabass_2023}
G.~Cabass, S.~Jazayeri, E.~Pajer and D.~Stefanyszyn, \emph{Parity violation in
  the scalar trispectrum: no-go theorems and yes-go examples},
  \href{https://doi.org/10.1007/jhep02(2023)021}{\emph{Journal of High Energy
  Physics} {\bfseries 2023} (2023) }.

\bibitem{arkanihamed2015cosmologicalcolliderphysics}
N.~Arkani-Hamed and J.~Maldacena, \emph{Cosmological collider physics},  2015.

\bibitem{Baumann:2025twist}
D.~Baumann, G.~Mathys, G.L.~Pimentel and F.~Rost, \emph{A new twist on spinning
  (a)ds correlators}, \href{https://doi.org/10.1007/JHEP01(2025)202}{\emph{J.
  High Energ. Phys.} {\bfseries 2025} (2025) 202}
  [\href{https://arxiv.org/abs/2408.02727}{{\ttfamily 2408.02727}}].

\bibitem{David:2019dSspinor}
A.~David, N.~Fischer and Y.~Neiman, \emph{Spinor-helicity variables for
  cosmological horizons in de sitter space},
  \href{https://doi.org/10.1103/PhysRevD.100.045005}{\emph{Phys. Rev. D}
  {\bfseries 100} (2019) 045005}
  [\href{https://arxiv.org/abs/1906.01058}{{\ttfamily 1906.01058}}].

\bibitem{Chowdhury:2025nnk}
C.~Chowdhury, A.~Lipstein, J.~Marshall and A.J.~Zhang, \emph{{On in-in
  correlators for spinning theories and their shadow formulation}},
  \href{https://arxiv.org/abs/2512.14694}{{\ttfamily 2512.14694}}.

\bibitem{DKKK}
S.~Das, D.~Karan, B.~Khatun and N.~Kundu, \emph{{Upcoming work}}, .

\bibitem{Pimentel:2025rds}
G.L.~Pimentel and C.~Yang, \emph{{Strongly Coupled Sectors in Inflation:
  Gapless Theories and Unparticles}},
  \href{https://arxiv.org/abs/2503.17840}{{\ttfamily 2503.17840}}.

\bibitem{Chakraborty:2023los}
T.~Chakraborty, J.~Chakravarty, V.~Godet, P.~Paul and S.~Raju,
  \emph{{Holography of information in de Sitter space}},
  \href{https://doi.org/10.1007/JHEP12(2023)120}{\emph{JHEP} {\bfseries 12}
  (2023) 120} [\href{https://arxiv.org/abs/2303.16316}{{\ttfamily
  2303.16316}}].

\bibitem{Chakraborty:2025izq}
T.~Chakraborty, A.~H and S.~Raju, \emph{{Cosmological correlators in
  gravitationally-constrained de Sitter states}},
  \href{https://arxiv.org/abs/2507.15926}{{\ttfamily 2507.15926}}.

\bibitem{Shukla:2016bnu}
A.~Shukla, S.P.~Trivedi and V.~Vishal, \emph{{Symmetry constraints in
  inflation, $\alpha$-vacua, and the three point function}},
  \href{https://doi.org/10.1007/JHEP12(2016)102}{\emph{JHEP} {\bfseries 12}
  (2016) 102} [\href{https://arxiv.org/abs/1607.08636}{{\ttfamily
  1607.08636}}].

\bibitem{Jain:2022uja}
S.~Jain, N.~Kundu, S.~Kundu, A.~Mehta and S.K.~Sake, \emph{{A CFT
  interpretation of cosmological correlation functions in
  {\ensuremath{\alpha}}{\ensuremath{-}}vacua in de-Sitter space}},
  \href{https://doi.org/10.1007/JHEP05(2023)111}{\emph{JHEP} {\bfseries 05}
  (2023) 111} [\href{https://arxiv.org/abs/2206.08395}{{\ttfamily
  2206.08395}}].

\bibitem{Chopping:2024bog}
A.J.~Chopping, C.~Sleight and M.~Taronna, \emph{Cosmological correlators for
  bogoliubov initial states},
  \href{https://doi.org/10.1007/JHEP09(2024)152}{\emph{JHEP} {\bfseries 2024}
  (2024) 152} [\href{https://arxiv.org/abs/2407.16652}{{\ttfamily
  2407.16652}}].

\bibitem{Ghosh:2024aqd}
D.~Ghosh, E.~Pajer and F.~Ullah, \emph{{Cosmological cutting rules for
  Bogoliubov initial states}},
  \href{https://doi.org/10.21468/SciPostPhys.18.1.005}{\emph{SciPost Phys.}
  {\bfseries 18} (2025) 005}
  [\href{https://arxiv.org/abs/2407.06258}{{\ttfamily 2407.06258}}].

\bibitem{Ansari:2024pgq}
A.~Ansari, P.~Banerjee, P.~Dhivakar, S.~Jain and N.~Kundu, \emph{{Inflationary
  non-Gaussianities in alpha vacua and consistency with conformal symmetries}},
  \href{https://doi.org/10.1007/JHEP10(2024)147}{\emph{JHEP} {\bfseries 10}
  (2024) 147} [\href{https://arxiv.org/abs/2403.10513}{{\ttfamily
  2403.10513}}].

\bibitem{Ghosh:2025pxn}
D.~Ghosh and F.~Ullah, \emph{{Cosmological cutting rules for Bogoliubov initial
  states: any mass and spin}},
  \href{https://doi.org/10.1088/1475-7516/2025/09/016}{\emph{JCAP} {\bfseries
  09} (2025) 016} [\href{https://arxiv.org/abs/2502.05630}{{\ttfamily
  2502.05630}}].

\bibitem{Loparco:2023rug}
M.~Loparco, J.~Penedones, K.~Salehi~Vaziri and Z.~Sun, \emph{{The
  K{\"a}ll{\'e}n-Lehmann representation in de Sitter spacetime}},
  \href{https://doi.org/10.1007/JHEP12(2023)159}{\emph{JHEP} {\bfseries 12}
  (2023) 159} [\href{https://arxiv.org/abs/2306.00090}{{\ttfamily
  2306.00090}}].

\bibitem{SalehiVaziri:2024joi}
K.~Salehi~Vaziri, \emph{{A non-perturbative construction of the de Sitter
  late-time boundary}},  \href{https://arxiv.org/abs/2412.00183}{{\ttfamily
  2412.00183}}.

\bibitem{DLMF}
F.W.J.~Olver, A.B.~{Olde Daalhuis}, D.W.~Lozier, B.I.~Schneider, R.F.~Boisvert,
  C.W.~Clark et~al., ``{NIST Digital Library of Mathematical Functions}.''
  Release 1.1.9 of 2023-03-15.

\bibitem{Goodhew:2023bcu}
H.~Goodhew, \emph{{The Cosmological Implications of Unitarity}}, Ph.D. thesis,
  Cambridge U., DAMTP, 2023.
\newblock 10.17863/CAM.99550.

\end{thebibliography}\endgroup

\end{document}